\documentclass[10pt,english,aps,manuscript,aps,superscriptaddress,nofootinbib]{revtex4}
\usepackage{float}
\usepackage{amsmath}
\usepackage{amssymb}
\usepackage{graphicx}
\usepackage{esint}

\makeatletter

\providecommand{\tabularnewline}{\\}

\@ifundefined{textcolor}{}
{%
 \definecolor{BLACK}{gray}{0}
 \definecolor{WHITE}{gray}{1}
 \definecolor{RED}{rgb}{1,0,0}
 \definecolor{GREEN}{rgb}{0,1,0}
 \definecolor{BLUE}{rgb}{0,0,1}
 \definecolor{CYAN}{cmyk}{1,0,0,0}
 \definecolor{MAGENTA}{cmyk}{0,1,0,0}
 \definecolor{YELLOW}{cmyk}{0,0,1,0}
}

\usepackage{multirow}\usepackage{float}\@ifundefined{definecolor}
 {\usepackage{color}}{}

\makeatother

\usepackage{babel}
\begin{document}

\preprint{Preprint SMU-13-25}

\title{A meta-analysis of parton distribution functions}

\author{Jun Gao}

\email{jung@smu.edu}

\selectlanguage{english}%

\author{Pavel Nadolsky}

\email{nadolsky@smu.edu}

\selectlanguage{english}%

\affiliation{Department of Physics, Southern Methodist University, Dallas, TX
75275-0181, USA}

\date{December 25, 2013}
\begin{abstract}
A ``meta-analysis'' is a method for comparison and combination of
nonperturbative parton distribution functions (PDFs) in a nucleon
obtained with heterogeneous procedures and assumptions. Each input
parton distribution set is converted into a ``meta-parametrization''
based on a common functional form. By analyzing parameters of the
meta-parametrizations from all input PDF ensembles, a combined PDF
ensemble can be produced that has a smaller total number of PDF
member sets than the original ensembles. The meta-parametrizations
simplify the computation of the PDF uncertainty in theoretical
predictions and provide an alternative to the 2010 PDF4LHC
convention for combination of PDF uncertainties. As a practical
example, we construct a META ensemble for computation of QCD
observables at the Large Hadron Collider using the
next-to-next-to-leading order PDF sets from CTEQ, MSTW, and NNPDF
groups as the input. The META ensemble includes a central set that
reproduces the average of LHC predictions based on the three input
PDF ensembles and Hessian eigenvector sets for computing the
combined PDF+$\alpha_s$ uncertainty at a common QCD coupling
strength of 0.118.
\end{abstract}

\keywords{Global QCD analysis, parton distributions, Large Hadron Collider}

\pacs{\noindent 14.65.Dw, 12.38.-t}

\maketitle

\section{Introduction\label{sec:intro}}
Parton distribution functions (PDFs) in the nucleon describe long-distance
nonperturbative effects in high-energy QCD processes. Their detailed
understanding is vital for the physics program at the Large Hadron
Collider (LHC). First years of the LHC operation have culminated in
the discovery of a Higgs scalar particle \cite{Aad:2012tfa,Chatrchyan:2012ufa}
and tested various Standard Model processes at the highest precision
ever attained. No direct indications for physics beyond
the Standard Model have been found so far, yet future plans
\cite{SnowmassHiggs2013}
envision reaching much higher accuracy in new physics searches at the
raised LHC energy in the next decade.
Toward this goal, many theoretical predictions
for the LHC must include next-to-next-to-leading order (NNLO) QCD
and next-to-leading order (NLO) electroweak contributions, as well
as parametrizations for nonperturbative PDFs of comparable
accuracy. Modern PDFs are determined by several groups from a statistical
analysis of multiple hadronic experiments utilizing
diverse methods. In this paper, we discuss estimation of the full uncertainty
in QCD predictions based on the combination of inputs from PDFs
by several groups. The combination of PDF uncertainties is non-trivial, given
a variety of considerations that go into their estimation, and
has important implications for those LHC measurements where the
PDF uncertainty results in a leading systematic error.

Recent NNLO PDF parametrizations include CT10 \cite{Gao:2013xoa},
MSTW 2008 \cite{Martin:2009iq}, and NNPDF 2.3 \cite{Ball:2012cx},
determined from the global analysis of hadronic data; ABM11
\cite{Alekhin:2012ig}, ABM12 \cite{Alekhin:2013nda}, and 
JR09 \cite{JimenezDelgado:2008hf,JimenezDelgado:2009tv},
determined from the analysis of deep-inelastic scattering (DIS)
and vector boson production; and HERAPDF 1.5
\cite{CooperSarkar:2011aa}, determined solely based on DIS.
Every group provides an ensemble of error PDF
sets, besides the central PDF set, to compute the PDF uncertainty
according to the Hessian~\cite{Pumplin:2001ct,Pumplin:2002vw}
or Monte-Carlo sampling
methods~\cite{Giele:1998gw,DelDebbio:2007ee}.\footnote{Throughout the
  paper, we draw a distinction between the ``PDF sets'' and ``PDF
  ensembles''. A ``PDF set''  is a group of parametrizations of PDFs
  for various flavors ($u(x,Q)$, $d(x,Q)$, etc.) corresponding to one
  point in the PDF parameter space. The ``PDF ensemble'' is a group of
  PDF sets from several points in the parameter space, such as in the
  ``CT10 ensemble of 51 PDF error sets''.}
The central PDFs and  PDF error bands at the 68\% confidence level (c.l.)
from CT10, HERAPDF1.5, MSTW'08, and NNPDF2.3 ensembles
are similar, but not exactly identical \cite{Ball:2012wy}. These PDF ensembles
may share some common ingredients, like using the same experimental data
sets or analogous statistical methods. They also
exhibit substantial differences, notably in
the choice of the QCD coupling strength, heavy-quark masses and heavy-quark
schemes, PDF functional forms, and error propagation. There are
also many other specialized PDFs, {\it e.g.}, CJ12~\cite{Owens:2012bv}, PDFs with
intrinsic charm~\cite{Pumplin:2007wg,Dulat:2013hea} and
photons~\cite{Martin:2004dh,Ball:2013hta}, SOM PDFs~\cite{Askanazi:2013ota},
and nuclear PDFs~\cite{Kovarik:2013sya,Hirai:2007sx,Eskola:2009uj,deFlorian:2011fp}.

When predicting a new QCD observable $X$, a user may estimate the
full uncertainty due to the PDFs by repeatedly computing $X$ for
every provided error PDF set and combining all such predictions
according to a prescribed formula. Such estimation may cause a
practical bottleneck if $X$ is a complex observable that must be
recomputed with hundreds of PDF parametrizations.

A related difficulty concerns combination of estimated PDF
uncertainties on $X$ using PDFs from several ensembles. Various
groups follow incongruous procedures when defining their PDF
uncertainties, which must be somehow reconciled in order for the
combination to proceed. The PDF4LHC study recommended in 2010
\cite{Botje:2011sn,Alekhin:2011sk} to find the combined 68\% c.l.
uncertainty for $X$ using NLO PDFs from three groups (CT, MSTW, and
NNPDF) by first calculating the 68\% c.l. interval independently for
each group and in accordance with the definition that the group had
adopted; and then taking an envelope of the three 68\% c.l.
intervals as the total uncertainty. Later, this procedure was
applied to the NNLO PDFs \cite{Ball:2012wy}. The resulting estimate
of the PDF uncertainty spans all individual confidence intervals
obtained under non-identical definitions. 

A related combination method \cite{Forte:2013wc,Forte:2010dt} (which will be
referred to as a 'replica combination method')
estimates the total PDF uncertainty by combining 
Monte-Carlo replicas (random PDF error sets) generated 
from the input PDF ensembles. To carry out this approach, 
each input ensemble utilizing the Hessian method is first converted
into a secondary Monte-Carlo ensemble with about 100 member sets 
that is then merged with the rest of the Monte-Carlo
ensembles. When CT10, MSTW'08, and NNPDF2.3 ensembles are combined, 
the total PDF uncertainty in the 
replica combination method is close to that derived according to the PDF4LHC
recommendation. In either the PDF4LHC or replica combination method,
the number of the final PDF error sets turns out to be  
the same or even larger than the number of the input sets, which is
in the range of hundreds when predictions from many groups are combined.  
Both methods estimate $X$ for multiply redundant 
PDF sets, many of which predict the $X$ values that are close 
to the central prediction and contribute little to the total uncertainty.

We will now outline an alternative approach called a
``meta-analysis'' in which the input PDF sets are combined
\emph{before} the observable $X$ is computed. Each member set of the
initial PDF ensembles is approximated at an energy scale $Q_{0}$ by
an intermediary functional form dependent on PDF parameters $\{a\}.$
The distribution of the member sets over the parameters $\{a\}$ is
analyzed, and a hypervolume spanning parameters of all PDF error
sets is determined. Finally, using the shared parametrization form,
we can construct a relatively small number of PDF eigenvector sets
spanning all 68\% c.l. error sets using the Hessian or Monte-Carlo
sampling method. When the input PDF ensembles are statistically
consistent with one another, the new PDF error sets reproduce the
average and total PDF uncertainty of the initial PDF sets, but the
number of the independent sets can be considerably reduced in the
new ensemble.

The final ensemble of META PDFs constructed this way serves the same
purpose as the PDF4LHC and replica combination methods, yet it
combines the PDFs directly
in the PDF parameter space while at the same time minimizing numerical
computation efforts. In general it is difficult to bring all
diverse PDFs into a common framework, for example because distinct
heavy-quark schemes are used at momentum scales of a few GeV. But
for purposes of the LHC studies, and focusing only the $x$ and $Q$
ranges corresponding to the LHC kinematics, most of the difficulties
can be circumvented by requiring the lowest scale $Q_{0}$ for the
PDF evolution to be above the bottom quark mass, where all PDF sets evolve
assuming the same number of active flavors. The effects of choosing
different heavy-quark schemes are then taken into account through
the boundary conditions at the low scale $Q_{0}$. A variety of checks
is performed to guarantee that the physics inputs and statistical
features of the input PDF sets are preserved by the META PDF ensemble.

In the next sections, we build a combined ensemble of NNLO META PDFs
for LHC studies from CT10, MSTW 2008 and NNPDF2.3 ensembles with 5
active quark flavors. These ensembles use compatible central values of
the QCD coupling ($\alpha_{s}$), similar data sets and procedures, and
are in quantitative agreement \cite{Ball:2012wy}. The combination
of such compatible sets is the simplest one to carry through, but
the framework of the meta-analysis does not limit the number of the PDF
ensembles that can be combined. We also derive and examine
the meta-parametrizations for the ABM11 and HERAPDF1.5
sets that were not included in the PDF4LHC combination. The ABM and
HERAPDF ensembles are quite distinct from the three global ensembles,
hence we cannot include them into the META ensemble yet using the
combination procedure that was chosen. We comment on the comparison of
the ABM and HERAPDF ensembles with the META PDFs in later sections.

The paper is organized as follows. Section~\ref{sec:bench}
introduces the meta-parametrizations to approximate the input PDFs
at energies relevant for the LHC. Section~\ref{sec:meta} constructs
a META PDF ensemble from CT10, MSTW'08, and NNPDF2.3 NNLO ensembles.
In Section~\ref{sec:lhc} we briefly discuss phenomenological
applications of the META PDFs. Section~\ref{sec:concl} contains a
short summary.

\section{Meta-parametrizations for input PDFs\label{sec:bench}}

\begin{table}

\begin{tabular}{|c|c|c|c|c|c|c|c|c|}
\hline
Input PDF ensemble & Ref. & Initial & $\alpha_{s}(M_{Z})$& Analysis & Error & Free & Error & Included\tabularnewline
 &  & scale & \quad  & type & type & parameters & sets & in META PDFs\tabularnewline
\hline
\hline
CT10 & \cite{Gao:2013xoa} & 1.3 GeV & 0.118 & Global & Hessian & 25 & 50 & Yes\tabularnewline
\hline
MSTW'08 & \cite{Martin:2009iq} & 1.0 GeV & 0.11706 & Global & Hessian & 20 & 40 & Yes\tabularnewline
\hline
NNPDF2.3 & \cite{Ball:2012cx} & 1.414 GeV & 0.118 & Global & MC & 259  & 100 & Yes\tabularnewline
\hline
\hline
ABM11($n_f=5$) & \cite{Alekhin:2012ig} & 3.0 GeV & 0.118 (0.1134) & DIS+DY & Hessian & 29 & 28 & No\tabularnewline
\hline
HERAPDF1.5 & \cite{CooperSarkar:2011aa} & 1.378 GeV & 0.1176 & DIS & Hessian & 14 & 28 & No\tabularnewline
\hline
\end{tabular}

\caption{Input NNLO PDF ensembles considered in the
  meta-analysis. In the ABM analysis, 28 eigenvector sets are provided with $\alpha_s(M_Z)$ varied around a best-fit
  value of 0.1134. The meta-parametrization is obtained for the
ABM set for $\alpha_s(M_Z)=0.118$ and compared to the other PDF ensembles at close $\alpha_s(M_Z)$ values.
    \label{tab:Input-PDF-ensembles}}
\end{table}

\subsection{Selection of input PDF ensembles \label{sec:frame}}

In a general-purpose PDF ensemble, parton distributions  are parametrized
by certain functions $\Phi_f(x,Q)$ of the partonic momentum fraction $x$ at an
initial scale $Q_{0}$ of order 1 GeV and found numerically at higher
scales $Q$ by solving QCD evolution equations. The PDF depends on the
flavor $f$ of the probed parton, but we omit the index $f$ in
$\Phi(x,Q)$ to simplify the notation, assuming that all considerations
apply to every parton flavor.

At low scales comparable to the charm and bottom masses, the PDFs from various
groups cannot be directly combined because of different implementations of
heavy-quark mass contributions.
In typical LHC applications, on the other hand, the hard factorization scale
$Q$ tends to be well above the bottom quark mass $m_{b}\approx4.8$
GeV. At such scales the PDFs from all groups evolve in the same way
according to the  DGLAP evolution equations with 5 active quark flavors. At
even higher scales, $Q^{2}\gg m_{b}^{2}$, masses of charm and bottom
quarks can be neglected in hard-scattering contributions.
We therefore can circumvent explicit treatment of heavy-quark effects
by approximating the PDFs  $\Phi(x,Q_{0})$ by
flexible meta-parametrizations $f(x,Q_{0};\{a\})$ at a scale $Q_{0}$
above $m_{b}$. For definiteness we choose $Q_{0}=8$ GeV and assume
9 independent PDF parametrizations at this scale, for $g$, $u$, $\overline{u}$,
$d$, $\overline{d}$, $s$, $\overline{s}$, $c=\overline{c}$, and
$b=\overline{b}$. We neglect small differences
between $c$ and $\bar{c}$, and $b$ and $\bar{b}$ PDFs.

The input PDF ensembles that will be considered in this analysis are
summarized in Table~\ref{tab:Input-PDF-ensembles}. They are obtained
at the next-to-next-to-leading order (NNLO) in the QCD coupling
strength $\alpha_{s}$. Every ensemble except ABM11 includes 28-100
PDF error sets probing the uncertainty due to the PDF
parametrization at a fixed $\alpha_s(M_Z)$ close to 0.118, as well
as additional best-fit PDF parametrizations for alternative
$\alpha_s(M_Z)$ to examine the magnitude of the $\alpha_s$
uncertainty. The 0.118 value is compatible with the world-average
QCD coupling $0.1184\pm0.0007$~\cite{Beringer:1900zz} and will be
used as a common $\alpha_s(M_Z)$ value in most comparisons. In the
ABM11 ensemble, the error PDFs are given for a lower central value
of $\alpha_s(M_Z)=0.1134$ and include the $\alpha_s(M_Z)$ variation
in the covariance matrix. An update of the ABM11 analysis tuned to the
LHC data, called ABM12, has been released
very recently~\cite{Alekhin:2013nda}. Only the ABM11 ensemble provides a
member set for $\alpha_s(M_Z)=0.118$ from their $\alpha_s$
series (but not the PDF error sets at this $\alpha_s$). Because the
ABM error sets correspond to a lower $\alpha_s(M_Z)$ value, 
 they should not be compared on the same
footing with the error PDFs for a fixed $\alpha_s(M_Z)\approx 0.118$
from the other groups. For this comparison,
we will examine the meta-parametrization
for the ABM11 member set at $\alpha_s(M_Z)=0.118$, 
as well as LHC predictions
for ABM12 using the central $\alpha_s(M_Z)=0.1132$.
Note that, for making theoretical predictions, 
the ABM group recommends to use their PDF sets 
corresponding to their best-fit $\alpha_s(M_Z)\approx 0.113$.  

When combining the PDF ensembles, one follows
two common methods used for estimating the PDF uncertainty, the 
Hessian method \cite{Pumplin:2001ct,Pumplin:2002vw} and the 
Monte Carlo (MC) sampling
method \cite{Giele:1998gw,DelDebbio:2007ee}. We summarize the core
relations of the two methods for completeness. 
Generally speaking, the Hessian approach
provides $2N_{eig}$ PDF error sets (called ``eigenvector sets'')
that only estimate a given confidence interval for the PDFs without
specifying the actual probability distribution. The MC method
returns $N_{rep}$ PDF error sets called ``replicas'' to estimate the
probability density for the desired QCD observable.

The following simplest statistics can be computed with either method.
\begin{enumerate}
\item \textbf{The central prediction} for a QCD observable $X$, such as
a cross section or the PDF itself, is the value $X_{C}^{H}$ given by
the most probable PDF in the Hessian approach (corresponding to the
lowest log-likelihood $\chi^{2}$ in the global fit); or the mean
value $X_{C}^{M}=\langle X\rangle_{replicas}$ of predictions from
all the replicas in the MC method.
\item \textbf{The PDF uncertainty} in the Hessian method estimates the boundaries
of the confidence interval on $X$, usually at the 68\% or 90\% c.l.
Both the symmetric \cite{Pumplin:2001ct,Pumplin:2002vw} uncertainty,
$\delta^{H}$, and asymmetric uncertainties
$\delta_{\pm}^{H}$~\cite{Nadolsky:2001yg} can be estimated:
\begin{eqnarray}
 &  & \delta^{H}(X)=\frac{1}{2}\sqrt{\sum_{i=1}^{N_{eig}}[X_{i}^{+}-X_{i}^{-}]^{2}}\ ,\label{HessianSymmetricError}\\
 &  & \delta_{+}^{H}(X)=\sqrt{\sum_{i=1}^{N_{eig}}[\max(X_{i}^{+}-X_{C}^H,\ X_{i}^{-}-X_{C}^H,\ 0)]^{2}},\label{HessianPlusError}\\
 &  & \delta_{-}^{H}(X)=\sqrt{\sum_{i=1}^{N_{eig}}[\max(X_{C}^H-X_{i}^{+},\ X_{C}^H-X_{i}^{-},\ 0)]^{2}},\label{HessianMinusError}
\end{eqnarray}
where $X_{i}^{\pm}$ are the values of $X$ corresponding to the upper
and lower boundary of the confidence interval for the parameter $a_i$.
These formulas are used by the CTEQ, HERA, and MSTW groups, while the 
ABM group uses a variant of the Hessian master formula that estimates 
the symmetric uncertainty from the differences 
between the predictions based on the central PDF and 
one error set for each eigenvector direction~\cite{Alekhin:2012ig}.

In the MC approach, the symmetric 68\% c.l. PDF uncertainty
is given by a standard deviation $\delta^{M}$:
\begin{eqnarray}
 &  & \delta^{M}(X)=\sqrt{\frac{1}{N_{rep-1}}\sum_{i=1}^{N_{rep}}[X_{i}-X_{C}^M]^{2}}.\label{MCSymmetricError}
\end{eqnarray}

\item \textbf{The tolerance (error) ellipse} is introduced to study correlations
between two observables, \emph{e.g.}, $X$ and $Y$. In the Hessian approach,
we first define the correlation angle $\Delta\varphi$ as \cite{Pumplin:2001ct,Nadolsky:2001yg,Nadolsky:2008zw}
\begin{equation}
\cos(\Delta\varphi)=\frac{1}{4\delta^{H}(X)\delta^{H}(Y)}\sum_{i=1}^{N_{eig}}(X_{i}^{+}-X_{i}^{-})(Y_{i}^{+}-Y_{i}^{-})\ .\label{eq:ell0}
\end{equation}
The tolerance ellipse is the projection onto the $XY$ plane of the
Hessian hypersphere spanned by the eigenvector sets
in the linear approximation for PDF dependence of $X$ and $Y$. The
ellipse can be found from the parametric equations
\begin{equation}
X(\theta)=X_{C}^{H}+\delta^{H}(X)\cos\theta,\quad Y(\theta)=Y_{C}^{H}+\delta^{H}(Y)\cos(\theta+\Delta\varphi)\label{eq:ell1}
\end{equation}
for $0\leq\theta<2\pi$.

In the MC approach we can also define the error ellipse by an equation
\begin{equation}
\frac{(X-X_{C}^{M})^{2}}{\delta^{M}(X)^{2}}+\frac{(Y-Y_{C}^{M})^{2}}{\delta^{M}(Y)^{2}}-\frac{2\rho(X-X_{C}^{M})(Y-Y_{C}^{M})}{\delta^{M}(X)\delta^{M}(Y)}=p_{0}^{2}(1-\rho^{2})\ ,\label{eq:ell2}
\end{equation}
where $\rho\equiv{\rm {\rm cov}}(X,\ Y)/\left(\delta^{M}(X)\,\delta^{M}(Y)\right)$
is the correlation of $X$ and $Y$, and
\begin{equation}
{\rm cov}(X,Y) = \frac{1}{N_{rep}-1}\sum_{i=1}^{N_{rep}}\left[X_i -
  X_C\right]\cdot\left[Y_i - Y_C\right].
\end{equation}
 Projections of the ellipse on the $X$ and $Y$ axes coincide
with the confidence intervals on $X$ and $Y$ determined by $p_{0}$,
for instance $p_{0}=1$ (1.64) for the 68 (90)\% c.l.
Eqs.~(\ref{eq:ell0}) and (\ref{eq:ell2}) delineate the regions of the
prescribed confidence when the distributions of $X$ and $Y$ are close
to Gaussian ones. The equivalence of the Hessian
and MC error ellipses in the Gaussian case
can be demonstrated by identifying $\rho=\cos(\Delta\varphi)$.
\end{enumerate}

\begin{figure}[tb]
\begin{centering}
\includegraphics[width=0.45\textwidth]{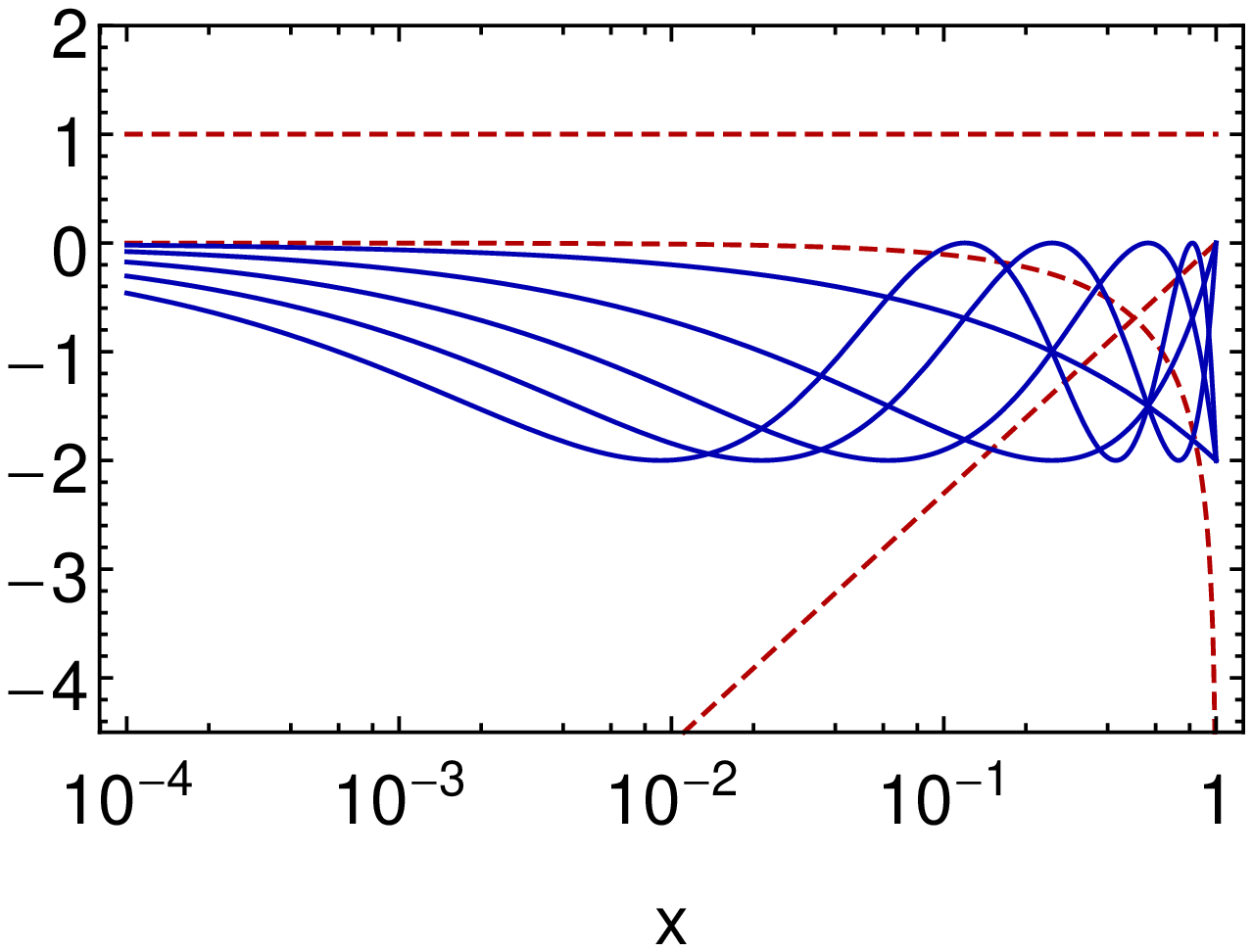}
\includegraphics[width=0.45\textwidth]{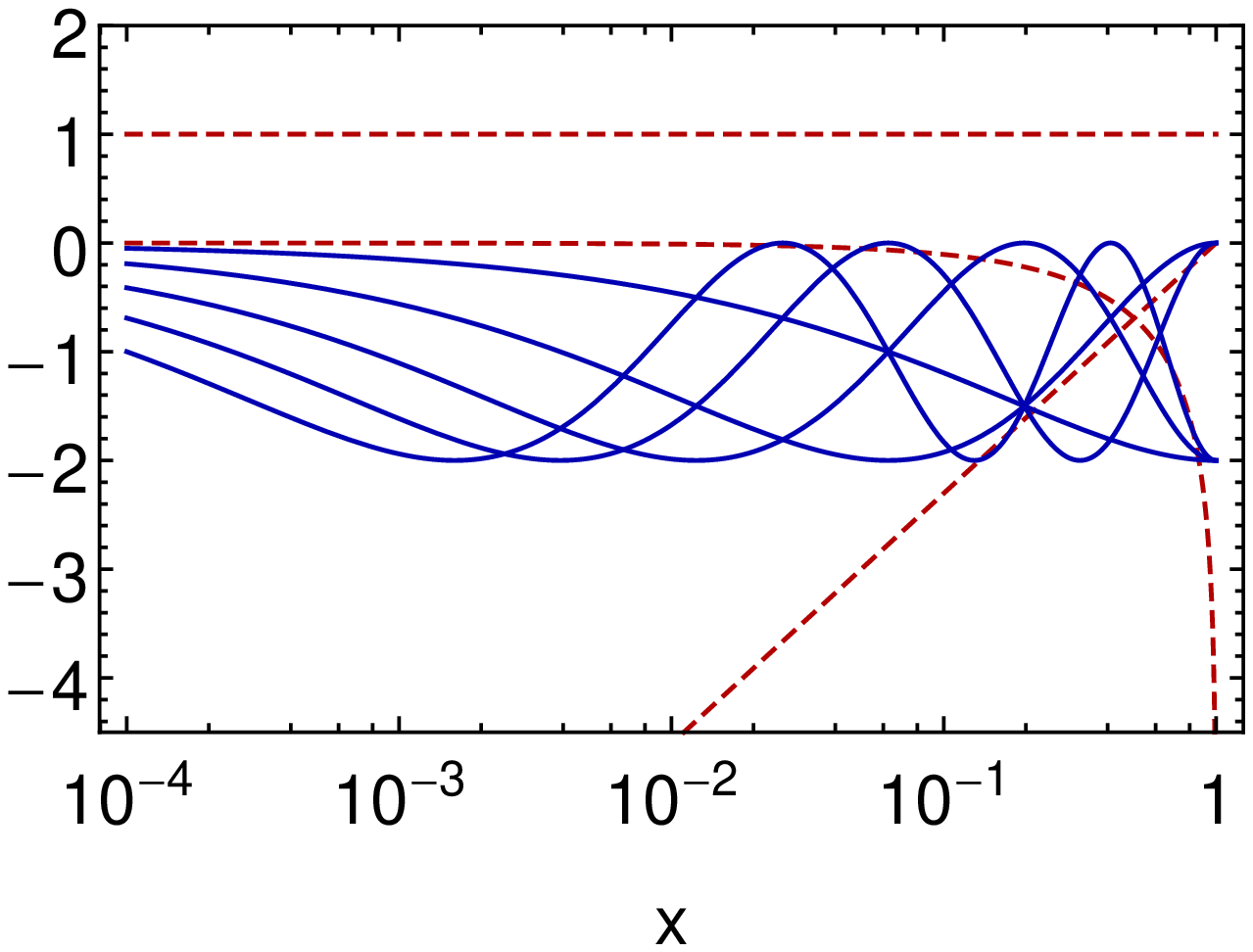}
\par\end{centering}

\vspace{-1ex}
 \caption{\label{fig:yx} $d\ln(f)/da_{i}$ for $a_{1,2,3}$ (red dashed curves)
and $a_{4,5,6,7,8}$ (blue solid curves) with $y(x)=1-2x^{1/2}$ (left
inset) and $y=\cos(\pi x^{1/4})$ (right inset).}
\end{figure}

\subsection{Parametrization and sum rules\label{sec:para}}

The functional forms for $\Phi(x,Q_0)$ are often assumed to behave
as $\sim x^{a_{2}}(1-x)^{a_{3}}$ in the $x\rightarrow0$ and
$x\rightarrow1$ limits on the basis of Regge and quark counting
arguments. They must satisfy the valence quark number and momentum
sum rules and predict positive cross sections. The free parameters
of $\Phi(x,Q_0)$ are found from a fit, their total number is chosen
so as to provide very flexible functional forms agreeing with the
data, but without overfitting the data. NNPDF2.3 uses a neural
network with 259 parameters to minimize the parametrization bias,
while the CT10 and MSTW'2008 NNLO ensembles keep 25 and 20 free
parameters, respectively. Another convenient form utilizes Chebyshev
orthogonal polynomials $T_{i}(y(x))$
\cite{Pumplin:2009bb,Glazov:2010bw,Martin:2012da}. This form is
particularly suited for constructing the meta-parametrizations and
will be employed in our analysis.

\begin{figure}[b]
\begin{centering}
\includegraphics[width=0.32\textwidth]{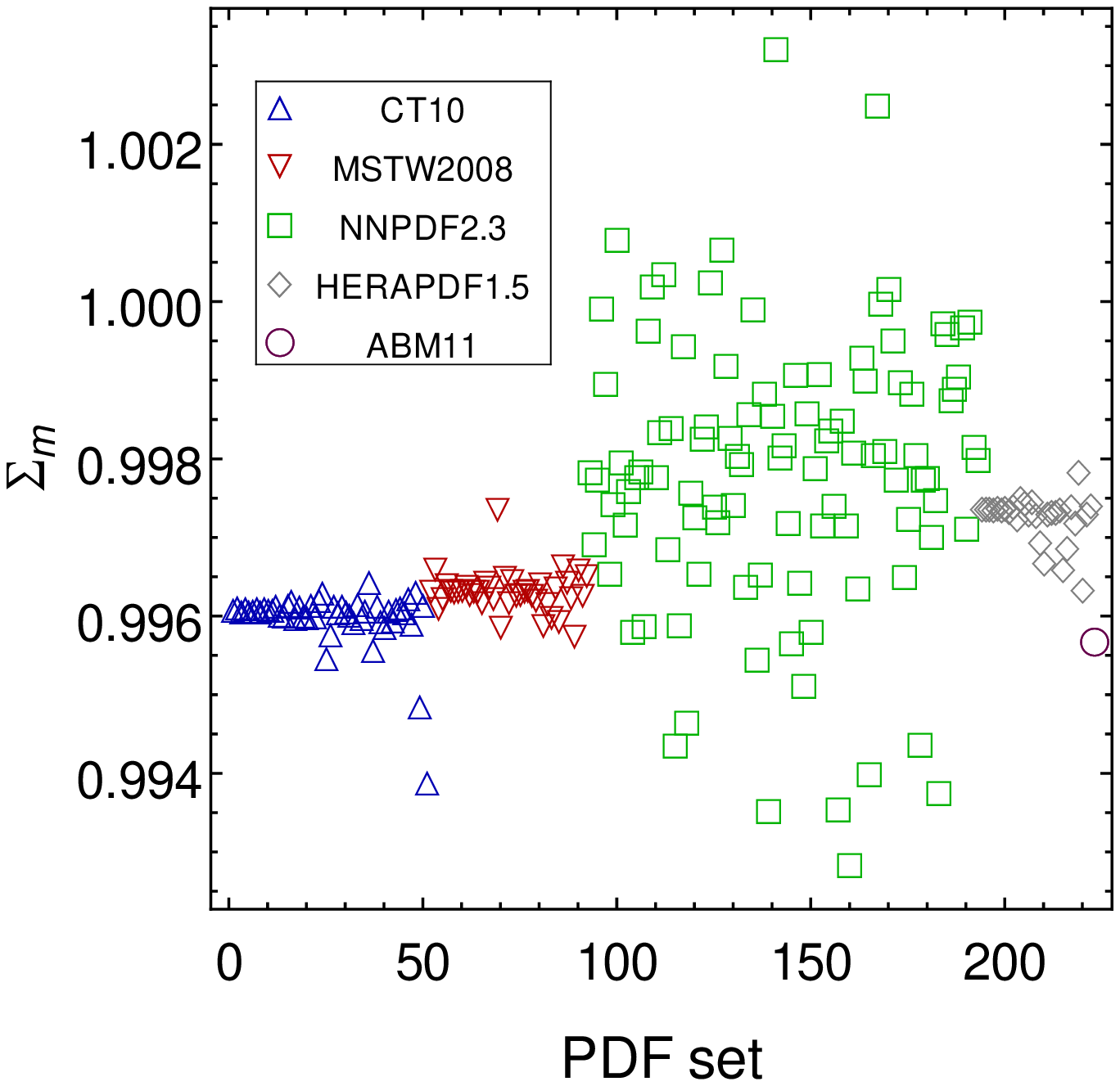}
\includegraphics[width=0.313\textwidth]{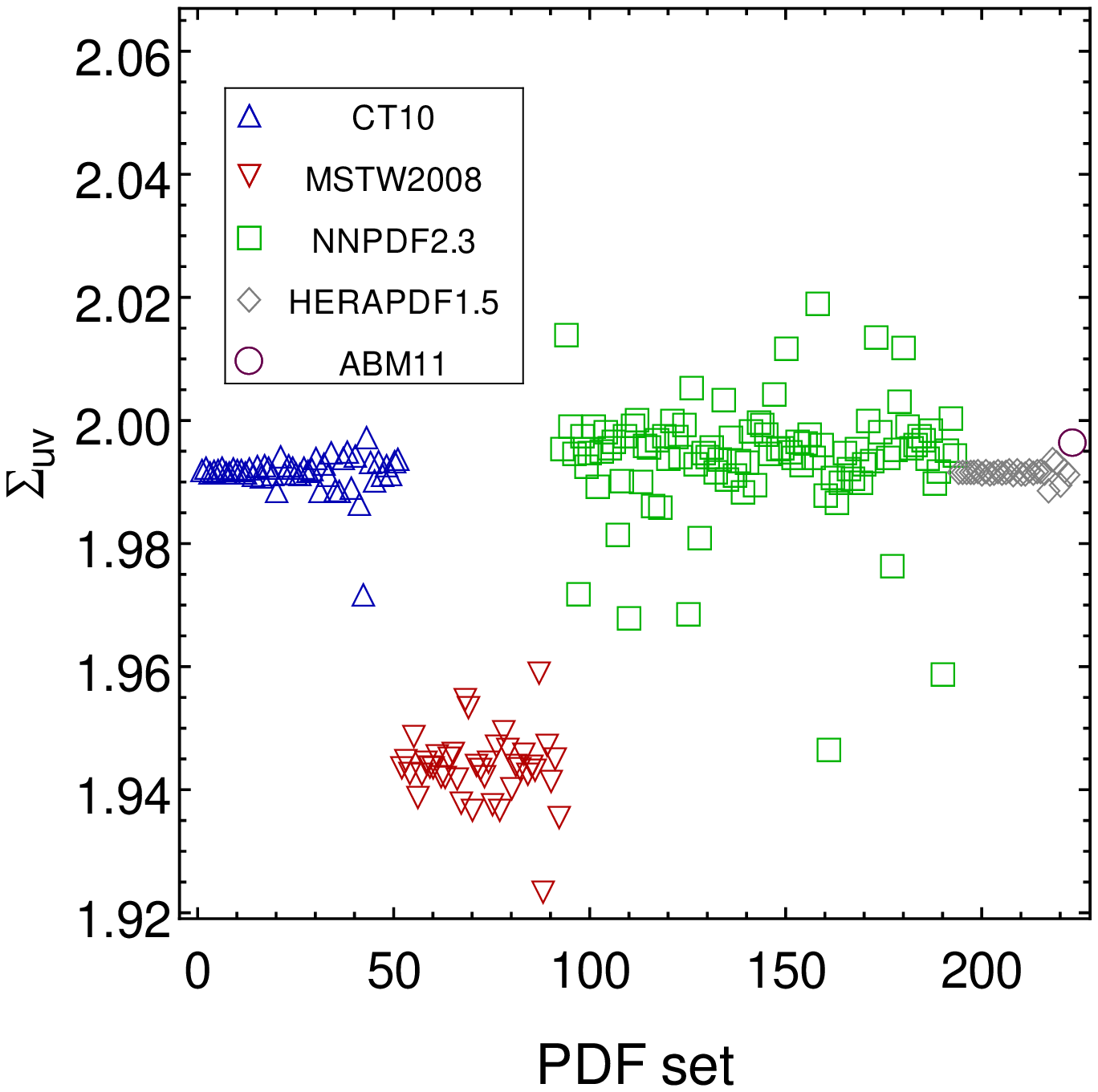}
\includegraphics[width=0.32\textwidth]{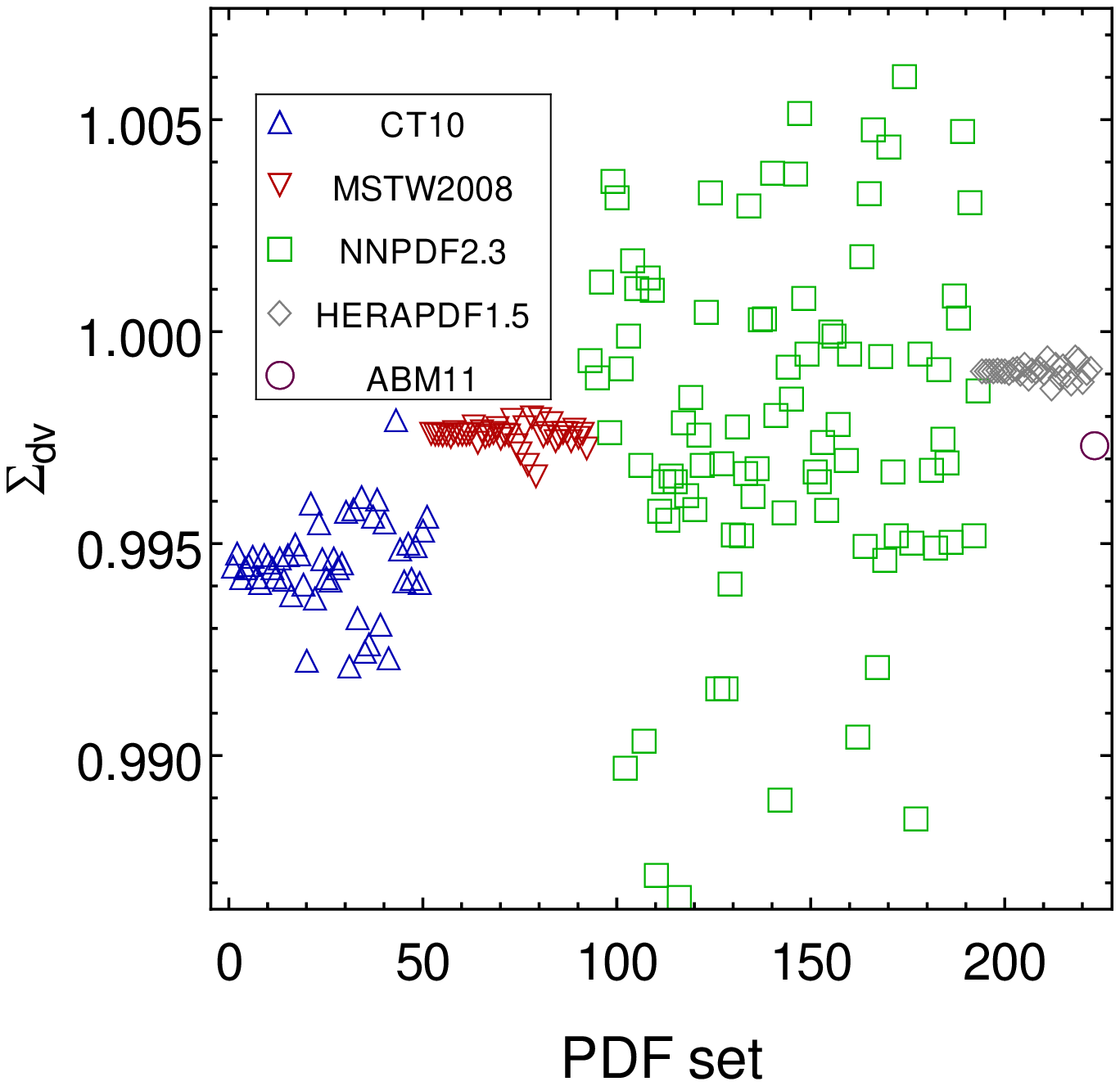}
\par\end{centering}

\vspace{-1ex}
 \caption{\label{fig:sum} Partial integrals over the fitted $x$ ranges
for the momentum sum rule, $u$-valence sum
rule, and $d$-valence sum rule at $Q_0=8\,{\rm GeV}$.
In each inset, the points from left to right
are found from CT10 (up triangles), MSTW2008 (down triangles),
NNPDF2.3 (squares), HERAPDF1.5 (diamonds), and ABM11 (circle) PDF sets.}
\end{figure}

In those $x$ ranges where the experimental data impose tight
constraints, the number of effective degrees of freedom is smaller
than in the full PDF ensemble, so a smaller number of parameters is
needed to approximate the acceptable PDF shapes. We will therefore
fit the effective parametrizations to the input PDFs only in the $x$
ranges where sufficient experimental constraints are available. The
boundaries of these ranges are taken to satisfy $x>3\cdot10^{-5}$
for all quark flavors; $x<0.8$ for the gluon and $u,d,c,b$ quarks; $
x<0.4$ for $\bar{u},\bar{d}$ quarks; and $x<0.3$ for $s,\bar{s}$
quarks. By construction, the uncertainty bands of the input PDFs and
their meta-parametrizations agree well in the fitted $x$ regions.
Outside these ranges, the meta-parametrizations are determined by
extrapolation and span a wide uncertainty band, which is close,
although not identical, to the original PDF uncertainty (which has a
large uncertainty of its own at such $x$). The PDFs in the outside
(unfitted) regions are hardly constrained at the moment and have
negligible contributions to most LHC observables even at 14 TeV.

\begin{figure}[h]
\begin{centering}
 \includegraphics[width=0.48\textwidth]{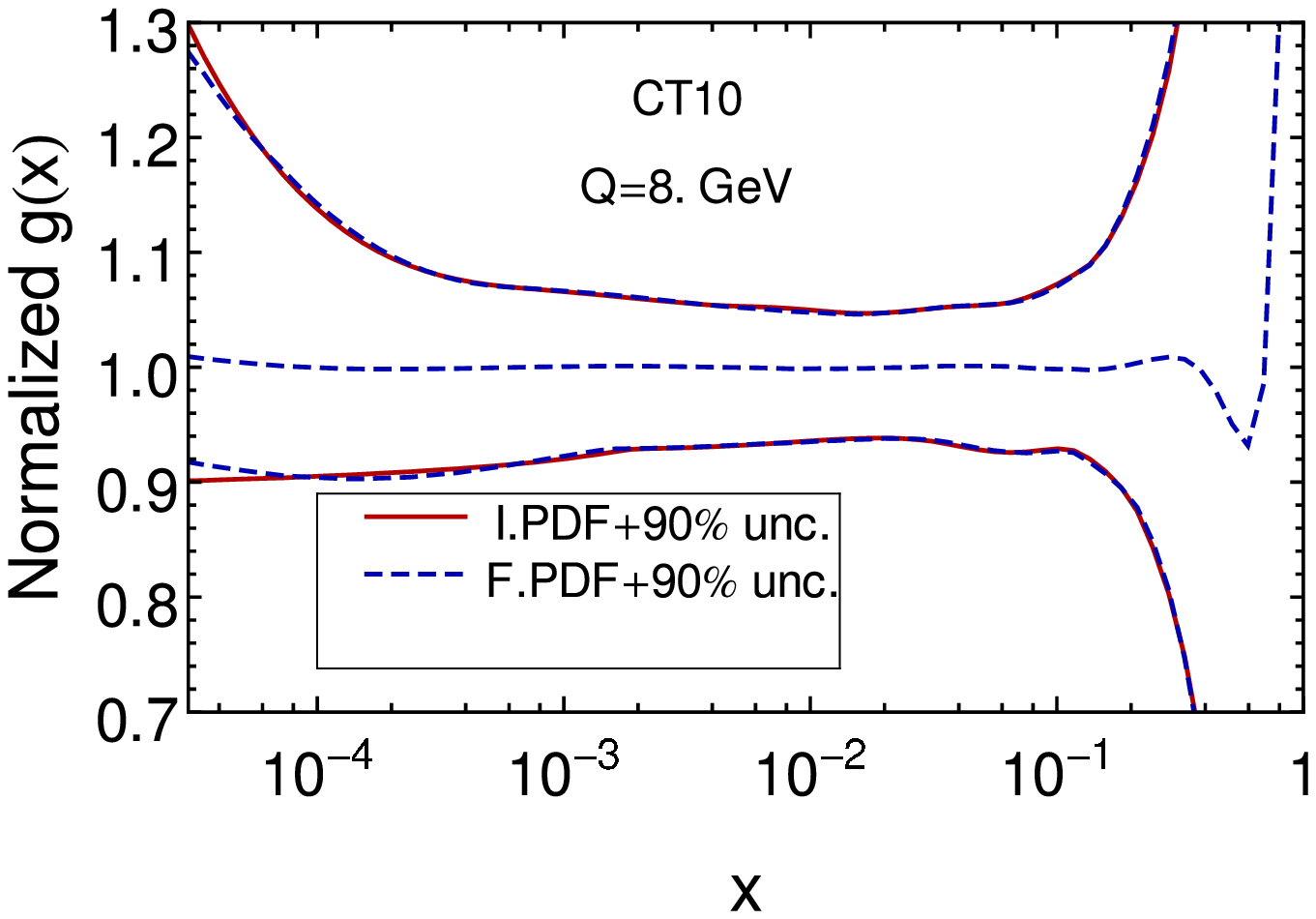} \hspace{10pt}
 \includegraphics[width=0.48\textwidth]{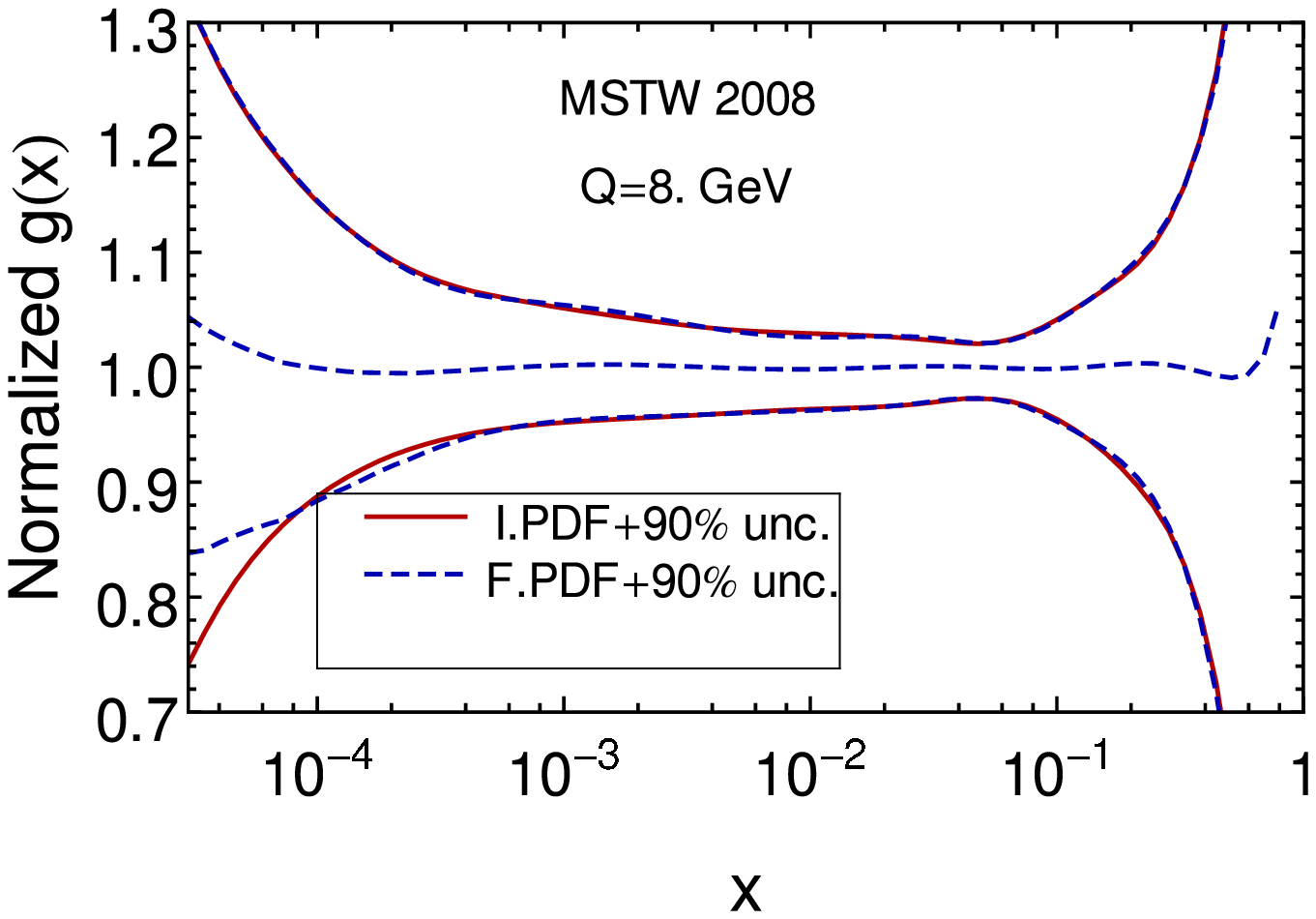}\\
 \includegraphics[width=0.48\textwidth]{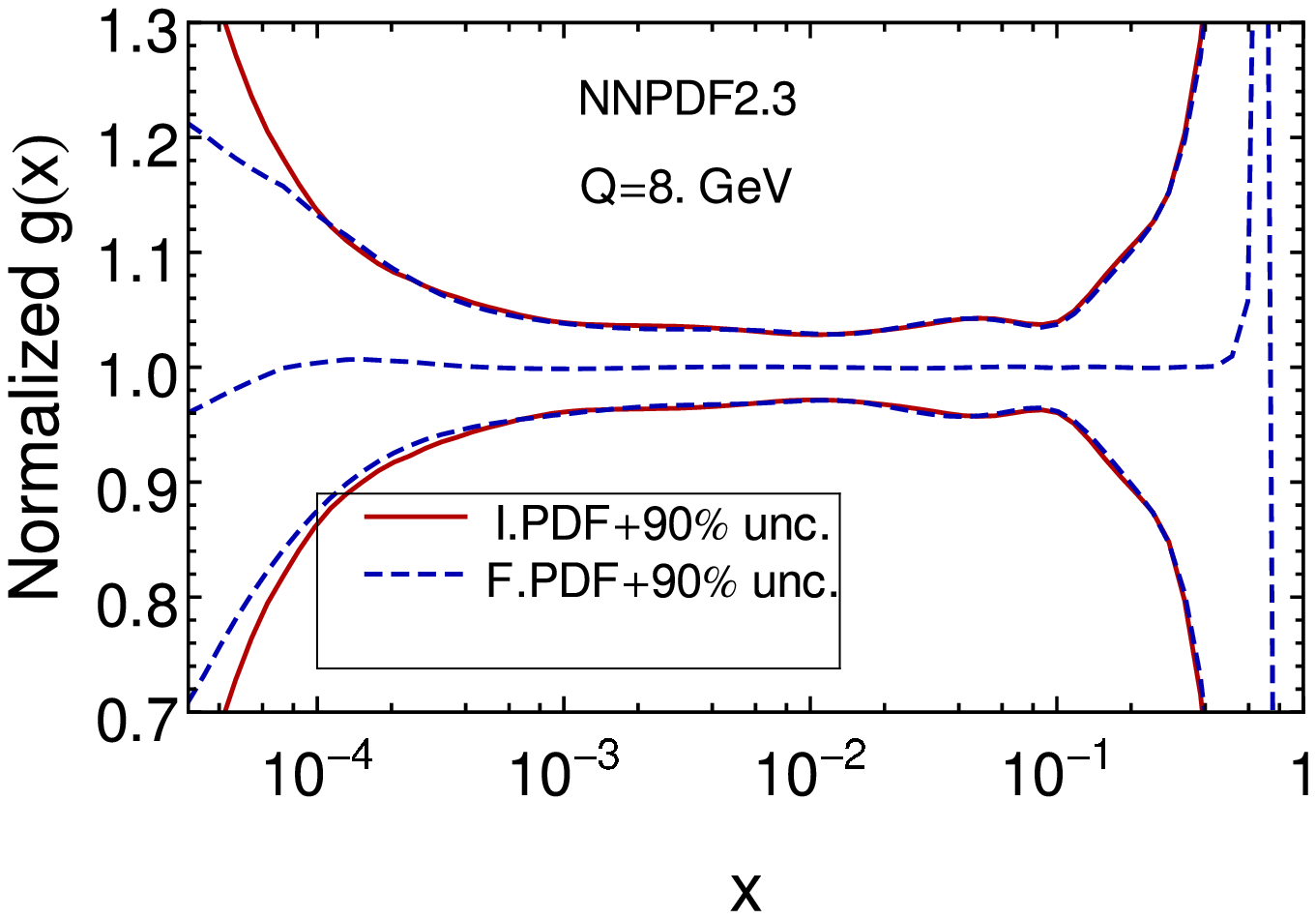} \hspace{10pt}
 \includegraphics[width=0.48\textwidth]{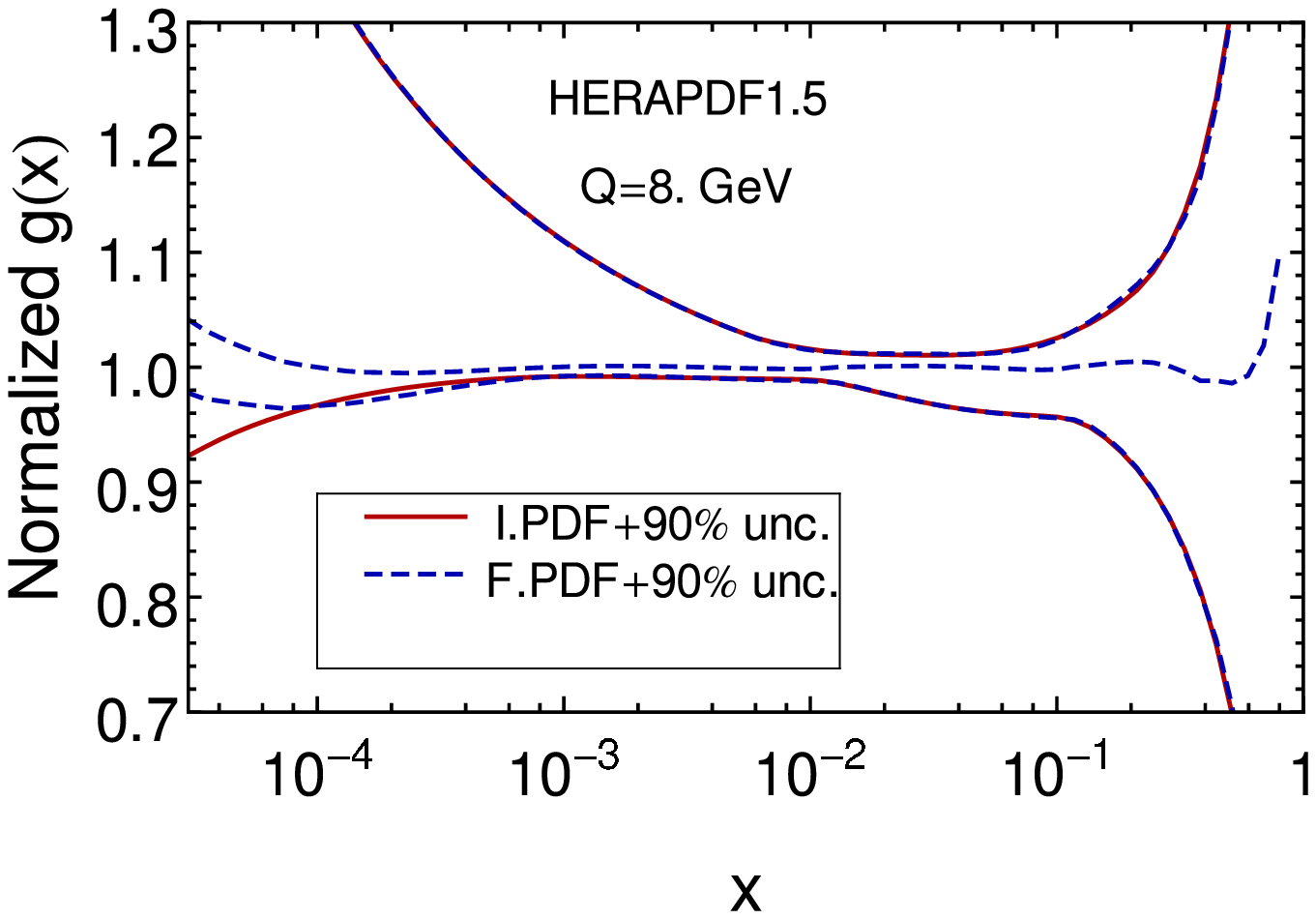}\\
 \includegraphics[width=0.48\textwidth]{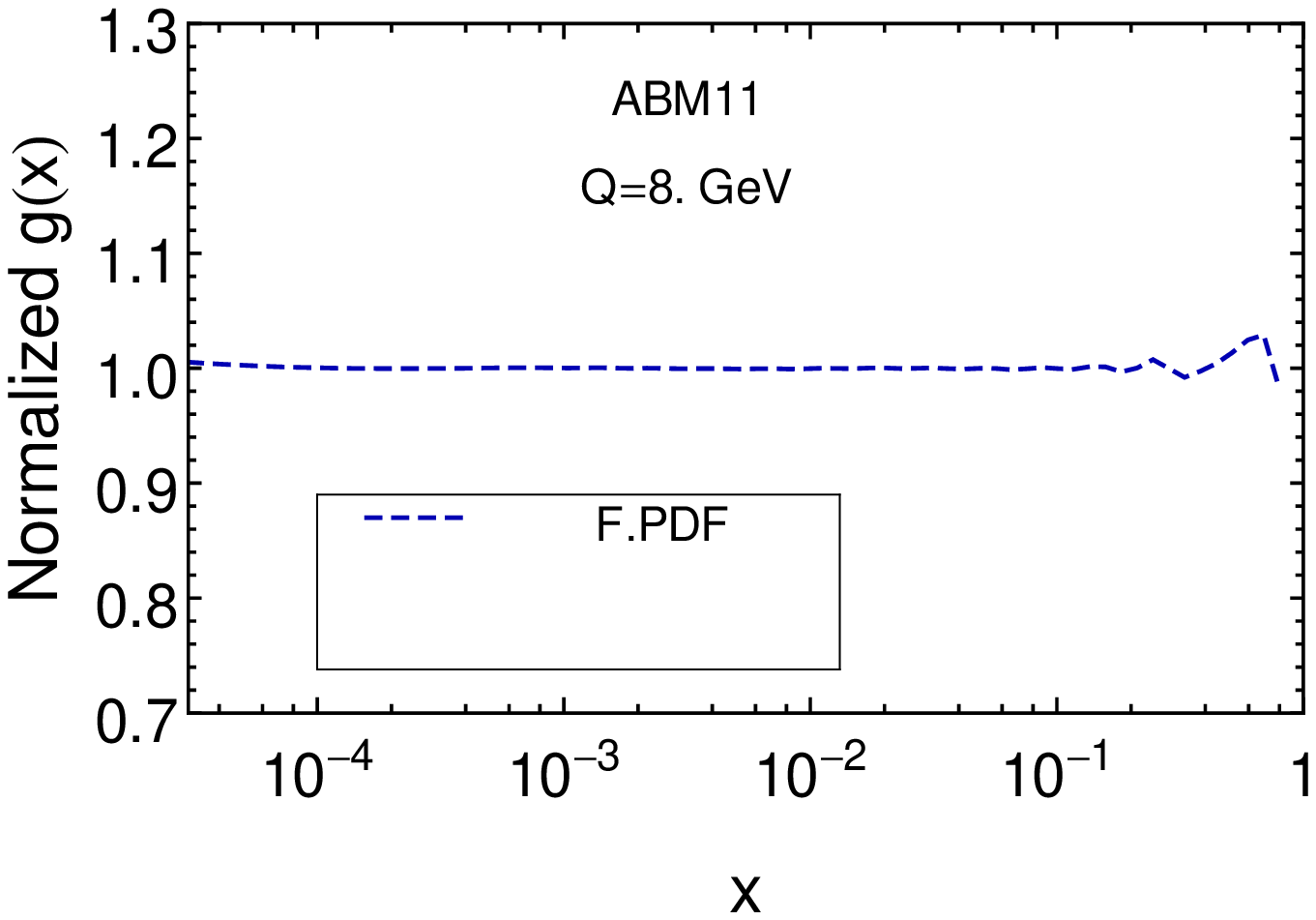}
\par\end{centering}

\vspace{-3ex}
 \caption{\label{fig:com1} Comparison of the input (I) and fitted
(F)  gluon PDFs and their 90\% c.l. uncertainties
at the initial scale $Q=8\,{\rm GeV}$, normalized
to the best-fit gluon PDF of the input ensemble.}
\end{figure}

\begin{figure}[h]
\begin{centering}
 \includegraphics[width=0.48\textwidth]{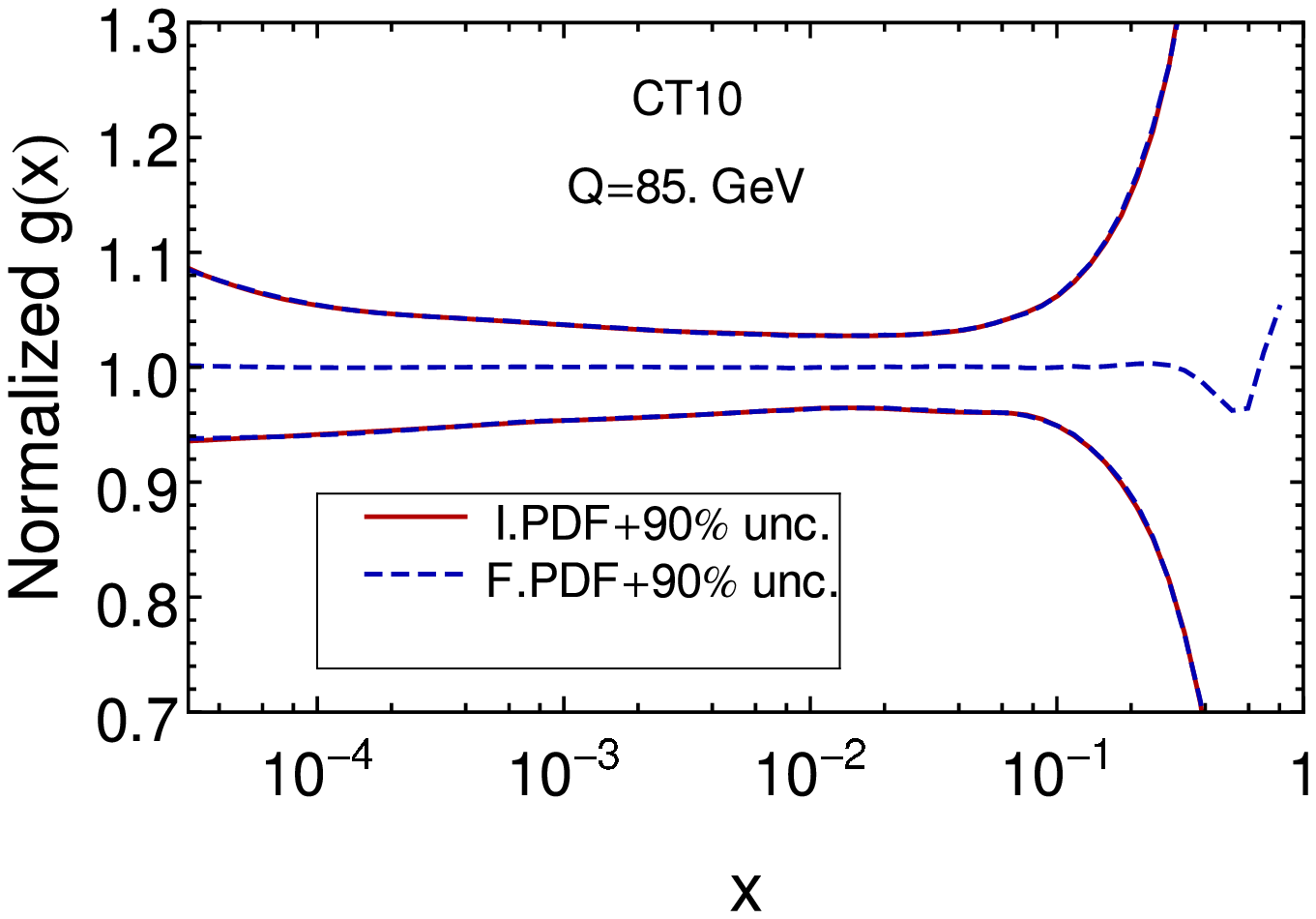} \hspace{10pt}
 \includegraphics[width=0.48\textwidth]{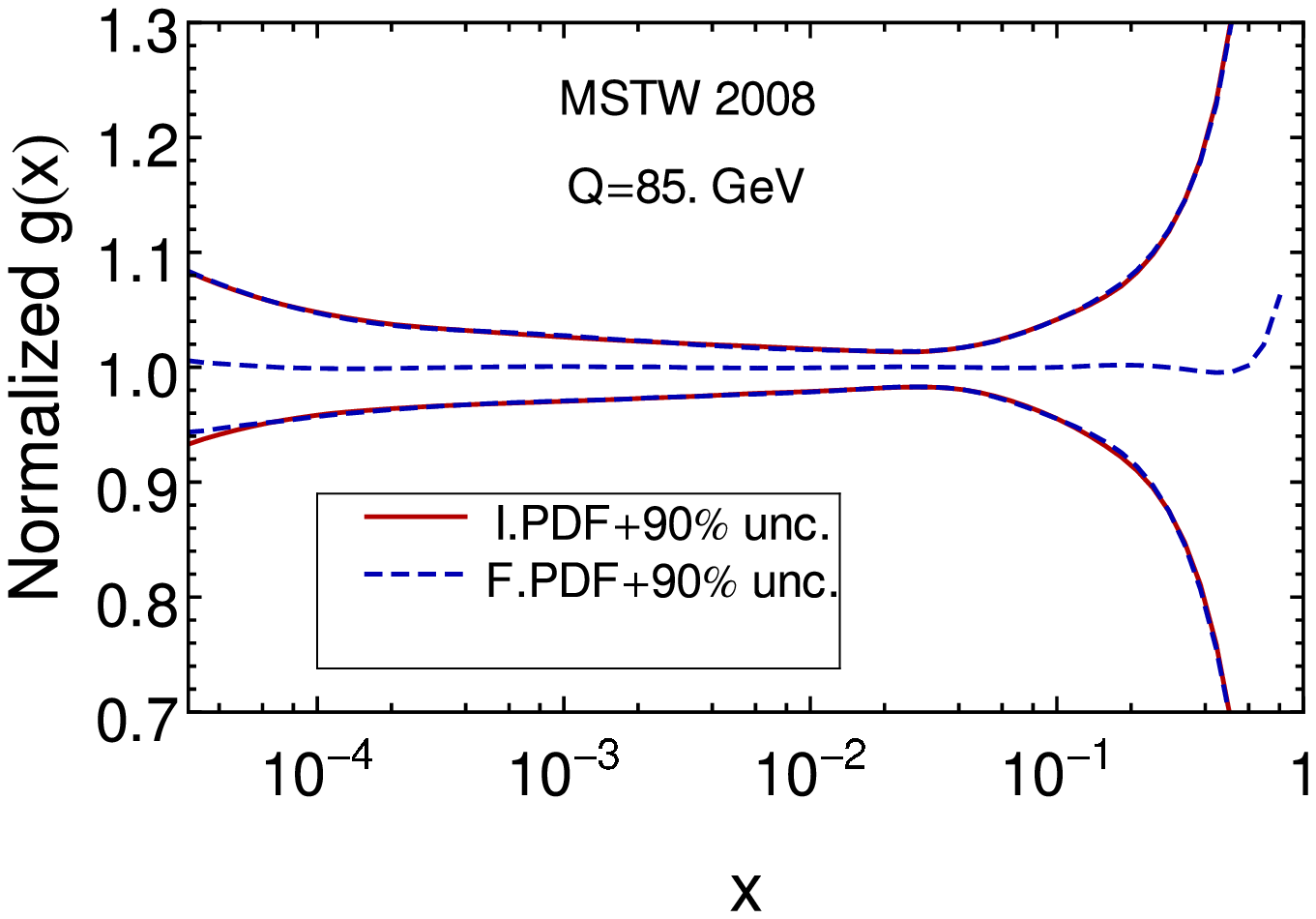}\\
 \includegraphics[width=0.48\textwidth]{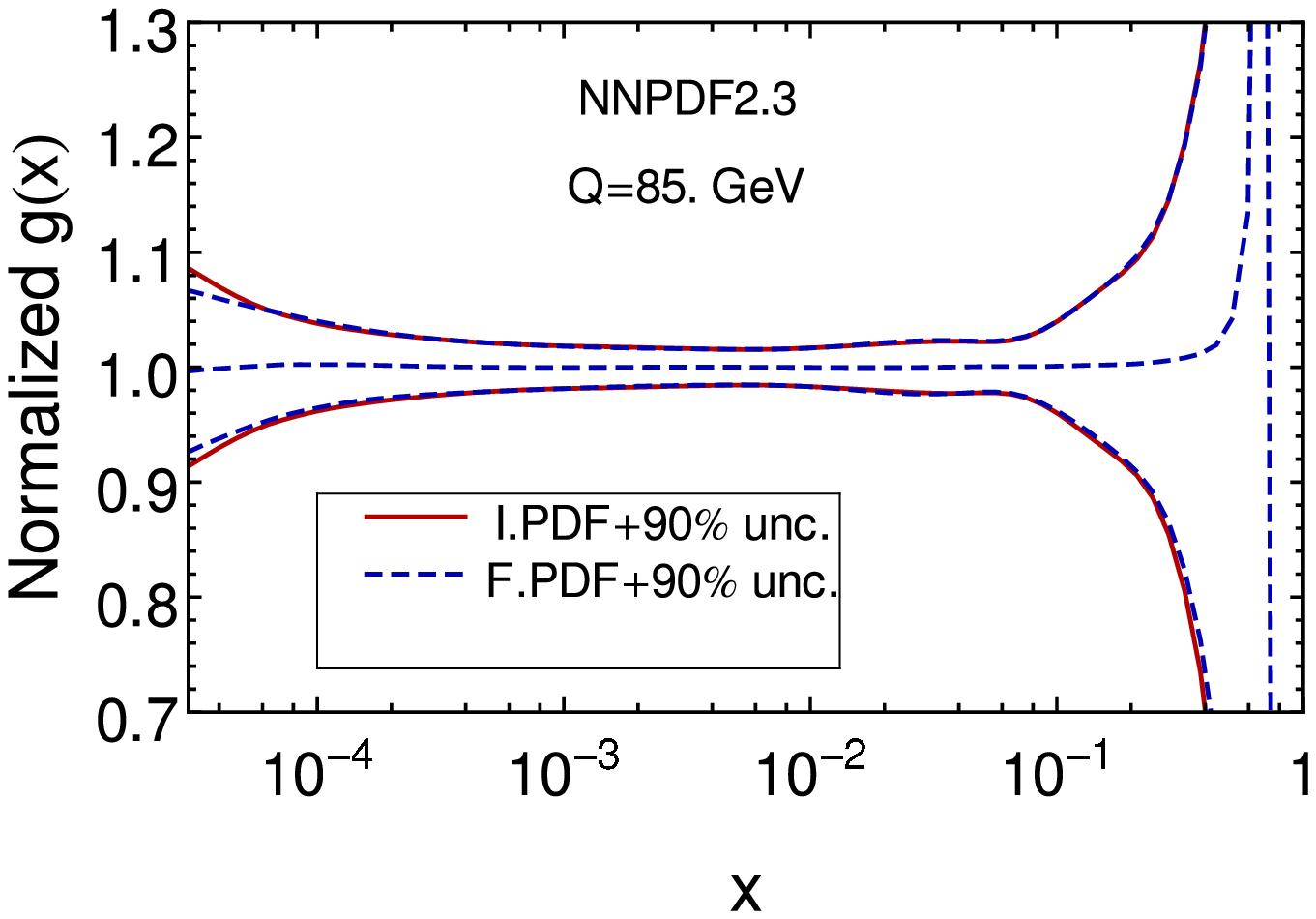} \hspace{10pt}
 \includegraphics[width=0.48\textwidth]{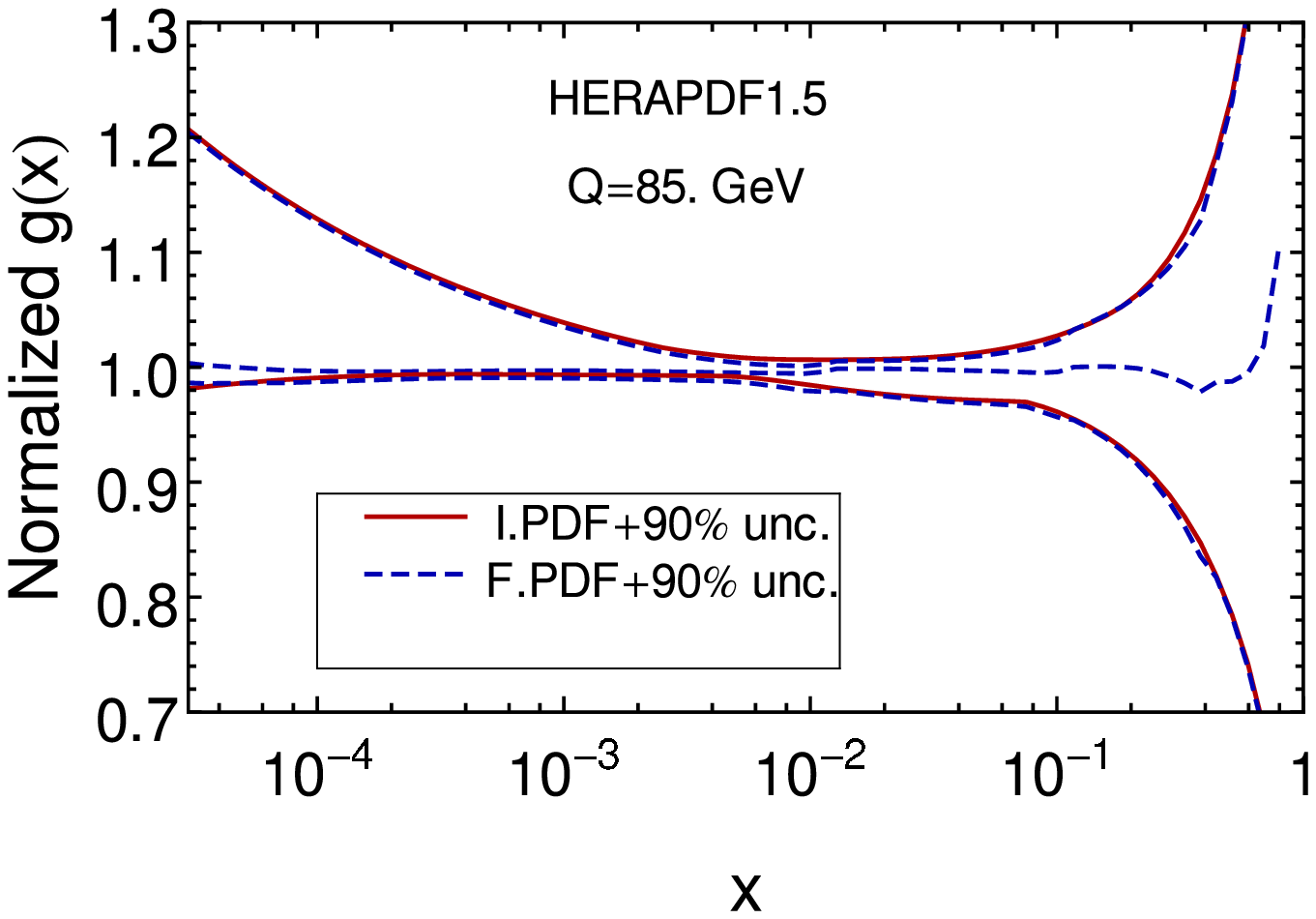}\\
 \includegraphics[width=0.48\textwidth]{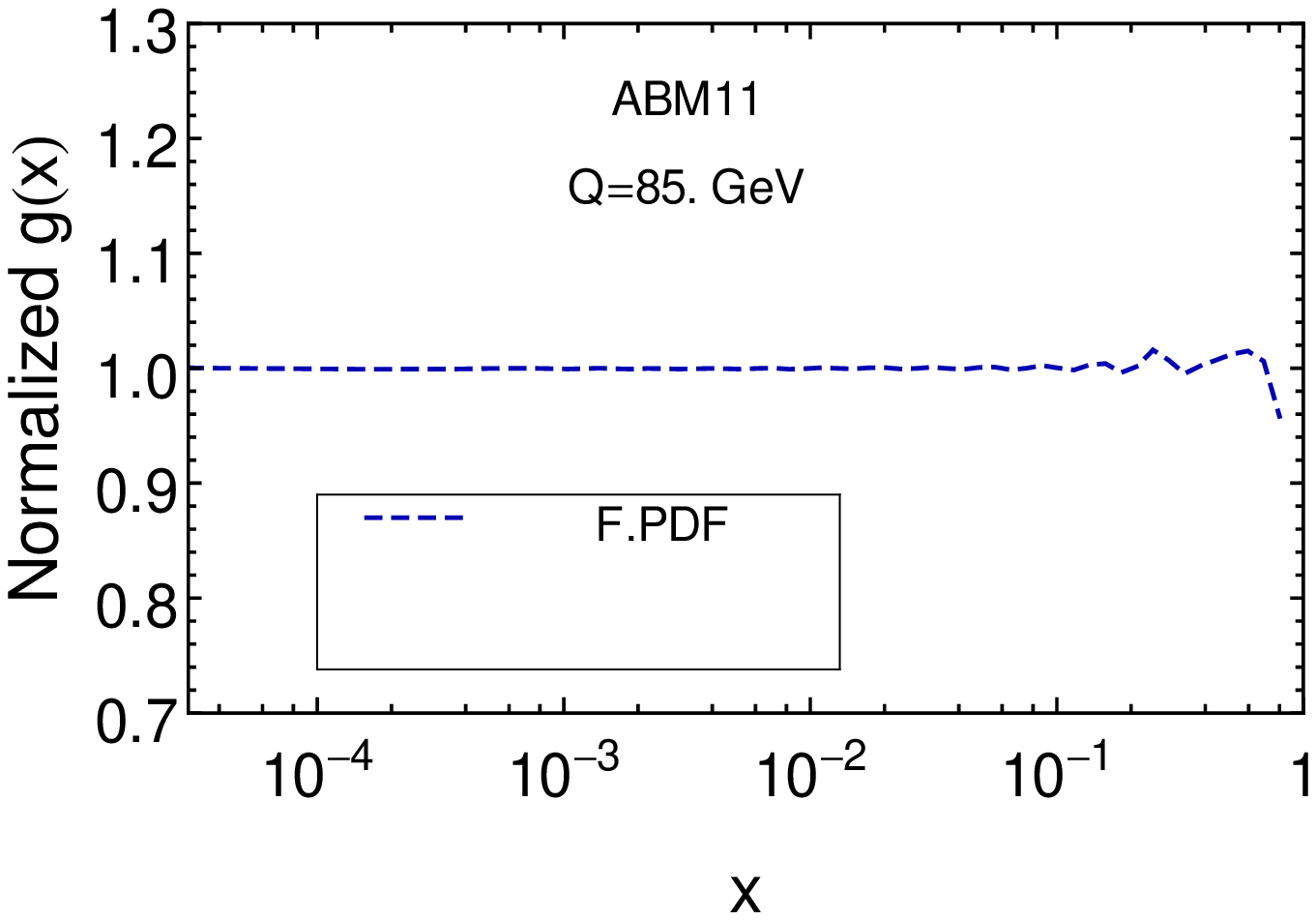}
\par\end{centering}

\vspace{-3ex}
 \caption{\label{fig:com2} Same as Fig.~\ref{fig:com1}, at
scale $Q=85$ GeV after evolution of the fitted PDFs.}
\end{figure}

The specific effective form that we choose at the scale $Q_{0}$ for
each flavor is
\begin{equation}
f(x,Q_{0};\{a\})=e^{a_{1}}x^{a_{2}}(1-x)^{a_{3}}e^{\sum_{i\geq4}a_{i}\,
  \bigl[T_{i-3}\left(y(x)\right)-1\bigr]}.\label{efun}
\end{equation}
It satisfies the above asymptotic behaviors, $\sim x^{a_{2}}$ at
$x\rightarrow0$ and $\sim(1-x)^{a_{3}}$ at $x\rightarrow1$, while
the detailed shape is regulated by the Chebyshev polynomials,
$T_{j}(y(x))$ with $j\geq1$, bound to lie between -1 to 1 for
$-1\leq y\leq1$. The positivity condition is automatically satisfied
for $0\leq x\leq1$, in accord with the input PDFs, which are
positive for all groups at the $x$ and $Q_{0}$ values we chose. The
function $y(x)$ maps the $0\leq x\leq1$ interval onto the $-1\leq
y\leq1$ interval. The form of $y(x)$ is selected so as to avoid
large cancellations between the coefficients of Chebyshev polynomials in
the $y\rightarrow\pm1$ limits, hence to reduce the number of
$T_{i}(y(x))$ needed for approximating $\Phi(x,Q_{0})$.

We select $y=\cos(\pi x^{\beta})$ with $\beta=1/4$ as a mapping
function that generally requires fewer Chebyshev polynomials than
the other tried form $y=1-2x^{\alpha}$ with $\alpha=1/2$ suggested
by ~\cite{Pumplin:2009bb}. The choice of $y(x)$ is illustrated in
Fig.~\ref{fig:yx}, where the logarithmic derivatives $d(\ln
f)/da_{i}$ are compared for $y=1-2x^{1/2}$ in the left subfigure and
$y=\cos(\pi x^{1/4})$ in the right subfigure. From Eq.~(\ref{efun})
we have
\begin{equation}
\frac{d\ln(f)}{da_{i}}=\bigl\{1,\,\ln x,\,\ln(1-x),\, T_{i-3}\left(y(x)\right)-1\bigr\},
\end{equation}
thus $\ln f$ is linear in the $a_{i}$ parameters.
The coefficients $d\ln f/da_{i}=\bigl\{T_{1}\left(y(x)\right)-1,\: T_{2}\left(y(x)\right)-1,...\bigr\}$
for $i\ge4$ are shown by blue solid lines. With
the choice $y=\cos(\pi x^{1/4})$ in the right inset,
the oscillations of the polynomials are stretched across a wider span of $x$,
resulting in a better approximation of the PDF shapes.

We evaluate $\Phi(x_{k},Q_{0})$ and $f(x_{k},Q_{0};\{a\})$
on a lattice of momentum fractions $\{x_{k}\}$ for each
flavor and fit $f$ to $\Phi$ by minimizing a metric function
\begin{equation}
E\left[\Phi,\ f(a)\right]=\sum_{{\rm flavors,}x{\rm \, grid}}\left[\frac{\ln\Phi(x_{k},Q_{0})-\ln f(x_{k},Q_{0};\{a\})}{\delta(\ln\Phi(x_{k},Q_{0}))}\right]^{2},
\end{equation}
where $\delta(\ln\Phi(x_{k},Q_{0}))\equiv
\delta(\Phi(x_{k},Q_0))/\Phi(x_{k},Q_0)$, and
$\delta(\Phi(x_{k},Q_0))$ is the symmetric PDF uncertainty of
$\Phi(x_{k},Q_{0})$. For each of the 9 flavors, $f(x,Q_0; {a})$
depends on at least three important parameters, $a_{1,2,3}$. The
number of additional Chebyshev polynomials varies depending on the
complexity of the PDF shape and is especially large for the NNPDF
parametrizations that oscillate. By trial and error, we found that
all current NNLO PDFs can be approximated by including up to order-5
polynomials $T_{j}$ for the gluon, $u$ and $d$ quarks; and up to
order-4 polynomials for other flavors. This will be our default
choice, rendering a total of 66 PDF parameters. The differences
between the input and fitted PDF uncertainty bands are much smaller
than the uncertainty bands in this case. The order of Chebyshev
polynomials is still low enough to avoid large cancellations between
their coefficients.

We checked that the sum rules are automatically preserved by the
approximate parametrizations. The integrals of the
meta-parametrizations must obey
\begin{align}
\Sigma_{m} & \equiv\sum_{flavors}\int_{0}^{1}x\, f(x,Q)\, dx=1,\\
\Sigma_{uv} & \equiv\int_{0}^{1}\left(u(x,Q)-\overline{u}(x,Q)\right)\, dx=2,\\
\Sigma_{dv} & \equiv\int_{0}^{1}\left(d(x,Q)-\overline{d}(x,Q)\right)\, dx=1.
\end{align}
In Fig.~\ref{fig:sum}, we evaluate partial contributions to these
integrals over the fitted $x$ regions at the scale $Q_0$. For all
input PDFs, the partial integrals render an average of 0.996 for the
momentum sum, 1.99 for the $u$-valence sum, and 0.997 for the
$d$-valence sum. Thus the sum rules are satisfied by the approximate
parametrizations even if they are not enforced by an explicit
condition, and if integration is only over the selected partial $x$
regions.

In Figs.~\ref{fig:com1} and \ref{fig:com2}, we compare the input
NNLO PDFs (solid lines) for 5 PDF ensembles to their respective
meta-parametrizations, referred to as ``fitted PDFs'' (dashed
lines). The comparisons are made at scales $Q=8$ and 85 GeV after
evolution of the fitted PDFs that will be explained a bit later. Due
to the limited space, we only show results for the gluon PDF
$g(x,Q)$ as a representative example. Analogous plots for all
flavors and PDF ensembles are available at~\cite{metapdfweb}.

In each subfigure, we see the ratio of the central fitted PDF to the
central input PDF for each ensemble, indicated by the dashed line at
the center.  The bands of the 90\% c.l. PDF uncertainties for the input
and fitted PDFs are also shown, normalized to the central input PDF.
 When the input ensemble provides only 68\% c.l. error sets,
the 90\% c.l. errors are obtained by multiplying the 68\% c.l. ones by 1.64.
For ABM11, only the best-fit
PDF with $\alpha_s(M_Z) = 0.118$
is presented, since
their PDF error sets correspond to a lower central $\alpha_s(M_Z)$ and
include the $\alpha_s(M_Z)$ uncertainty in the covariance matrix.

Apart from not too consequential differences at very small or large
$x$ values, the agreement between the input and fitted PDFs is good,
especially for CT10. For MSTW at $Q=8$ GeV and, similarly, for
HERAPDF, only for $x$ below $10^{-4}$ the fitted PDFs show a
systematically upward shift due to the double pole structure of the
MSTW gluon PDF~\cite{Martin:2009iq} that is absent in our functional
form. It is even more interesting to examine the fit to NNPDF2.3,
which originally has a much more flexible parametrization with a
total of 259 PDF parameters. From Fig.~\ref{fig:com1}, we find that
the fitted PDFs show almost the same statistical features as the
original NNPDF except for the PDF uncertainties in the region with
$x<10^{-4}$. The differences between the input and fitted PDF errors
are further reduced at higher scales, such as $Q=85$ GeV in
Fig.~\ref{fig:com2}. These small differences outside of the typical $x-Q$
region of the LHC will hardly matter in practice.

In principle, above the initial scale $Q_{0}=8\ {\rm GeV}$, the
selected NNLO PDFs and associated QCD coupling strengths
$\alpha_{s}$ follow identical evolution equations based on 3-loop
QCD splitting functions with 5 active
flavors~\cite{Moch:2004pa,Vogt:2004mw}, since $Q_{0}$ is above the
PDF transition threshold from 4 to 5 flavors used in the variable
flavor number schemes. On the other hand, some numerical differences
can't be excluded initially in numerical implementation of these
equations by various groups. To benchmark the $\alpha_{s}$ running
and PDF evolution above $Q_{0}$, we compared the tabulated $Q$
dependence of the PDFs and $\alpha_s(Q)$ from the five ensembles
available in the LHAPDF library \cite{lhapdfweb} to the explicit
evolution of their respective input parametrizations done with the
computer code HOPPET~\cite{Salam:2008qg}. The benchmarking
comparison is summarized in the appendix. Acceptable agreement
between the tabulated and HOPPET evolution at better than 1-2\% is
observed for all PDF ensembles and at practically all $x$ and $Q$.

\begin{figure}[b]
\begin{centering}
 \includegraphics[width=0.46\textwidth]{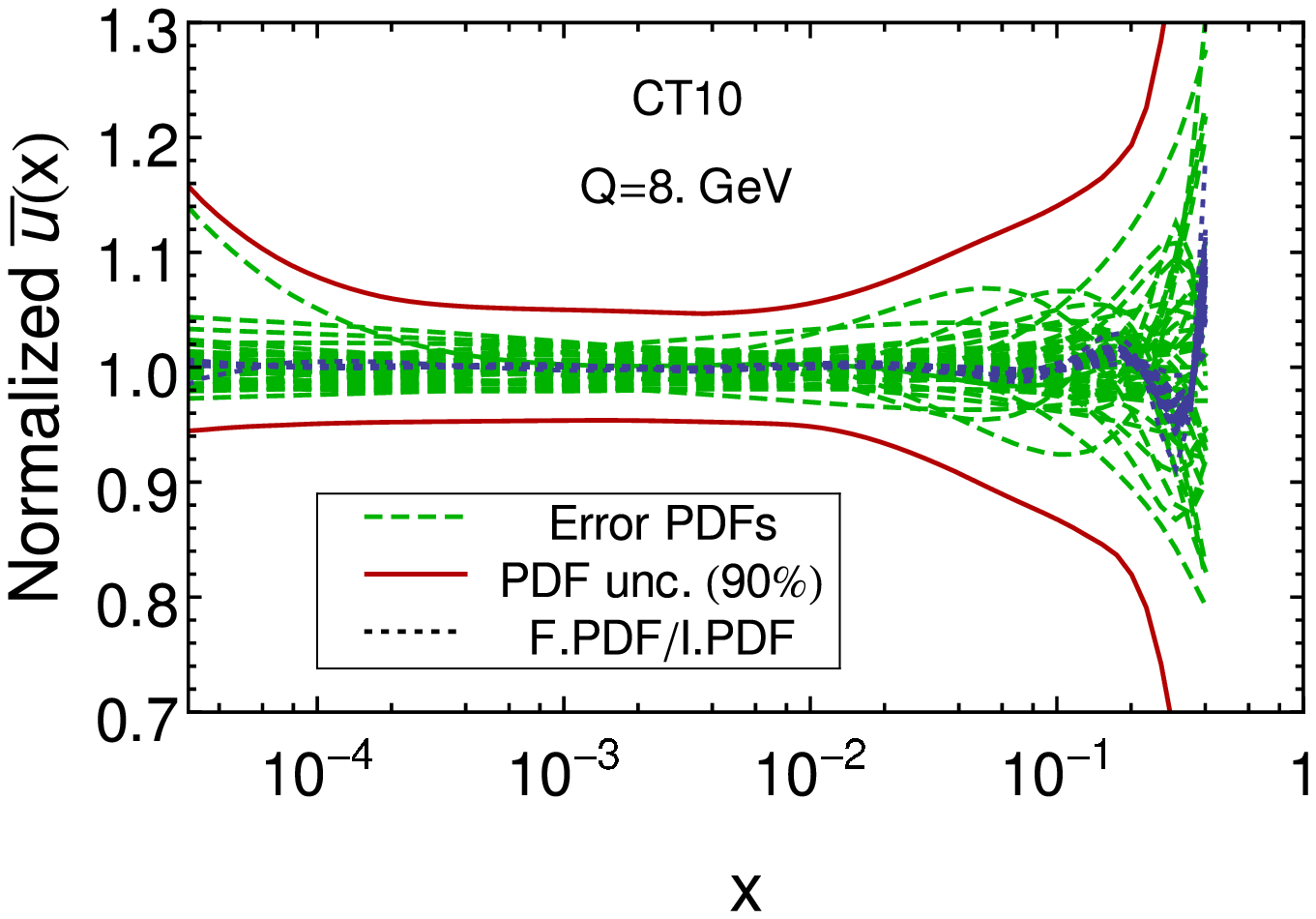} \hspace{10pt}
 \includegraphics[width=0.46\textwidth]{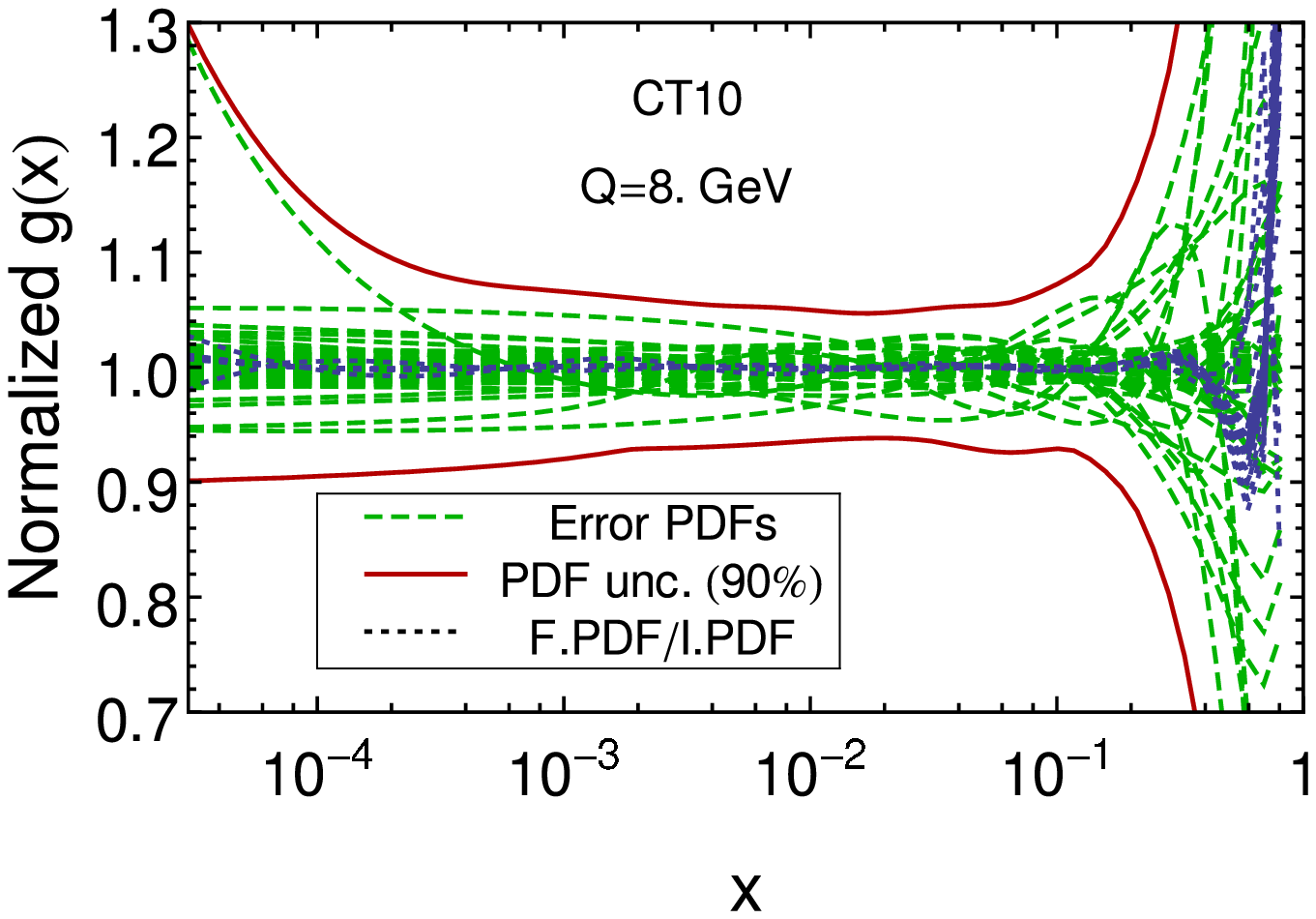} \\ 
 \includegraphics[width=0.46\textwidth]{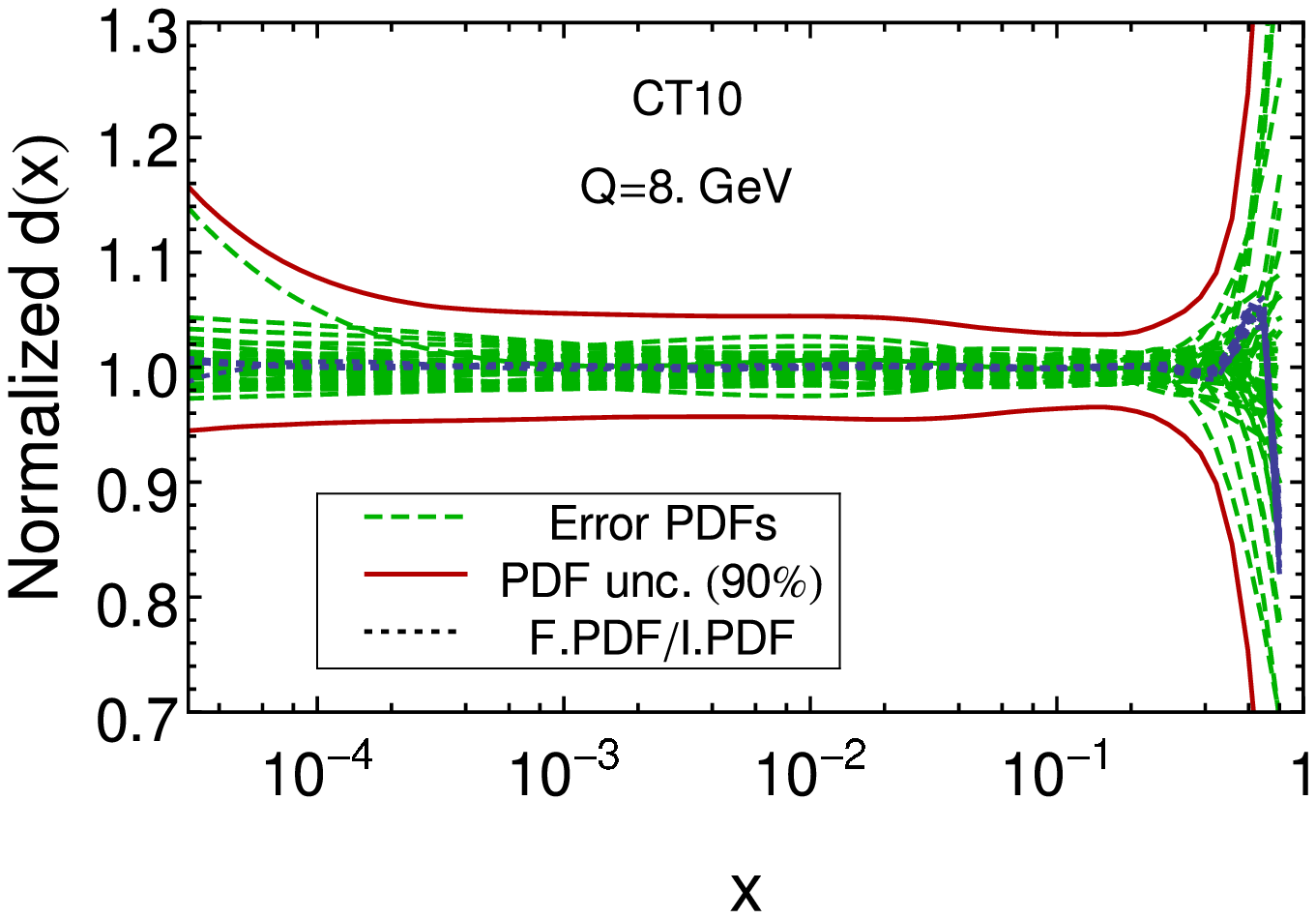} \hspace{10pt}
 \includegraphics[width=0.46\textwidth]{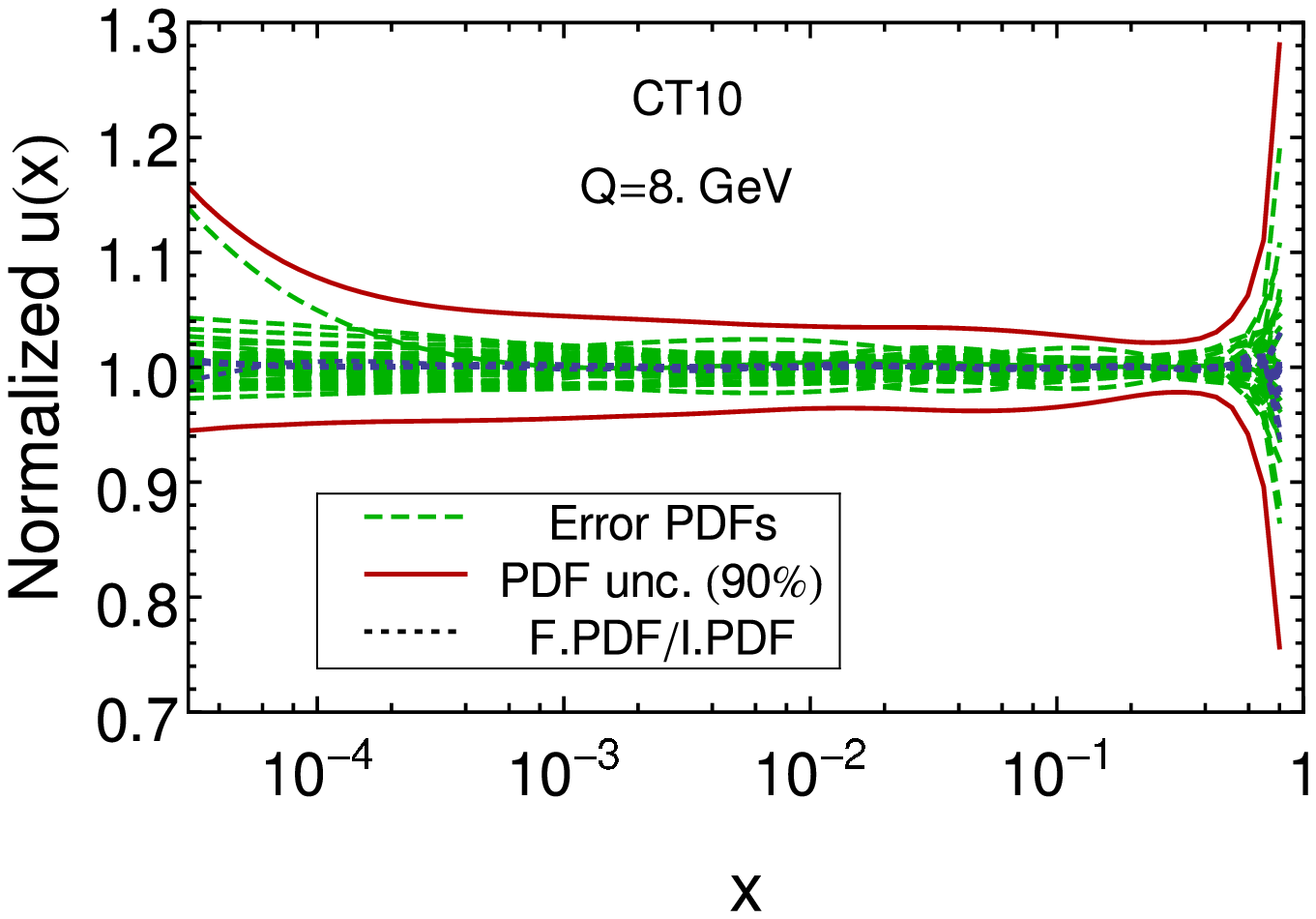} \\
\par\end{centering}

\vspace{-1ex}
 \caption{\label{fig:RelDiffCT10}
The CT10 NNLO error band at 90\% c.l. (outer solid lines) is compared
against individual error sets of the CT10 NNLO ensemble (long-dashed
lines), all normalized to the central CT10 NNLO set, as well as
ratios of the input and fitted
parametrizations for each error set (short-dashed lines).
The scale
$Q=8$ GeV is assumed.}
\end{figure}

\begin{figure}[tb]
\begin{centering}
\includegraphics[width=0.46\textwidth]{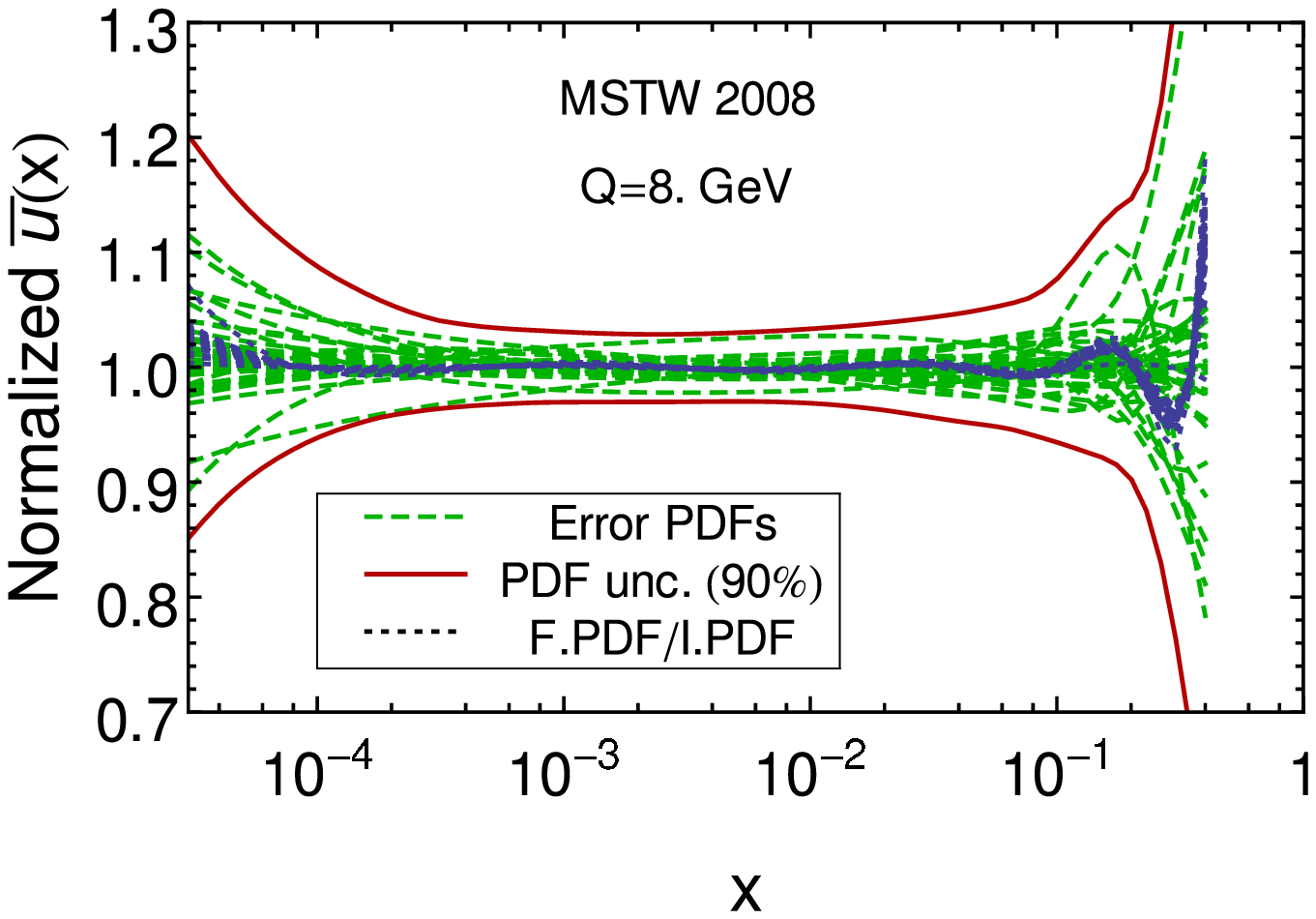} \hspace{10pt}
\includegraphics[width=0.46\textwidth]{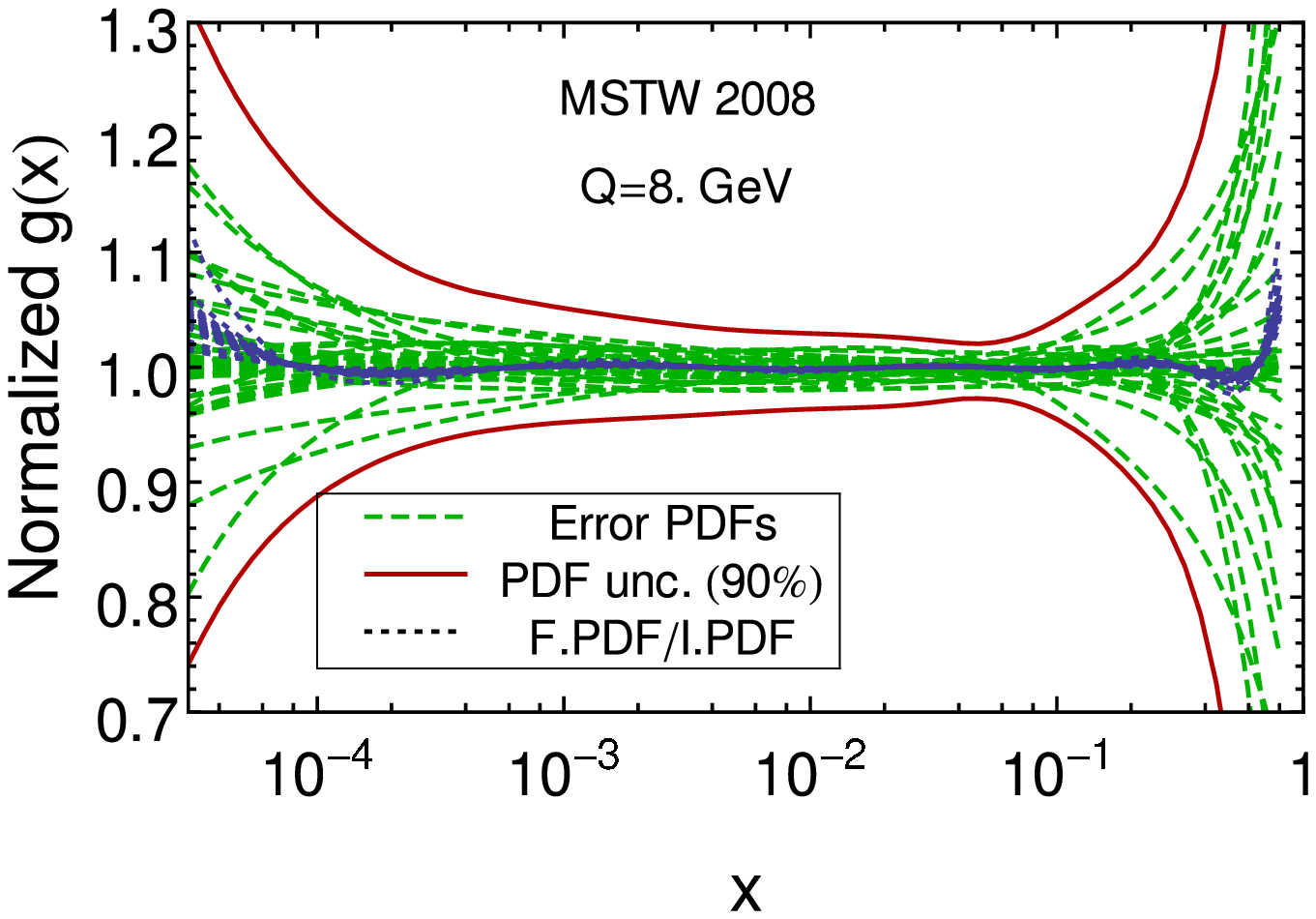} \\
\includegraphics[width=0.46\textwidth]{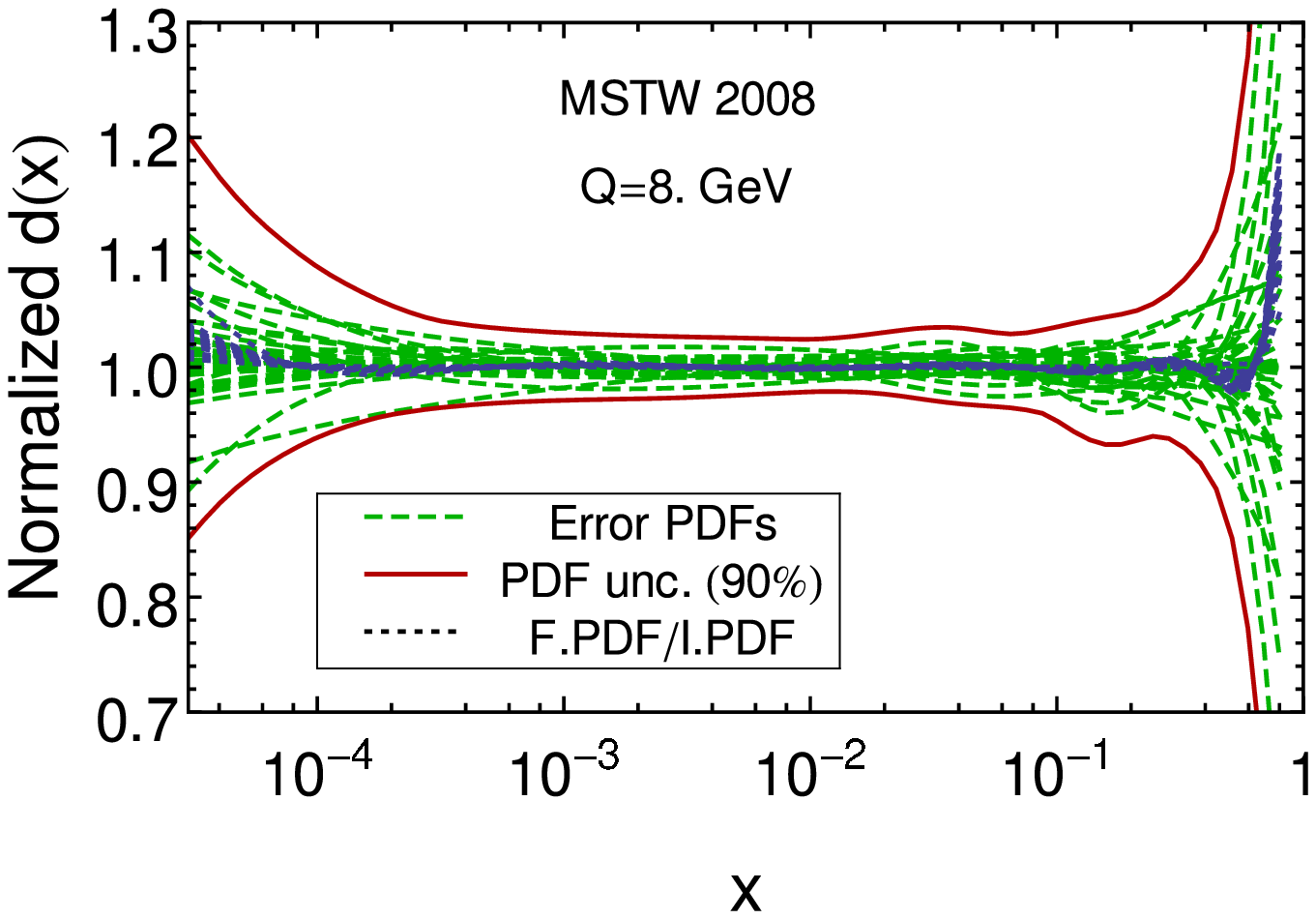} \hspace{10pt}
\includegraphics[width=0.46\textwidth]{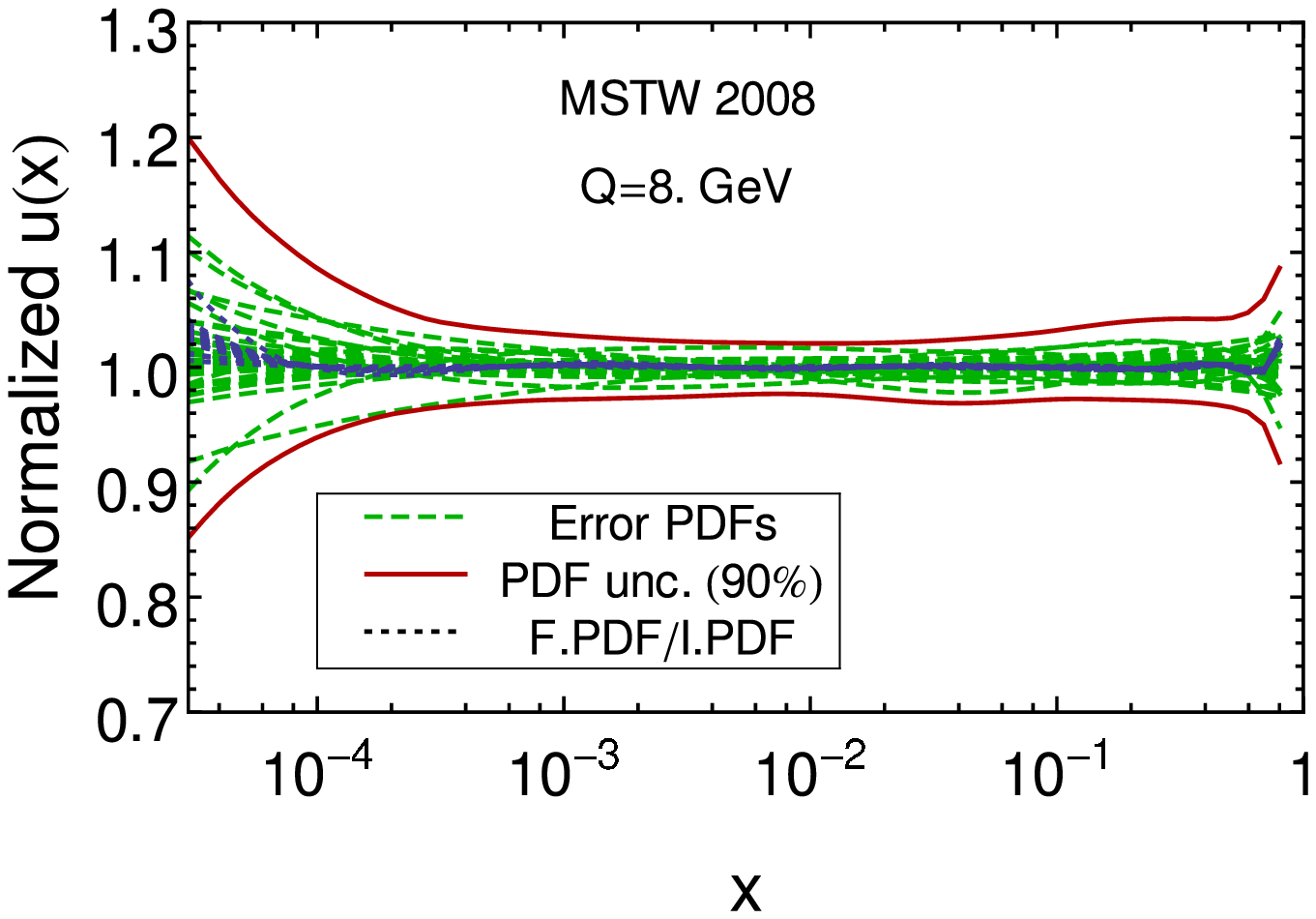} \\
\par\end{centering}
\vspace{-1ex}
 \caption{\label{fig:RelDiffMSTW} Same as Fig.~\ref{fig:RelDiffCT10}, for
the MSTW'08 NNLO ensemble.}
\end{figure}

\begin{figure}[p]
\begin{centering}
 \includegraphics[width=0.46\textwidth]{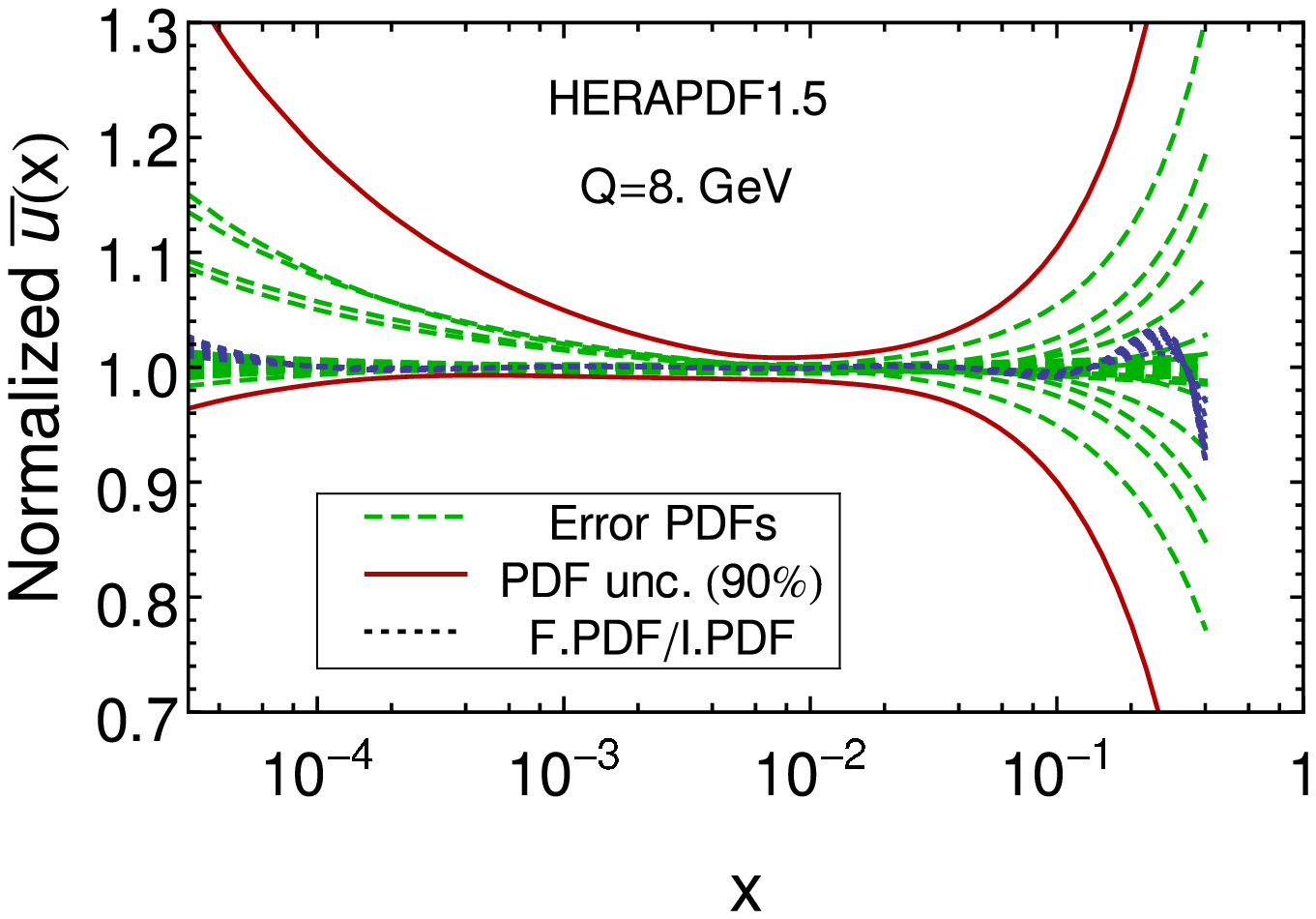} \hspace{10pt}
 \includegraphics[width=0.46\textwidth]{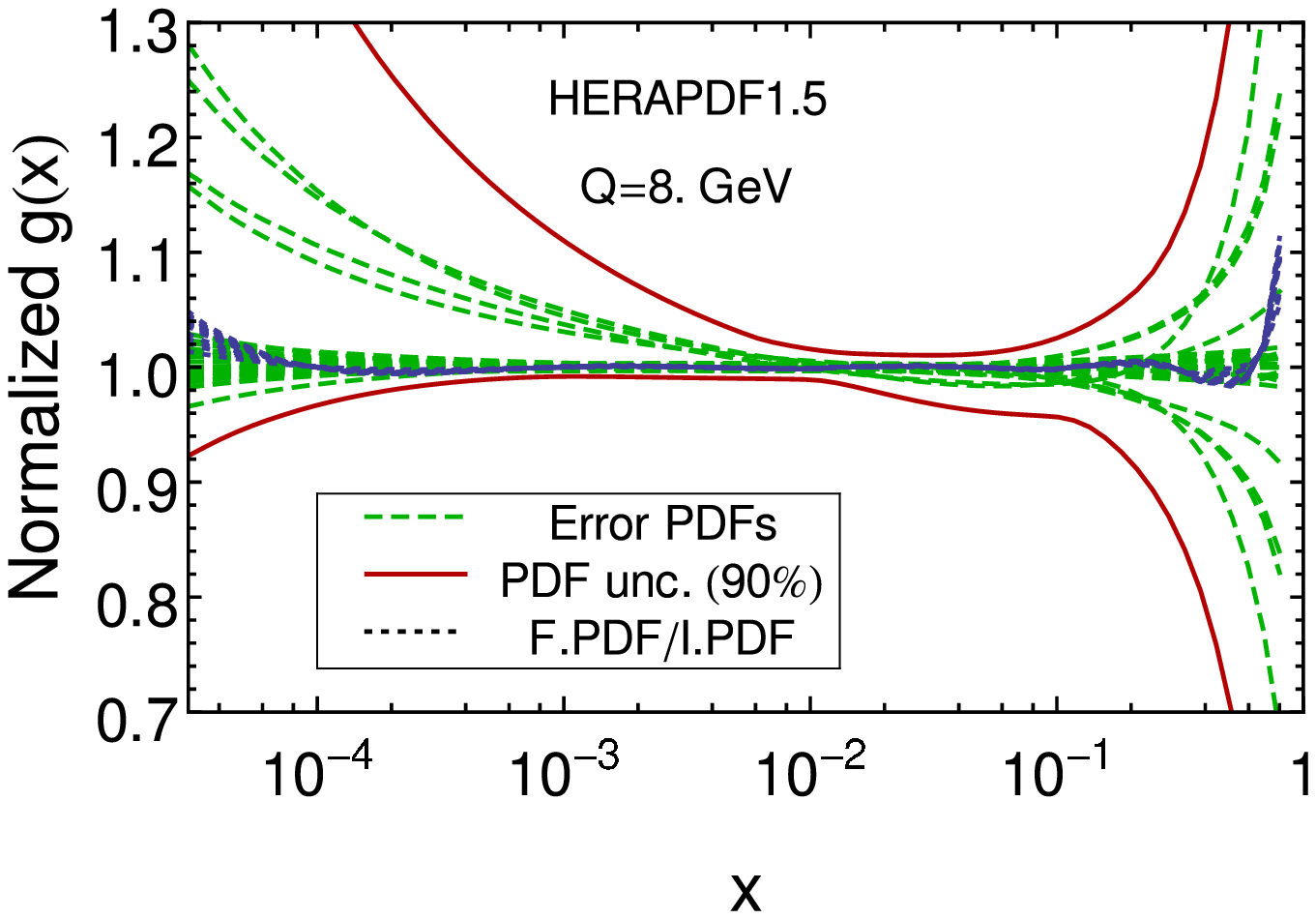} \\
 \includegraphics[width=0.46\textwidth]{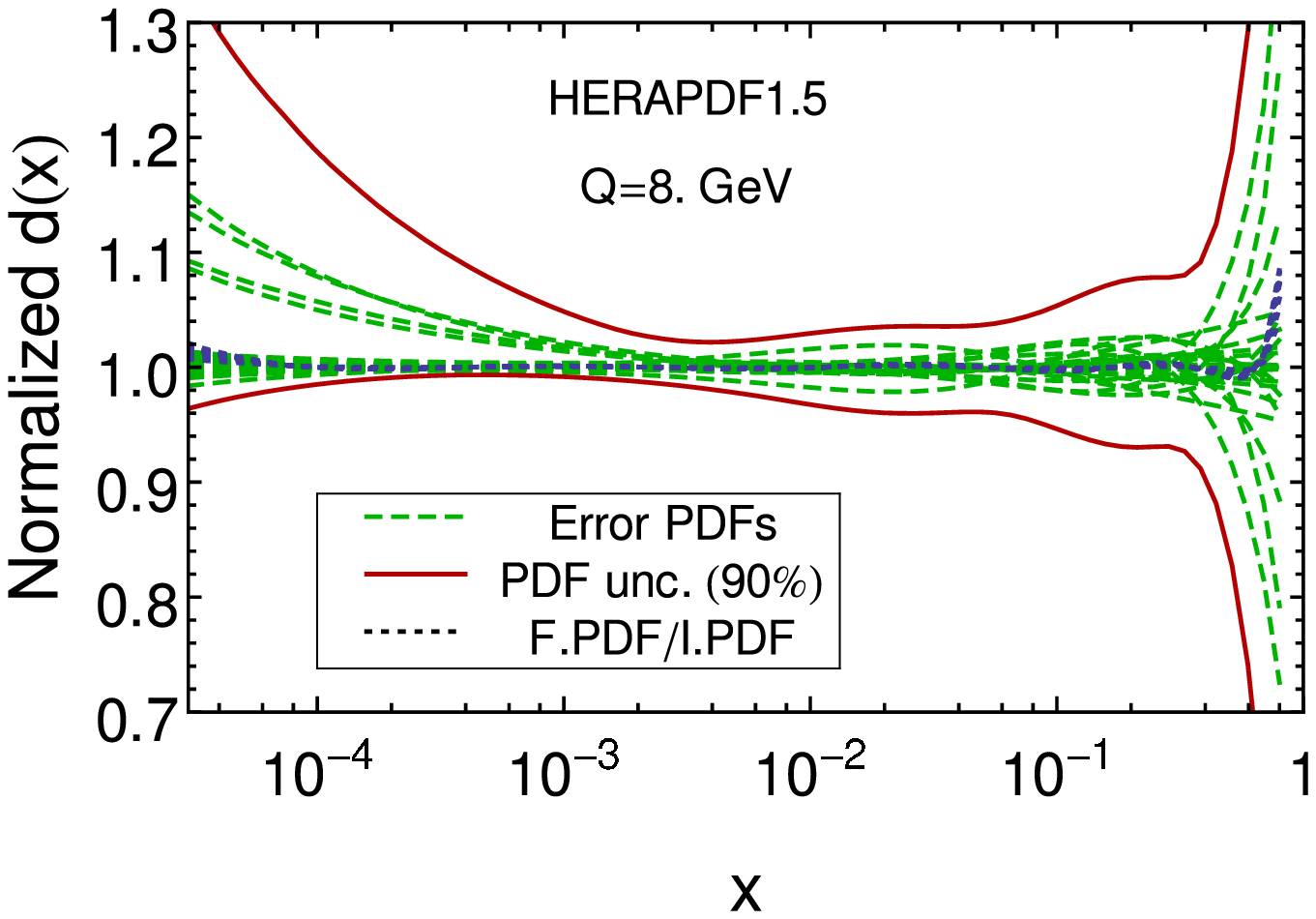} \hspace{10pt}
 \includegraphics[width=0.46\textwidth]{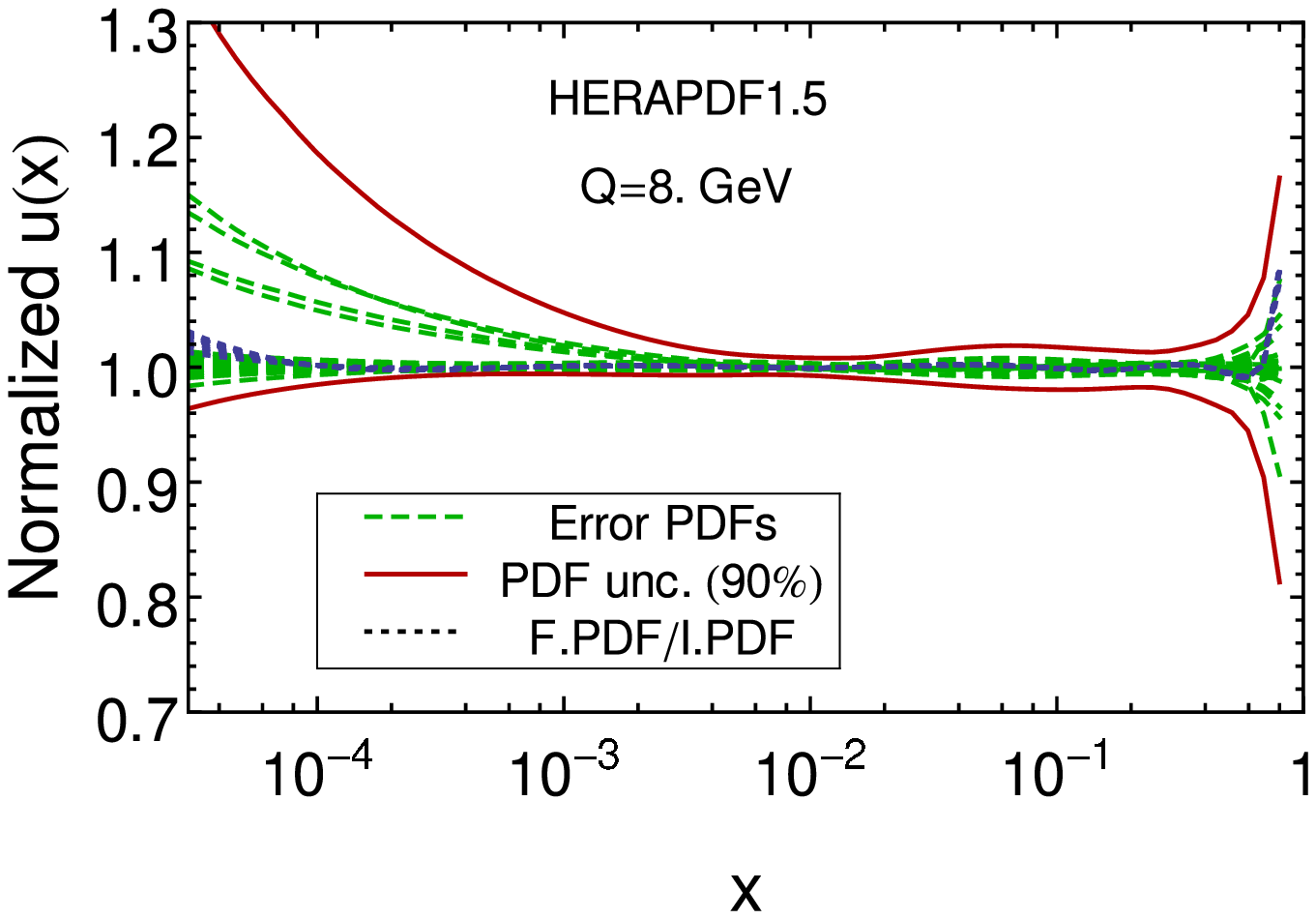} \\
\par\end{centering}

\vspace{-1ex}
 \caption{\label{fig:RelDiffHERA} Same as Fig.~\ref{fig:RelDiffCT10}, for
the HERAPDF1.5 NNLO ensemble.}
\end{figure}

\begin{figure}[p]
\begin{centering}
 \includegraphics[width=0.46\textwidth]{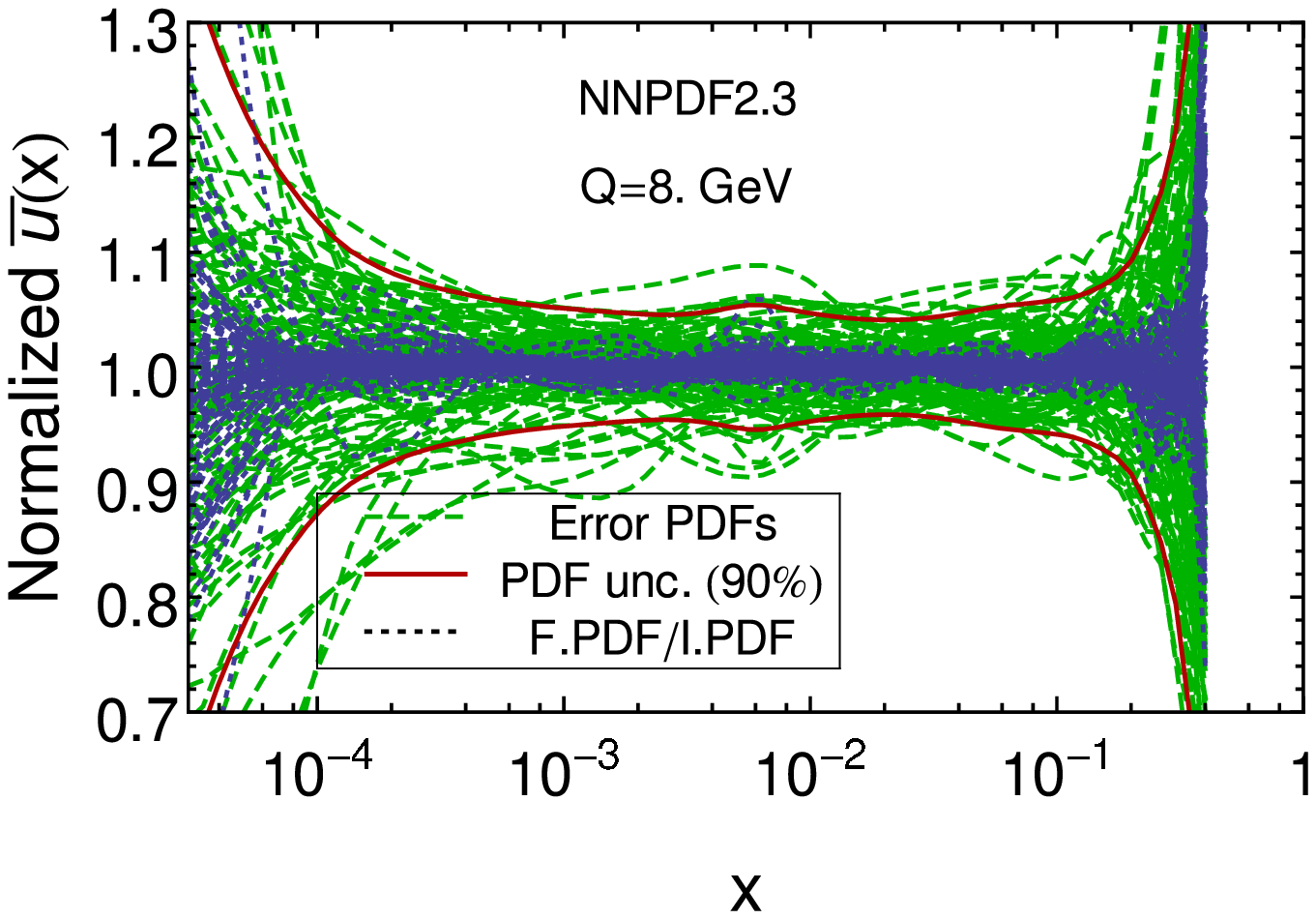} \hspace{10pt}
 \includegraphics[width=0.46\textwidth]{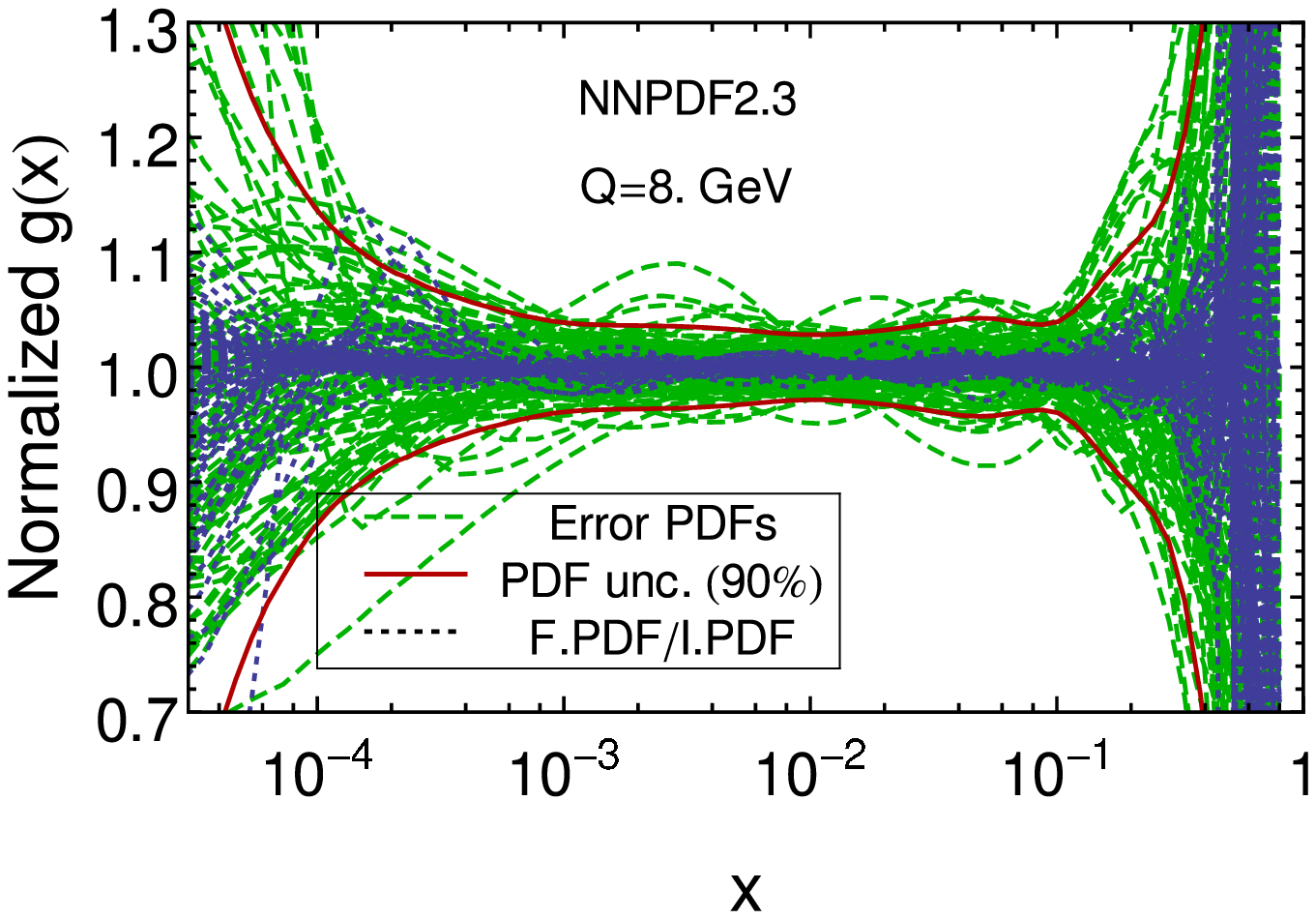} \\
 \includegraphics[width=0.46\textwidth]{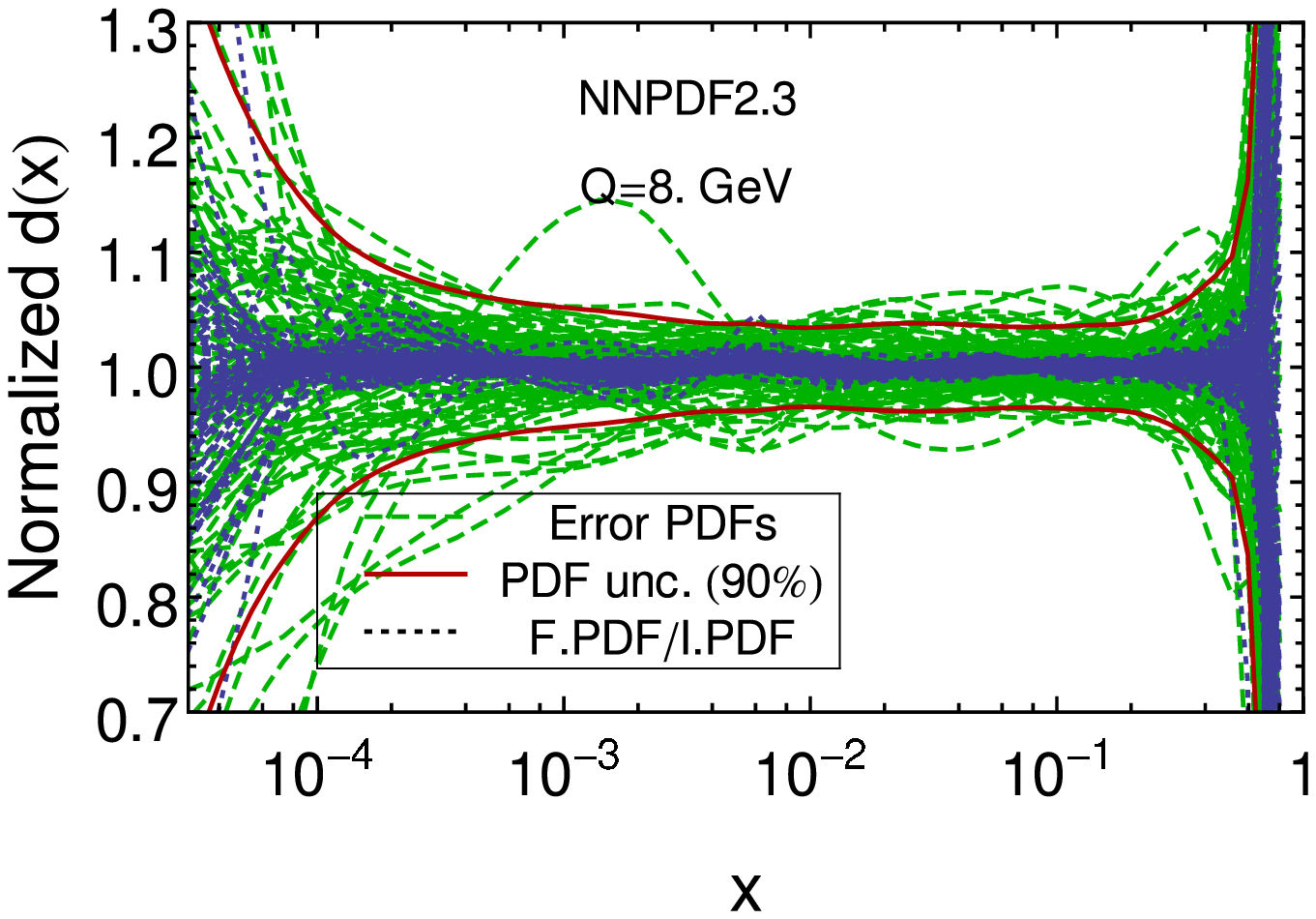} \hspace{10pt}
 \includegraphics[width=0.46\textwidth]{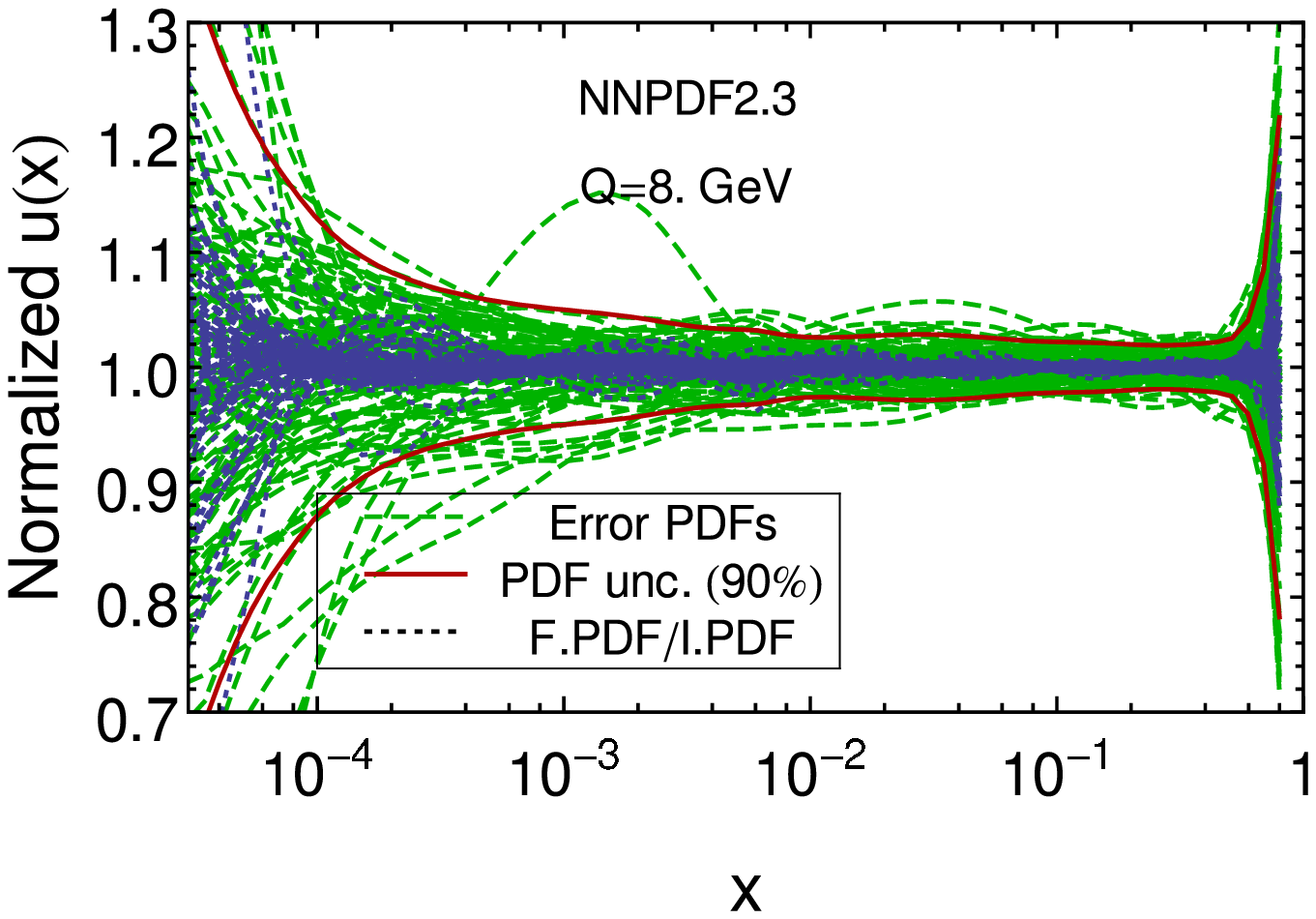} \\
\par\end{centering}

\vspace{-1ex}
 \caption{\label{fig:RelDiffNNPDF} Same as Fig.~\ref{fig:RelDiffCT10}, for
the NNPDF2.3 NNLO ensemble.}
\end{figure}

We can also examine the quality of fits to individual error sets
from each group, as is illustrated in
Figs.~\ref{fig:RelDiffCT10}-\ref{fig:RelDiffNNPDF} for some 
PDF flavors. For instance, in
Fig.~\ref{fig:RelDiffCT10} the green long-dashed lines indicate the
ratios of 50 PDF eigenvector sets in CT10 to the central set. The
blue short-dashed lines indicate the ratio of the
meta-parametrization to the input parametrization for each
eigenvector set. The differences between the input and
meta-parametrizations are below 2\% in most cases, much smaller than
the spread of the error PDFs. Similar level of agreement is reached
for the MSTW and HERA PDFs in Figs.~\ref{fig:RelDiffMSTW} and
\ref{fig:RelDiffHERA}, where slightly larger deviations (up to about
5\%) are observed at $x<10^{-4}$.
Finally, for NNPDF in Fig.~\ref{fig:RelDiffNNPDF}, the meta-parametrizations
tend to have fewer oscillations than the input error PDF sets because
of the functional form we chose. The differences between the
individual error PDFs of the input and fitted NNPDF ensembles
are larger in this case, but cancel well
in the full PDF uncertainty, which comes to be about the same in the
input and fitted ensembles as discussed above.

\subsection{Comparison of meta-PDF parameters from various
ensembles}

\begin{figure}[htb]
\begin{centering}
\par\end{centering}
\includegraphics[width=0.32\textwidth]{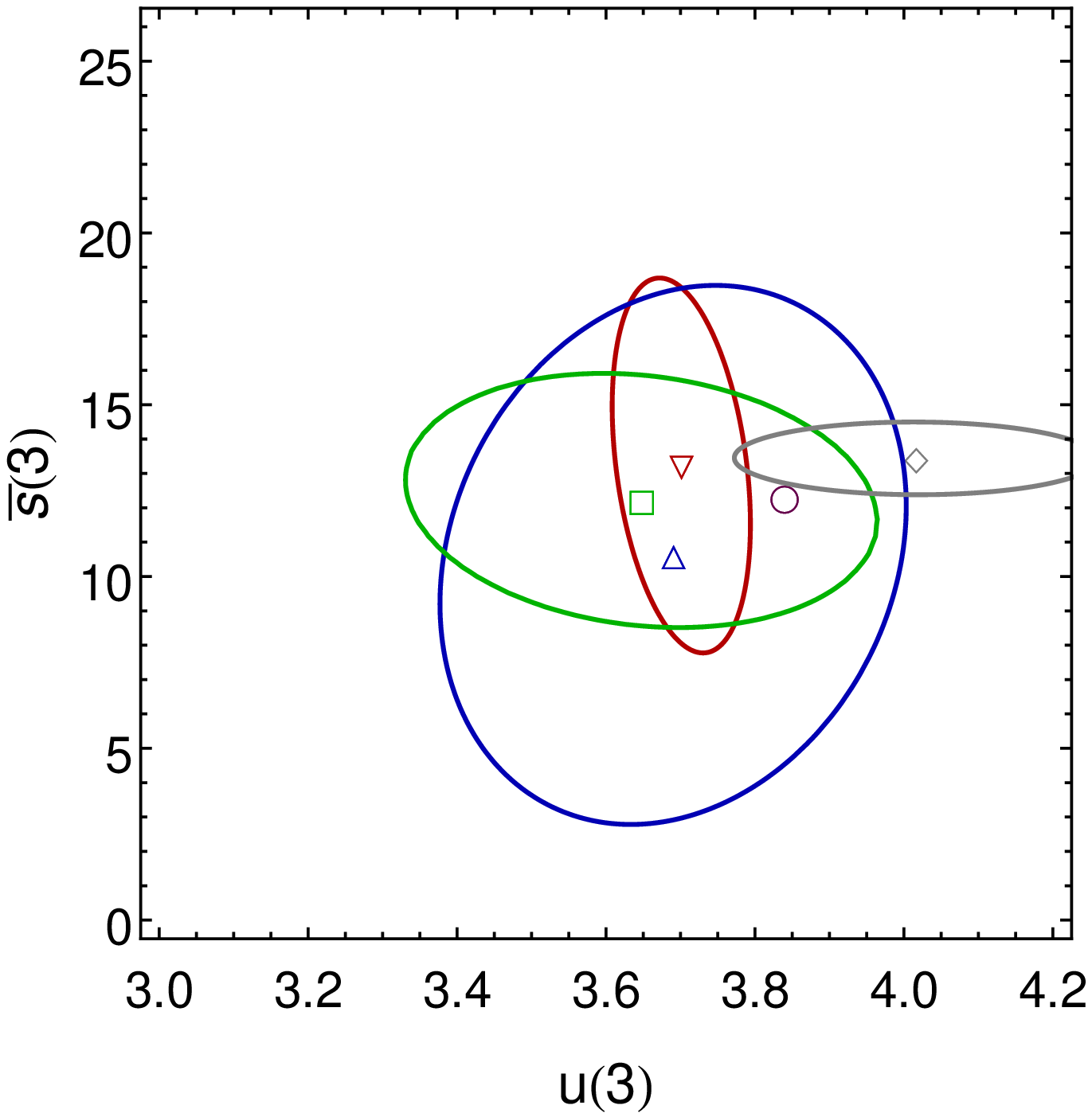} 
\includegraphics[width=0.32\textwidth]{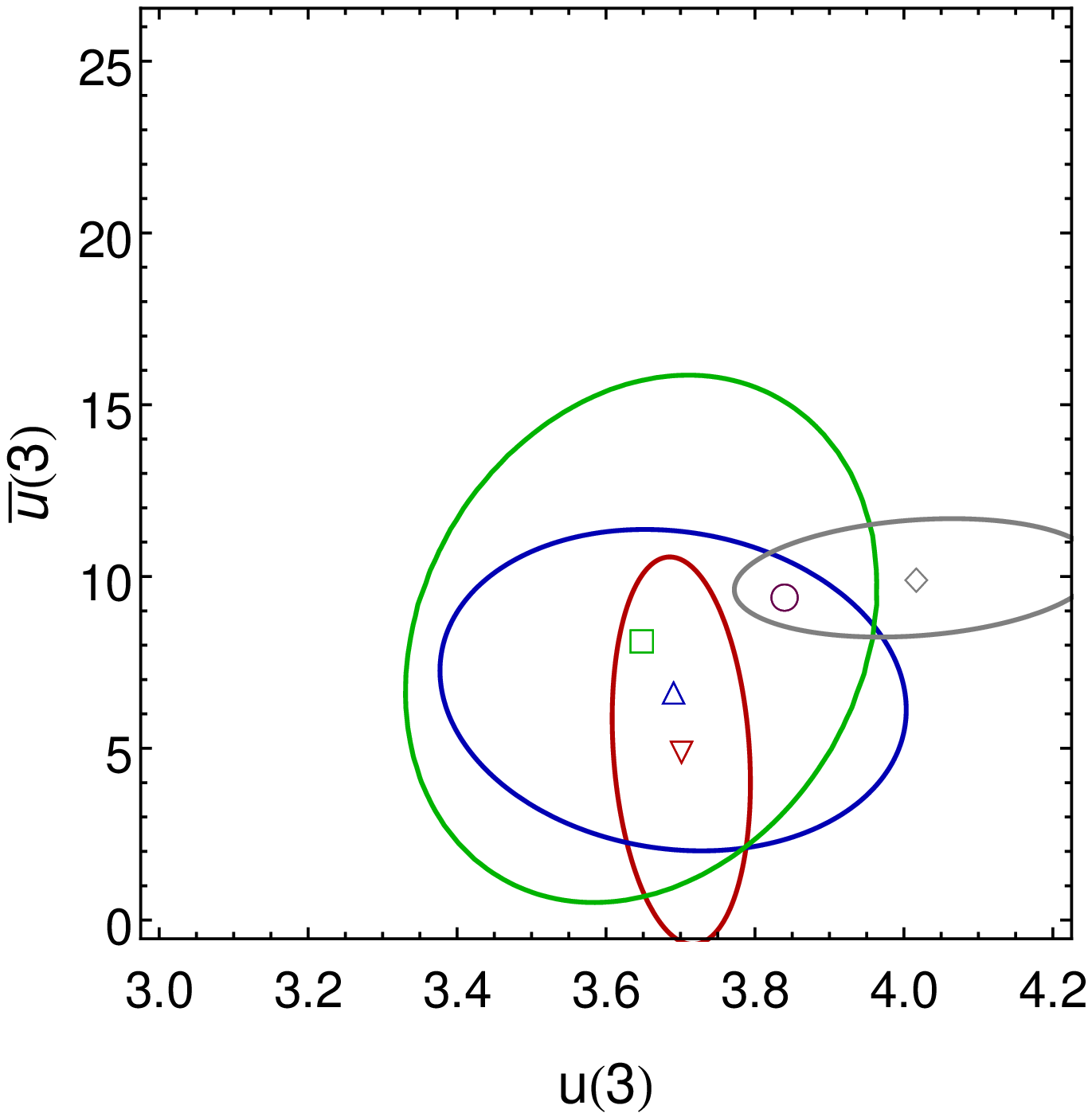}
\includegraphics[width=0.32\textwidth]{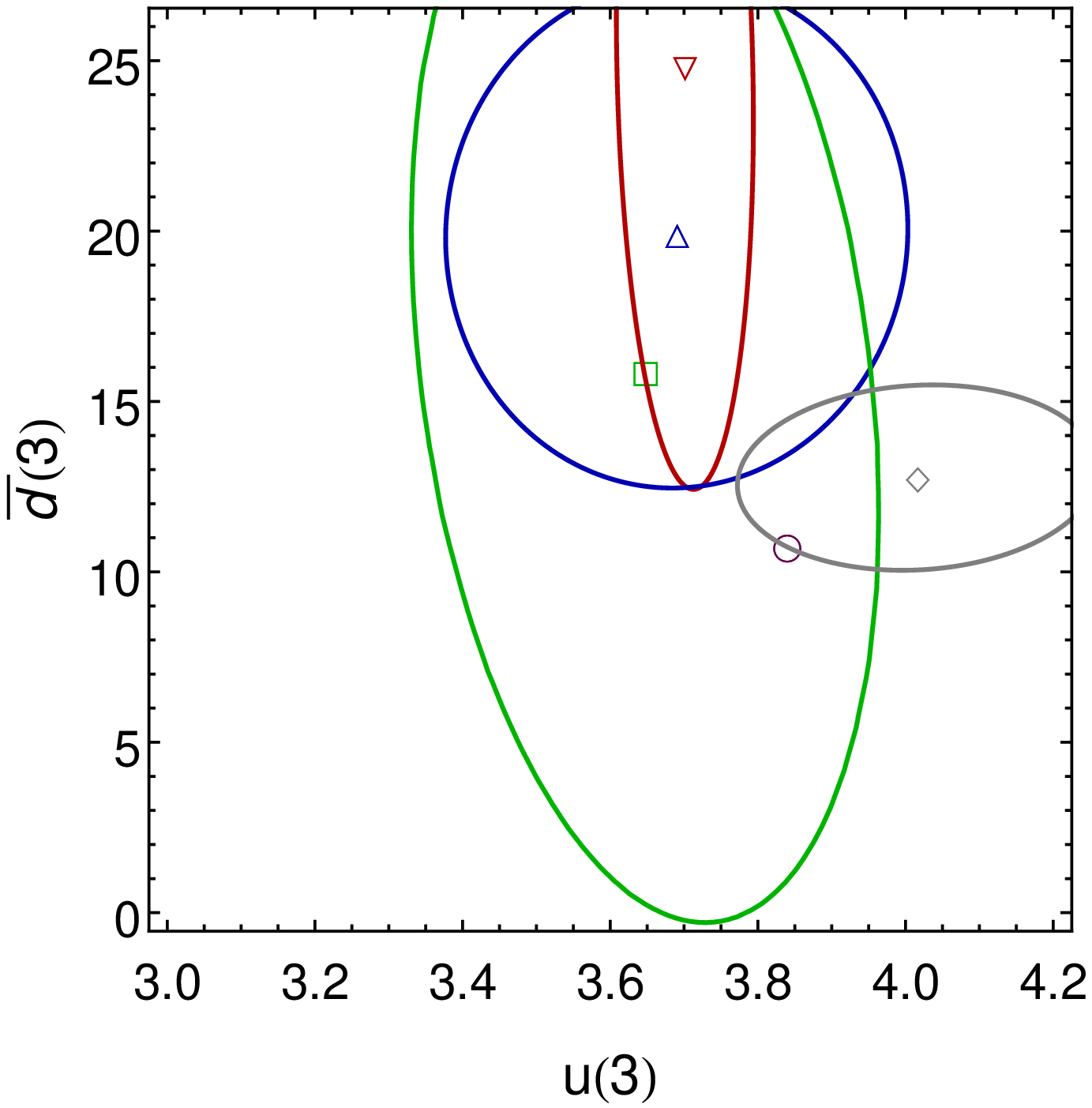}\\
 \includegraphics[width=0.32\textwidth]{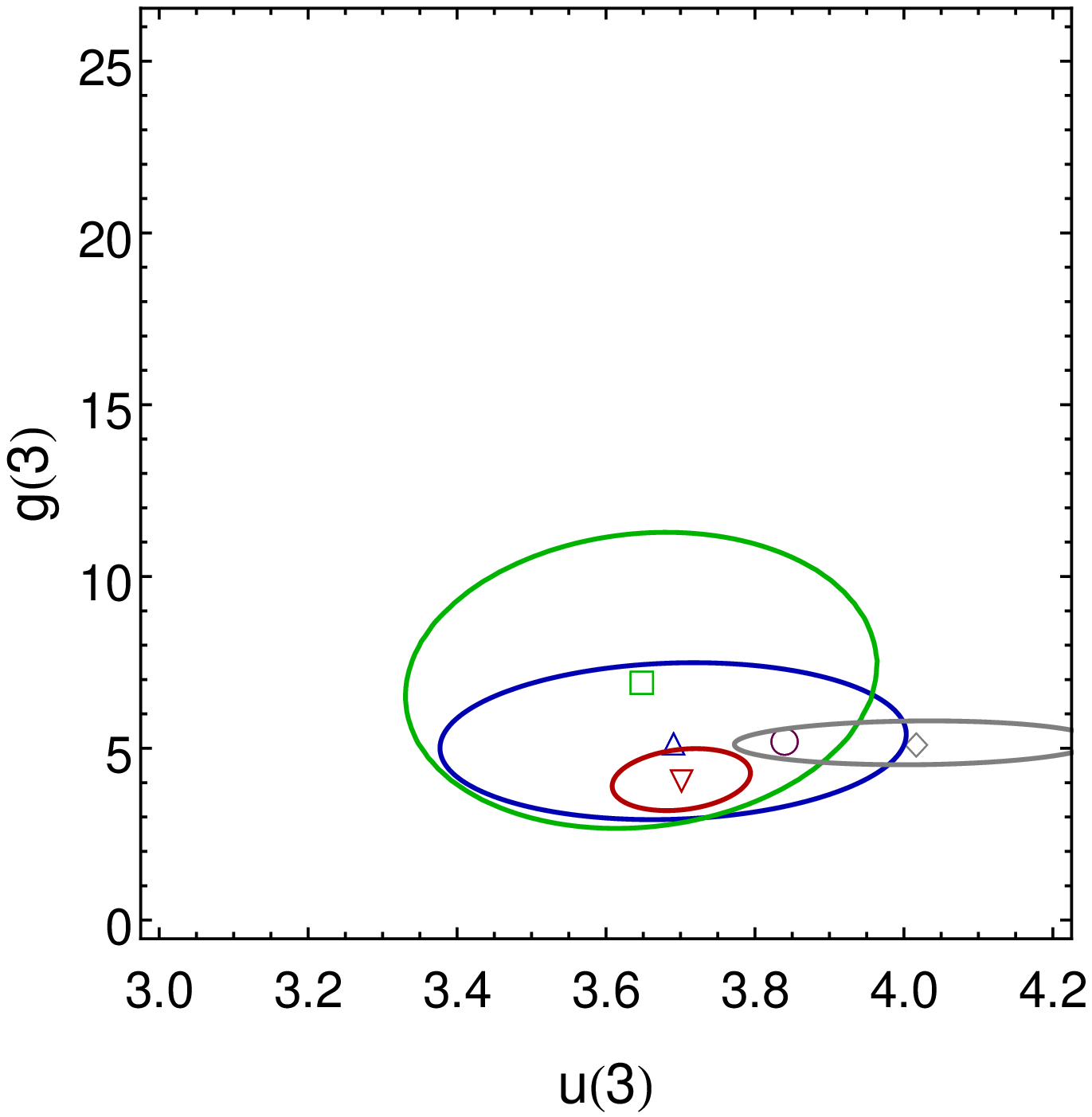} 
 \includegraphics[width=0.32\textwidth]{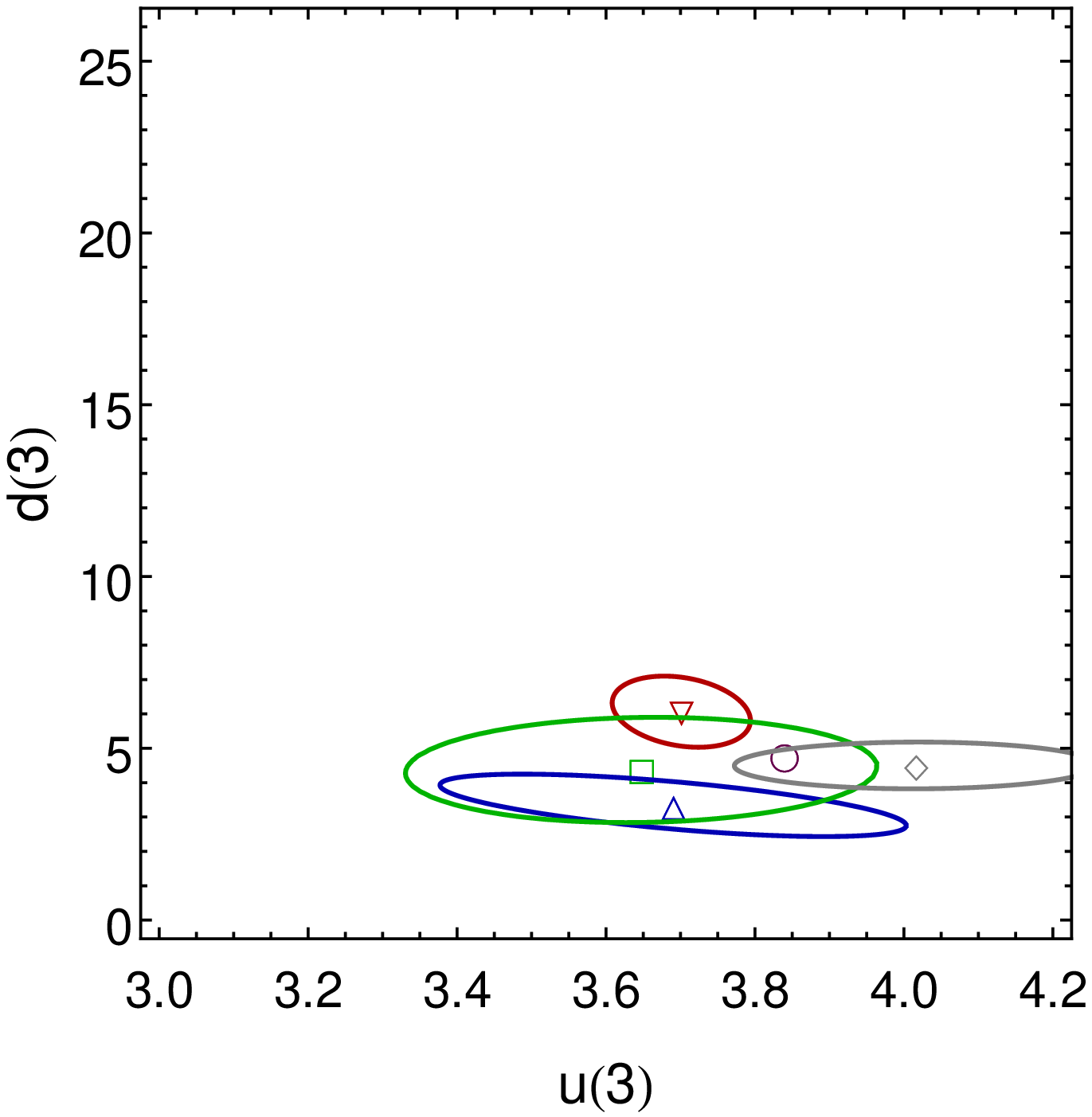}
 \includegraphics[width=0.32\textwidth]{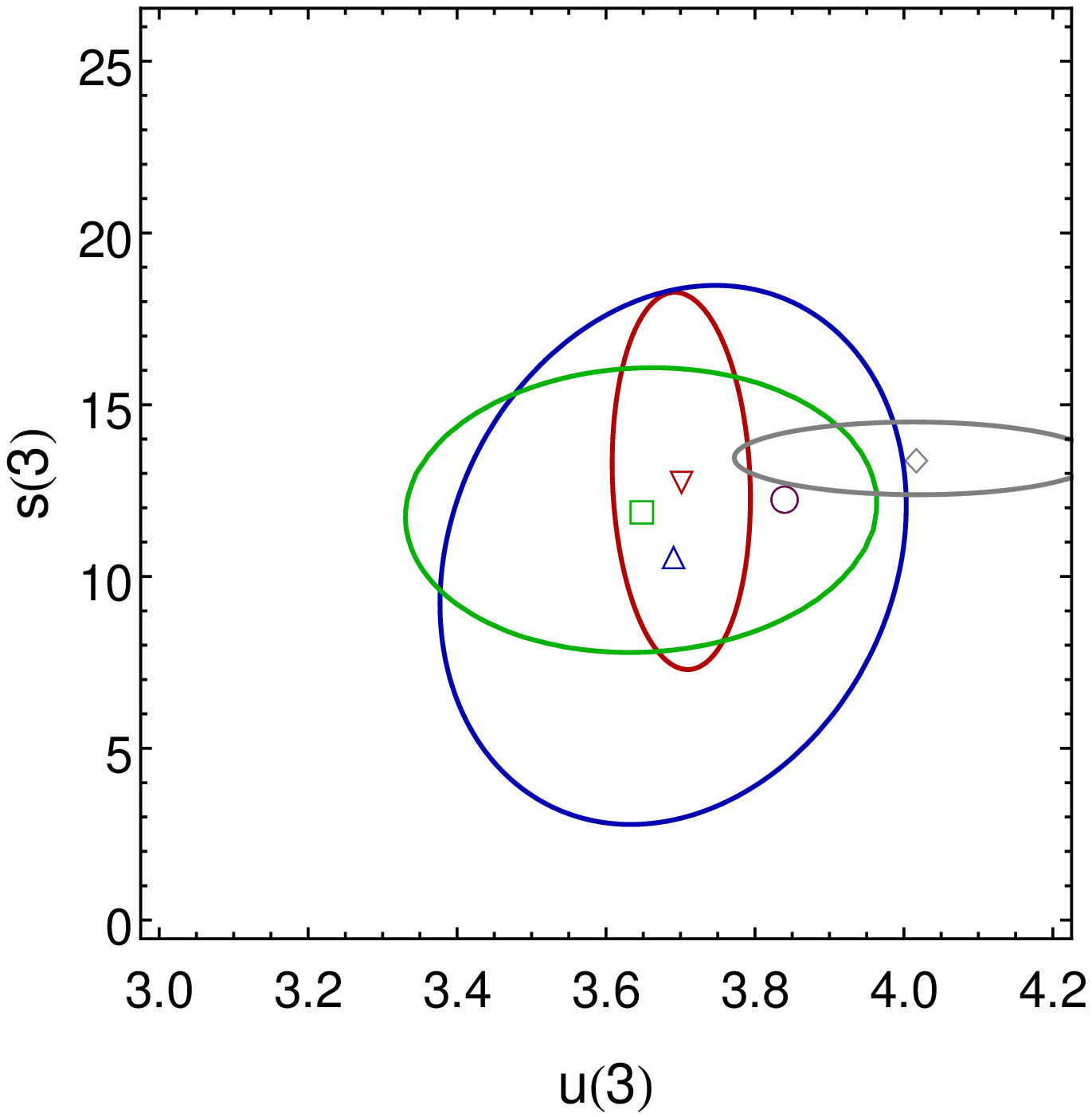}\\ 
 \includegraphics[width=0.32\textwidth]{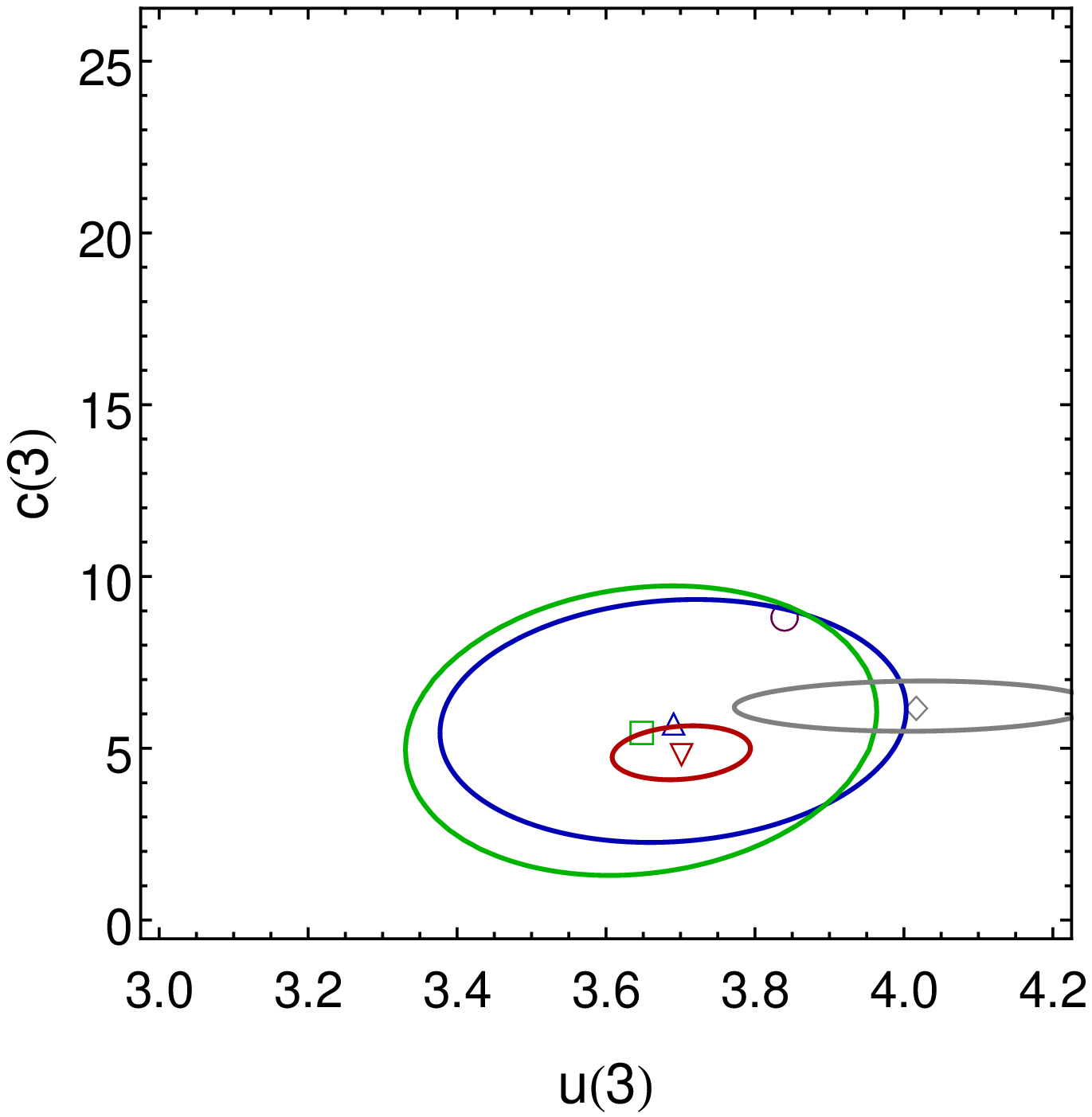}
\includegraphics[width=0.32\textwidth]{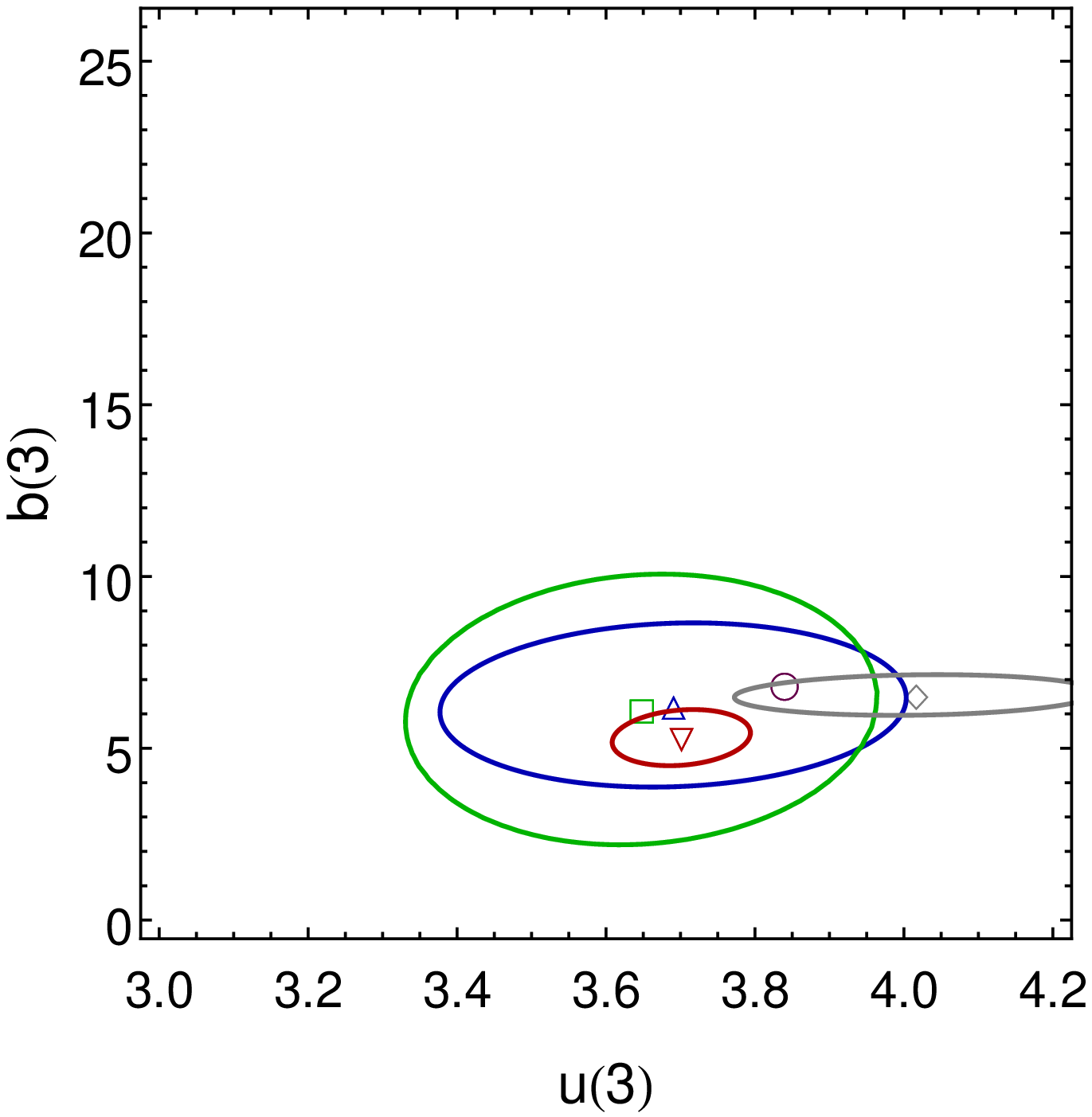}
\vspace{-2ex}
 \caption{\label{fig:metapar1} Fitted PDF parameters and 90\%
   c.l. ellipses for CT10 (blue up triangle), MSTW08
(red down triangle), NNPDF2.3 (green square), HERAPDF1.5 (gray diamond) and ABM11 (magenta circle).}
\end{figure}

\begin{figure}[htb]
\begin{centering}
\includegraphics[width=0.24\textwidth]{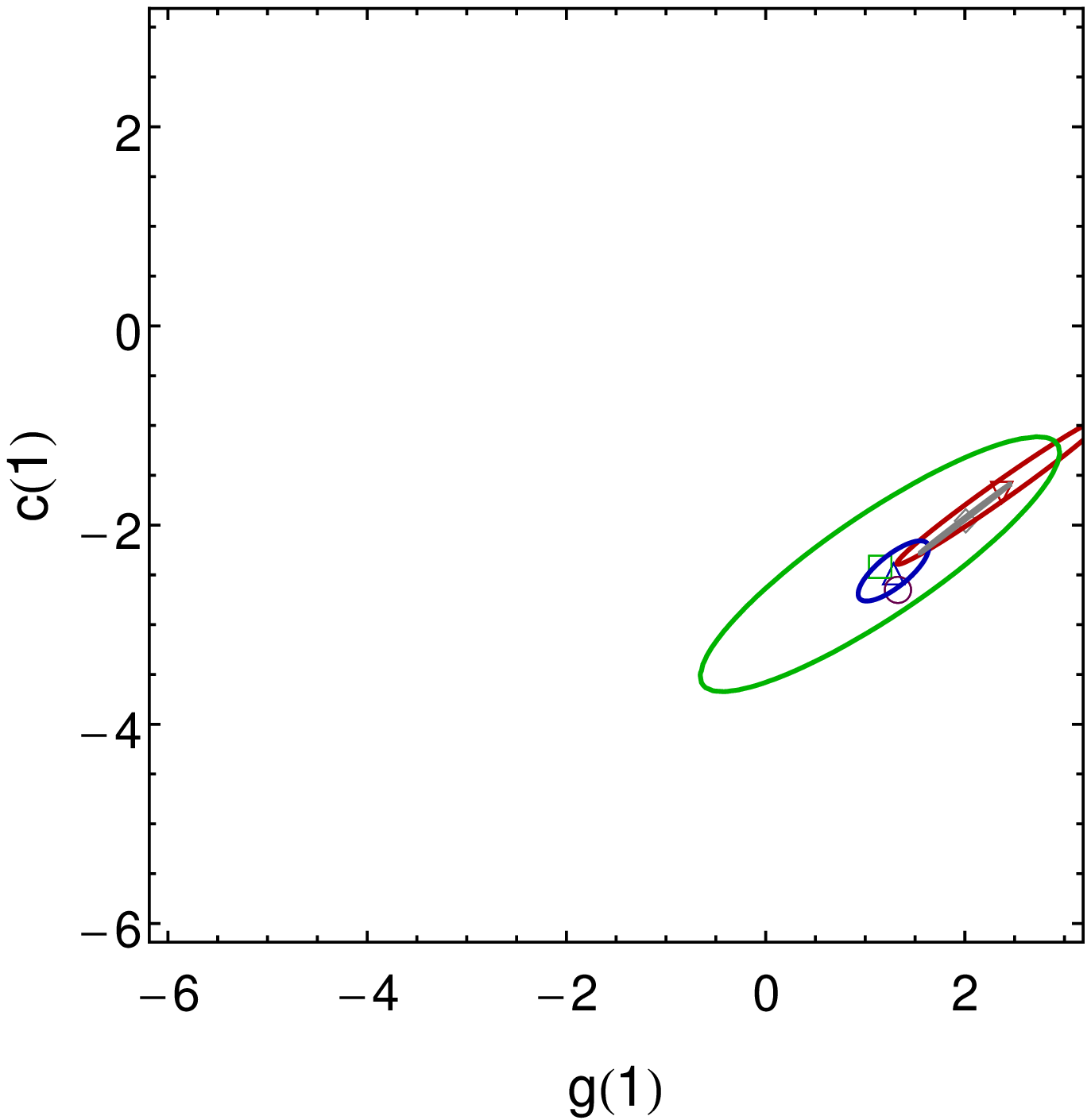} \includegraphics[width=0.24\textwidth]{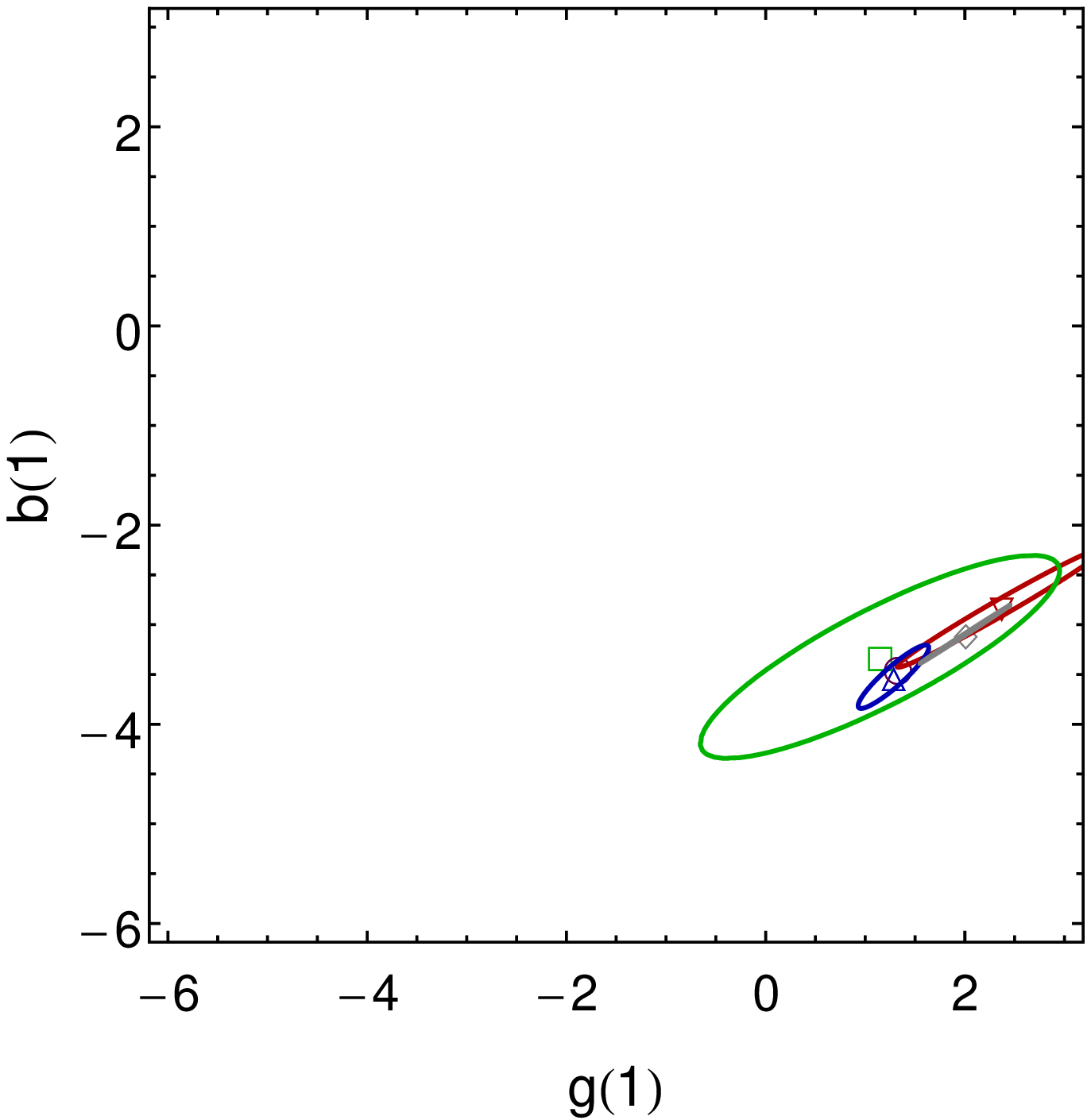}
\includegraphics[width=0.24\textwidth]{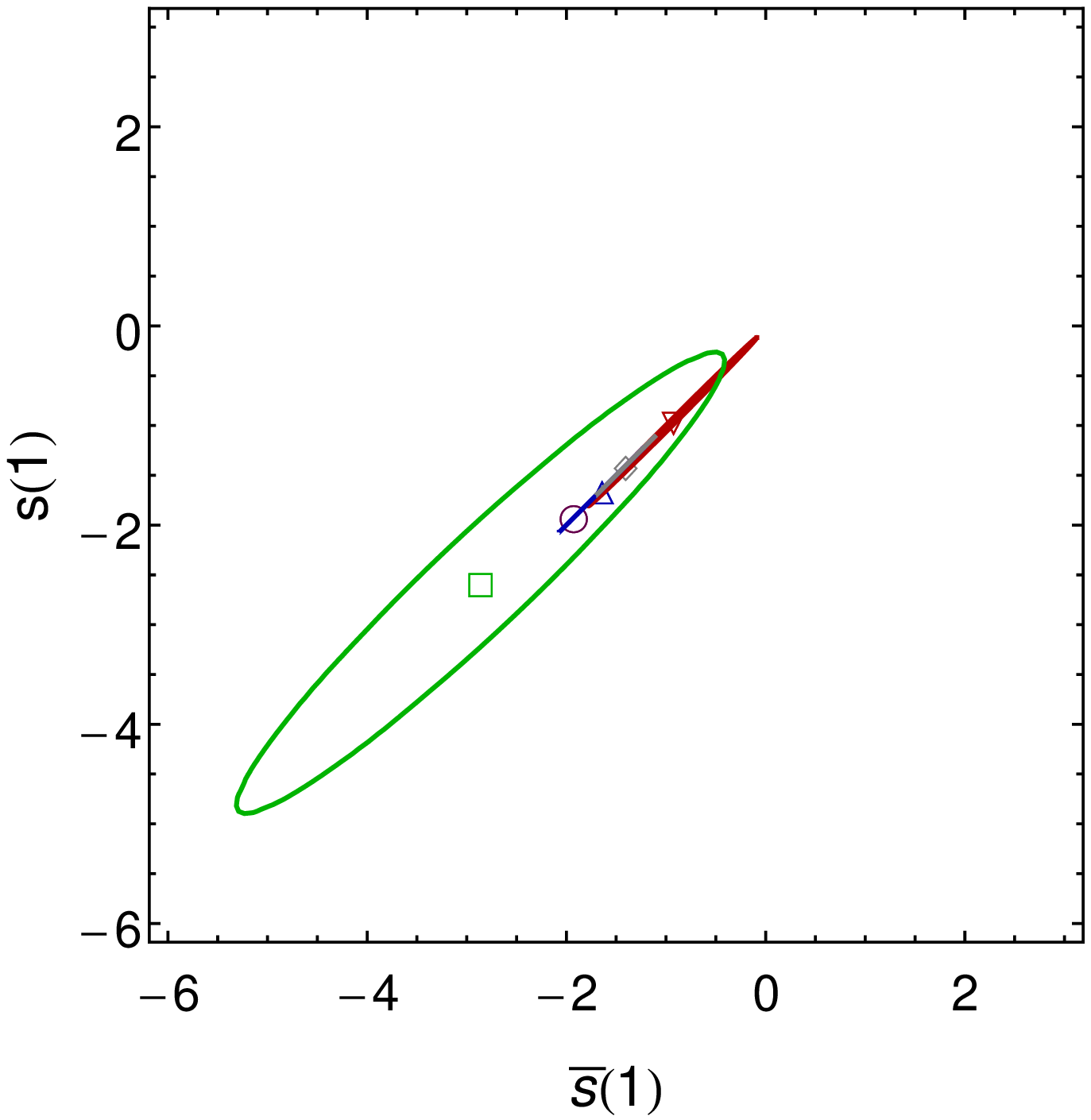} \includegraphics[width=0.24\textwidth]{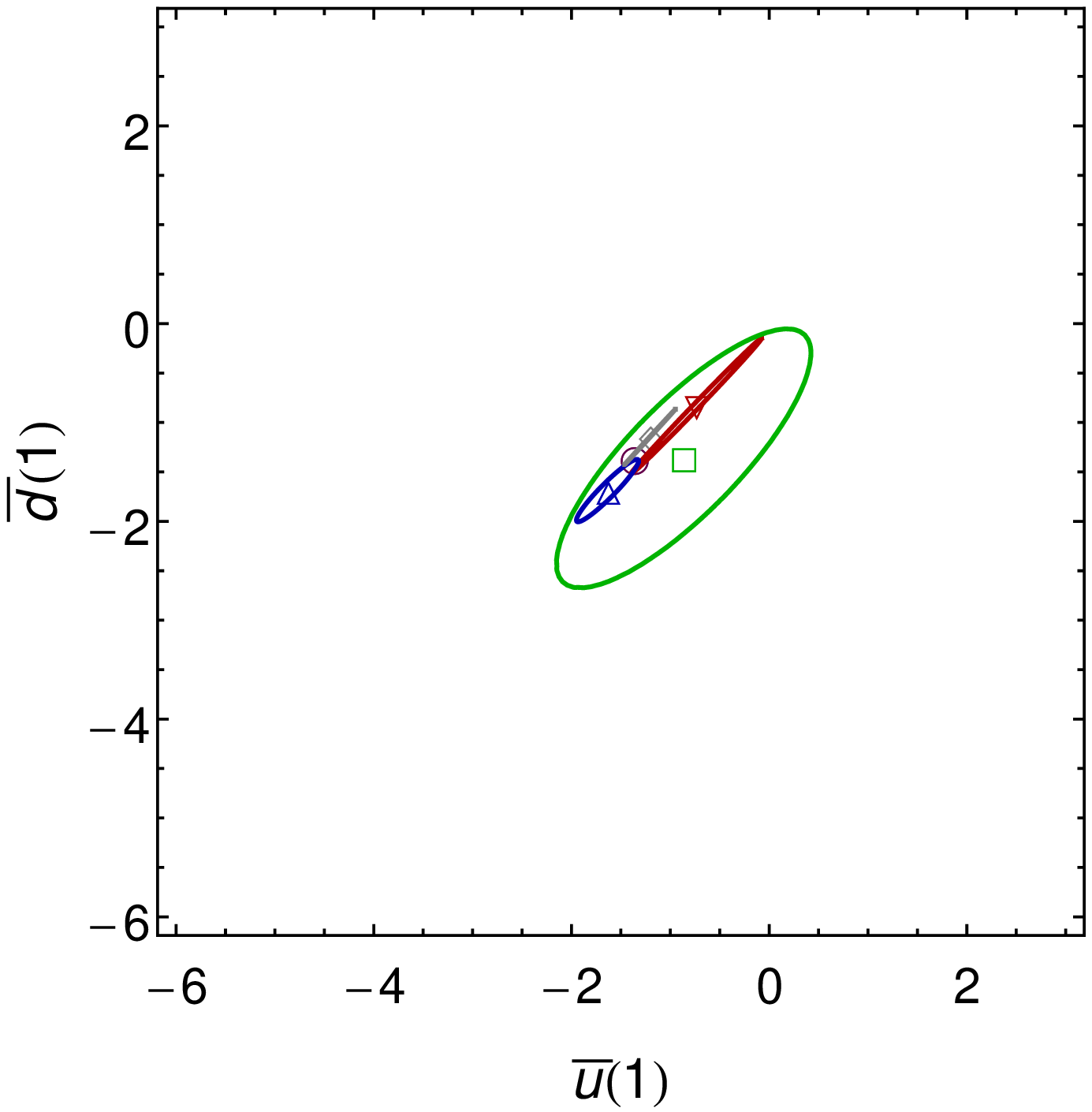}\\
 \includegraphics[width=0.24\textwidth]{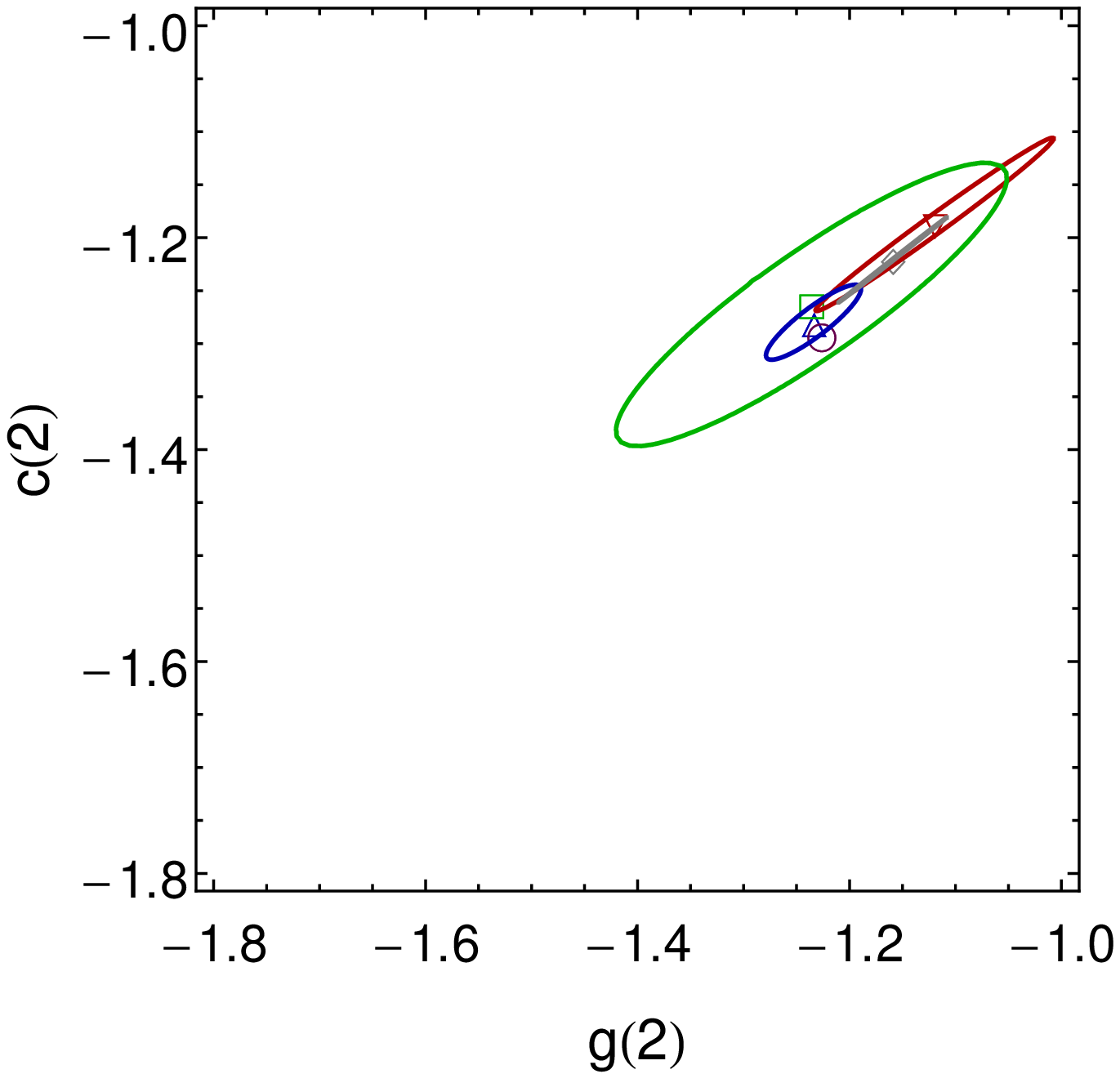} \includegraphics[width=0.24\textwidth]{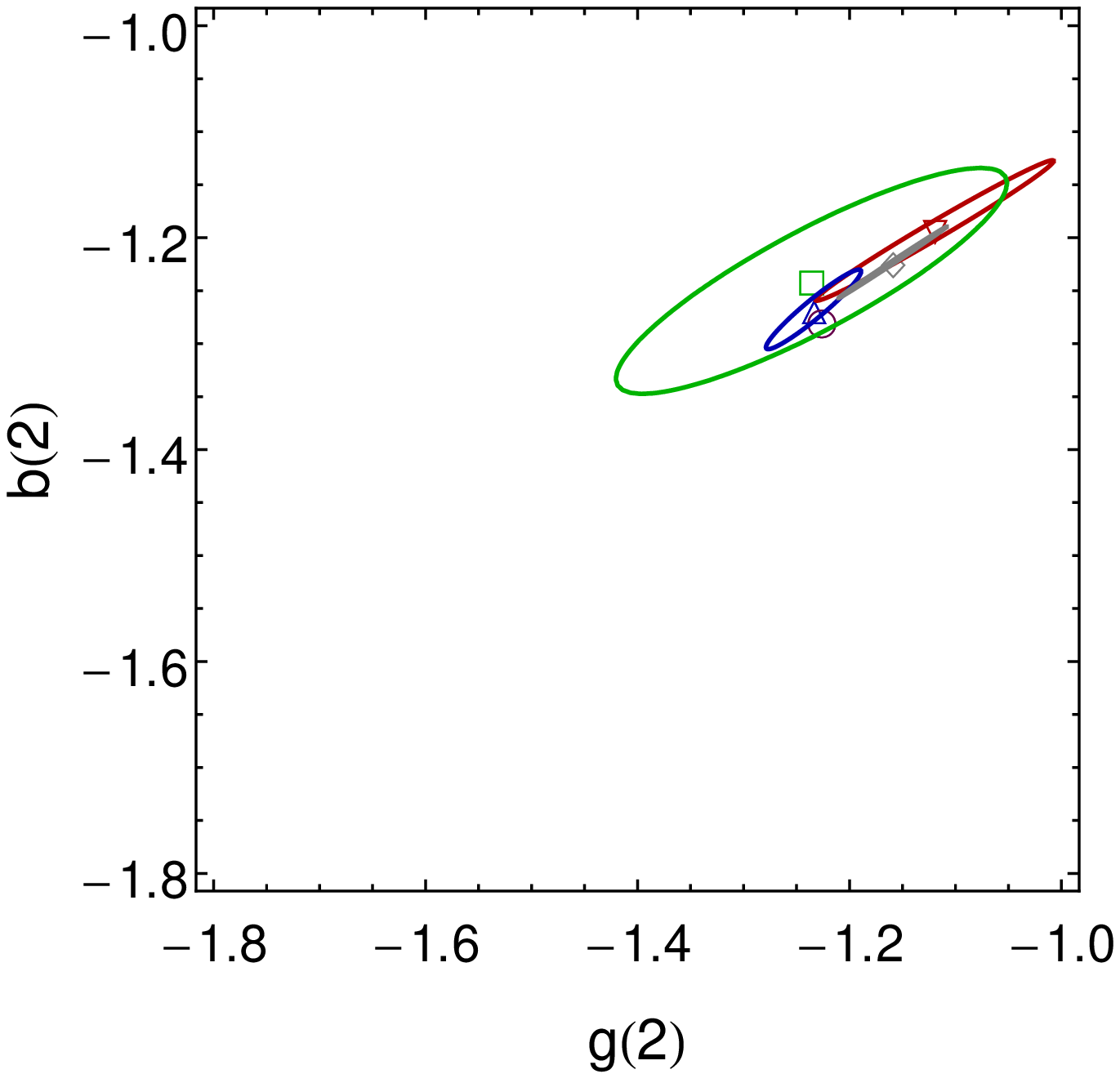}
\includegraphics[width=0.24\textwidth]{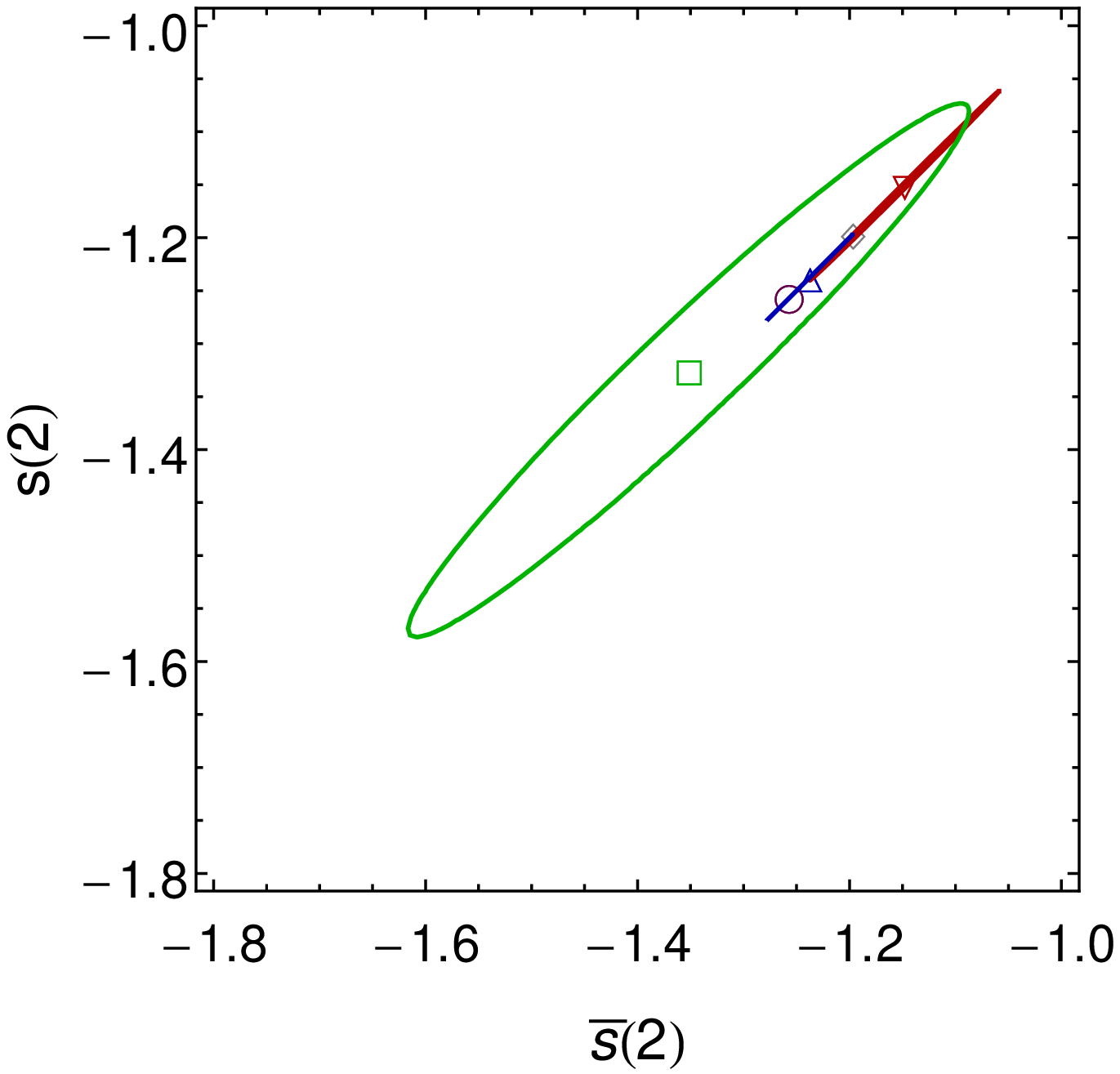} \includegraphics[width=0.24\textwidth]{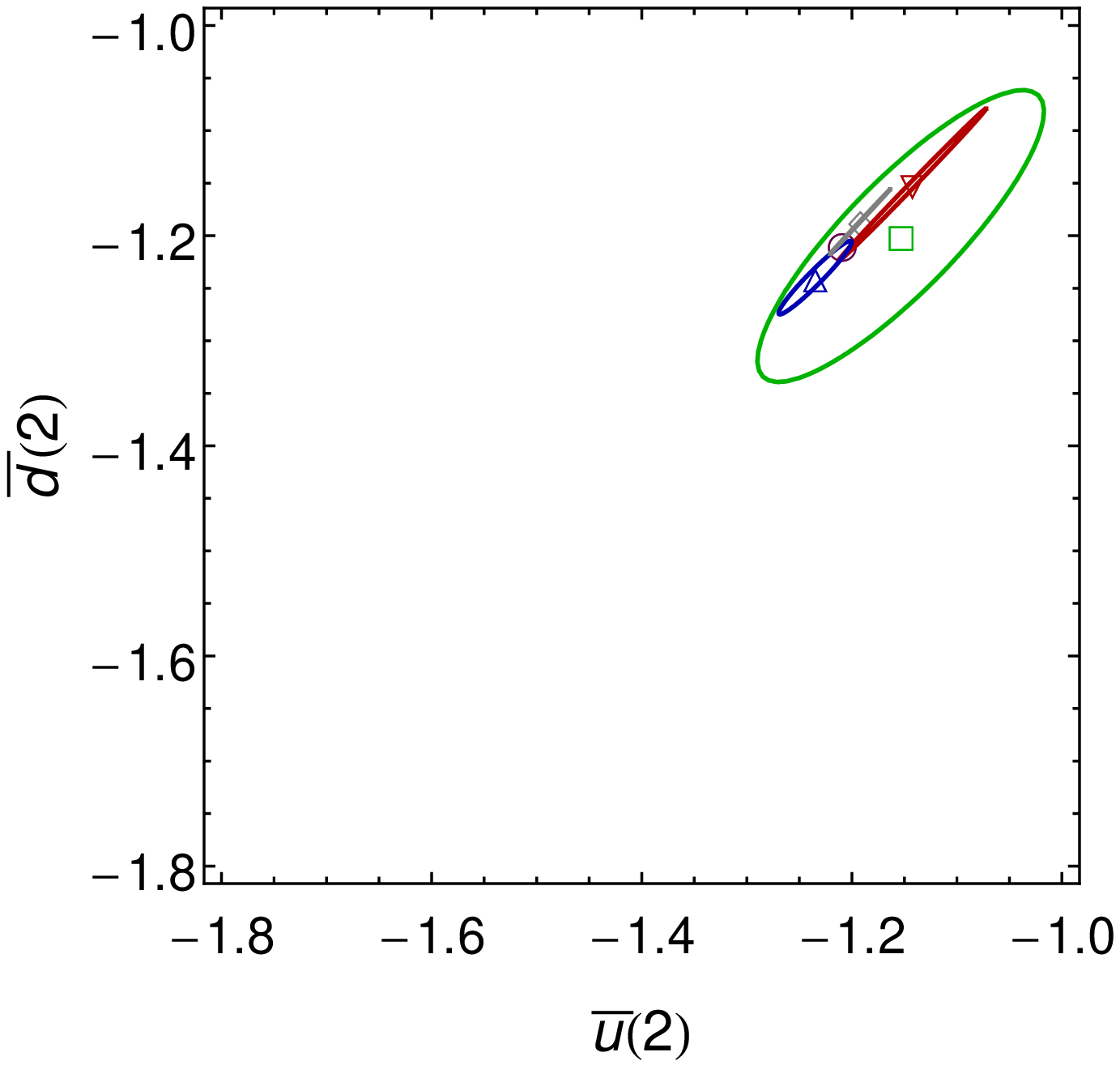}
\includegraphics[width=0.24\textwidth]{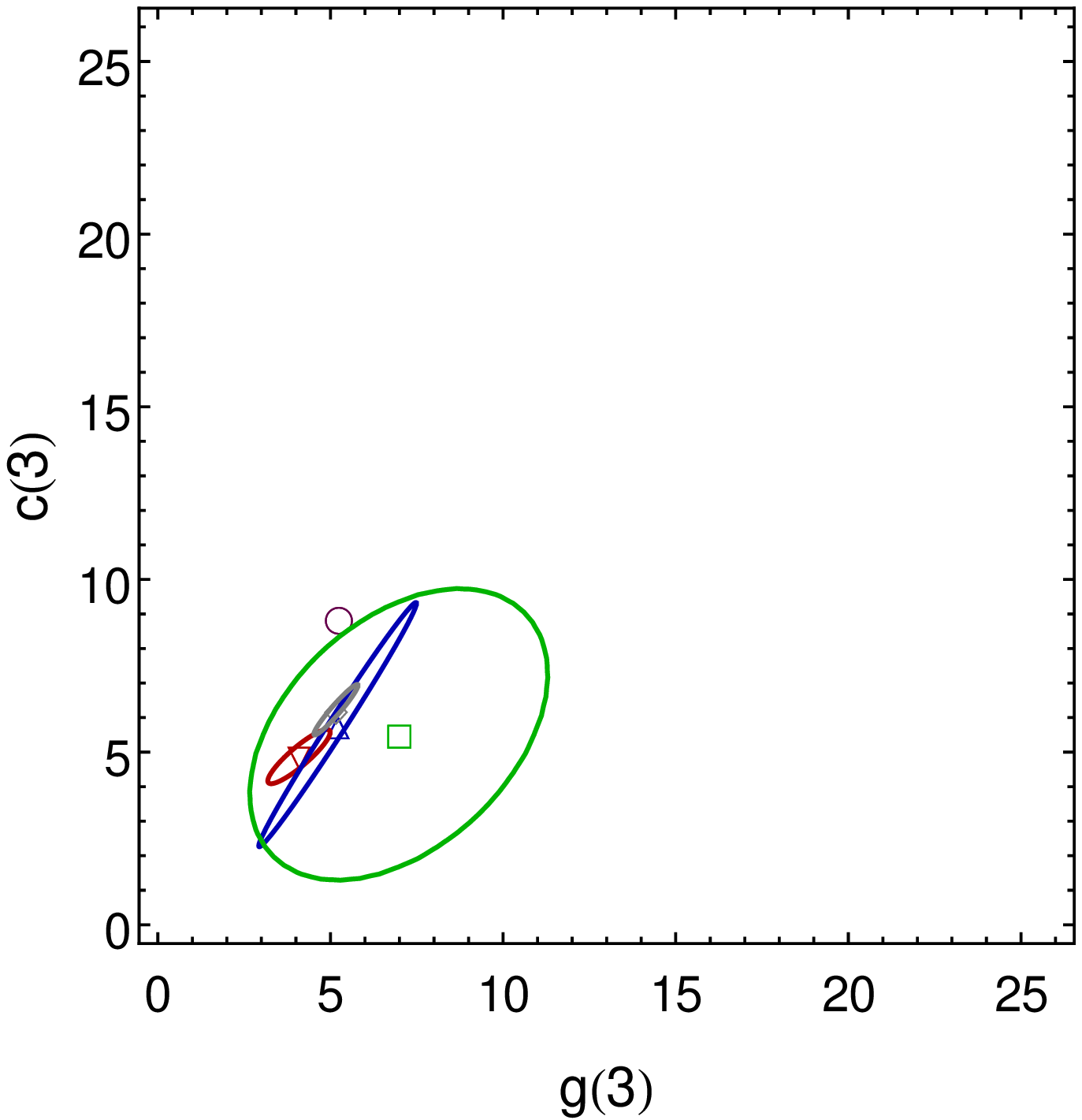} \includegraphics[width=0.24\textwidth]{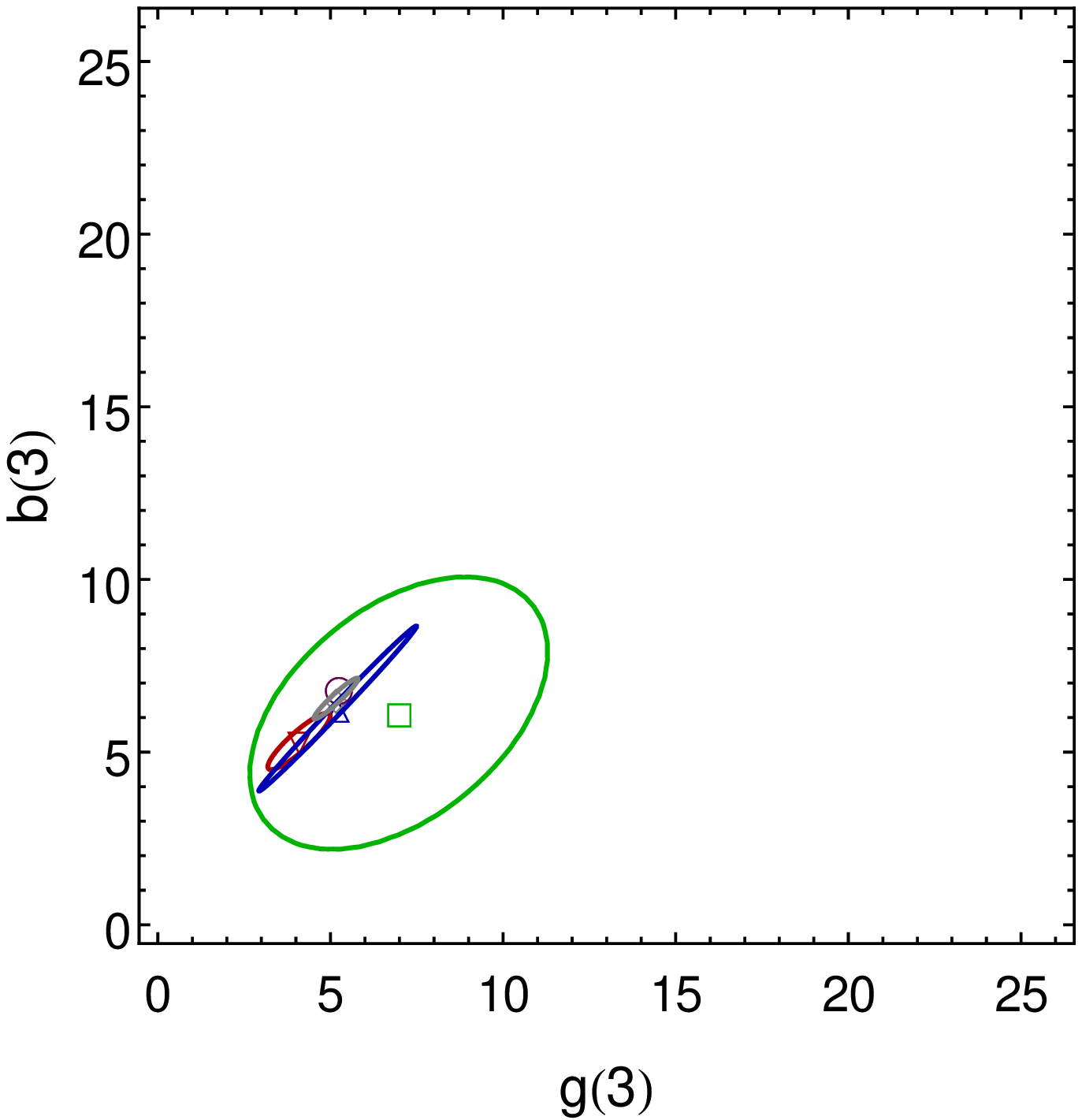}
\includegraphics[width=0.24\textwidth]{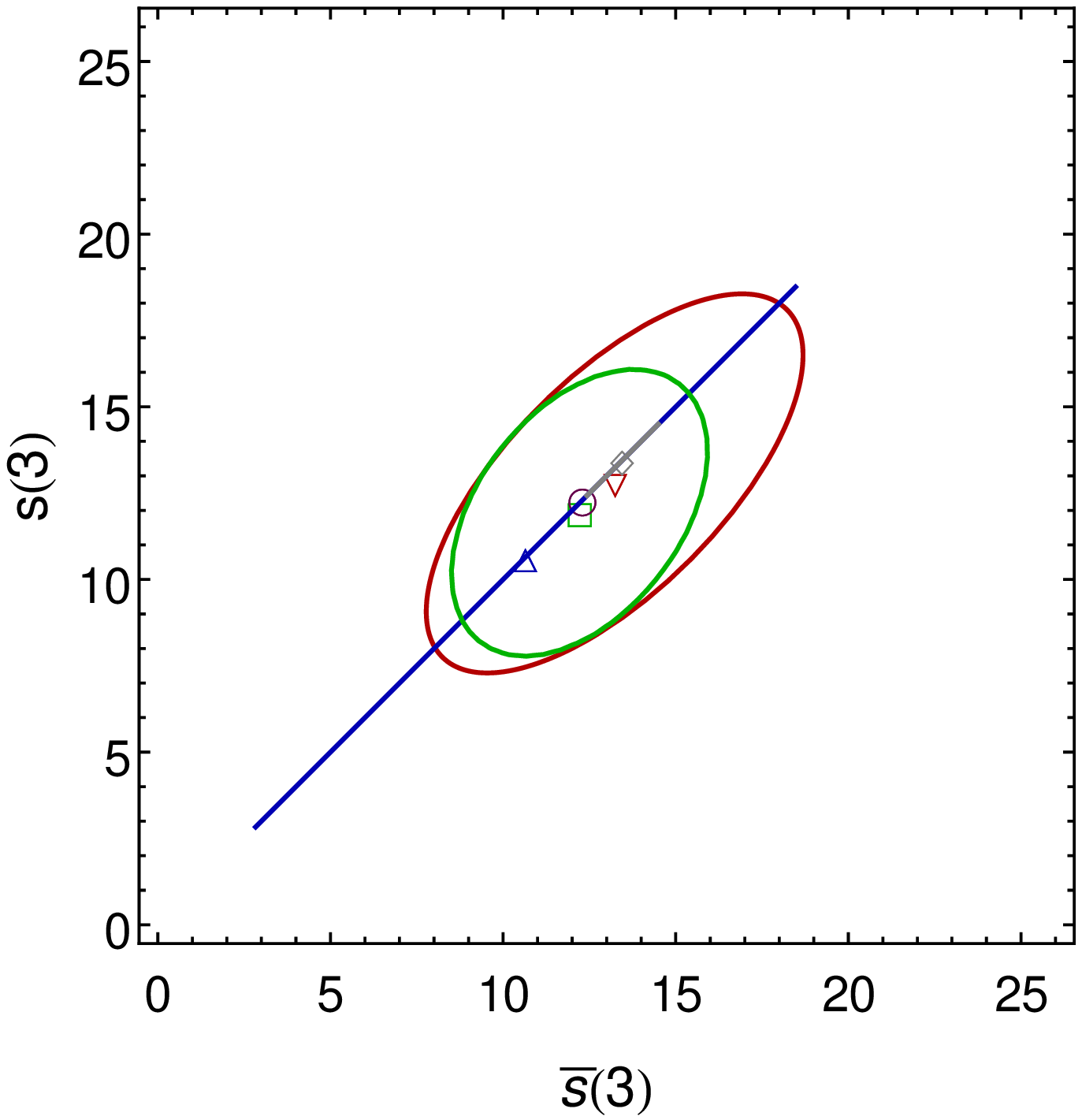} \includegraphics[width=0.24\textwidth]{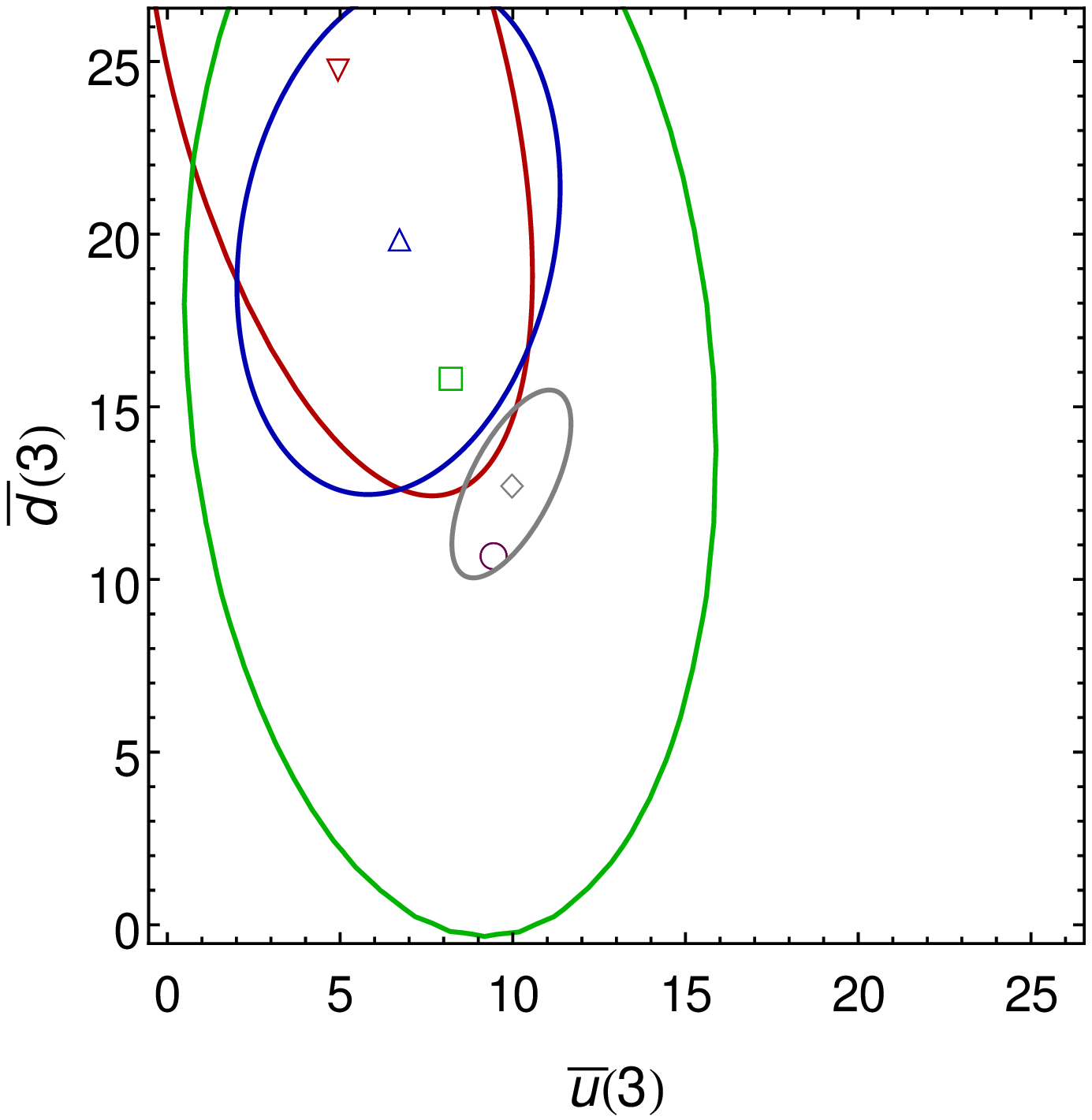}\\
 \includegraphics[width=0.24\textwidth]{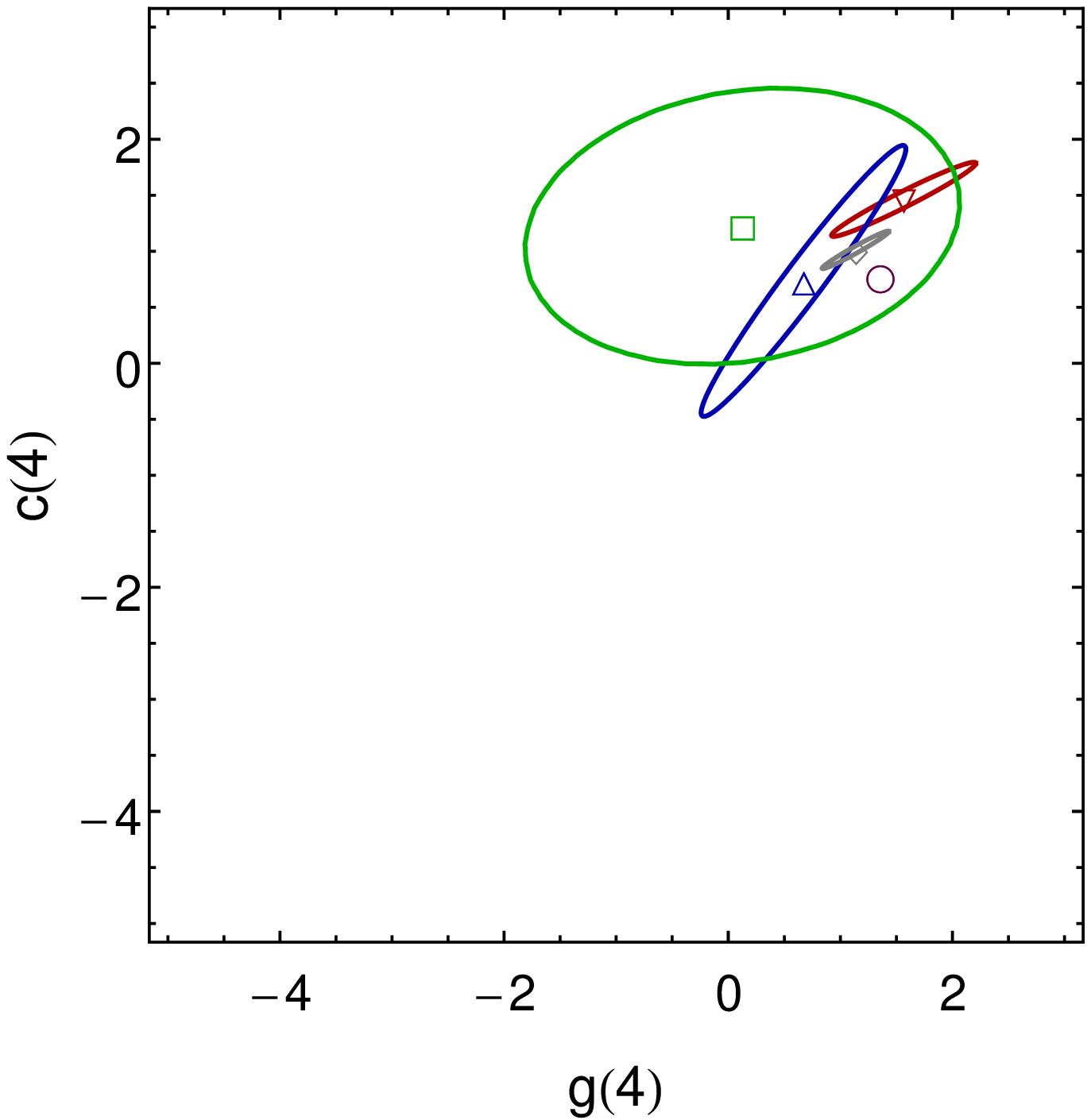} \includegraphics[width=0.24\textwidth]{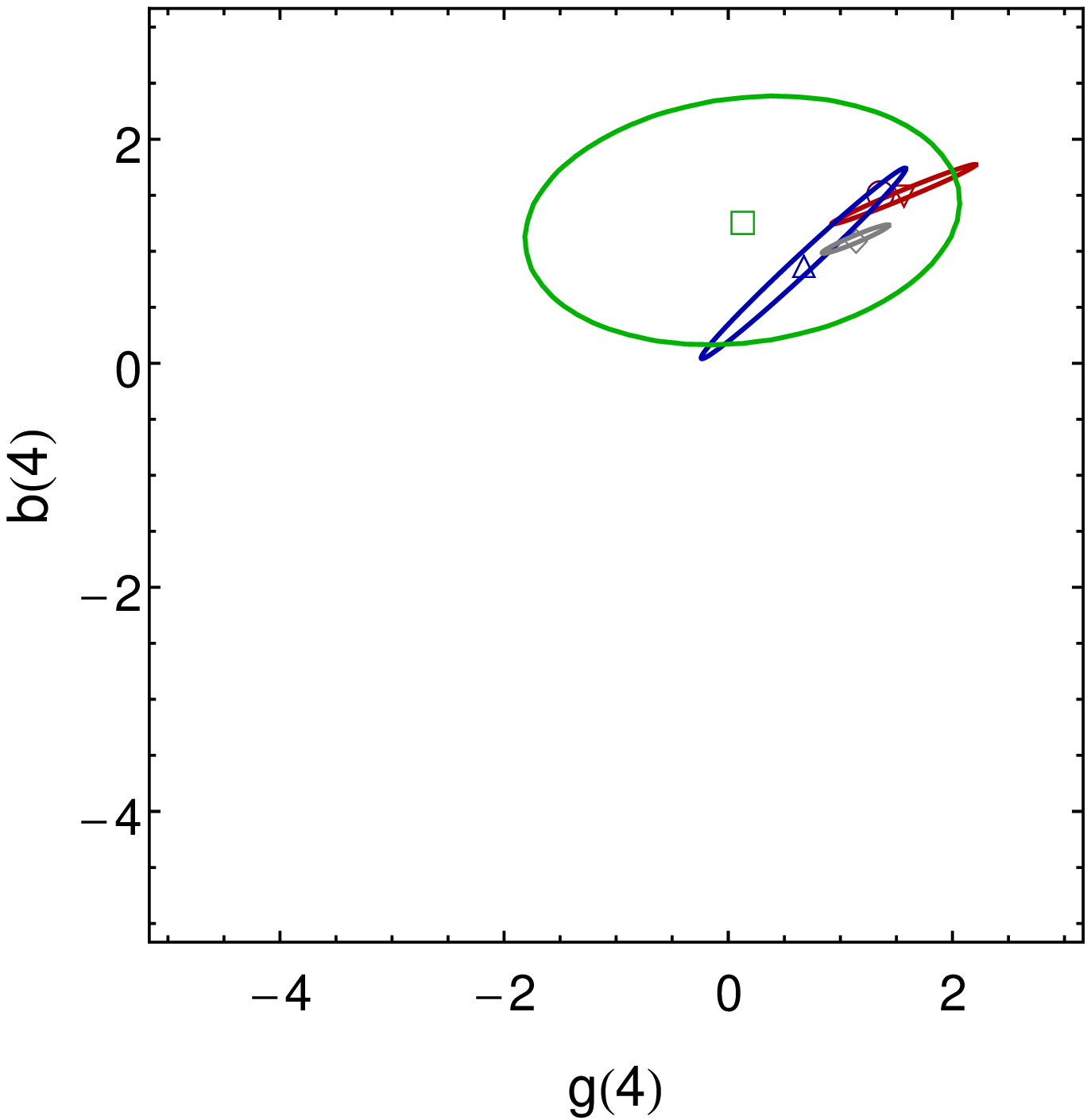}
\includegraphics[width=0.24\textwidth]{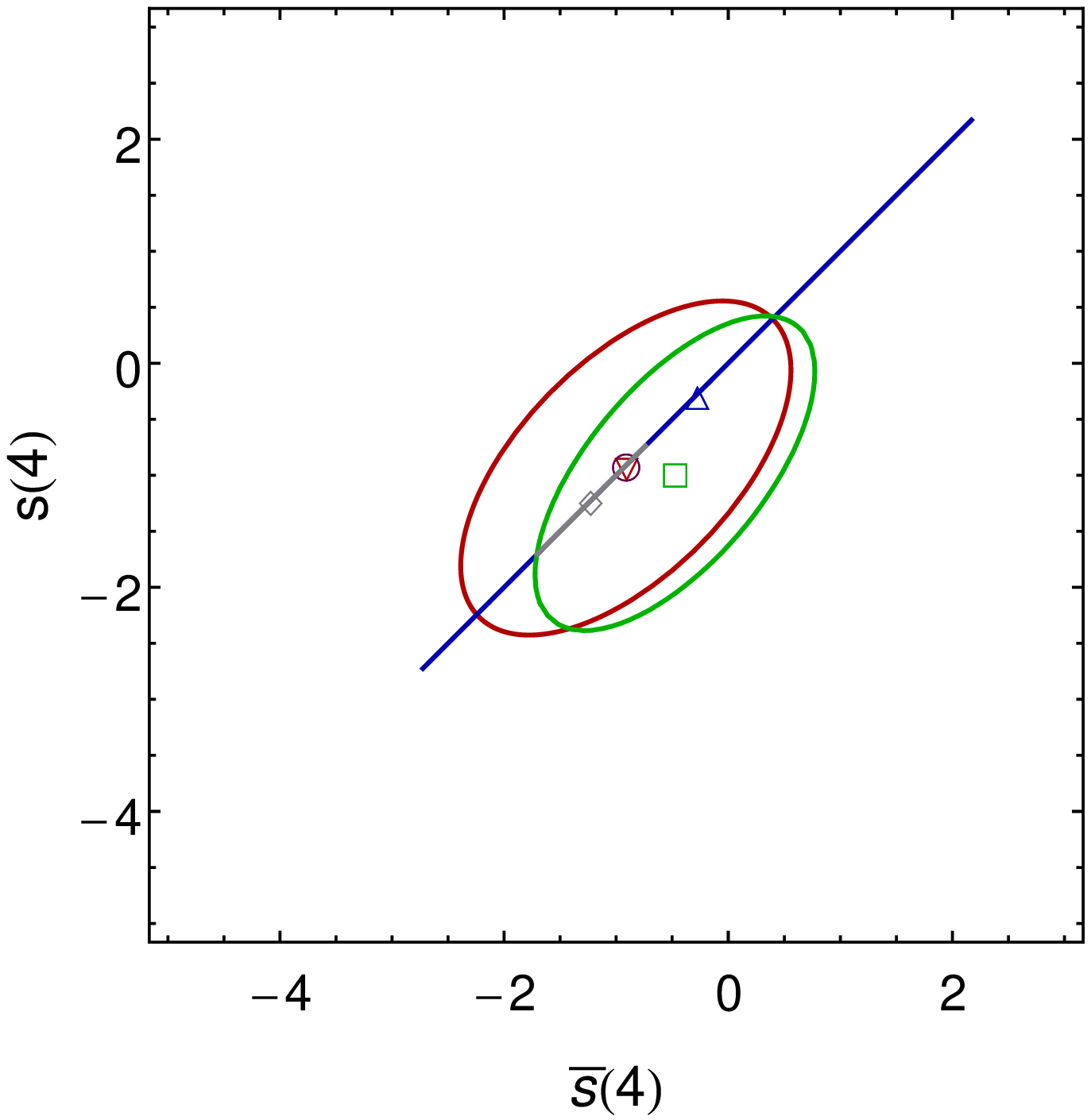} \includegraphics[width=0.24\textwidth]{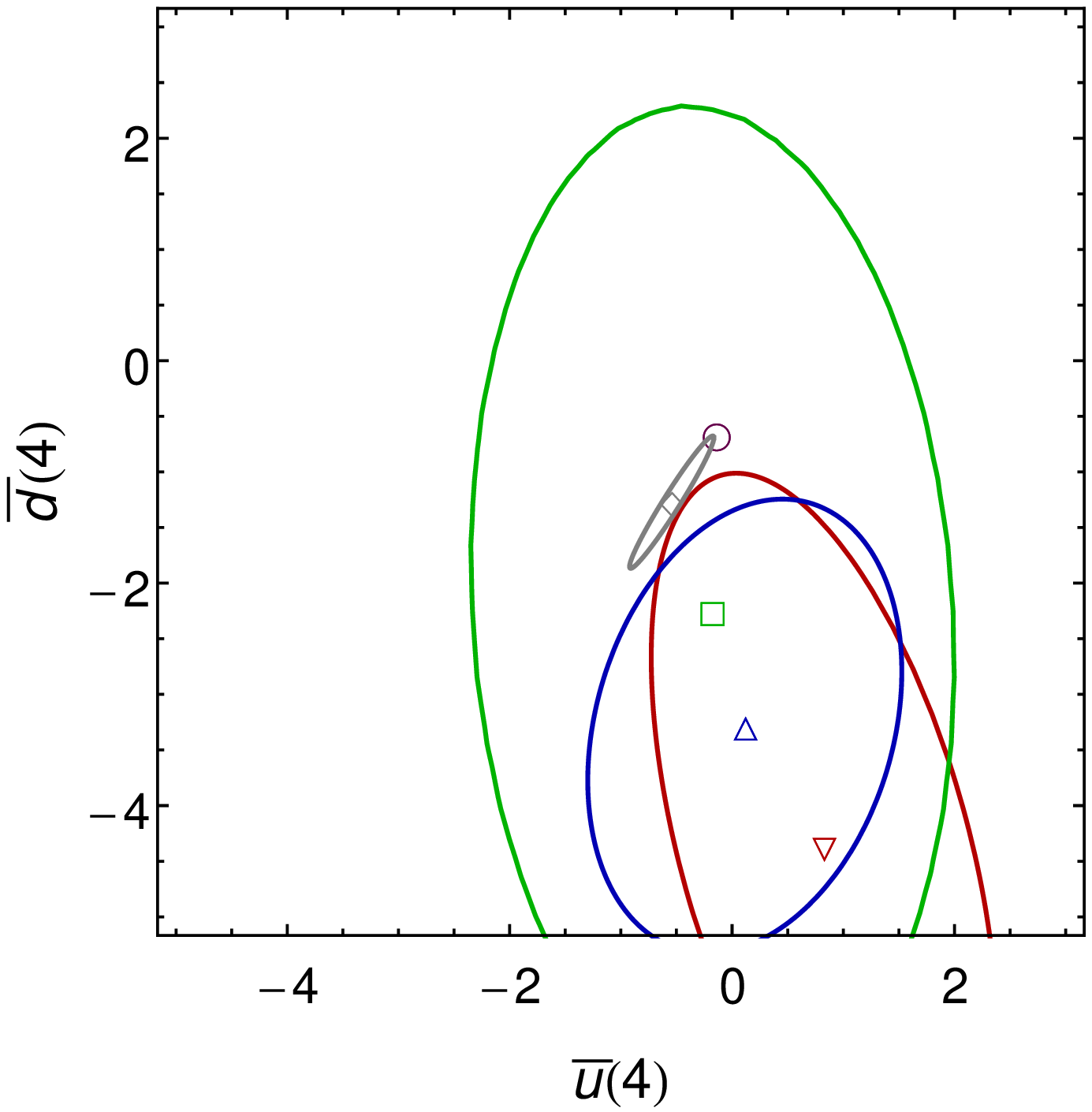}
\par\end{centering}

\vspace{-1ex}
 \caption{\label{fig:metapar2} Correlations
between representative pairs of meta-PDF parameters, for the same PDF ensembles as in Fig.~\ref{fig:metapar1}.}
\end{figure}

Once all input ensembles are converted into a shared functional
form, it becomes possible to directly examine and compare their
meta-parameters as an alternative to the usual comparisons of the
PDF shapes in the $x$ space. The distributions of the meta-parameters
follow a variety of trends~\cite{metapdfweb}, some of each are
illustrated in Figs.~\ref{fig:metapar1} and \ref{fig:metapar2}. Here
we show the probability distributions for select pairs of the
meta-parameters from the five fitted ensembles, with $f(i)$ in the
axis labels indicating the parameter $a_{i}$ in the
meta-parametrization (\ref{efun}) for flavor $f=\bar s, \bar u, $
etc. The discrete markers and lines indicate the central values and
90\% c.l. error ellipses computed according to
Eqs.~(\ref{eq:ell0})-(\ref{eq:ell2}) for each fitted ensemble. Blue,
red, green, gray, and magenta colors correspond to the CT, MSTW, NNPDF,
HERAPDF, and ABM ensembles, respectively.

In general, the figures indicate good agreement between the recent
NNLO PDFs. While the CT, MSTW, and NNPDF parameters are the most
compatible among themselves, the ABM and HERAPDF parameters are more
likely to deviate from the averages of the three global sets, as is
illustrated by some insets in Figs.~\ref{fig:metapar1} and
\ref{fig:metapar2}, as well as later in Figs.~\ref{fig:fit1} and
\ref{fig:fit2}.

Fig.~\ref{fig:metapar1} shows distributions of the parameter $a_{3}$
that controls the PDFs in the $x \to 1$ limit. The values of $a_{3}$
are well constrained for $u$ and $d$ quarks, but the spread of $a_3$
for sea flavors is much wider. [The axis ranges are chosen to
coincide in most subfigures to visually compare the spread of the
$a_3$ parameters for various flavors.] For the HERA ensemble, the
central $a_3$ value for $u$ is outside of the 90\% error ellipses of
the global ensemble, and similarly, the central $a_3$ value of ABM
for $c$ quark is barely on the boundary of the 90\% ellipses of the
global ensembles. It is also remarkable that the $a_3$ parameters
for many flavors are not correlated: the axes of the corresponding
error ellipses are oriented along the axes of the plot, rather than
along the diagonals. On the other hand, in Fig.~\ref{fig:metapar2}
we observe strong correlations between the PDF parameters of
different flavors that are due to the basic physical properties of
the PDFs, e.g., close connections between the heavy quarks and
gluon, strangeness and anti-strangeness, $\bar{u}$ and $\bar{d}$.

\subsection{Predictions based on the input and fitted PDFs for the LHC
\label{sec:LHCpredictions}}

As another test of consistency between the input and fitted PDF
ensembles, we compared their predictions for several common
observables at the LHC. We examined a group of observables that probe
various combinations of PDFs in the typical LHC measurements, as
discussed
in \cite{Gao:2013xoa}. They include the total cross sections for
$W$ and $Z$ boson production (with a decay into a lepton pair) at
NNLO \cite[Vrap]{Anastasiou:2003ds}; SM Higgs boson production via
$gg$ and $b\bar b$ fusion, computed at NNLO~\cite[iHixs1.3]{Anastasiou:2011pi}; $t\bar{t}$
production with $m_{t}=173.5\,{\rm GeV}$ at partial NNLO \cite{Baernreuther:2012ws,Czakon:2012zr,Czakon:2012pz,Czakon:2011xx}
and including resummed contributions \cite{Beneke:2009rj,Czakon:2009zw,Cacciari:2011hy}, implemented
in TOP++1.5;
as well as differential distributions of single-inclusive jet production
at NLO \cite[FastNLO2.0]{Wobisch:2011ij} for ATLAS kinematic
bins~\cite{Aad:2011fc}.

\begin{figure}[htb]
\begin{centering}
 \includegraphics[width=0.32\textwidth]{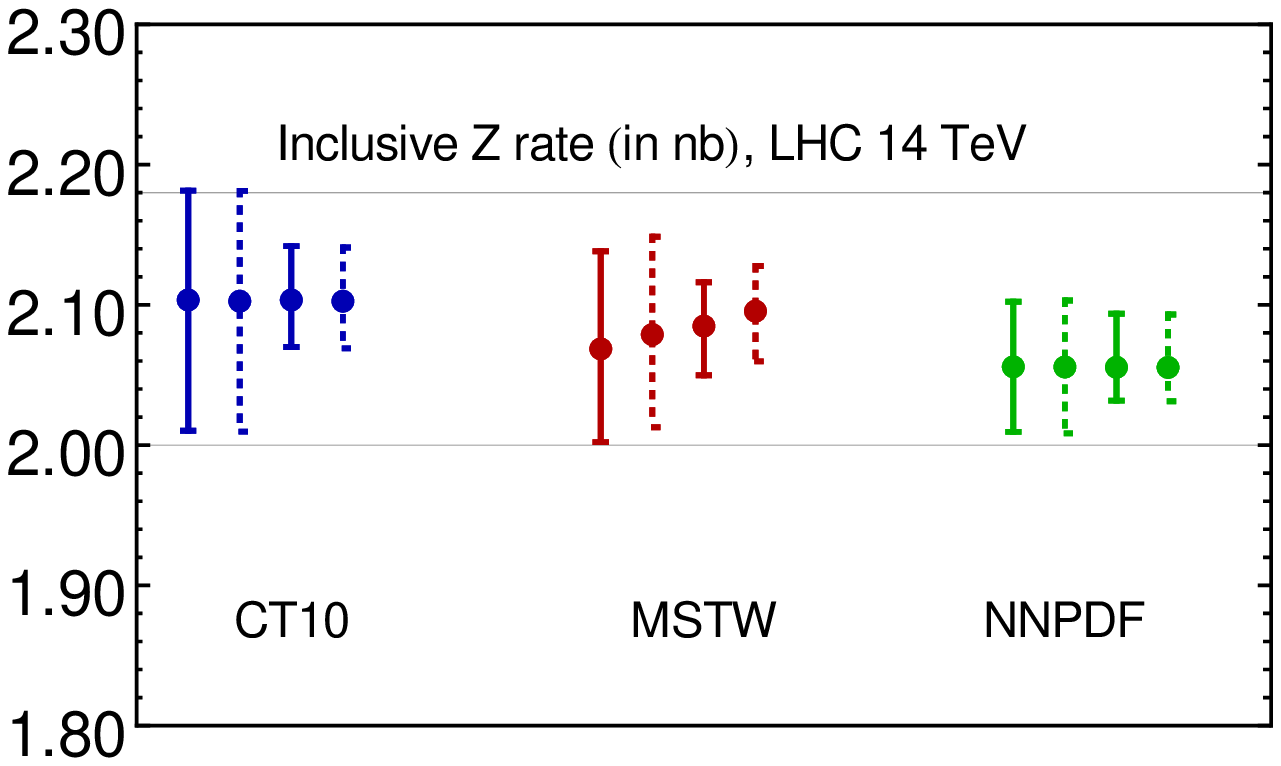} 
 \includegraphics[width=0.32\textwidth]{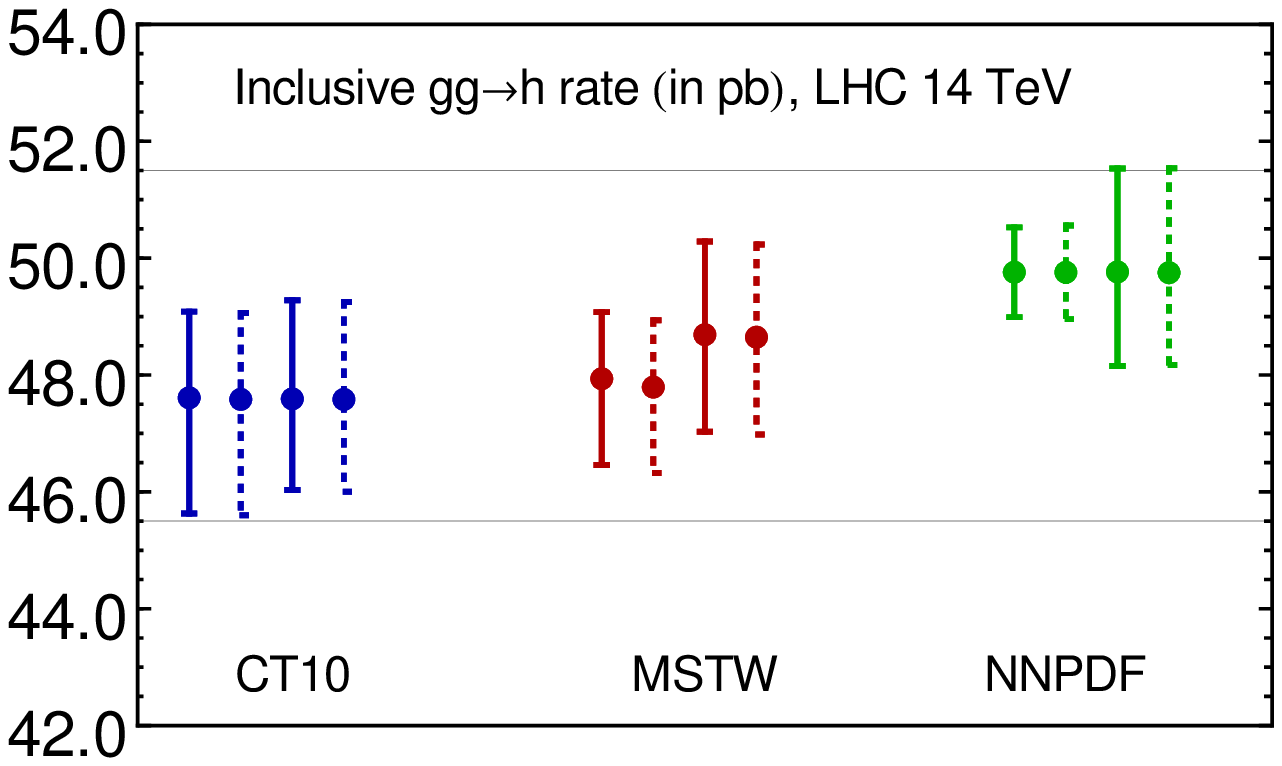}
\includegraphics[width=0.32\textwidth]{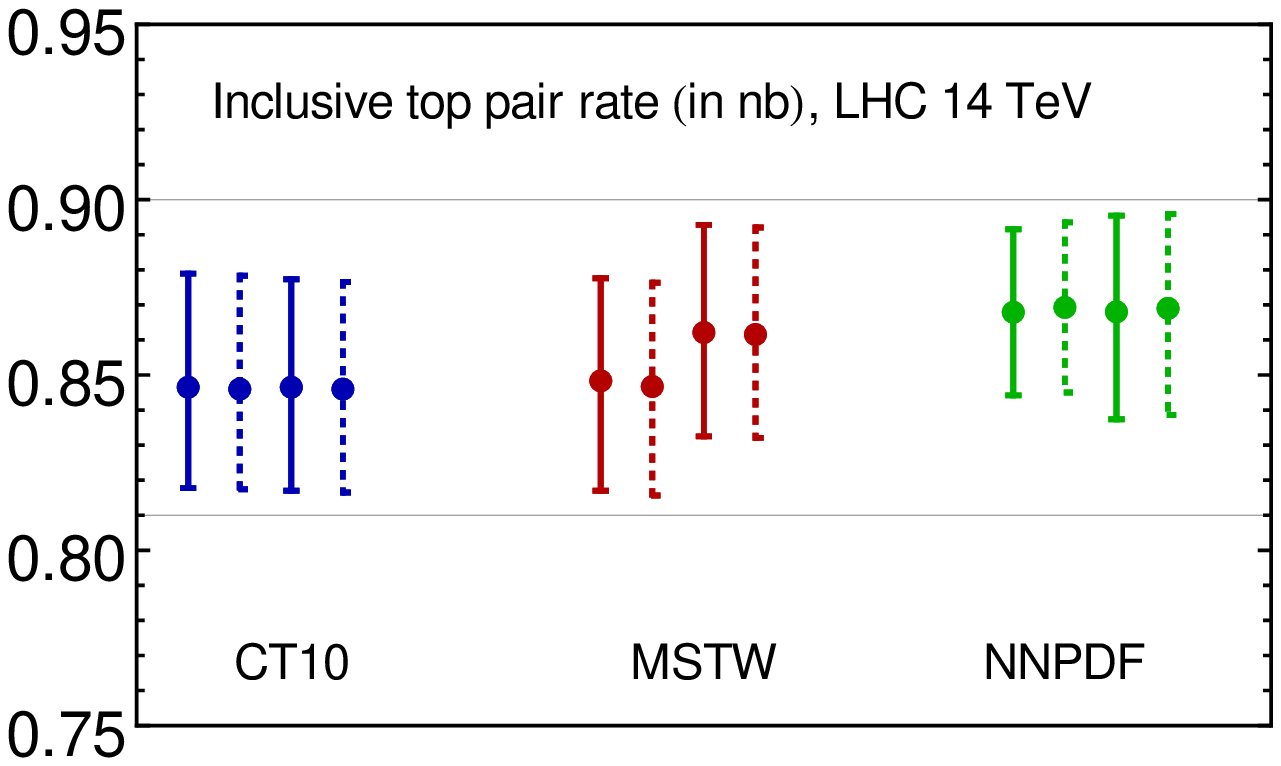}
\par\end{centering}
\vspace{-2ex}
 \caption{\label{fig:xsec1} Comparison of inclusive $Z$, SM Higgs, and
   $t\bar t$ production cross 
sections at NNLO using the input (solid) and fitted (dotted) PDFs.
The first (second) pair of error bars for each ensemble indicates the PDF
($\alpha_{s}$) uncertainty. The renormalization and
factorization scales are set equal to $M_{Z}$, $M_H$, and $m_t$, respectively.}
\end{figure}

\begin{figure}[htb]
\begin{centering}
\includegraphics[width=0.32\textwidth]{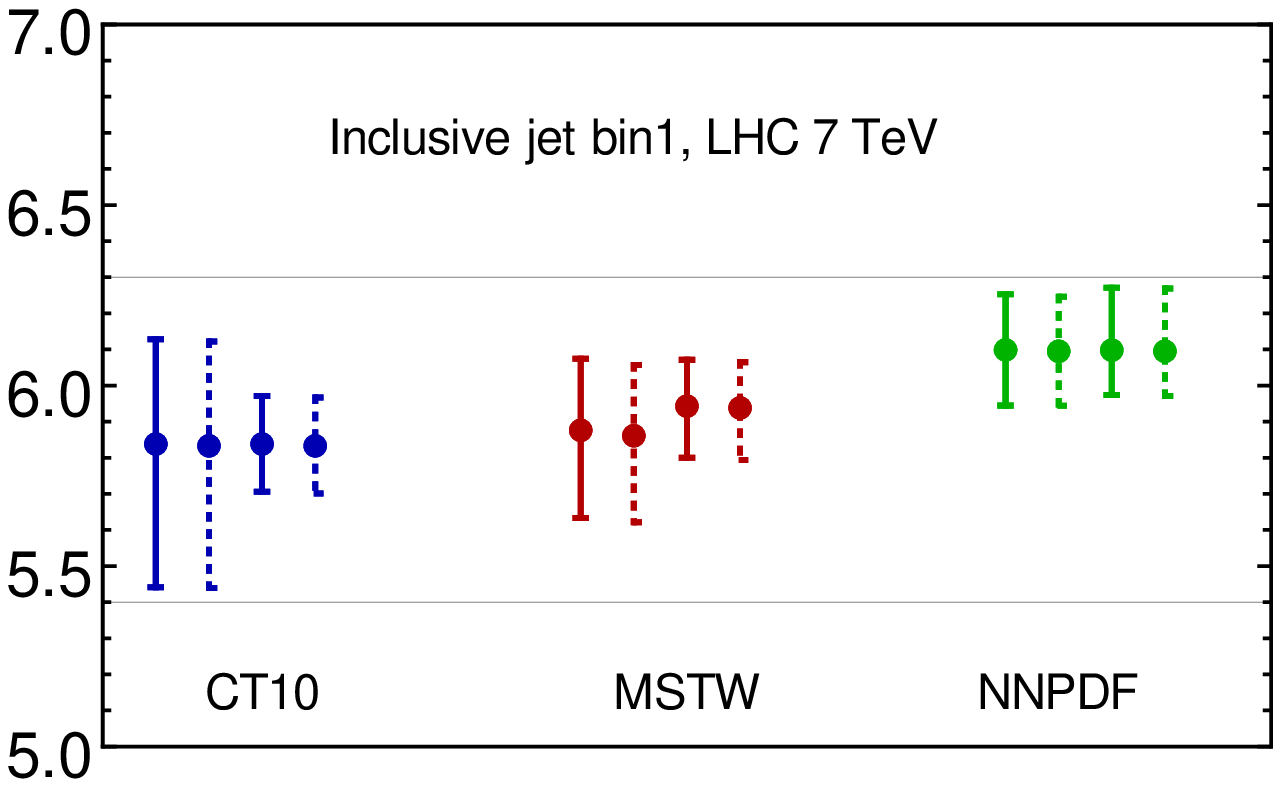} 
\includegraphics[width=0.32\textwidth]{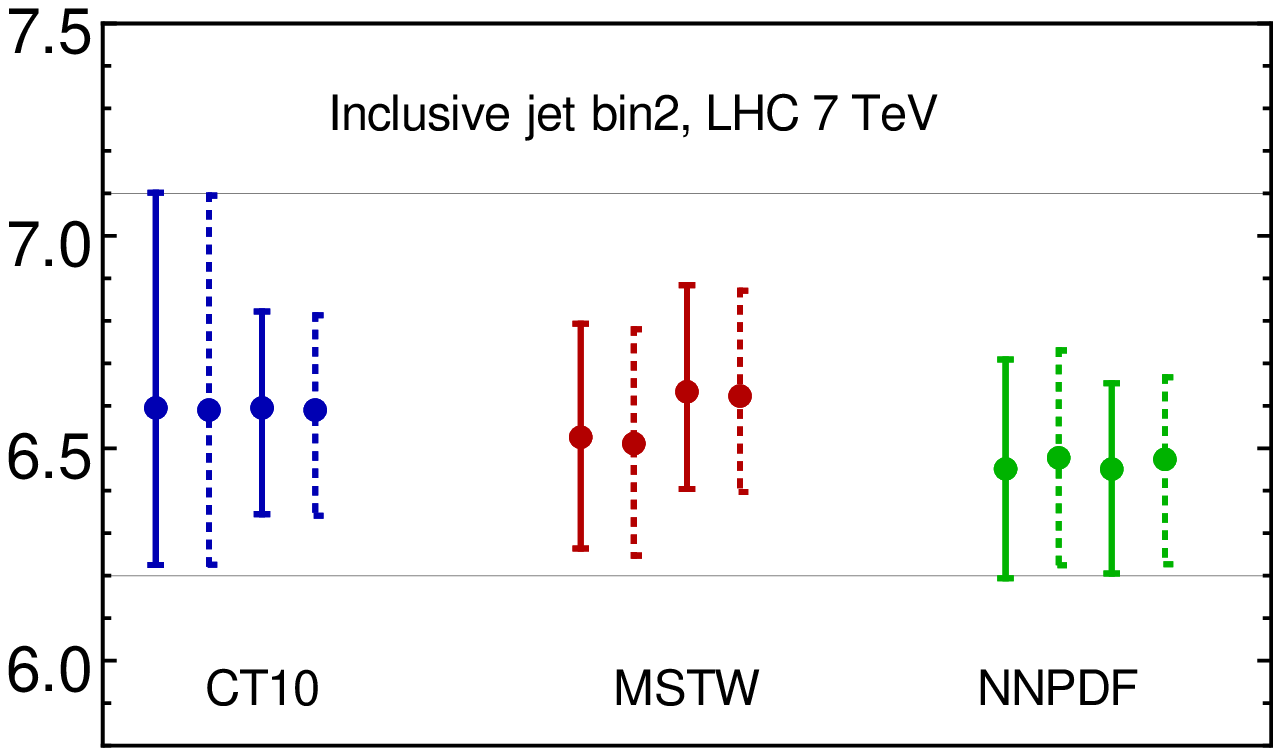}
\includegraphics[width=0.32\textwidth]{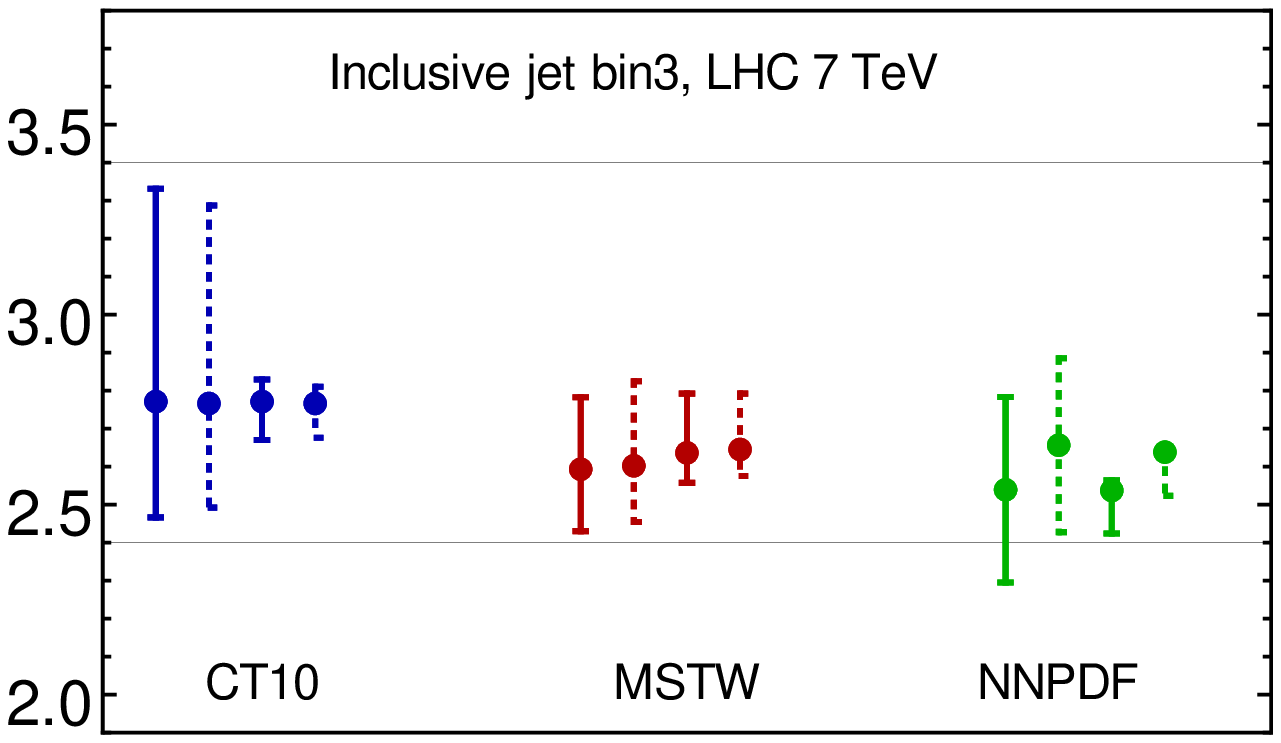}\\
 \includegraphics[width=0.32\textwidth]{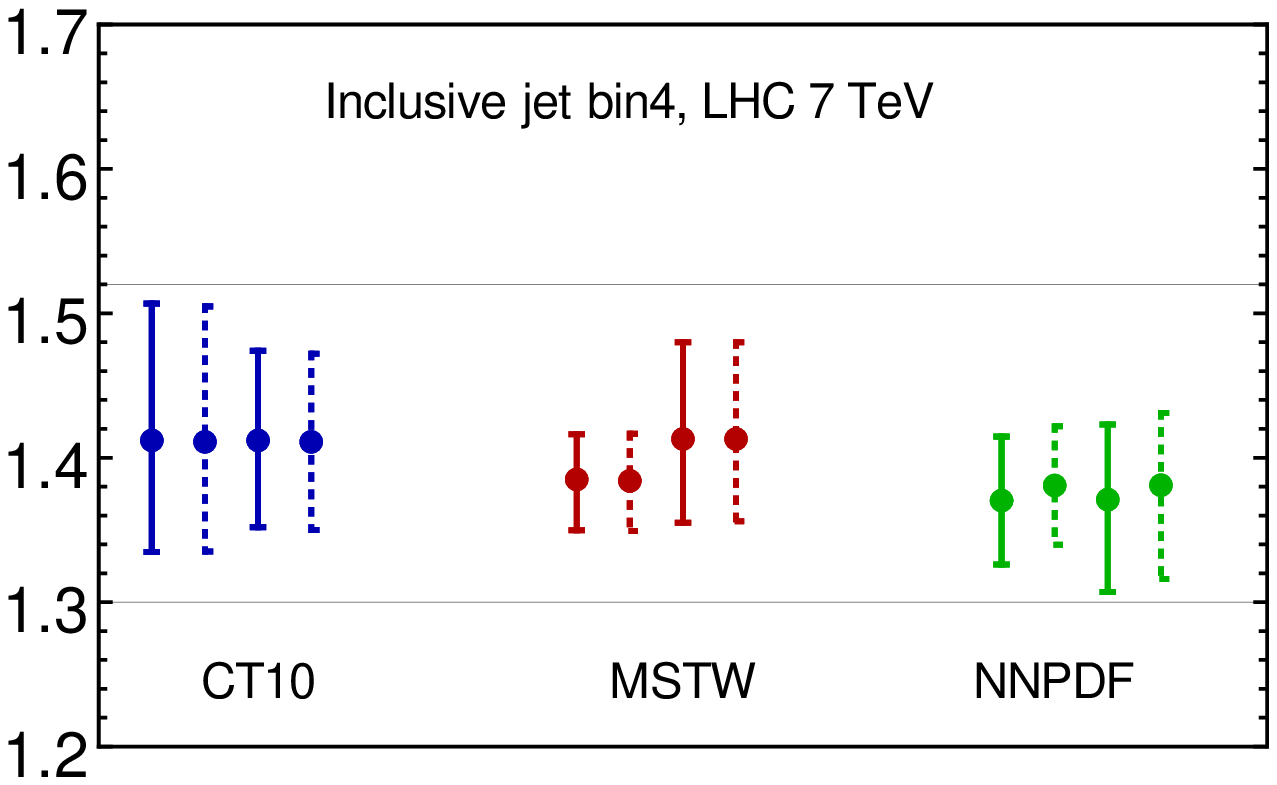} 
 \includegraphics[width=0.32\textwidth]{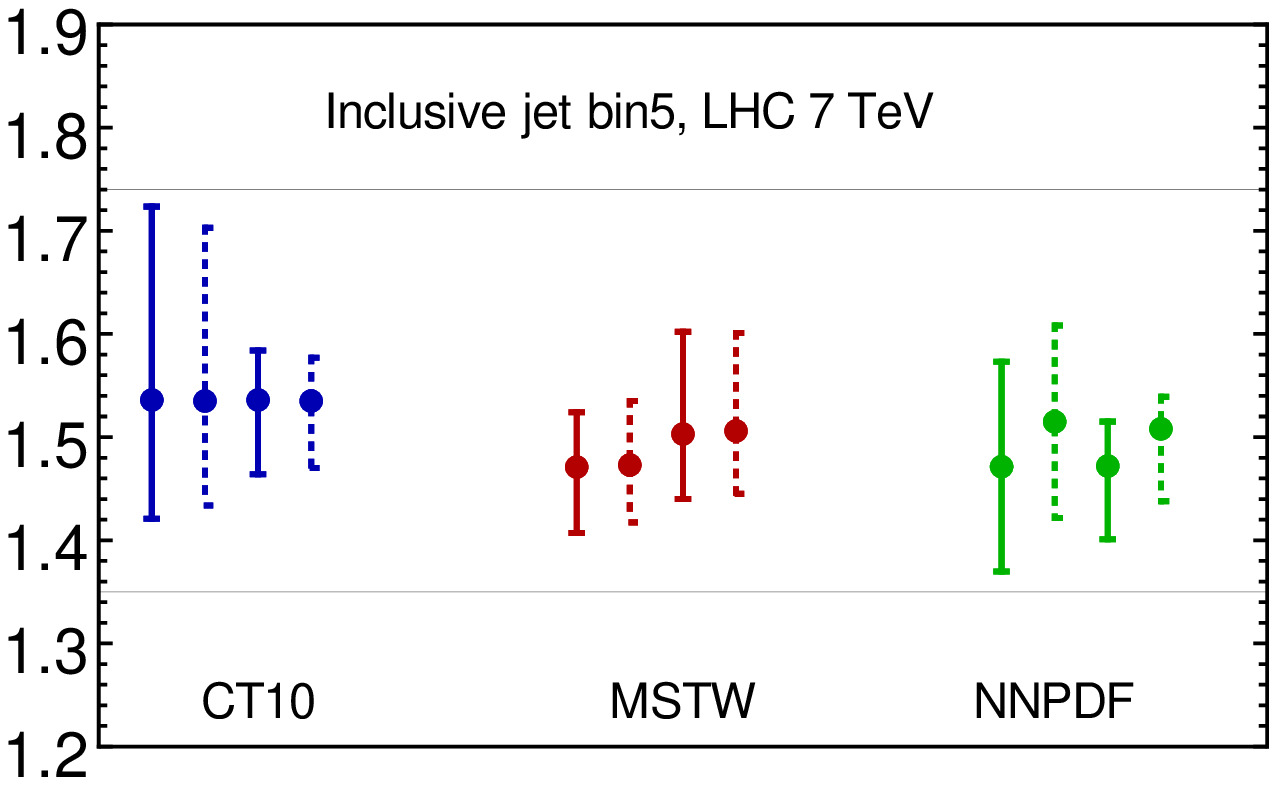}
\includegraphics[width=0.32\textwidth]{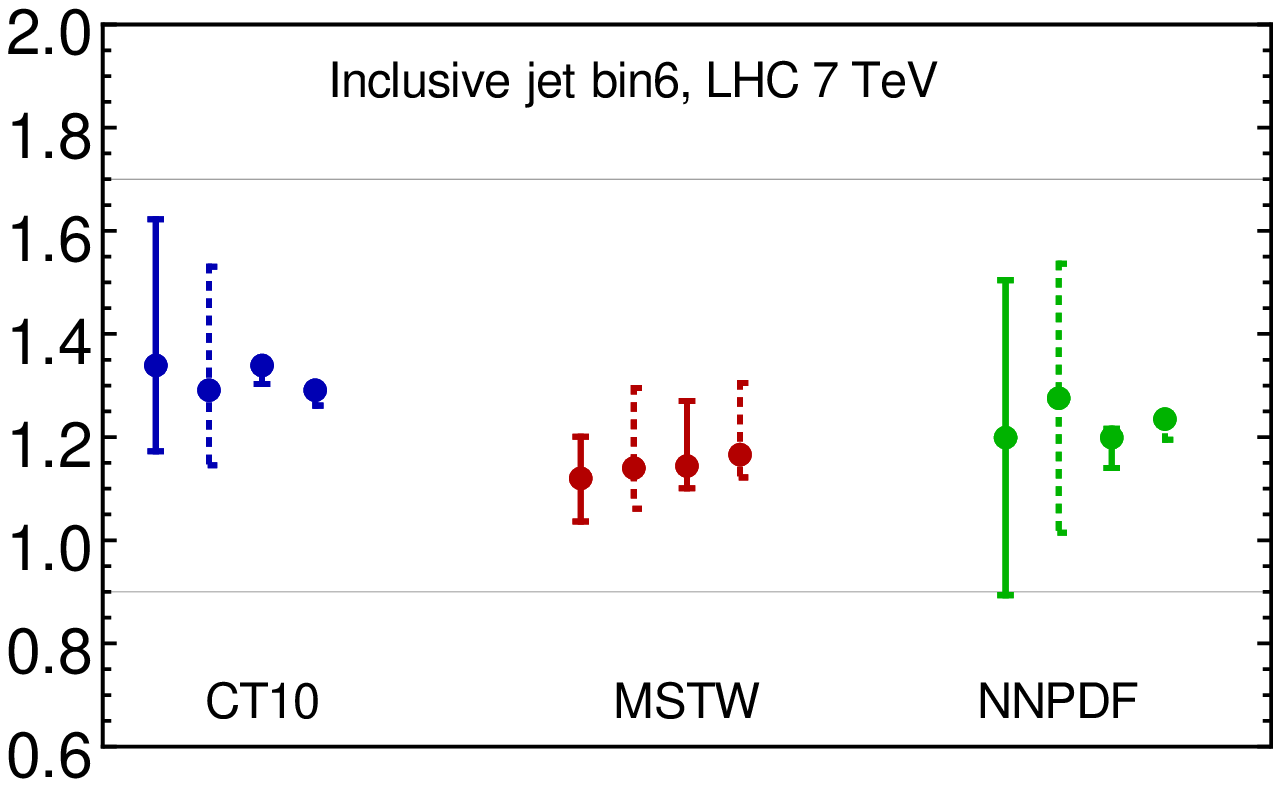}
\par\end{centering}

\vspace{-2ex}
 \caption{\label{fig:xsec3} Same as Fig.~\ref{fig:xsec1}, for inclusive jet
production cross sections (in arbitrary unit) at NLO at the LHC 7 TeV.
The upper row corresponds, from left to right, to the central
jet rapidity region $[0,\,0.3]$ in $p_{T}$ bins of $[20,\,30]$,
$[310,\,400]$, and $[1200,\,1500]$ GeV.  The lower row corresponds
to the rapidity $[4.0,\,4.4]$ and $p_{T}$ bins of $[30,\,45]$,
$[60,\,80]$, and $[110,\,160]$ GeV. The renormalization and factorization
scales are equal to the $p_{T}$ of the individual jet. }
\end{figure}

Figs.~\ref{fig:xsec1}-\ref{fig:xsec3} illustrate the level of agreement observed in these comparisons. Here  the central predictions and their PDF uncertainties using the input (fitted) PDFs are indicated by solid
(dotted) error bars, for CT10, MSTW'08, and NNPDF2.3 at the LHC $14$ TeV . Similar results for other processes and 
the LHC $8$ TeV are available at~\cite{metapdfweb}.
The PDF uncertainties at 90\% c.l.
are computed according to Eqs.~(\ref{HessianPlusError})-(\ref{HessianMinusError})
for the CT10 and MSTW ensembles, and Eq.~(\ref{MCSymmetricError}) for
the NNPDF2.3 ensemble. The central predictions are for the default
$\alpha_s(M_Z)$ values of each ensemble listed in Table~\ref{tab:Input-PDF-ensembles}.
In each group of four predictions, the first pair of error
bars corresponds to the PDF uncertainties, while the second
pair is the $\alpha_{s}$ uncertainty obtained with a central value
of 0.118 for $\alpha_s(M_Z)$ and variations of $\pm 0.002$.
\footnote{To compute the $\alpha_{s}$ uncertainty, we also fit the $\alpha_{s}$
PDF series of all the groups.}

The agreement between the input and fitted PDFs (both for 
$\alpha_s(M_Z)=0.118$ and the
$\alpha_{s}$ series) is nearly perfect for CT10 and NNPDF2.3. Minor
differences can be noticed upon closer examination, but they are
always smaller than the PDF uncertainties. For some MSTW2008
predictions, there is an overall shift ($<1\%$) due to the small
difference between their tabulated and the HOPPET numerical DGLAP
evolution mentioned in the appendix.

\begin{figure}[ht]
\begin{centering}
 \includegraphics[width=0.24\textwidth]{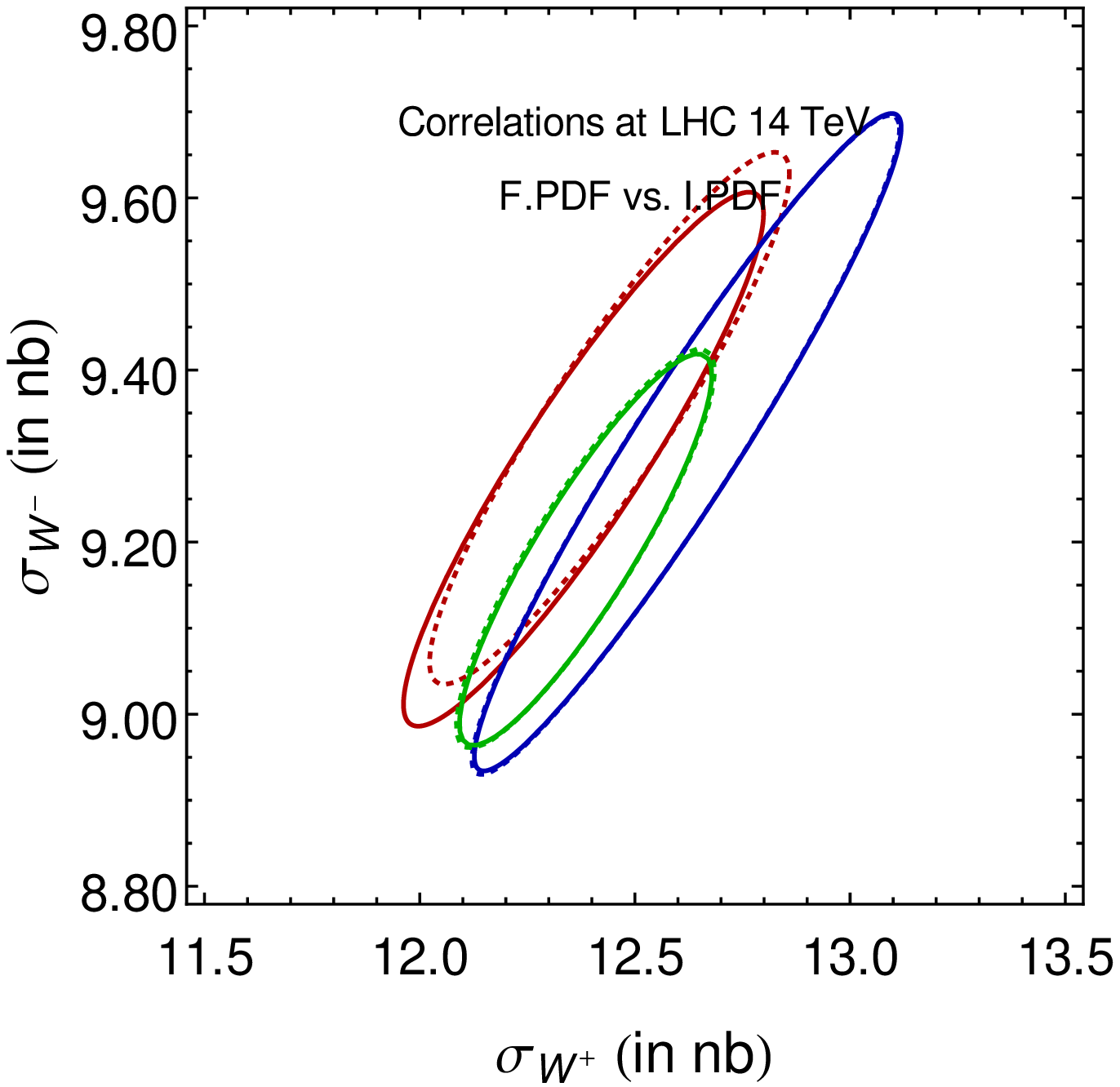} \includegraphics[width=0.24\textwidth]{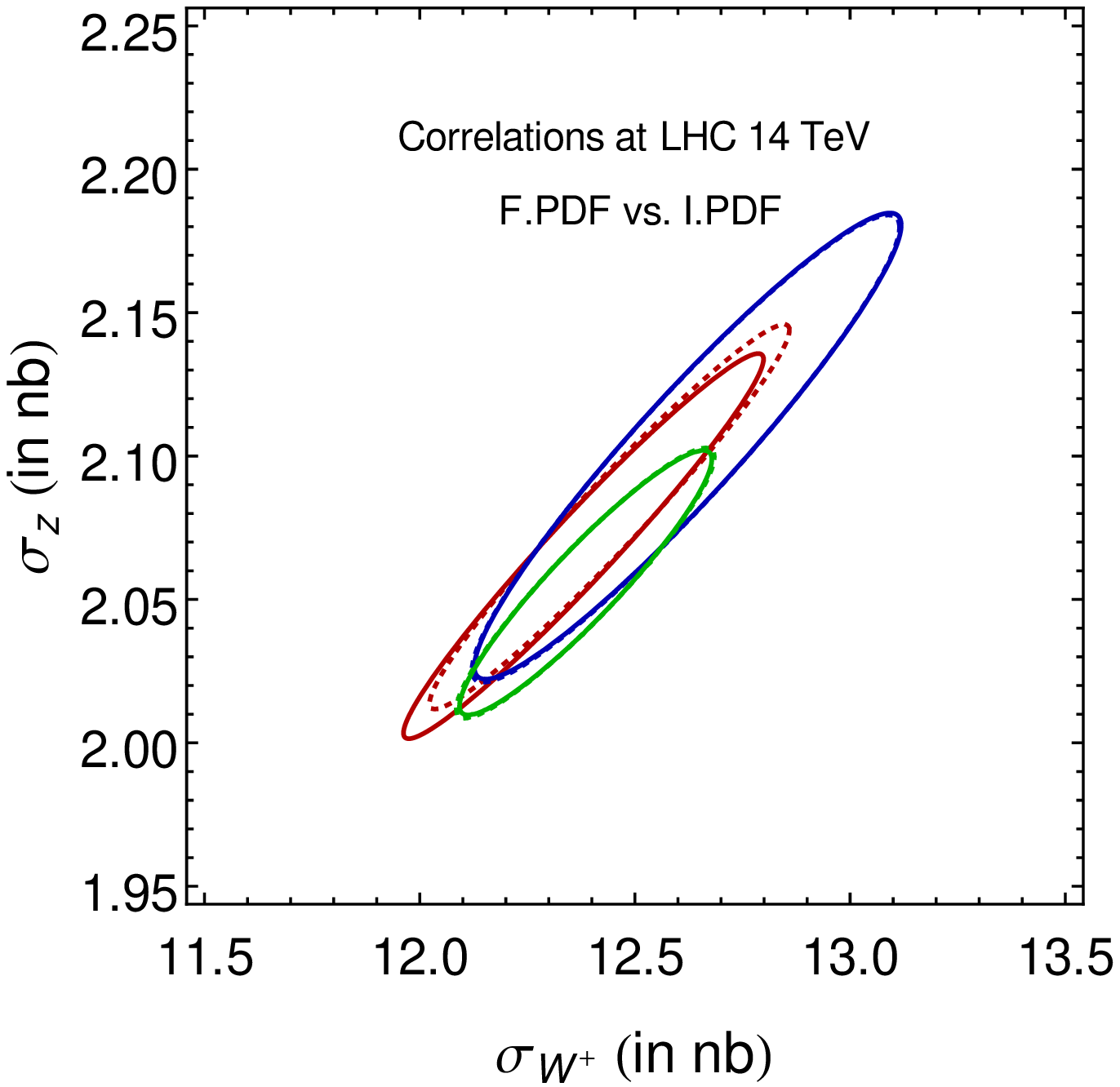}
\includegraphics[width=0.24\textwidth]{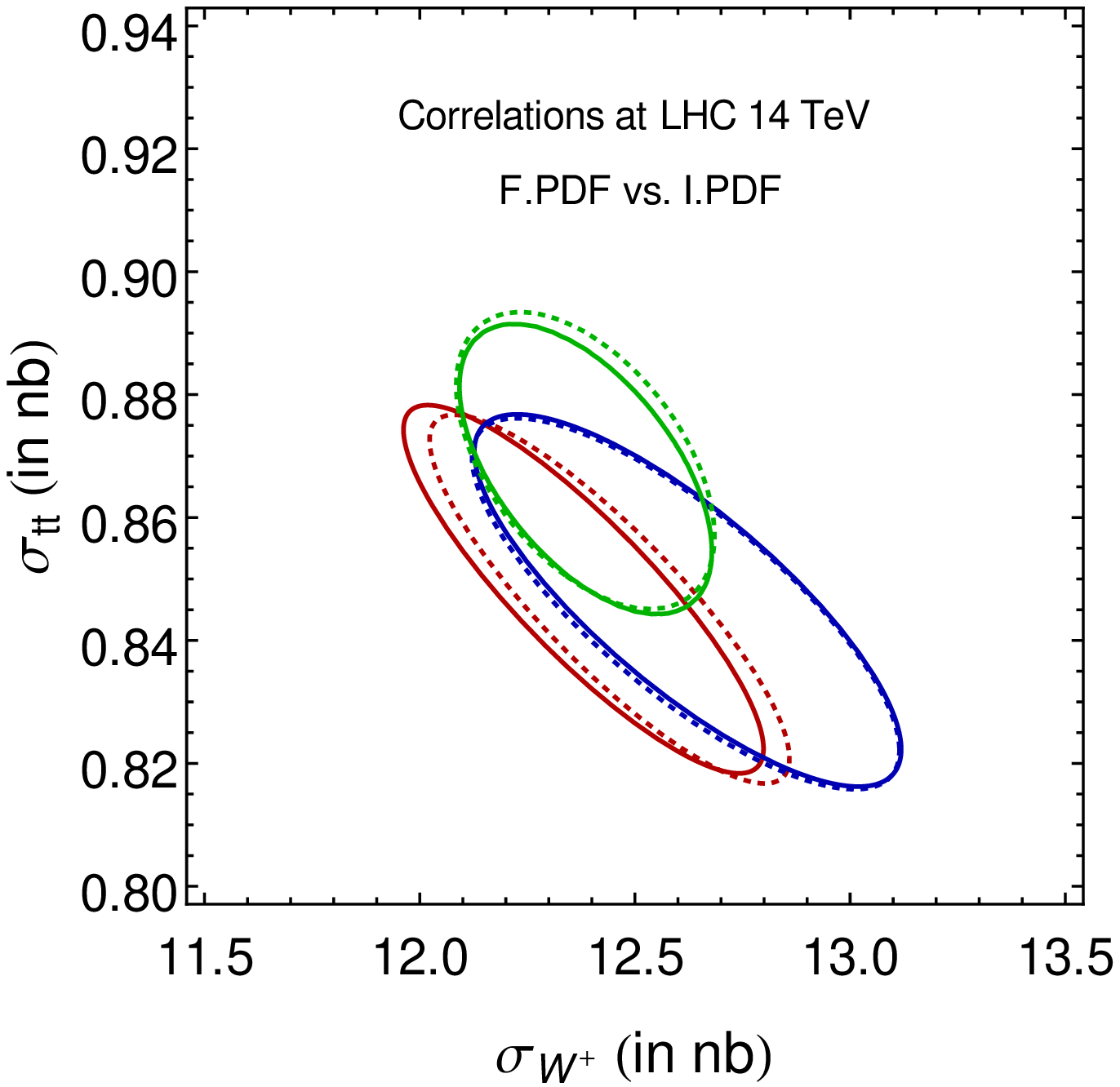} \includegraphics[width=0.24\textwidth]{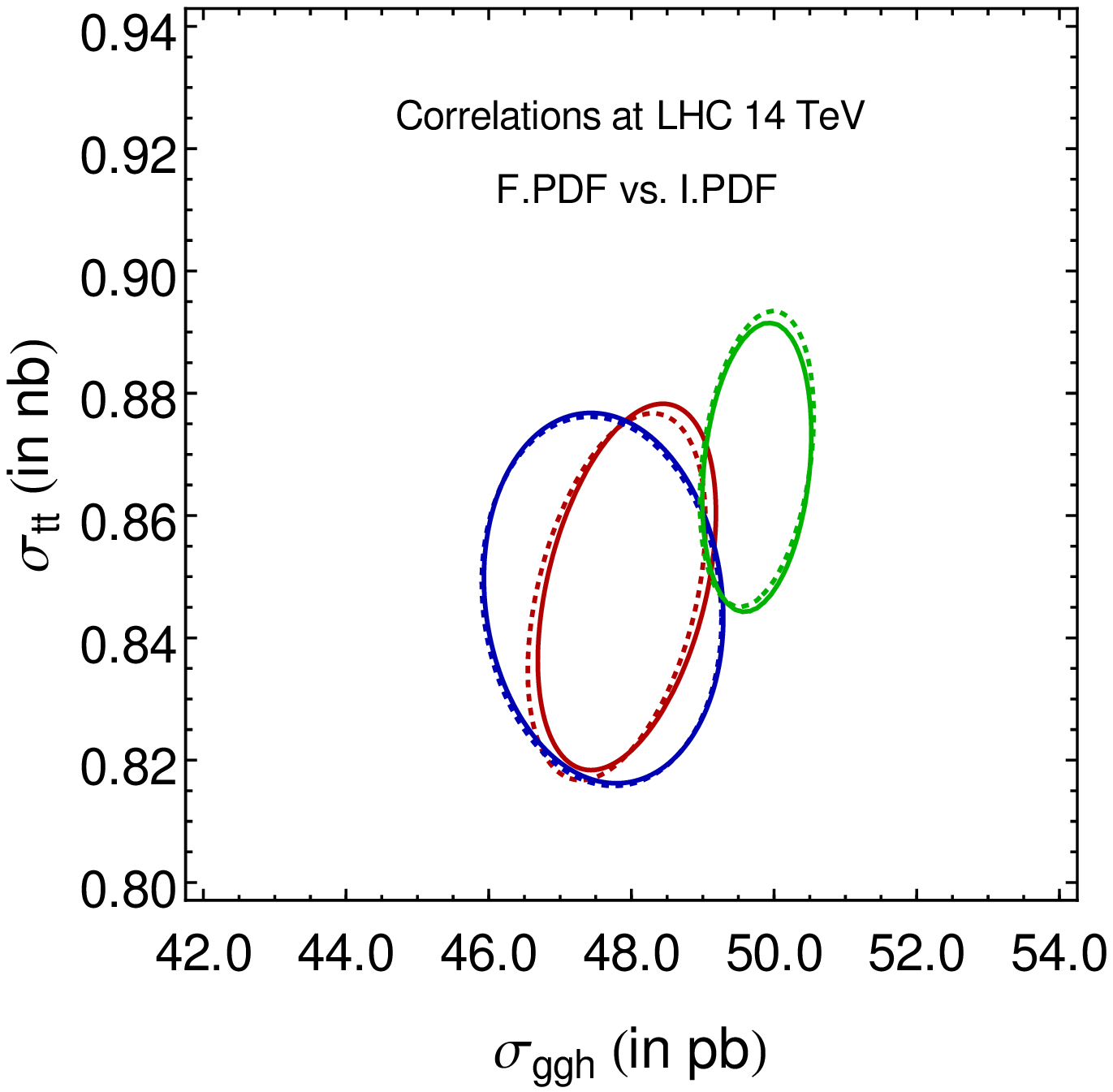}
\par\end{centering}

\vspace{-1ex}
 \caption{\label{xsec4} Comparison of correlations of NNLO cross sections using
the input (solid) and fitted (dotted) PDFs.}
\end{figure}

The 90\% c.l. error ellipses,
computed according to Eqs.~(\ref{eq:ell0})-(\ref{eq:ell2})
for the above LHC cross sections, are plotted in Fig.~\ref{xsec4}.
As before, the solid (dotted) ellipses represent predictions based
on the input (fitted) PDFs for CT10 (in blue), MSTW (in red), and
NNPDF (in green). The locations and shapes of the original error ellipses
are well preserved by the meta-parametrizations. From these comparisons
we conclude that the chosen meta-parametrization form with 66 parameters
in Eq.~(\ref{efun})
reproduces well the central LHC cross sections, their PDF uncertainties,
and correlated flavor dependence of all input PDF sets. We can now
proceed to the combination of the error PDFs into a META ensemble.

\clearpage

\section{Combining PDFs into a META ensemble \label{sec:meta}}

\subsection{Overview}

In the previous section, we have shown that, by using a shared PDF
parametrization form, we can closely approximate NNLO error PDFs from all groups
at the initial scale of $Q_{0}=8\,{\rm GeV}$. Even more, if numerical
evolution by HOPPET is adopted, the fitted PDFs also agree with the
input PDFs at all scales above $Q_{0}$.

Using these meta-parametrizations, we can create a META ensemble of
PDFs from all groups that would serve similar purposes as the
PDF4LHC recommendation~\cite{Botje:2011sn}. For the combination to
proceed, the PDFs must use compatible values of $\alpha_{s}$,
heavy-quark masses, number of active flavors, and other physics
inputs. Furthermore, it is desirable to have about the same number
of experimental points fitted in each input ensemble to be allowed
to use unweighted averages when combining them, as we do below.

Thus we choose to combine the meta-parametrizations from CT10,
MSTW2008, and NNPDF2.3 with $\alpha_{s}(M_{Z})=0.118$, 0.11707, and
0.118, which satisfy these requirements. In the META PDF ensemble we
will assume a common $\alpha_{s}(M_{Z})=0.118$. The available error
sets from MSTW correspond to a lower $\alpha_{s}(M_{Z})=0.11707$,
but MSTW also provides an additional central set corresponding to
$\alpha_{s}(M_{Z})=0.118$. We estimate the meta-parameters of all
MSTW error PDF sets shifted to 0.118 by adding the differences
between the parameters of MSTW central sets for 0.118 and 0.11707 to
the parameters of the MSTW error PDF sets for 0.11707. In the
future, we hope that the different groups will make their PDF
ensembles available for the same $\alpha_{s}(M_{Z})$ to facilitate
their combination into META PDFs.

Before proceeding, we should warn that the definitions of the PDF
uncertainties, as well as the procedures for implementing them, differ
significantly from group to group. It is beyond the scope of our study
to reconcile all pertinent differences among the individual input
PDF sets. For now, we will assume that the PDF error sets that \emph{nominally}
correspond to the same confidence level can be combined.

In the Hessian approach~\cite{Pumplin:2001ct,Pumplin:2002vw}
adopted by CT10 and MSTW, a
small number of independent PDF eigenvector sets spans only the boundary
of the 68\% or 90\% c.l. hypervolume in their respective PDF parameter
space. In the Monte-Carlo (MC) approach of NNPDF~\cite{DelDebbio:2007ee}, the
same hypervolume is populated by discrete unweighted PDF replicas
that can be used to reconstruct the probability density in space of
PDF parameters. To combine the Hessian and MC error sets, we
first generate MC replicas from the CT10 and MSTW2008 meta-sets using
a prescription that is close to the one in~\cite{Watt:2012tq}. The
distribution of replicas from all ensembles will approximate the \emph{a
priori} unknown probability distribution in space of 66 meta-PDF parameters.
With a sufficient number of replicas generated from the original sets,
this probability distribution can be reconstructed with increasing
accuracy.

We will follow a simplistic but not unreasonable thinking that the
parameters of the META ensemble roughly obey a 66-dimensional
Gaussian distribution. We will also symmetrize the PDF errors and
include all PDF ensembles with the same weights to simplify many
considerations.

The combined META PDF uncertainties can be estimated using either
the Hessian or MC approach. The latter could follow the strategy
that is similar to the combination of Monte-Carlo replicas from
various ensembles discussed in \cite{Forte:2013wc,Forte:2010dt}.
The Hessian approximation is at least as
attractive, as we found that only about 100 eigenvector sets are required to
estimate the combined confidence levels regardless of the number of
the input error sets that we tried. Furthermore, if we are interested in
the uncertainty of a specified observable, the number
of needed Hessian META sets can be reduced to a few by the procedure
dubbed 'data set diagonalization'~\cite{Pumplin:2009nk}, the
possibility that we investigated in a follow-up study \cite{GaoMETA2014}.

\subsection{Generation of MC replicas from Hessian ensembles}

The first step in the derivation of the META ensemble consists in
generating the MC replicas for the CT and MSTW ensembles. Initially
each of them contains $2N_{eig}$ Hessian eigenvector sets distributed \emph{on
the boundary of} their respective hypervolume, corresponding to 90\%
c.l. for both CT and MSTW. The MC replicas
for each ensemble are produced by generating
groups of $N_{eig}$ random numbers $R_{j}$,
each sampled independently from a standard normal
distribution in the interval $-\infty<R_{j}<+\infty$. For every group
of $R_{j}$, we construct a pair of MC replicas $f_{MC}(x,Q_{0})$ for each
flavor by
\begin{align}
f_{MC}(x,Q_{0}) & =f_{0}(x,Q_{,0})\pm\sum_{j=1}^{N_{eig}}\frac{R_{j}}{r}\cdot\frac{f_{j,+}(x,Q_{0})-f_{j,-}(x,Q_{0})}{2},
\end{align}
where $f_{0}(x,Q_{0})$ indicates the central PDF, and $f_{j,\pm}(x,Q_{0})$
are two Hessian error sets associated with the positive and negative
displacements along the $j$-th eigenvector.
The factor $r$ is 1.64 to convert from the 90\% to 68\% c.l.,
assuming the Gaussian distribution of the PDF parameters.\footnote{If
  the MC replicas are generated from a 68\% c.l. eigenvector set, $r=1$.}
This formula symmetrizes
the displacements of the MC replicas from the start, as indicated
by ``$\pm$''. It is equivalent to the generation of MC replicas
for Hessian PDF sets in \cite{Watt:2012tq}, but operates with symmetrically distributed
replicas.\footnote{In Ref.~\cite{Watt:2012tq}, the MC replicas are generated either
for PDF sets or for QCD observables. We follow the first option, by generating
the MC replicas for the PDFs themselves.}

By doing this we can convert any Hessian eigenvector set into an
arbitrary number $N_{rep}$ of MC replicas representing the
probability density in $\{a\}$ space. The underlying assumption
(which holds well in the $\{x,Q\}$ regions with sufficient
kinematic constraints) is that the input probability distributions
are close to the Gaussian ones, not too asymmetric, and compatible
with one another. By trial and error, we found that $N_{rep}=100$ of
replicas per ensemble is sufficient for the combination of errors.
This is the number of MC replicas that was included from each PDF
ensemble.

\subsection{Constructing META PDFs}
Given $N_{rep}$ values $a_i(k)$ of a parameter $a_i$ on a group $g$
of PDF sets, we can find the expectation value $\langle a_i
\rangle_g$, sample covariance ${\rm
  cov}(a_i,a_j)_g$, and standard deviation $(\delta a_i)_g$ of $a_i$ on $g$:
\begin{eqnarray}
\langle a_i\rangle_{g} & = &\frac{1}{N_{rep}}\sum_{k=1}^{N_{rep}}a_i(k),\label{aveg}\\
{\rm  cov}(a_i,a_j)_g & =
&\frac{N_{rep}}{N_{rep}-1}\langle\left(a_i-\langle
a_i\rangle_g\right)\cdot \left(a_j-\langle
a_j\rangle_g\right)\rangle_g,\label{covg}\\
 (\delta a_i)_g &=& \sqrt{{\rm cov}(a_i,a_i)_g}.\label{stdevg}
\end{eqnarray}
When  $N_{g}$ groups of PDF replicas ($N_g=3$: CT10,
MSTW'08, and NNPDF2.3) are combined into a META ensemble,
the expectation value $\langle a\rangle_{META}$ of
$a$ on the whole ensemble of $N_{rep}\cdot N_{g}$ replicas
can be found by averaging the expectation values on each group:
\begin{equation}
\langle a_i\rangle_{META}=\frac{1}{N_{g}}\sum_{g}\langle a_i\rangle_{g}.\label{ExpectationValue}
\end{equation}
The summation over ``g'' runs over all PDF groups. The sample
covariance and standard deviation on the META ensemble are similarly
related to their values on the individual groups by (assuming
$N_{rep}\gg 1$)
\begin{equation}
{\rm cov}(a_i,\,a_j)_{META} =\frac{1}{N_{g}}  \sum_{g}{\rm cov}(a_i,\,a_j)_{g}+\frac{1}{N_{g}}
\sum_{g}\langle a_i\rangle_{g}\langle a_j\rangle_{g}-\langle
a_i\rangle_{META}\langle a_j\rangle_{META},
\label{Covariance}
\end{equation}
and
\begin{equation}
\left(\delta a_i\right)^2_{META}  =\frac{1}{N_{g}}
\sum_{g}\left(\delta a_i\right)^2_g +\frac{1}{N_{g}}
\sum_{g}\langle a\rangle_{g}^{2}-\langle a\rangle_{META}^{2}.\label{Variance}
\end{equation}

Given these equations, the expectation value and standard deviation
for any parameter or observable $a$ on the META ensemble can be
derived either by averaging on the full META ensemble according to
Eqs.~(\ref{aveg})-(\ref{stdevg})  (and identifying the group $g$
with the whole META ensemble); or by averaging the expectation
values and standard deviations on the individual input groups with
the help of Eqs.~(\ref{ExpectationValue}) and (\ref{Variance}).
However, since much of the information contained in 300 replicas is
redundant, we can reduce the number of needed error PDFs by
constructing a smaller Hessian eigenvector ensemble that will still
be sufficient for estimating the confidence intervals.

Toward this goal, we select a new basis $a'_i= O_{ij}\cdot
(a_j-\langle a_j\rangle_{META})$ that diagonalizes the covariance
matrix, $O_{ki}\cdot {\rm cov}(a_i,\, a_j)_{META}\cdot
O^T_{jl}=\lambda_k \delta_{kl}$. Each new parameter $a'_i$ follows
an independent one-dimensional Gaussian distribution with an
expectation value 0 and standard deviation $\sqrt \lambda_i$. Here
$O_{ij}$ is an orthogonal matrix, and $\lambda_i$ are the
eigenvalues of the covariance matrix. The eigenvalues $\lambda_i$ in
the PDF analysis span many orders of magnitude
\cite{Pumplin:2001ct}. Those directions in $\{a'\}$ space that are
associated with very small $\lambda_i$ have negligible contributions
to the PDF uncertainties, so that the corresponding parameters
$a'_{i}$ can be fixed at their central values, thus reducing the
number of independent eigenvectors. For each remaining eigenvector
direction $a_{i}'$, we construct two meta-PDF sets corresponding to
the lower and upper boundaries of the respective 68\% c.l. interval.

In the end, we obtain a set of Hessian eigenvector PDFs that spans
the 68\% c.l. hypersurface on the ensemble of META PDF replicas. We
can repeat the same procedure to construct the 90\% c.l. META PDF
replicas.

In the current analysis, we found that a Hessian set with 50
eigenvectors (or 100 eigenvector sets) is sufficient for estimation
of the combined PDF uncertainty. The PDF parameter associated with
each eigenvector direction follows an independent Gaussian
distribution, thus we can compute the PDF uncertainties and
correlation angles for arbitrary observables according to the master
formulas of the Hessian formalism:
\begin{eqnarray}
 &  & \delta^H(X)=\frac{1}{2}\sqrt{\sum_{i=1}^{N_{eig}}[X_{i}^{+}-X_{i}^{-}]^{2}}\ ,\nonumber\\
 &  & \delta^H_{+}(X)=\sqrt{\sum_{i=1}^{N_{eig}}[\max(X_{i}^{+}-X_{C}^H,\ X_{i}^{-}-X_{C}^H,\ 0)]^{2}},\nonumber \\
 &  & \delta^H_{-}(X)=\sqrt{\sum_{i=1}^{N_{eig}}[\max(X_{C}^H-X_{i}^{+},\ X_{C}^H-X_{i}^{-},\ 0)]^{2}},\label{HessianMinusError2}
\end{eqnarray}
and
\begin{equation}
\cos(\Delta\varphi)=\frac{1}{4\delta(X)\delta(Y)}\sum_{i=1}^{N_{eig}}(X_{i}^{+}-X_{i}^{-})(Y_{i}^{+}-Y_{i}^{-}).\label{Correlation2}
\end{equation}

\begin{figure}[H]
\begin{centering}
\includegraphics[width=0.24\textwidth,height=120pt]{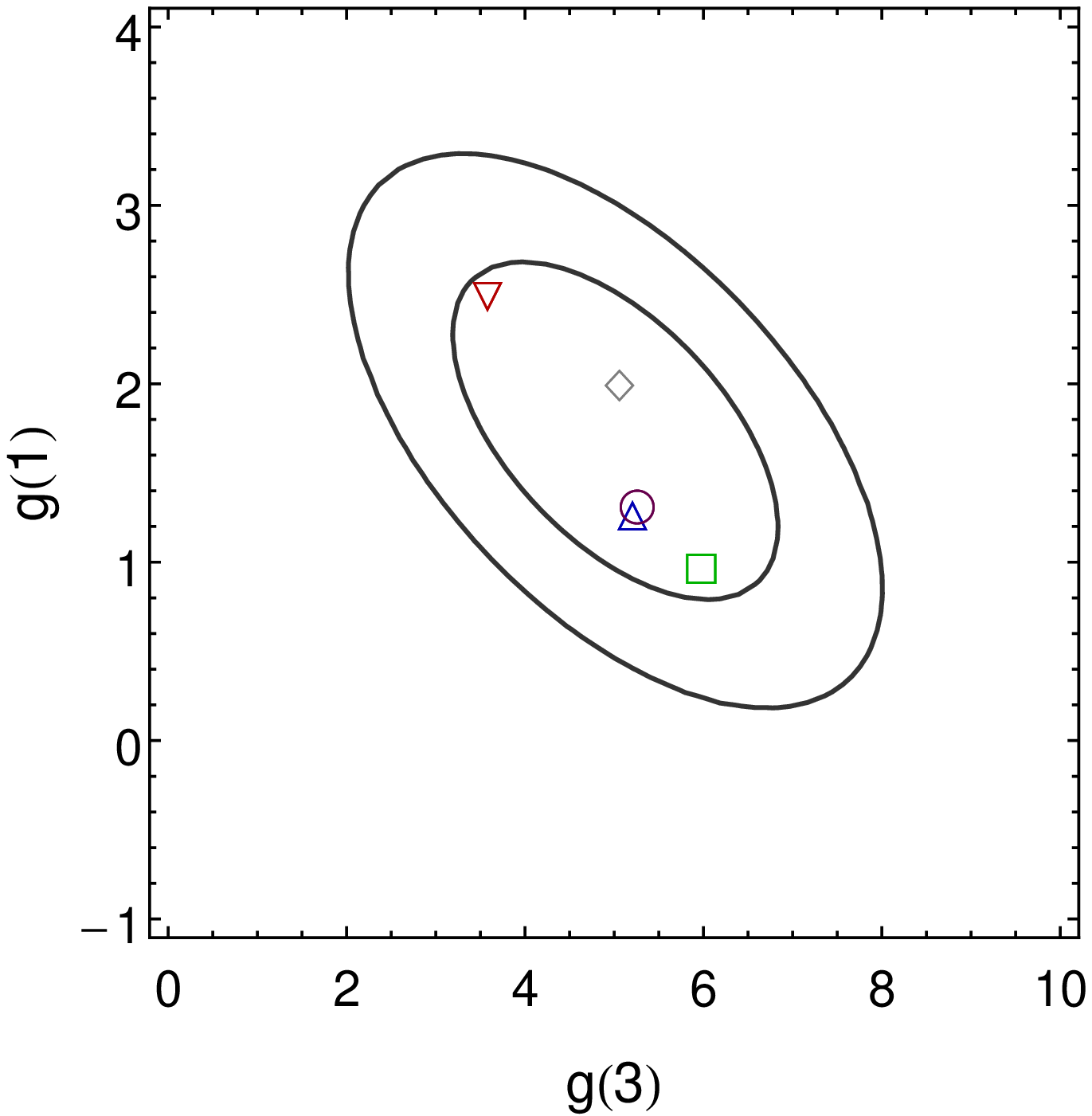} \includegraphics[width=0.24\textwidth,height=120pt]{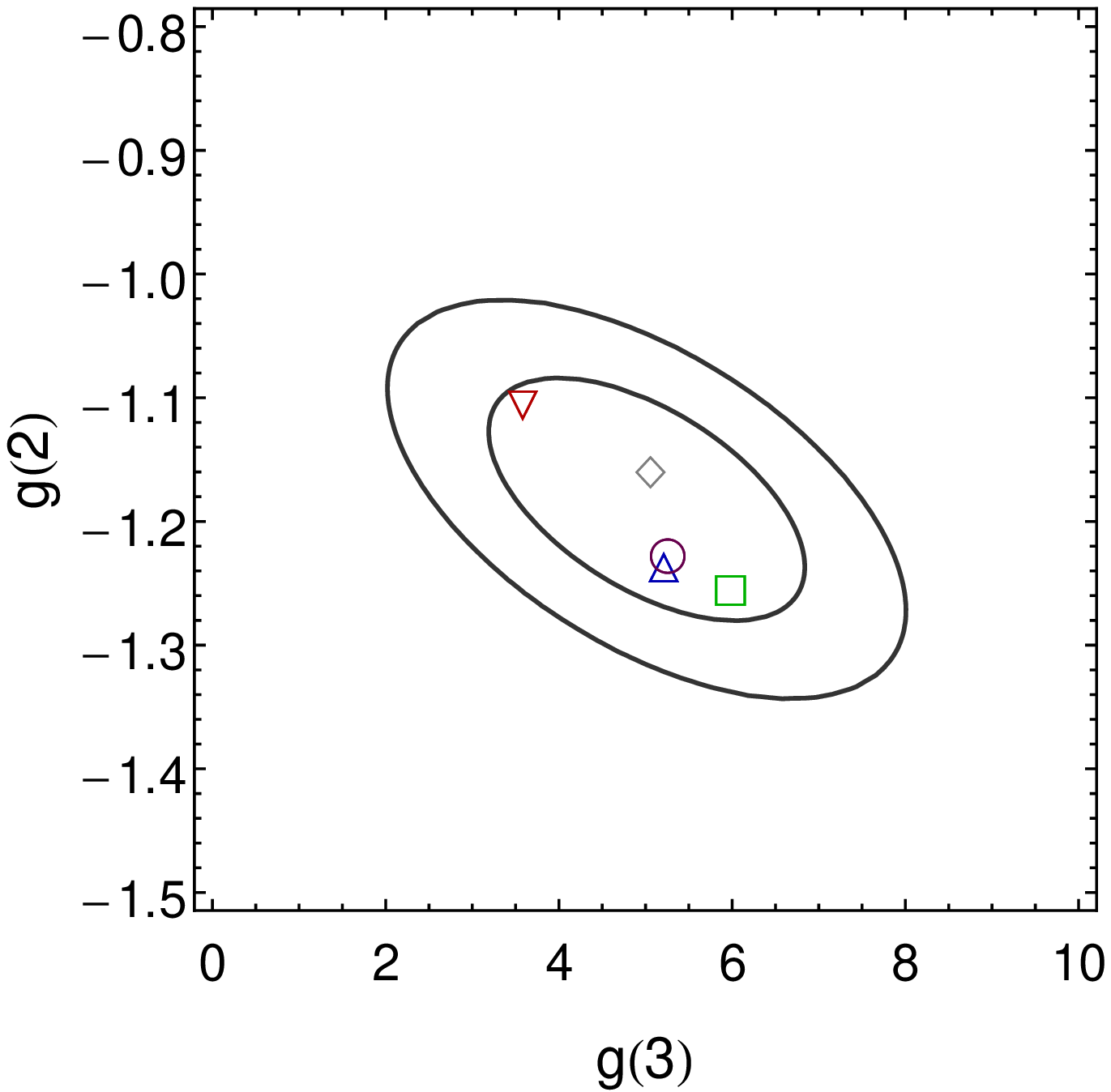}
\includegraphics[width=0.24\textwidth,height=120pt]{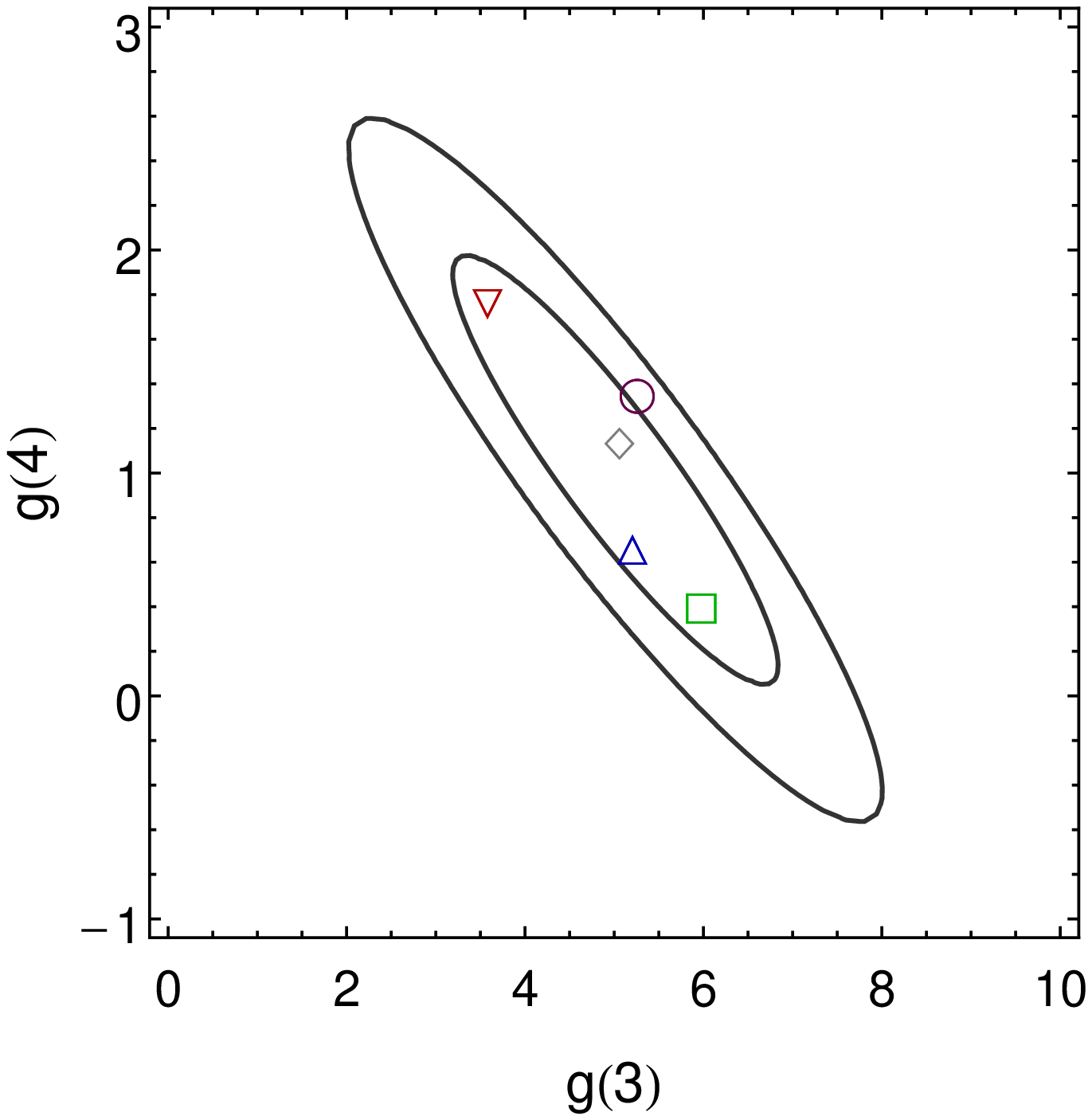} \includegraphics[width=0.24\textwidth,height=120pt]{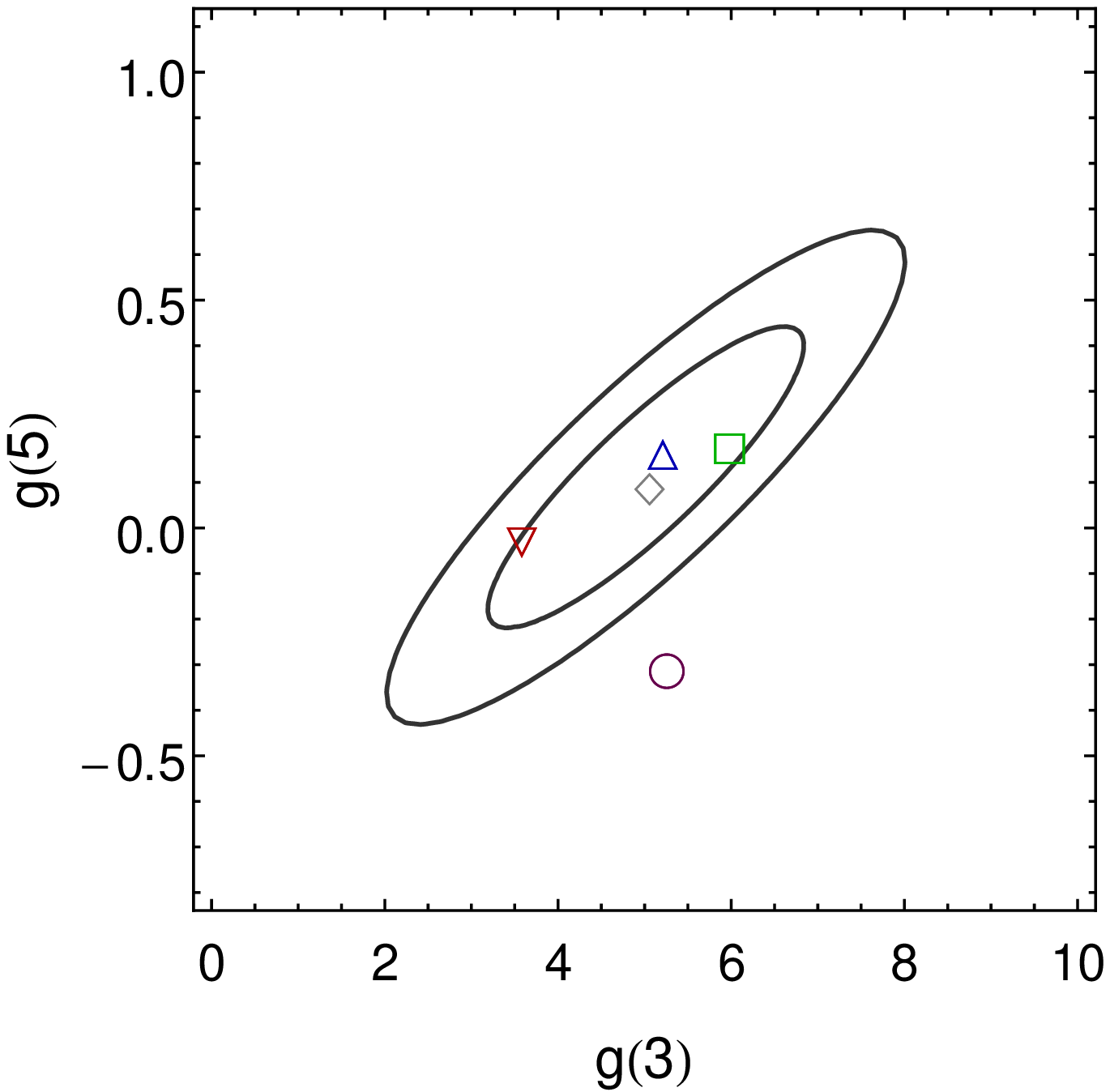}
\par\end{centering}

\vspace{-3ex}
 \caption{\label{fig:fit1} Comparison of META PDF confidence intervals with
central NNLO PDFs of the input PDF ensembles in space of
meta-parameters $a_{1-5}$ for the gluon PDF.
Up triangle, down triangle, square, diamond,
and circle correspond to the best-fit PDFs from CT10, MSTW, NNPDF,
HERAPDF, and ABM respectively. The ellipses correspond to 68 and 90\%
c.l. ellipses of META PDFs.}
\end{figure}

\begin{figure}[H]
\begin{centering}
\includegraphics[width=0.24\textwidth,height=120pt]{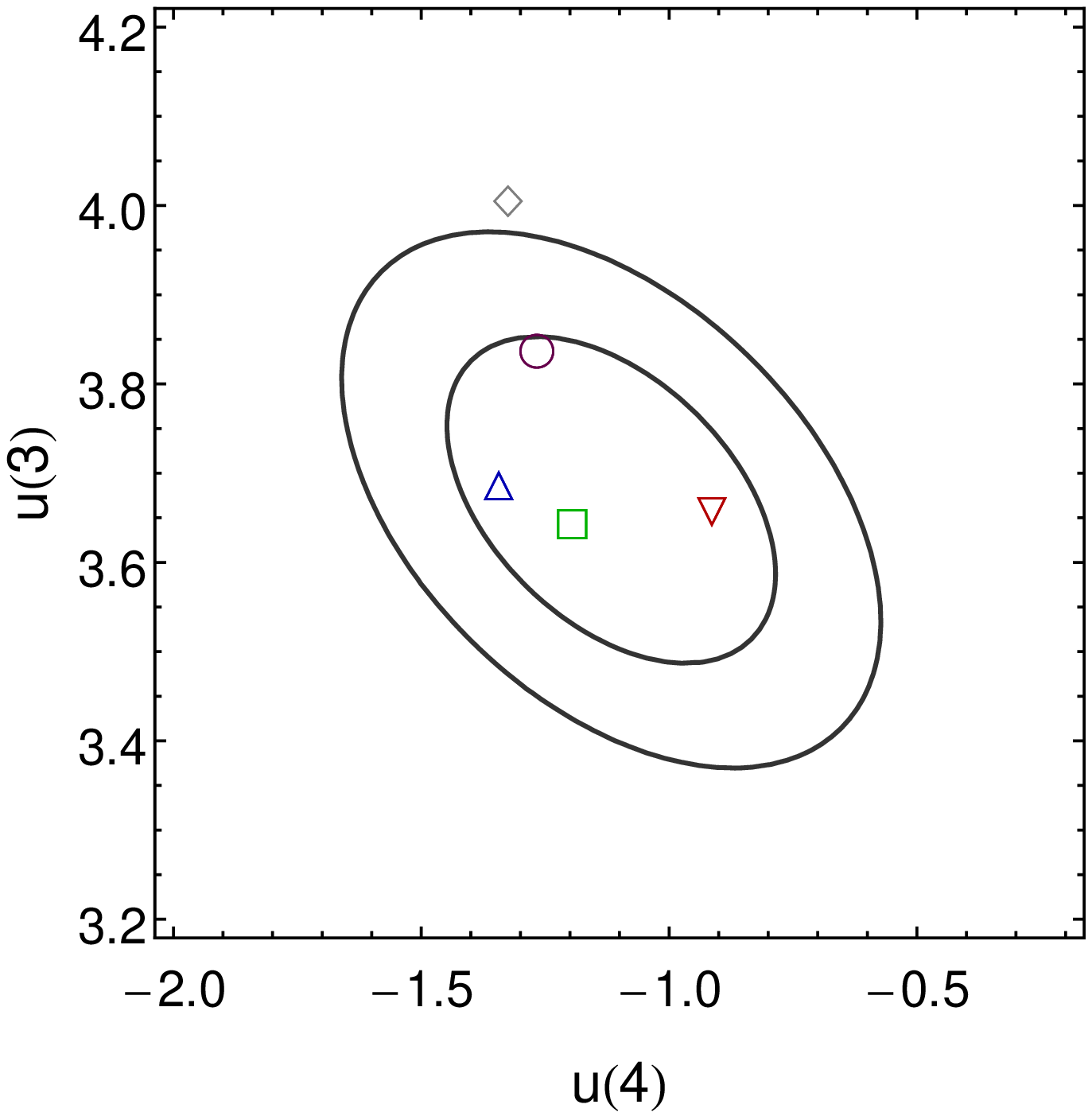} \includegraphics[width=0.24\textwidth,height=120pt]{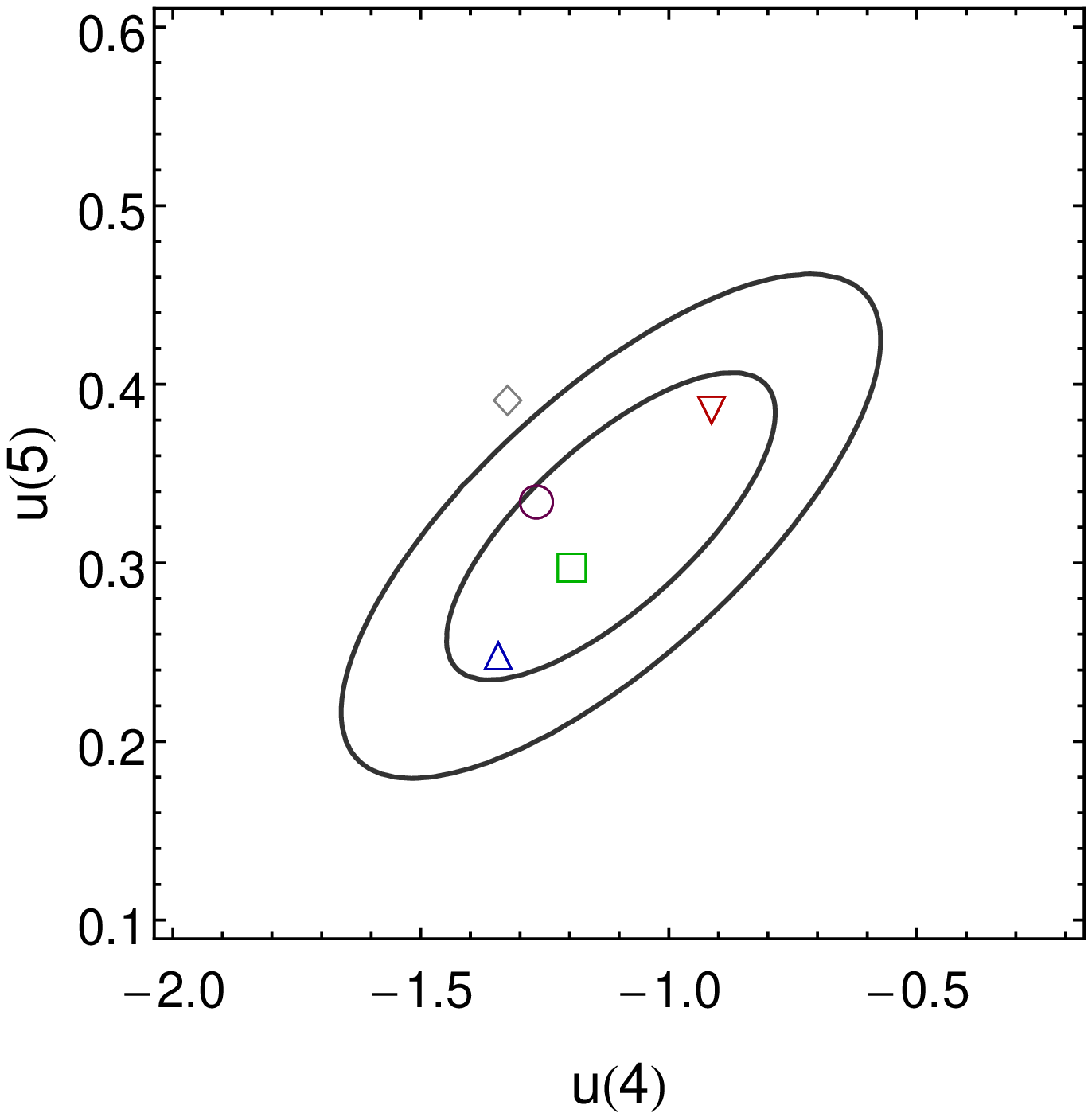}
\includegraphics[width=0.24\textwidth,height=120pt]{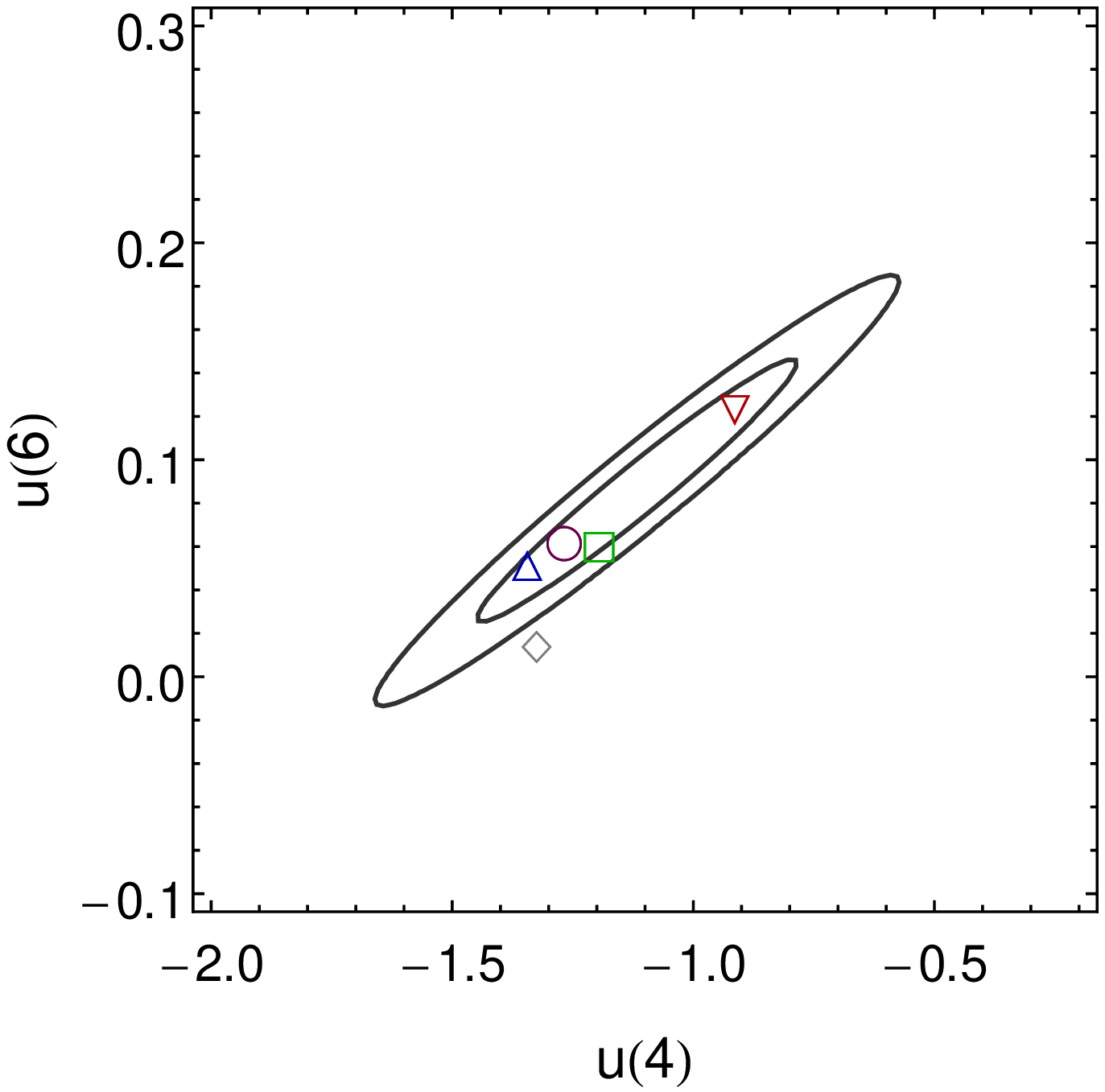} \includegraphics[width=0.24\textwidth,height=120pt]{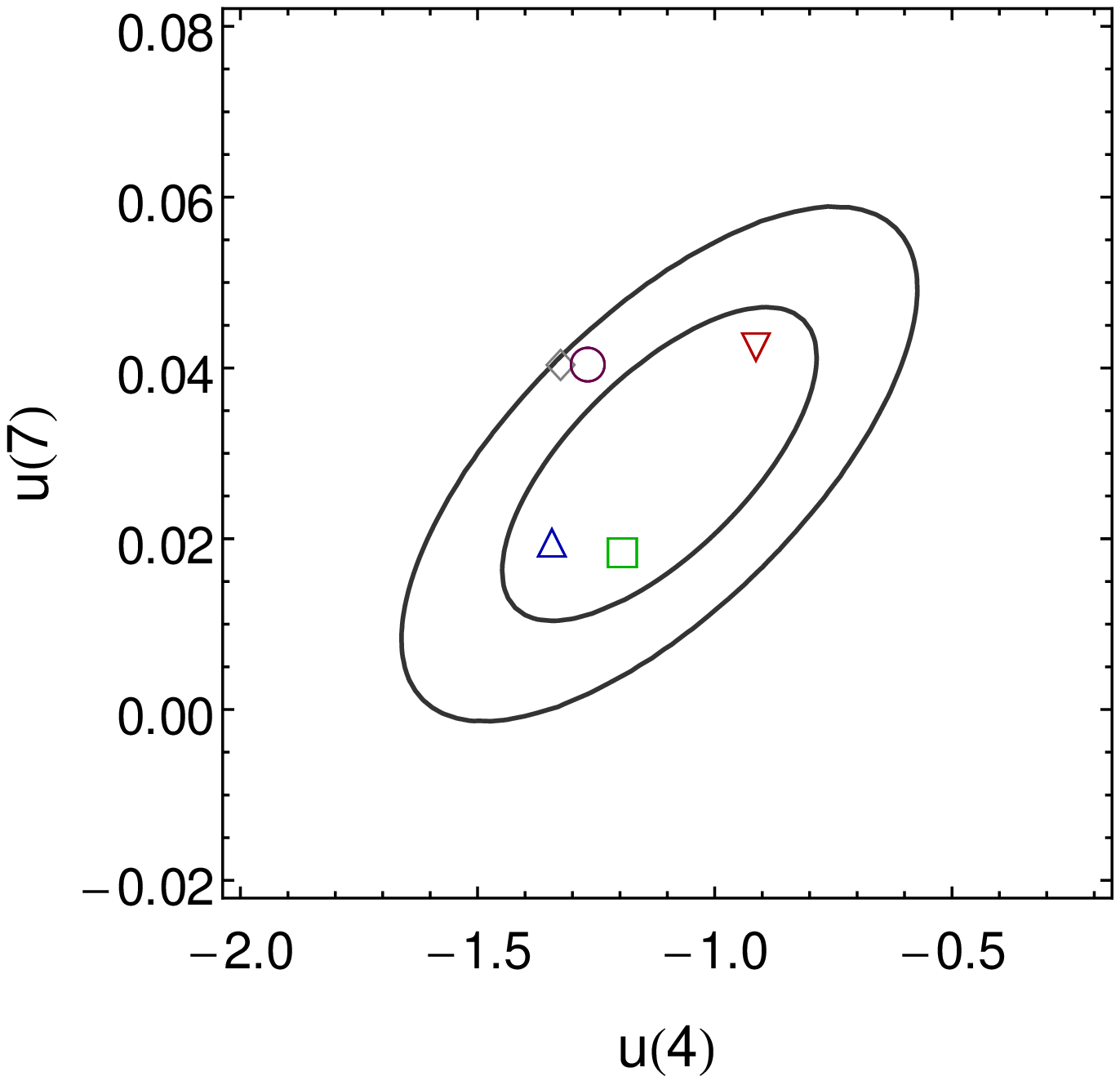}
\par\end{centering}

\vspace{-3ex}
 \caption{\label{fig:fit2} Same as Fig.~\ref{fig:fit1}, for $a_{3-7}$ of the $u$ quark PDF. }
\end{figure}

The 68\% or 90\% c.l. hypervolumes of the META ensemble enclose the
central predictions of all PDF groups, as is illustrated on the
example of some gluon or $u$ quark PDF parameters in
Figs.~\ref{fig:fit1} or \ref{fig:fit2}. As before, these comparisons
are obtained with a common $\alpha_{s}(M_{Z})$ value of 0.118. The
smaller (larger) ellipse corresponds to the 68\% (90\%) confidence
region. The markers indicate parameter combinations for the best-fit
PDFs from different groups. They are localized inside the 90\% error
ellipse of the META prescription in general. Almost all the best-fit
PDFs are within the 90\% c.l. ellipse of the META ensemble, except
for some parameters of the ABM and HERA meta-PDFs.

\begin{figure}[htb]
\begin{centering}
\includegraphics[width=0.32\textwidth]{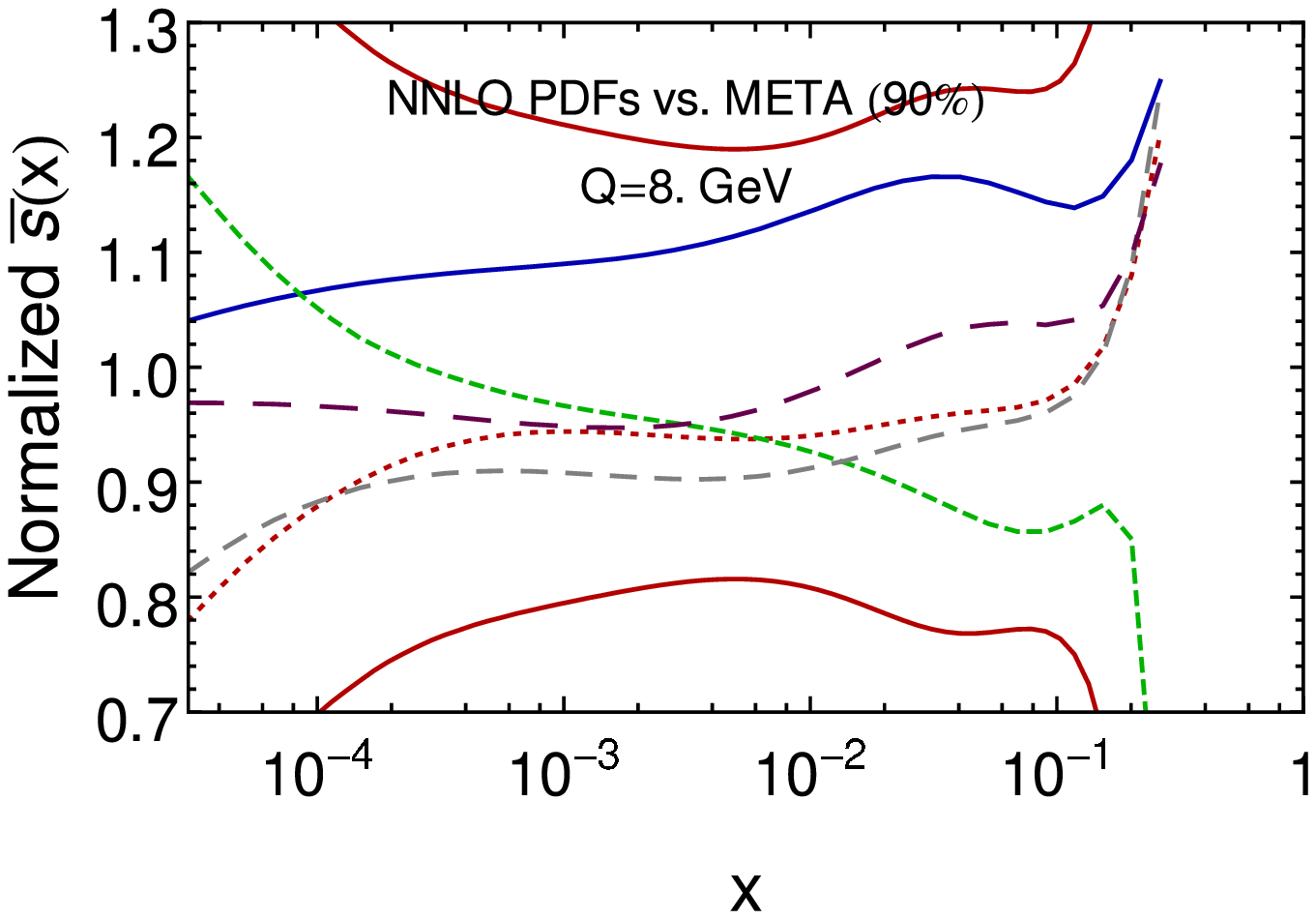} \includegraphics[width=0.32\textwidth]{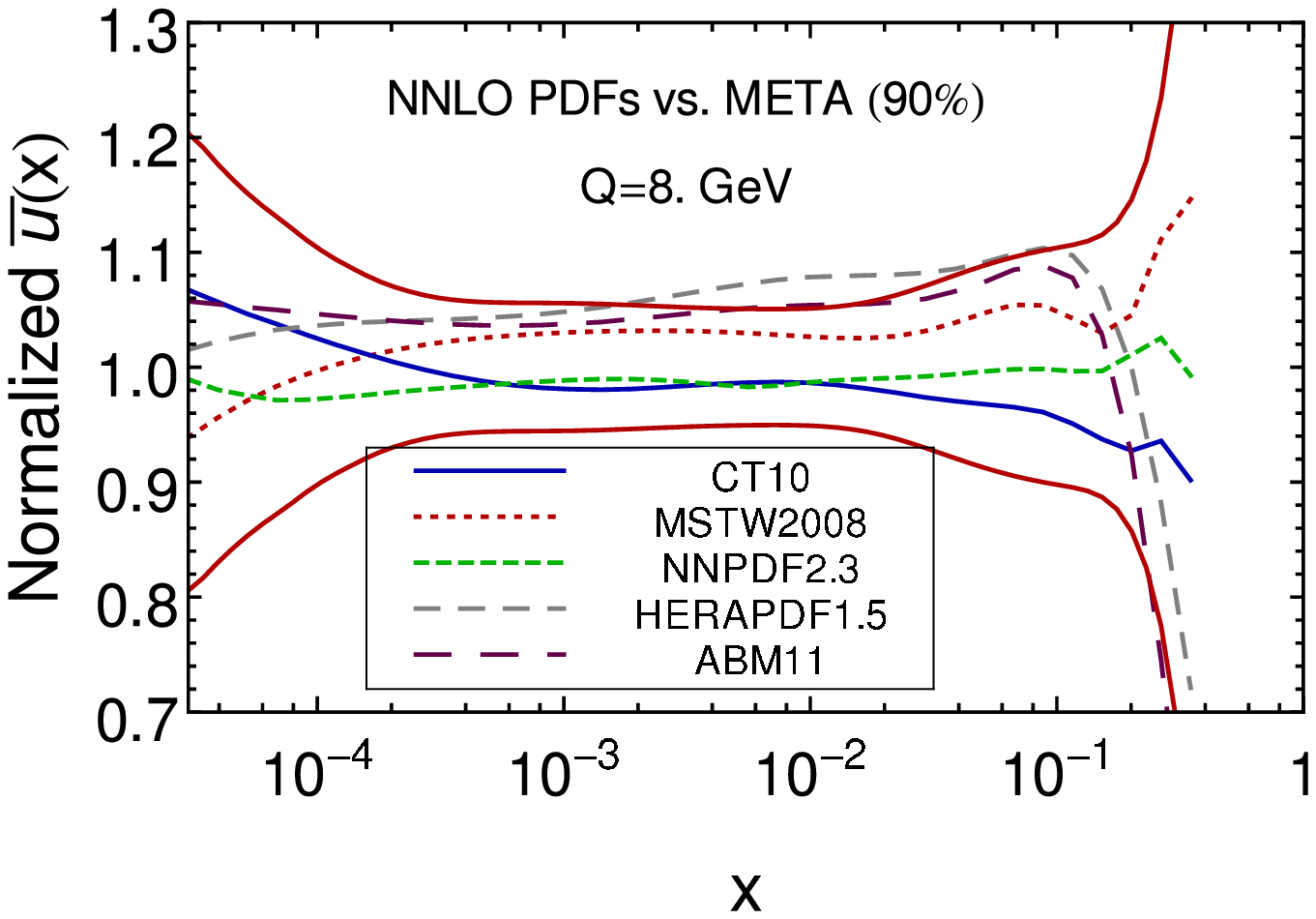}
\includegraphics[width=0.32\textwidth]{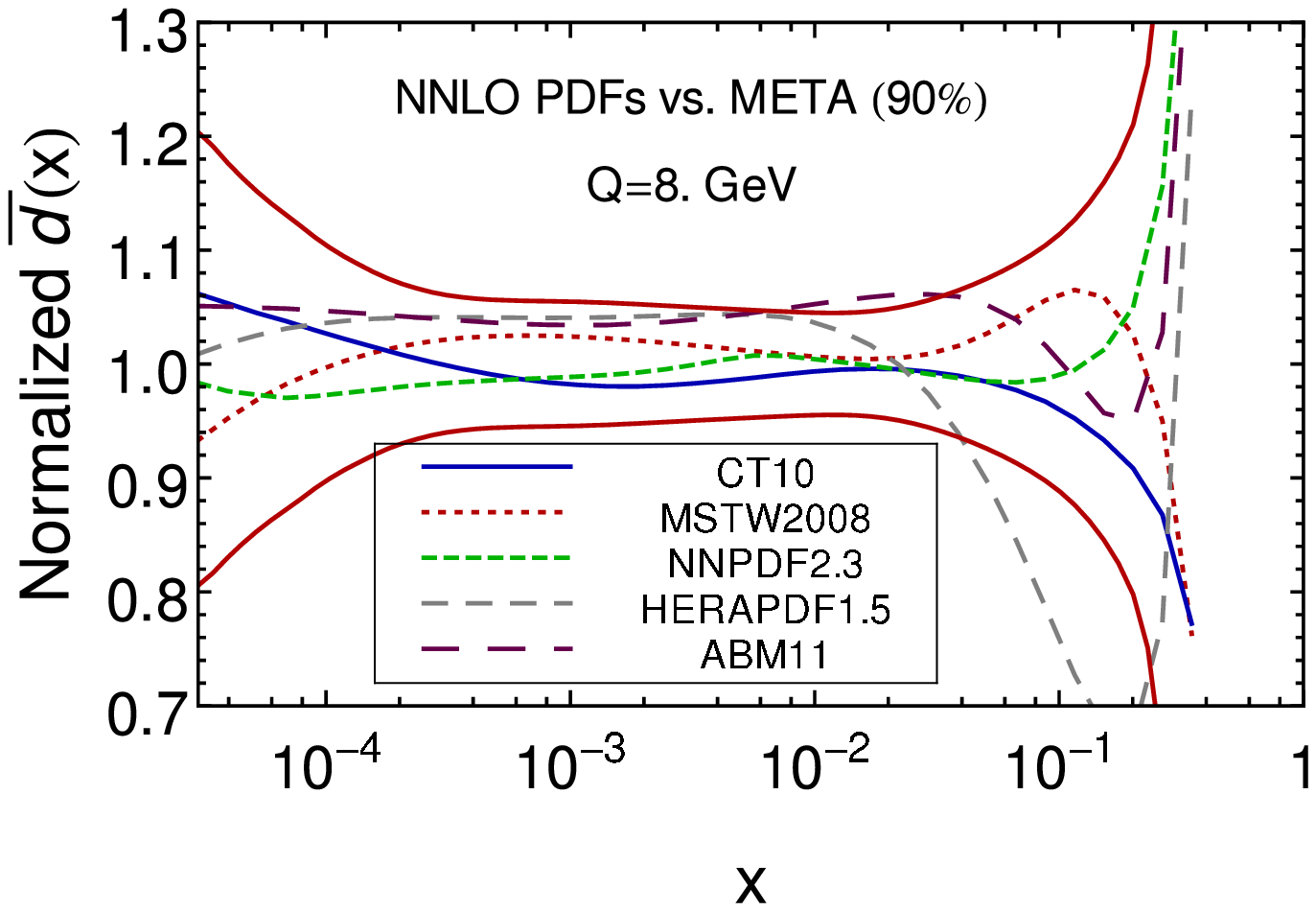} \\
 \includegraphics[width=0.32\textwidth]{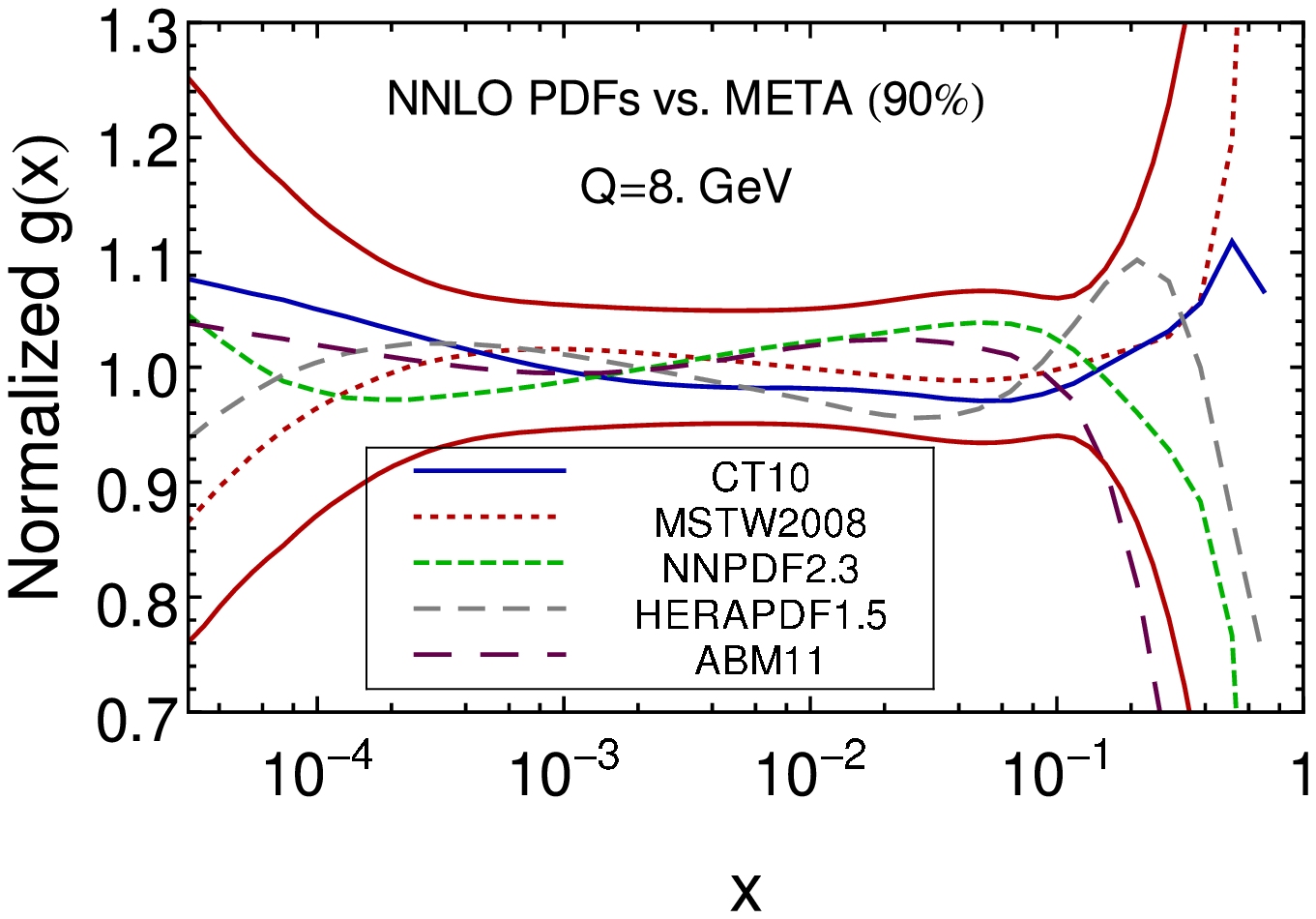} \includegraphics[width=0.32\textwidth]{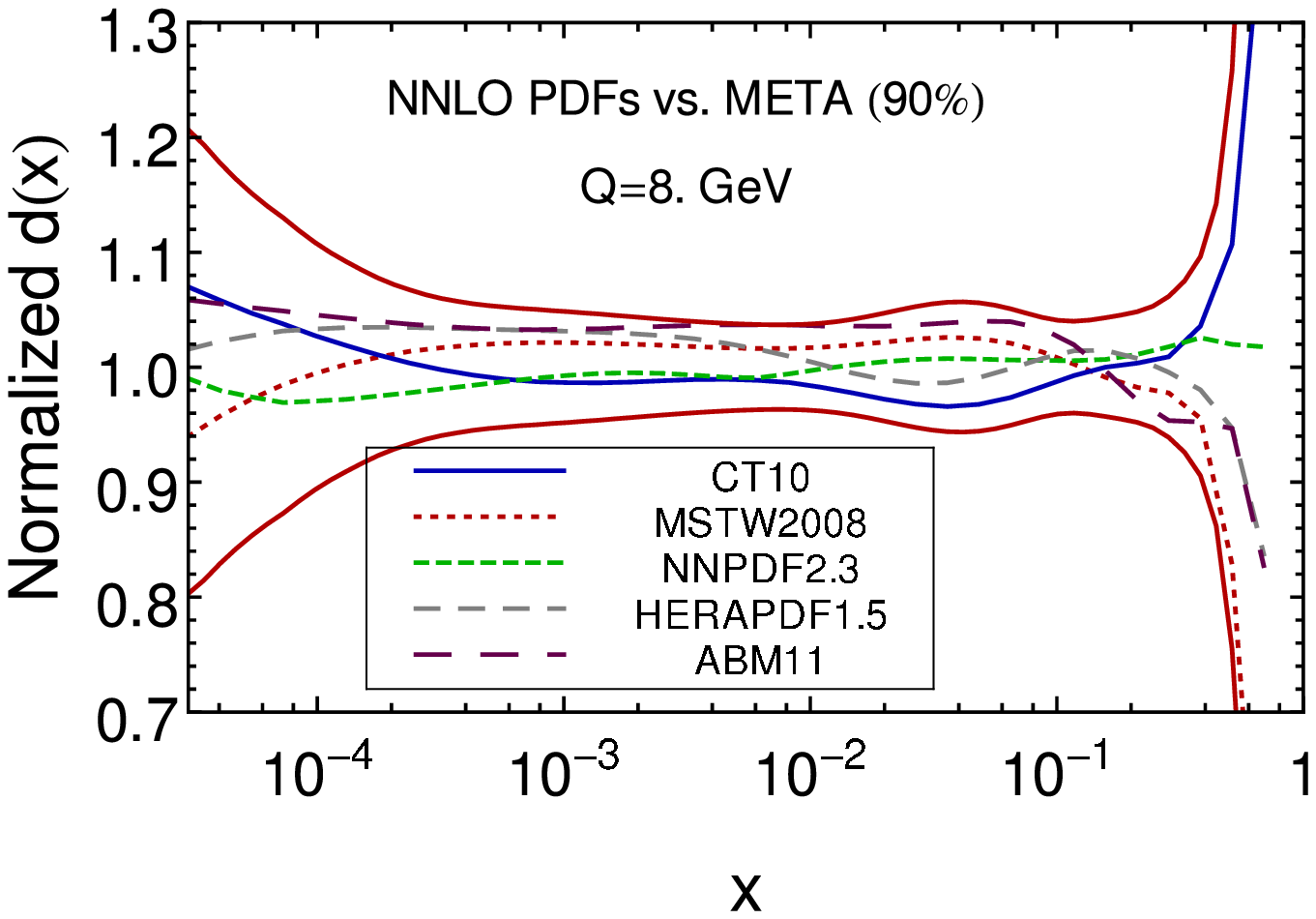}
\includegraphics[width=0.32\textwidth]{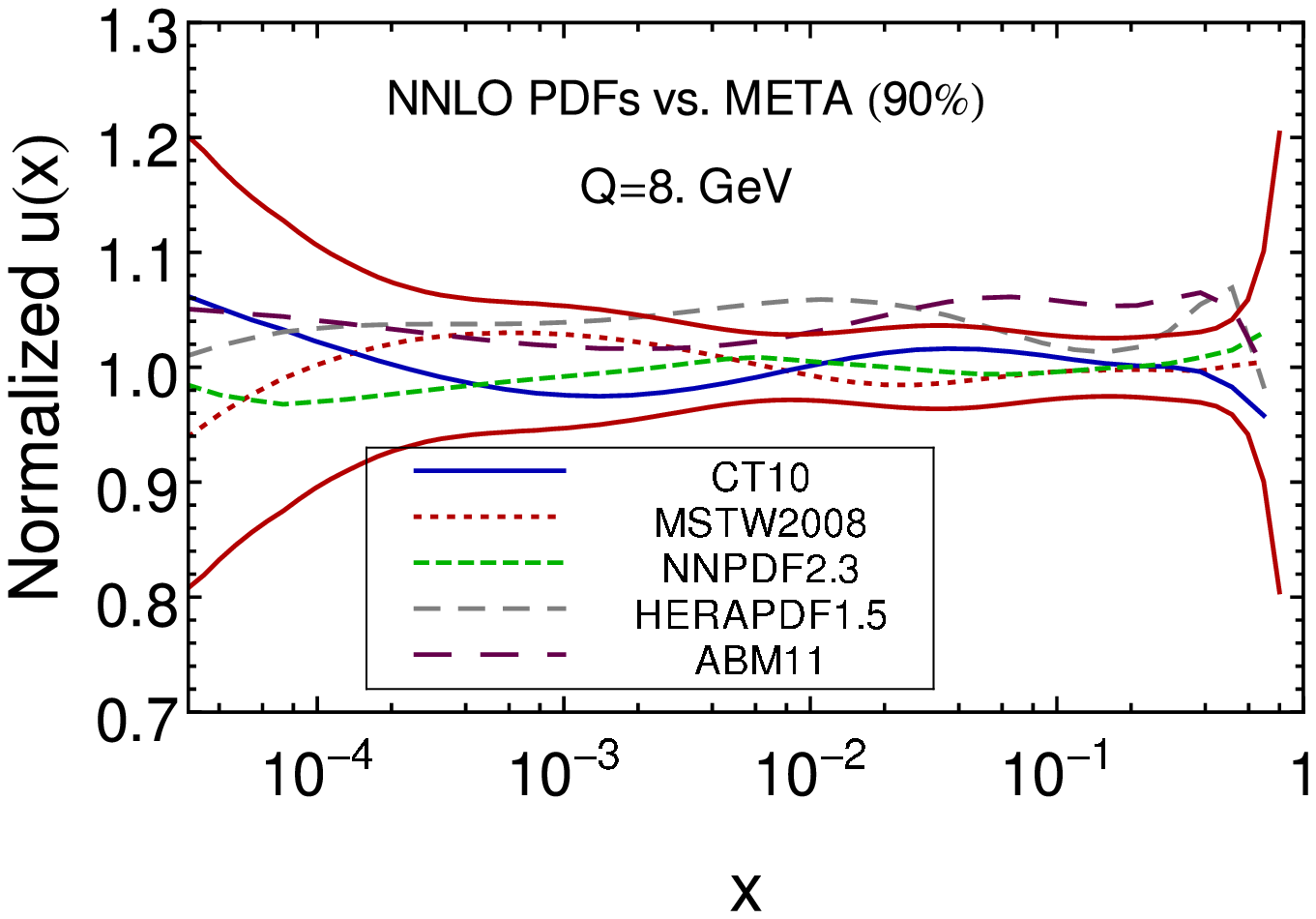} \\

\par\end{centering}

\begin{centering}
\includegraphics[width=0.32\textwidth]{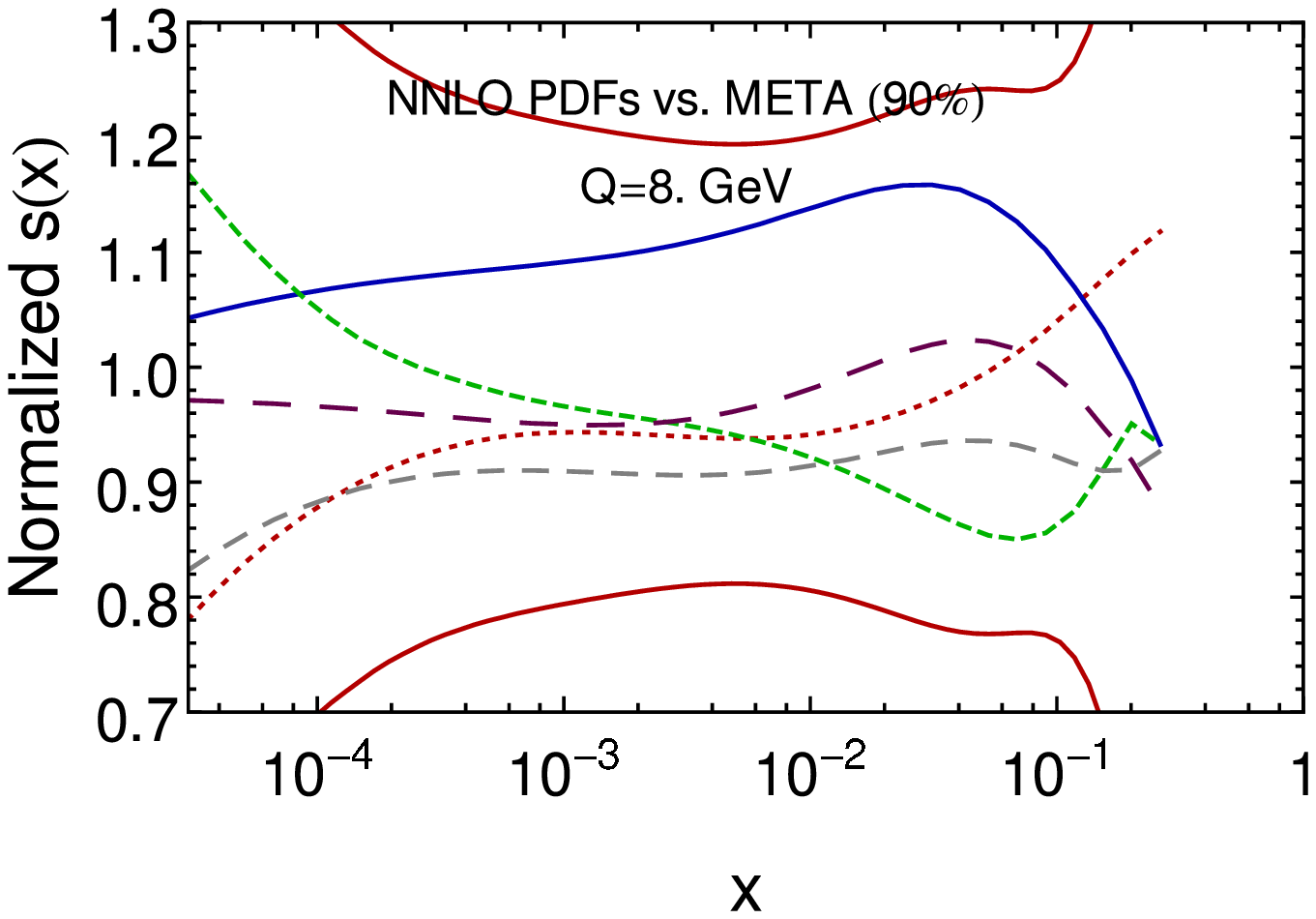} \includegraphics[width=0.32\textwidth]{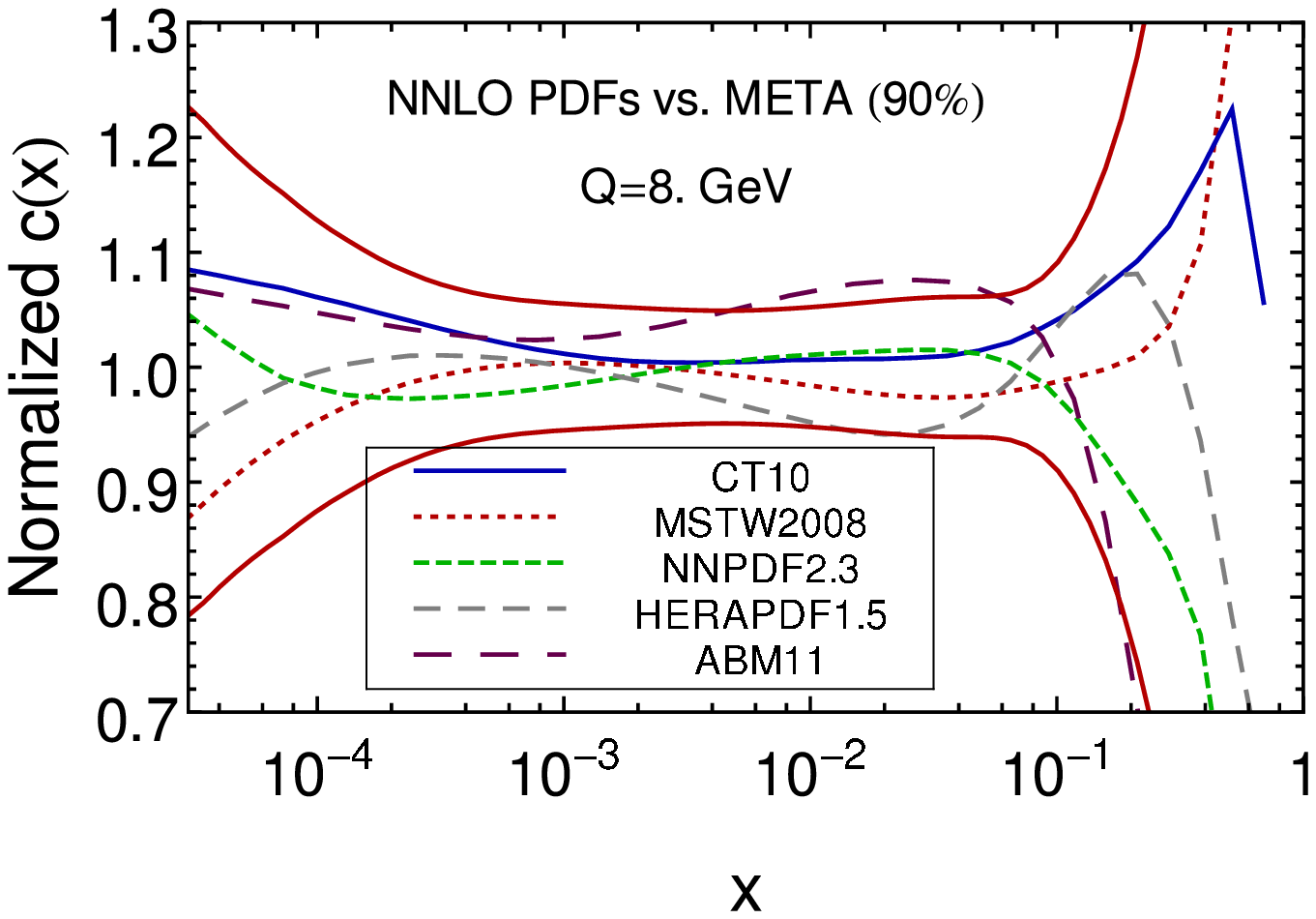}
\includegraphics[width=0.32\textwidth]{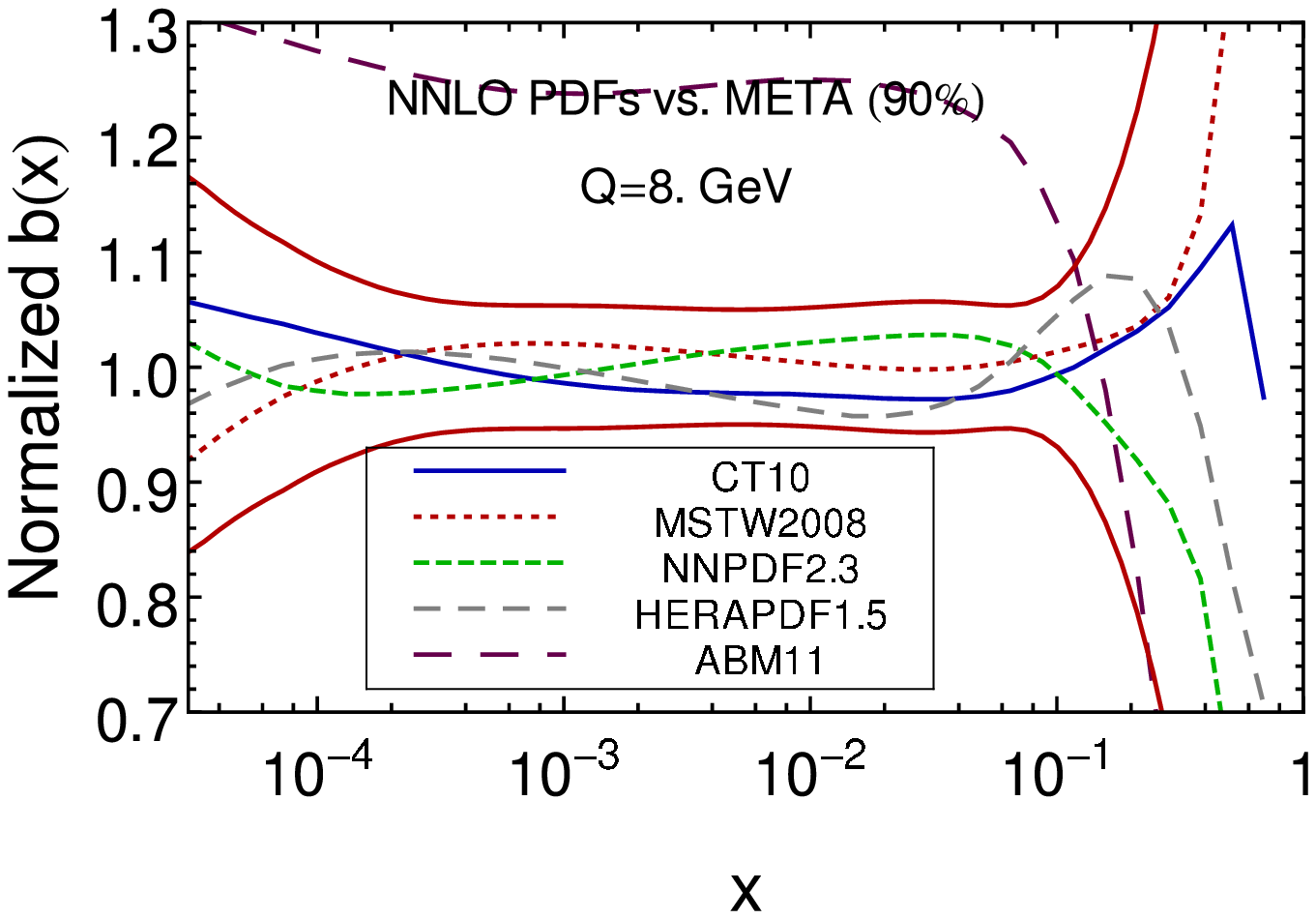}
\par\end{centering}

\vspace{-1ex}
 \caption{\label{fig:bench2} Comparison of the META PDFs and all the best-fit
NNLO PDFs with $\alpha_{s}(M_{Z})=0.118$ at a common scale $Q=8\ {\rm GeV}$.}
\end{figure}

\begin{figure}[htb]
\begin{centering}
\includegraphics[width=0.32\textwidth]{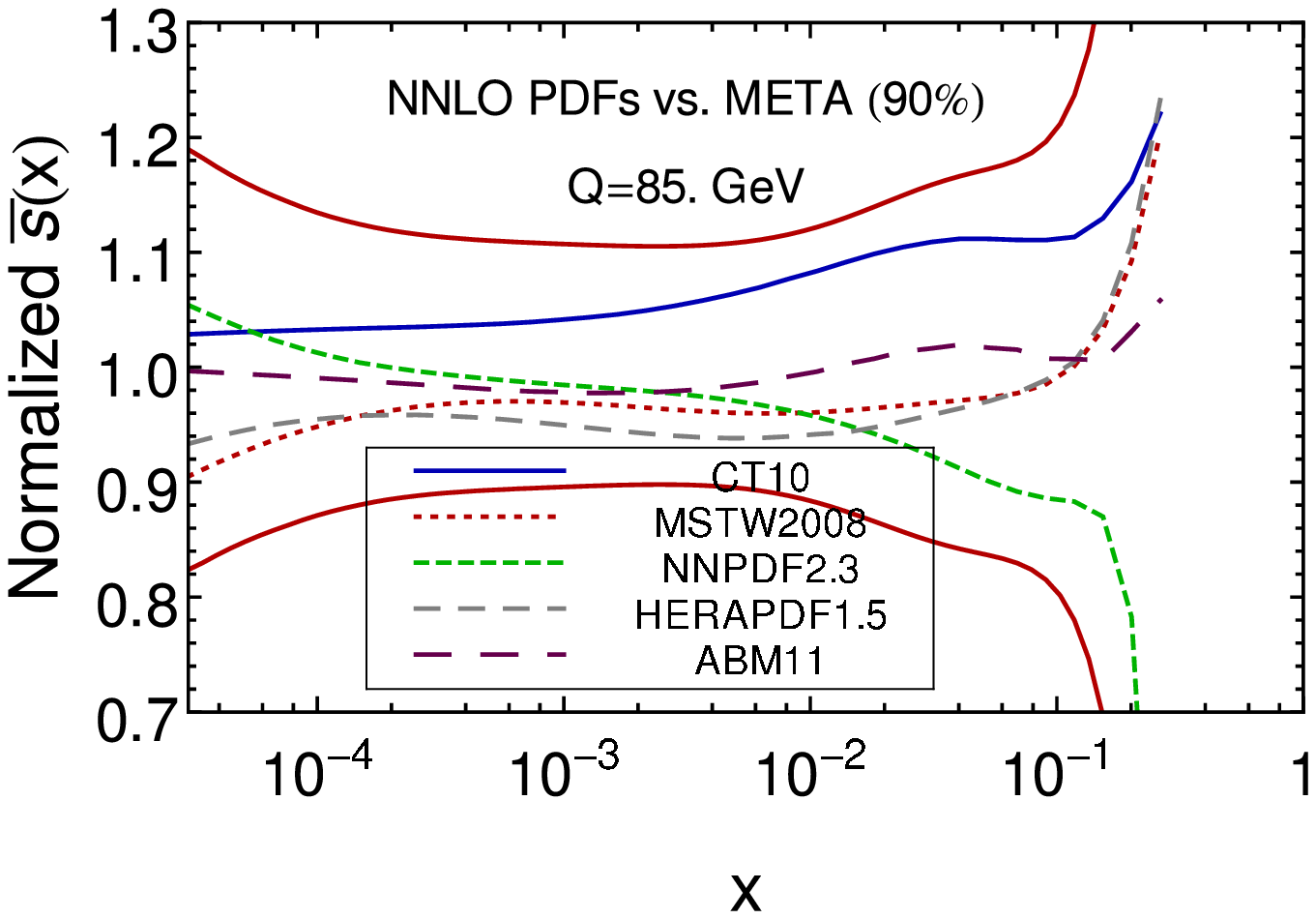} \includegraphics[width=0.32\textwidth]{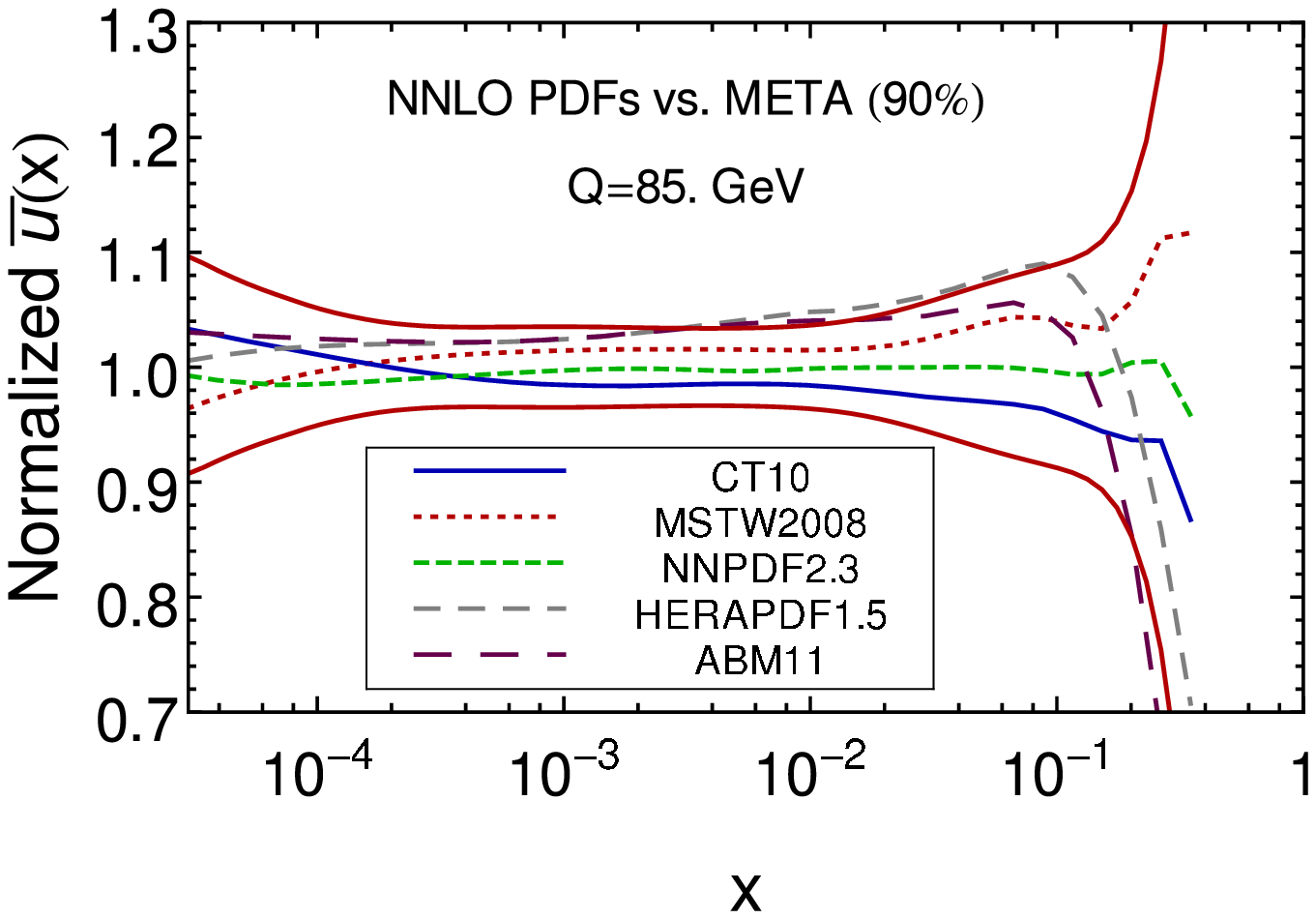}
\includegraphics[width=0.32\textwidth]{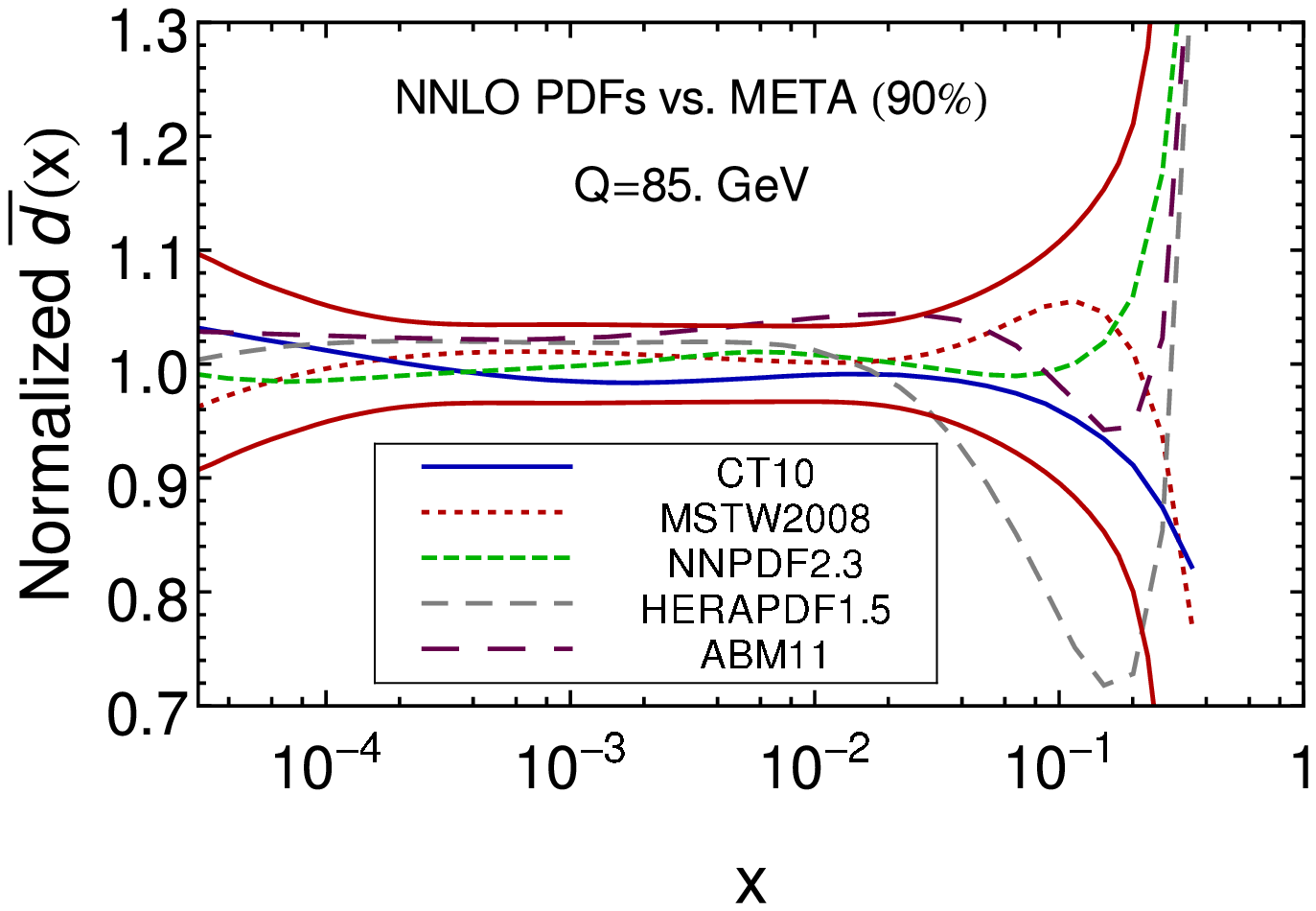} \\
 \includegraphics[width=0.32\textwidth]{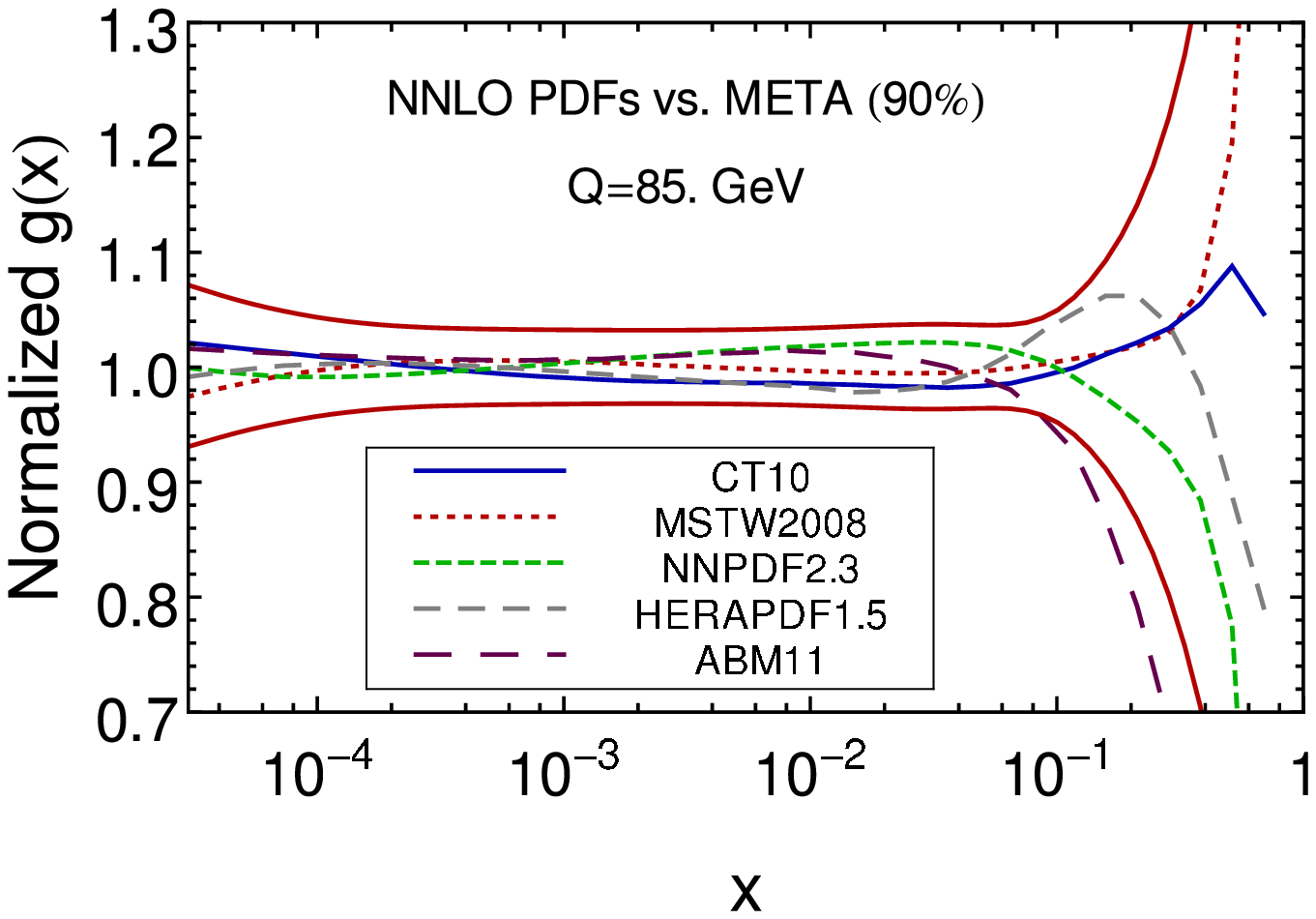} \includegraphics[width=0.32\textwidth]{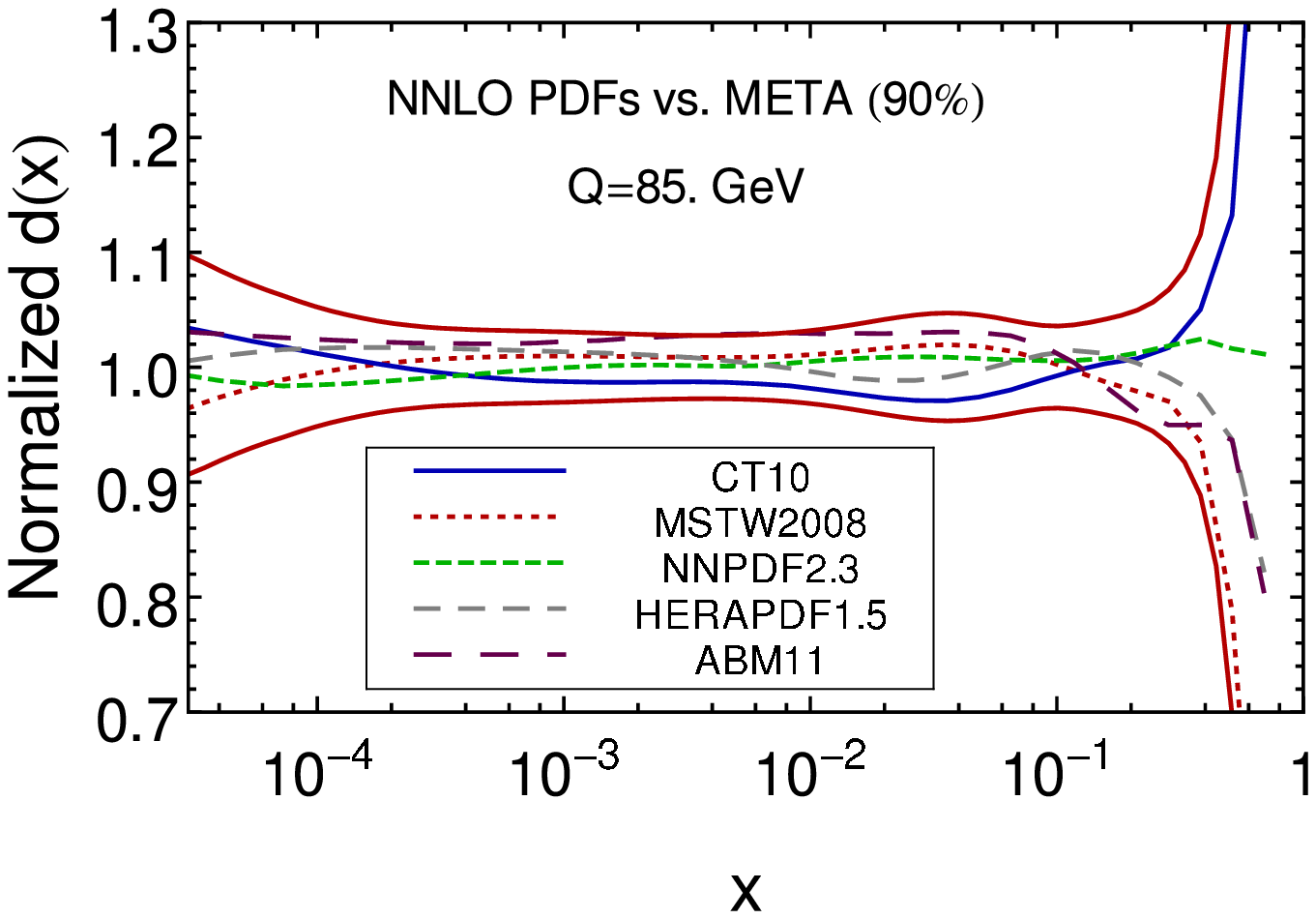}
\includegraphics[width=0.32\textwidth]{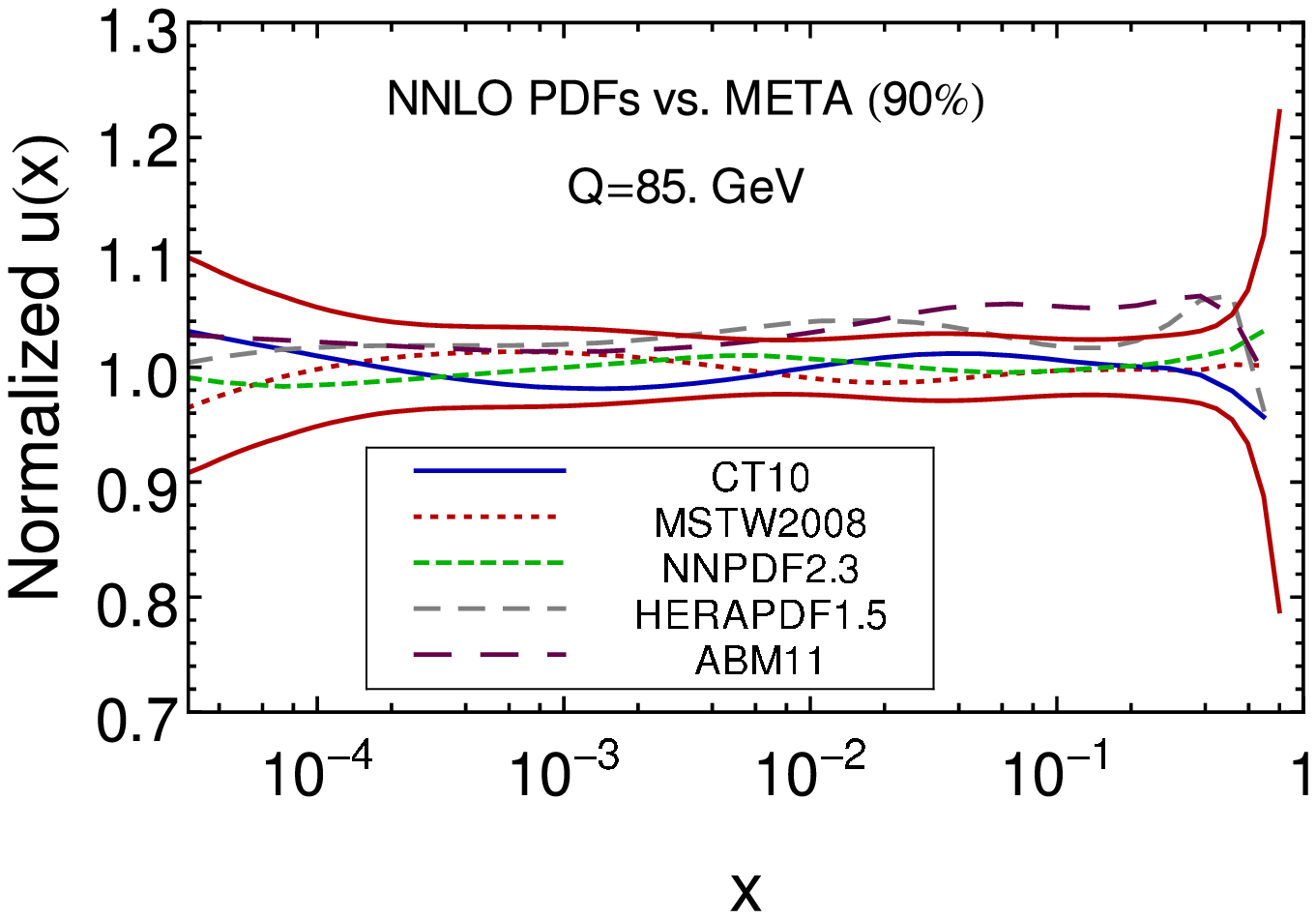} \\

\par\end{centering}

\begin{centering}
\includegraphics[width=0.32\textwidth]{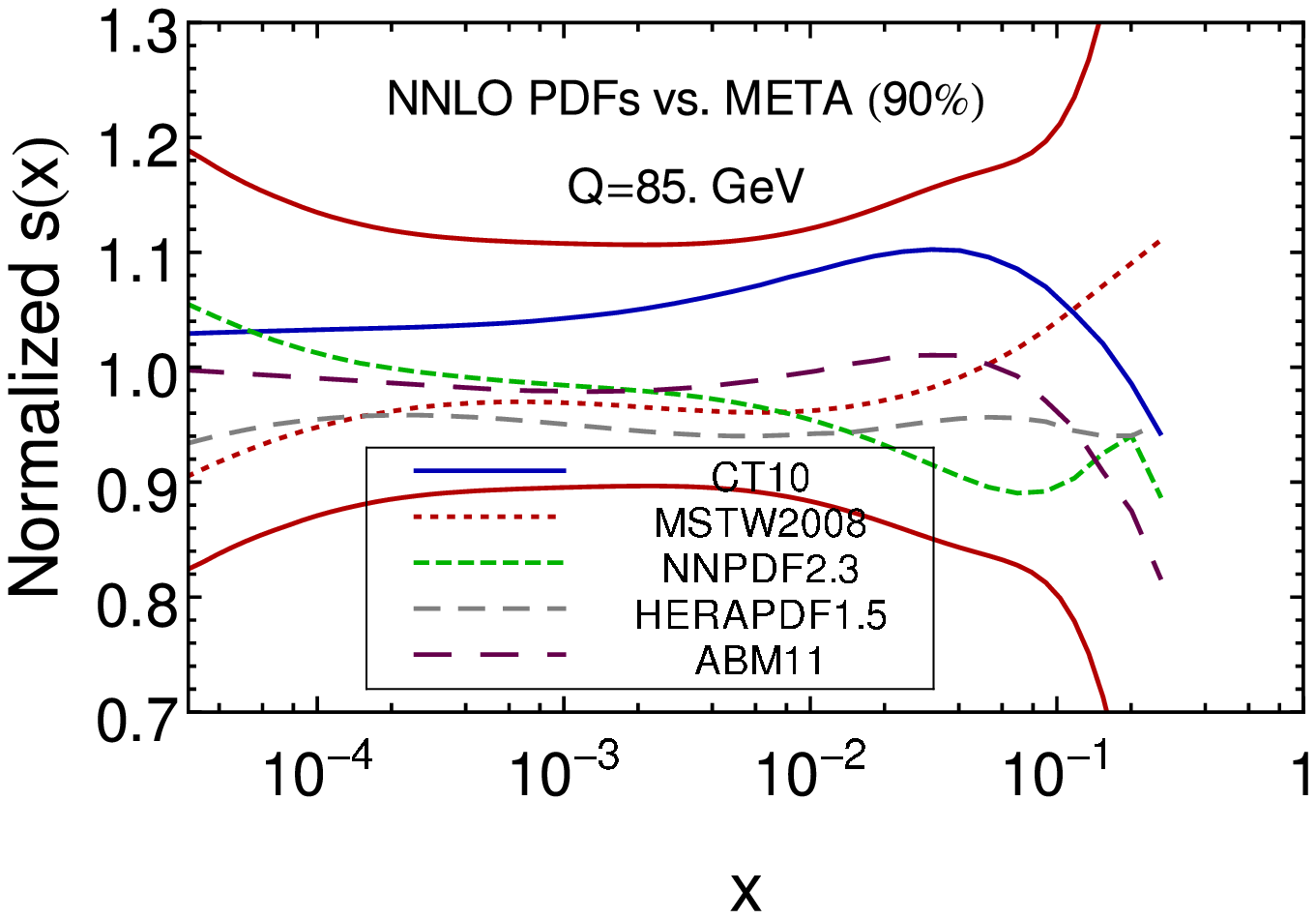} \includegraphics[width=0.32\textwidth]{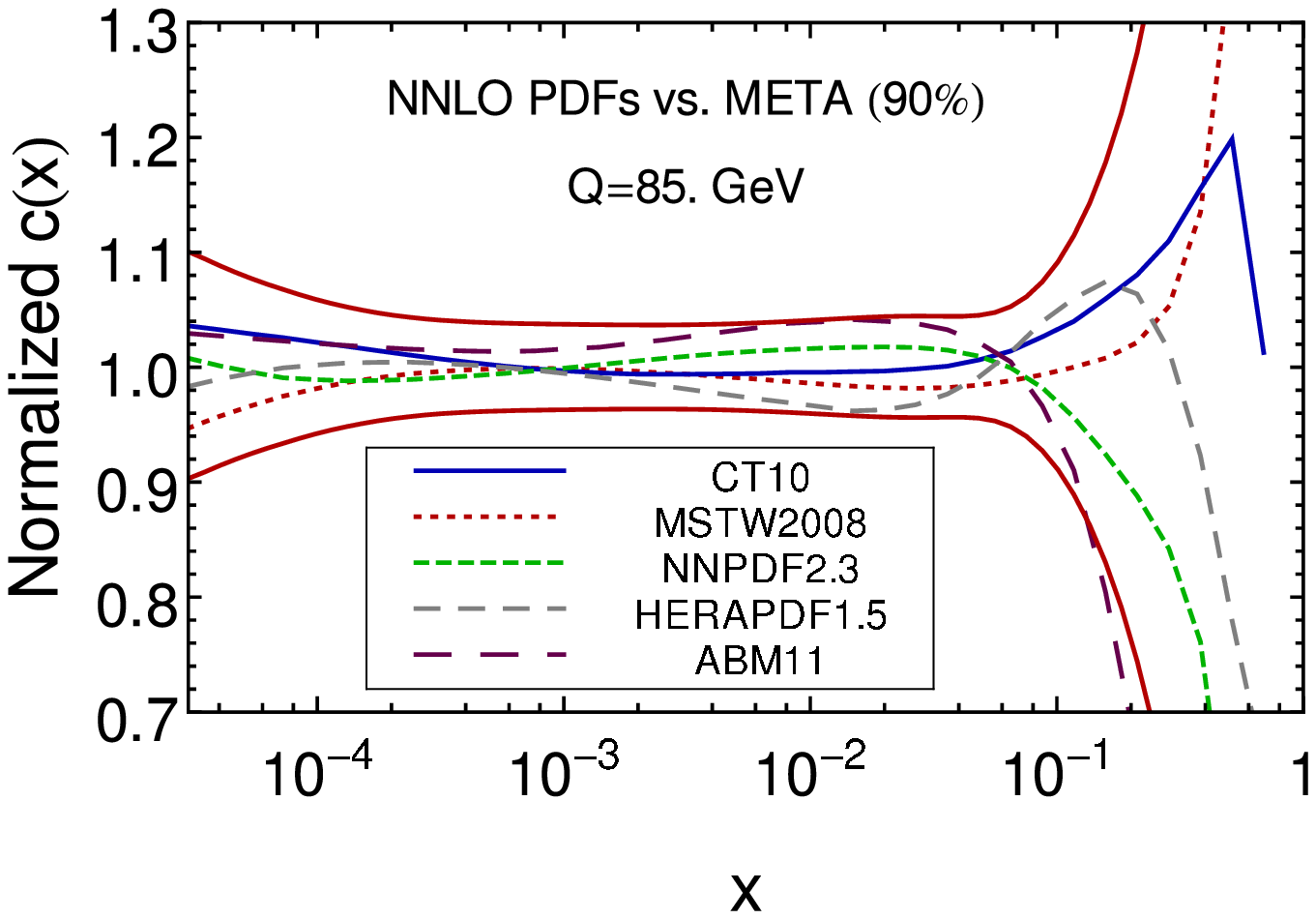}
\includegraphics[width=0.32\textwidth]{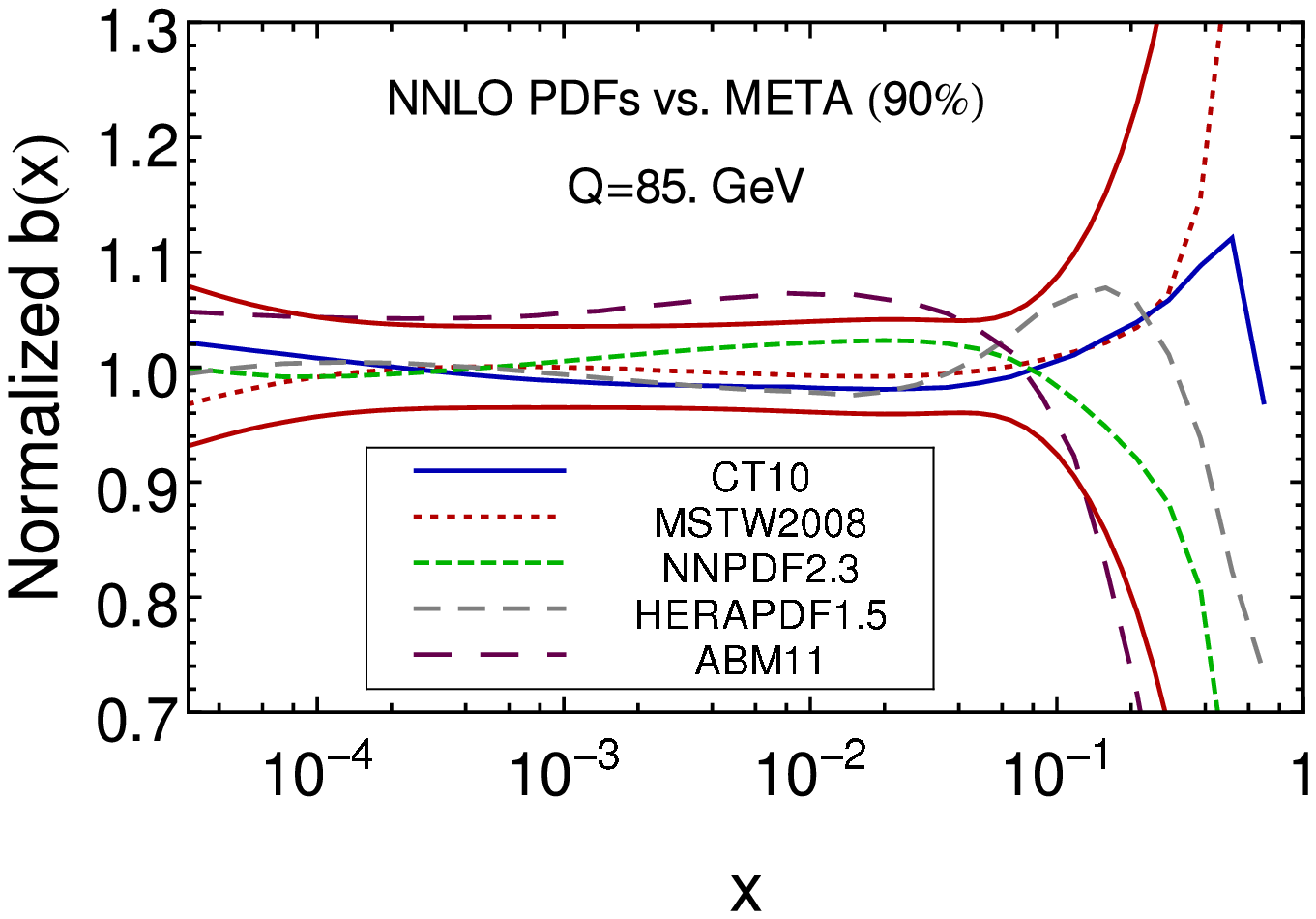}
\par\end{centering}

\vspace{-1ex}
 \caption{\label{fig:bench3} Same as Fig.~\ref{fig:bench2}, for $Q=85\ {\rm GeV}$.}
\end{figure}

Going back to the $x$ space representation, in Figs.~\ref{fig:bench2} and \ref{fig:bench3}
we compare the 90\% c.l. error band of META PDFs with the
central PDFs with $\alpha_{s}(M_{Z})=0.118$ from five groups, all
normalized to the central META
PDF. Again, the central PDFs of three global ensembles lie within the 90\%
c.l. uncertainty bands of the META PDF. The ABM and HERA PDFs for
$u$, $\bar u$, $\bar d$, $g$ flavors can be outside of the META
bands in some $x$ regions. The largest deviation is observed in the
bottom quark PDF for ABM11, which is very different at the scale 8
GeV compared to the other ensembles due to the different treatment
of heavy quarks. This difference of the bottom distribution is largely
reduced as going to higher scales like 85 GeV.

\subsection{Combination with the $\alpha_{s}$ uncertainty}

To estimate the PDF+$\alpha_{s}$ uncertainty using the META PDF
ensemble, we provide a series of additional meta-sets for
$\alpha_{s}(M_{Z})\neq 0.118$, obtained by taking the average of the
corresponding PDFs from the three groups. Depending on the PDF
analysis, $\alpha_{s}(M_{Z})$ is either treated as an external
parameter determined by its world average (as in CT10
\cite{Lai:2010nw}) or fitted together with the PDFs (as in
MSTW~\cite{Martin:2009bu} or NNPDF~\cite{Ball:2011us}). In the
latter case, the best-fit value of $\alpha_{s}(M_{Z})$ from the PDF
fit at NNLO (0.1171 from MSTW, 0.1173 from NNPDF, and 0.1132 from ABM12) tends to be
lower than the latest world average of $0.1184\pm0.0007$ at 68\%
c.l.~\cite{Beringer:1900zz}. The PDF4LHC
recommendation~\cite{Botje:2011sn,Alekhin:2011sk} suggests a
slightly larger error of $0.0012$ at 68\% c.l. than in the world
average. Note that the PDF4LHC recommendation also takes different
central values of $\alpha_s(M_Z)$ for different PDF ensembles, which
enlarges the $\alpha_s$ uncertainty in the end.

Given these options, one can come up with several ways for
calculating the combined PDF+$\alpha_s$ uncertainty, depending on
the interpretation and statistical confidence assigned to the
$\alpha_s$ input. The combined uncertainty may vary significantly
depending on the prescription, especially for the QCD observables
that are directly sensitive to $\alpha_s$ or the gluon PDF. Here we
suggest several practical possibilities for estimating the
PDF+$\alpha_s$ uncertainty using the META PDFs.

\begin{enumerate}

\item We may take the world-average $\alpha_s$ value
as an external input for the PDF analysis and assume that the
$\alpha_s$ uncertainty is decoupled from the systematic errors in
the fit due to missing higher-order effects. In this approach, the
68\% c.l. error on $\alpha_s(M_Z)$ is given by the world average of
$\pm0.0007$. We vary $\alpha_s$ in this range when estimating the
$\alpha_s$ uncertainty on a QCD observable and add it to the PDF
uncertainty in quadrature, which produces the PDF+$\alpha_s$
uncertainty with full correlations, as has been discussed by CTEQ
group~\cite{Lai:2010nw}.

\item An alternative prescription is to replace the
input $\alpha_s(M_Z)$ error by $0.0012$ at 68\% c.l. as in the
PDF4LHC recommendation~\cite{Botje:2011sn,Alekhin:2011sk}, and add
the $\alpha_s$ and PDF uncertainties in quadrature. The resulting
PDF+$\alpha_s$ uncertainty will be smaller than the one from the
PDF4LHC recommendation, as a single value of 0.118, rather than the
PDF4LHC envelope of central $\alpha_s$ values from all ensembles, is
used in the META case.

\item Lastly, we can enlarge the input $\alpha_s(M_Z)$ error
to $0.002$ at 68\% c.l. and add the $\alpha_s$ and PDF uncertainties
linearly. This prescription fully covers the preferred $\alpha_s$ values by
different groups and is numerically close to the PDF+$\alpha_s$
uncertainty based on the PDF4LHC envelope.

\end{enumerate}

The first two conventions predict smaller
$\alpha_s$ and PDF+$\alpha_s$ uncertainties than the envelope
prescription used by the 2010 PDF4LHC study
\cite{Botje:2011sn,Alekhin:2011sk} and 2012 PDF benchmarking
study \cite{Ball:2012wy}.
In these cases the PDF uncertainty always dominates in the combined
uncertainty. The third convention
turns out to be numerically close to the envelope prescription. In
our remaining comparisons the PDF+$\alpha_s$ uncertainties are estimated
based on the third convention.

\section{Phenomenological applications \label{sec:lhc}}

The META PDF ensemble incorporates measurements from HERA,
fixed-target experiments, and Tevatron. Although the NNPDF2.3 ensemble
used here includes some of the early LHC data in the fit, like jet
production cross sections~\cite{Aad:2011fc}, $W$ and $Z$ rapidity
distributions~\cite{Aad:2011dm,Aaij:2012vn}, and $W$ electron charge
asymmetry~\cite{Chatrchyan:2012xt}, we do not expect these
relatively weak LHC constraints to impact strongly the META PDFs'
behavior. Thus the META ensemble can be used for any LHC predictions
based on non-LHC measurements.

Before discussing the LHC applications, we would like to review two important
features of the META PDFs. First, as mentioned earlier, the sum rule
constraints are not enforced directly in the meta-fit. After obtaining
our final META PDFs, we examined the partial integrals for the
momentum and valence sums in the same
fashion as in Fig.~\ref{fig:sum}. 
The sum rules are obeyed by the META PDFs at the accuracy that is
similar, or better than, the individual PDF ensembles
\cite{metapdfweb}. 

\begin{figure}[htb]
  \begin{center}
  \includegraphics[width=0.48\textwidth]{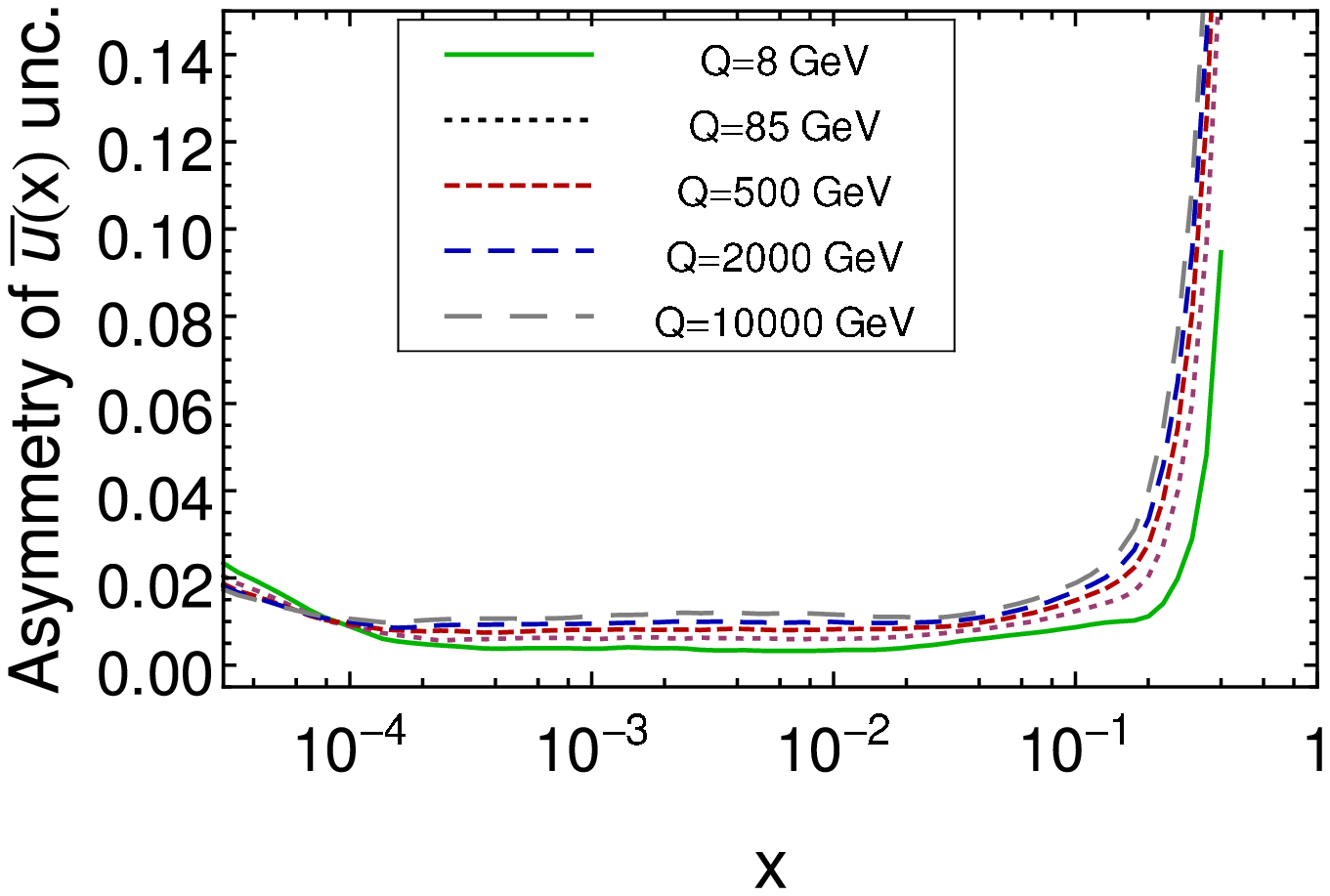}\hspace{10pt}
  \includegraphics[width=0.48\textwidth]{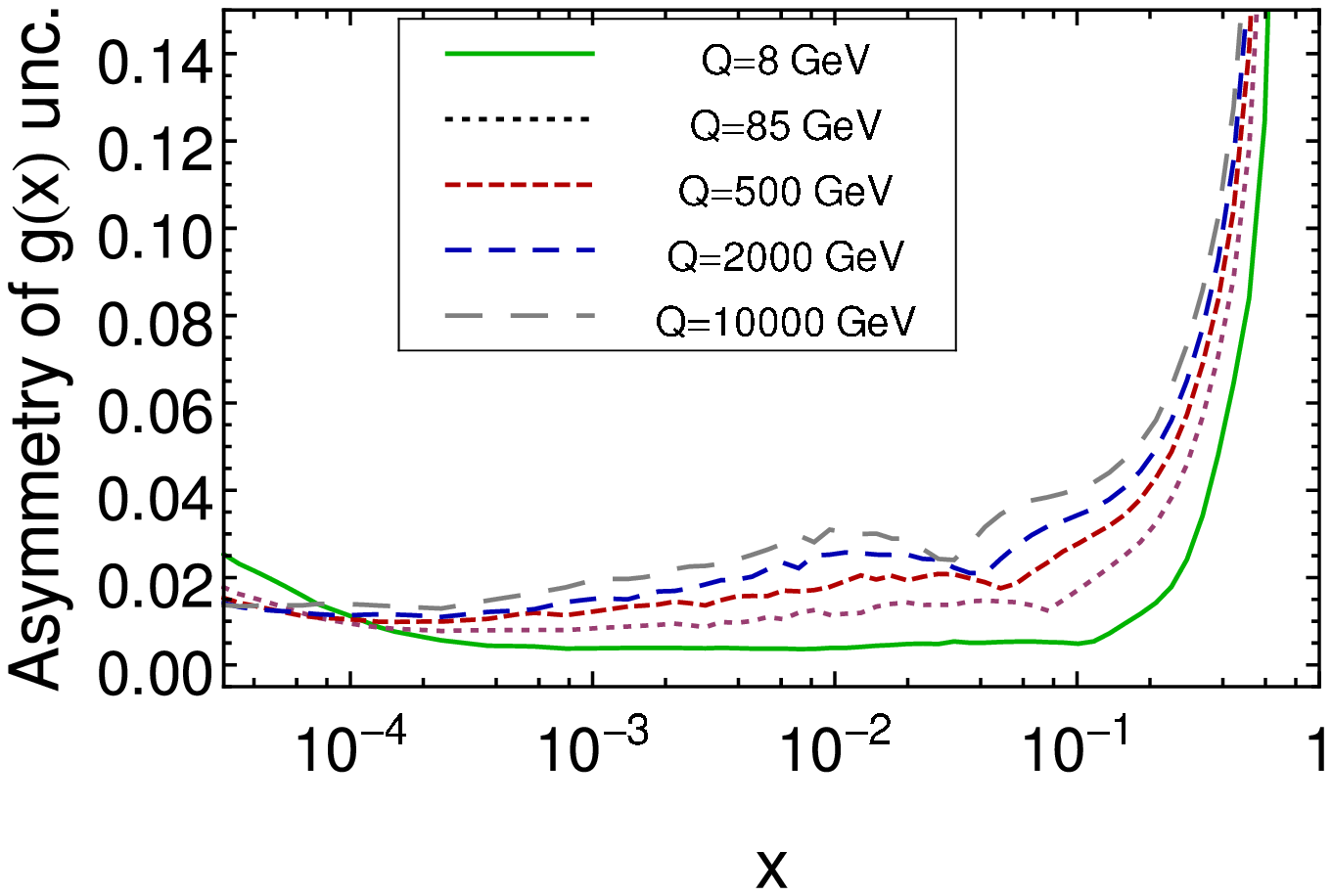}\\
  \includegraphics[width=0.48\textwidth]{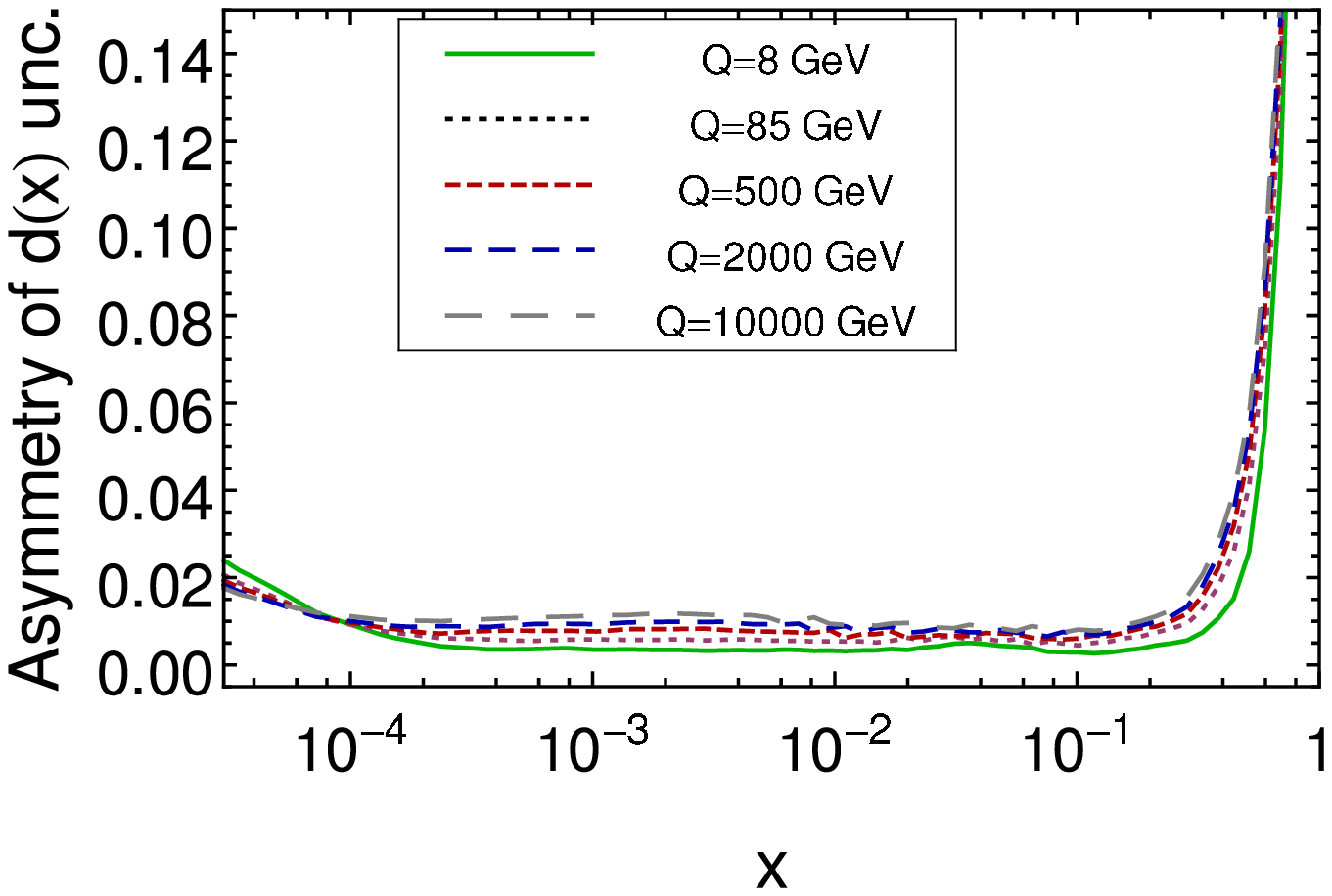}\hspace{10pt}
  \includegraphics[width=0.48\textwidth]{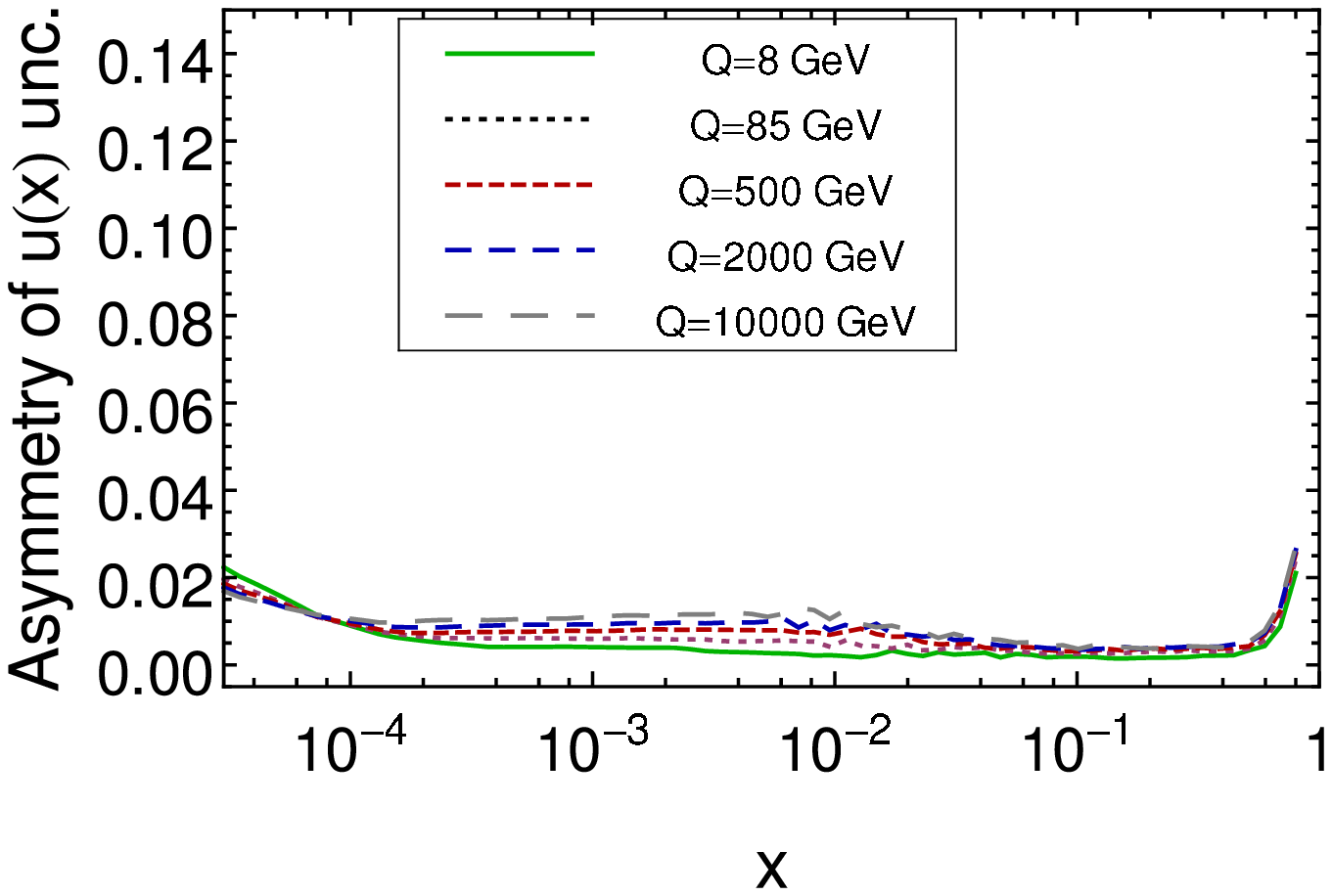}
  \end{center}
  \vspace{-1ex}
  \caption{\label{fig:lin1} Asymmetry of PDF errors of different flavors for
  the META PDFs at different $Q$ values.}
\end{figure}

\begin{figure}[htb]
  \begin{center}
  \includegraphics[width=0.48\textwidth]{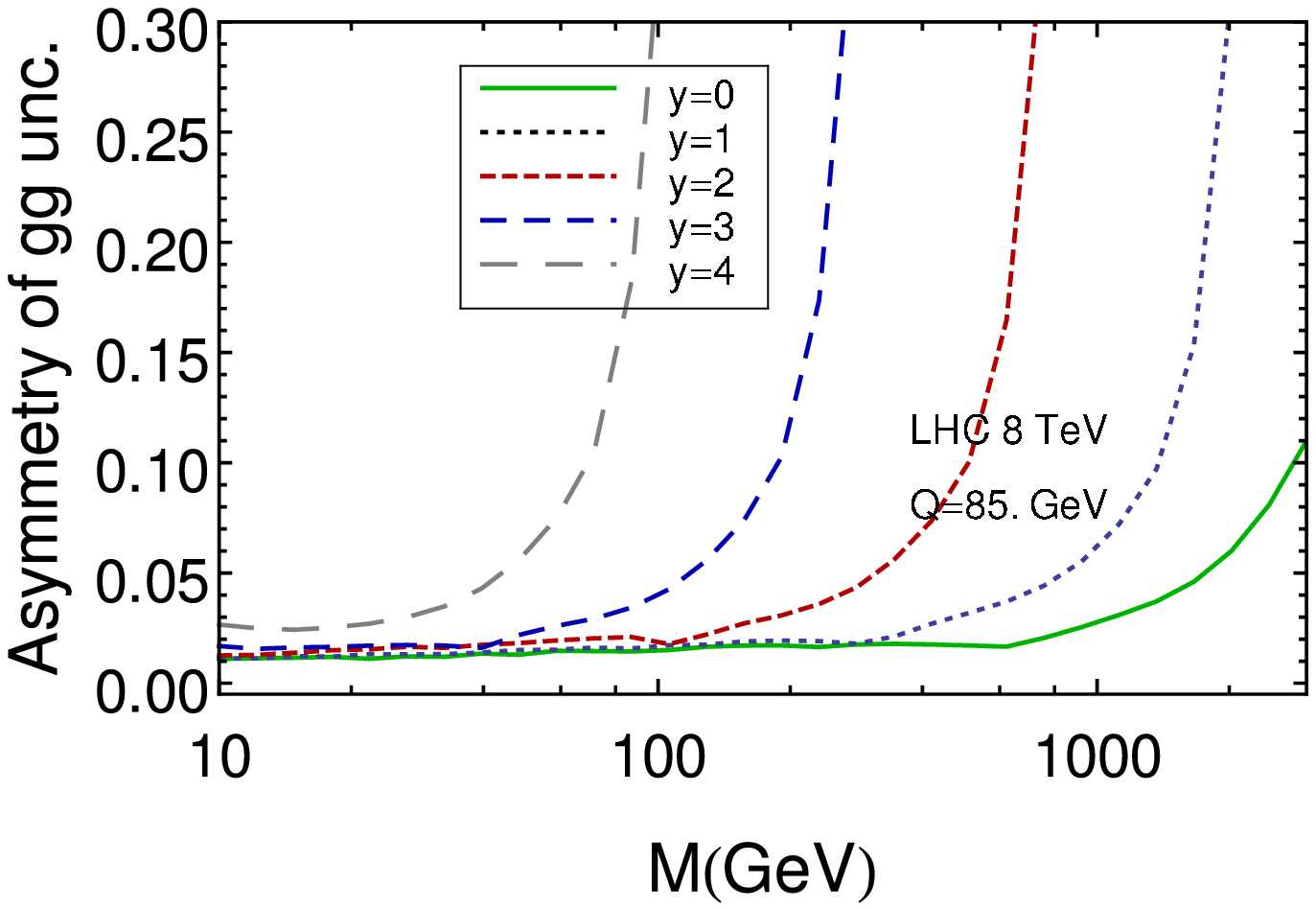}\hspace{10pt}
  \includegraphics[width=0.48\textwidth]{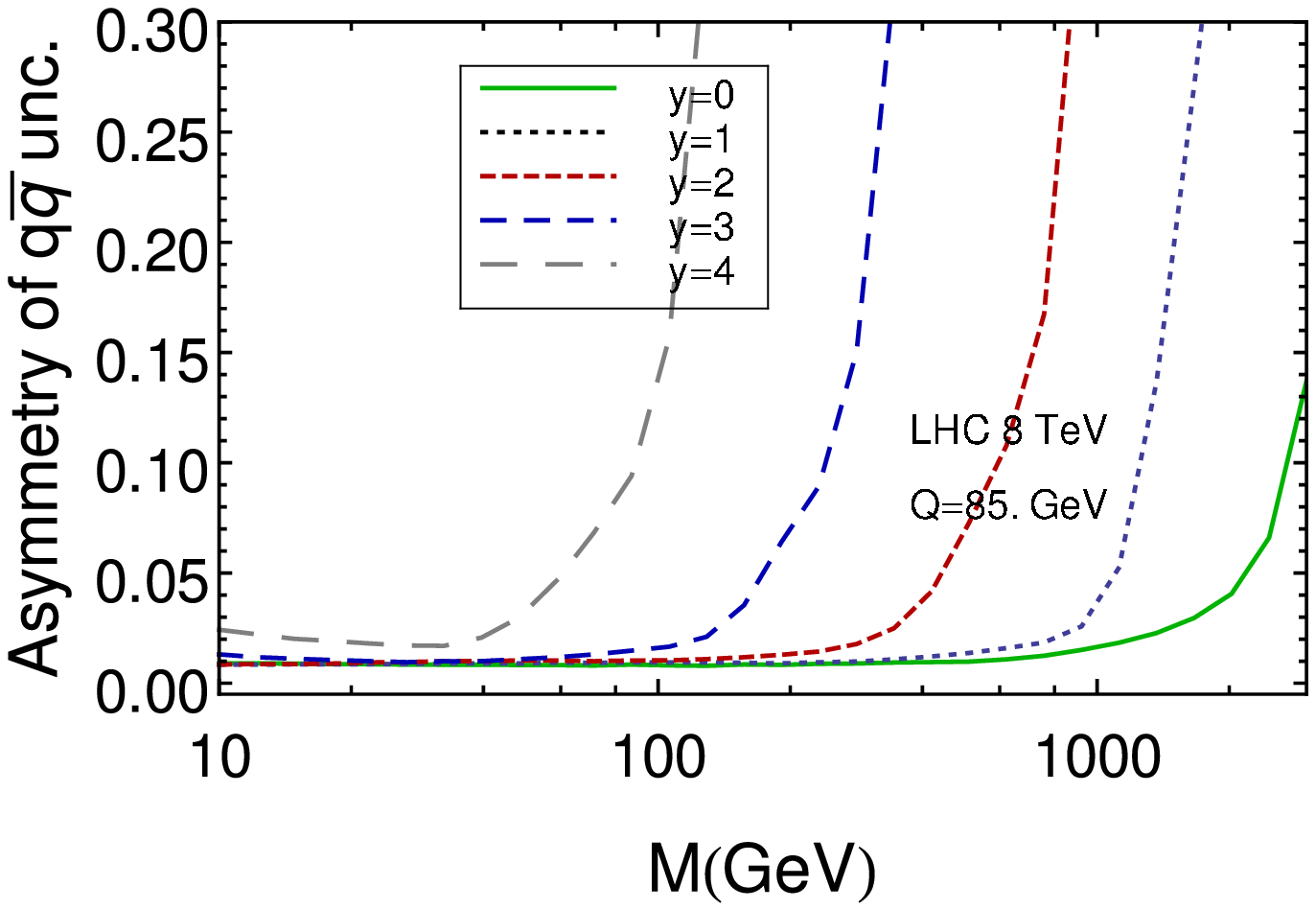}\\
  \includegraphics[width=0.48\textwidth]{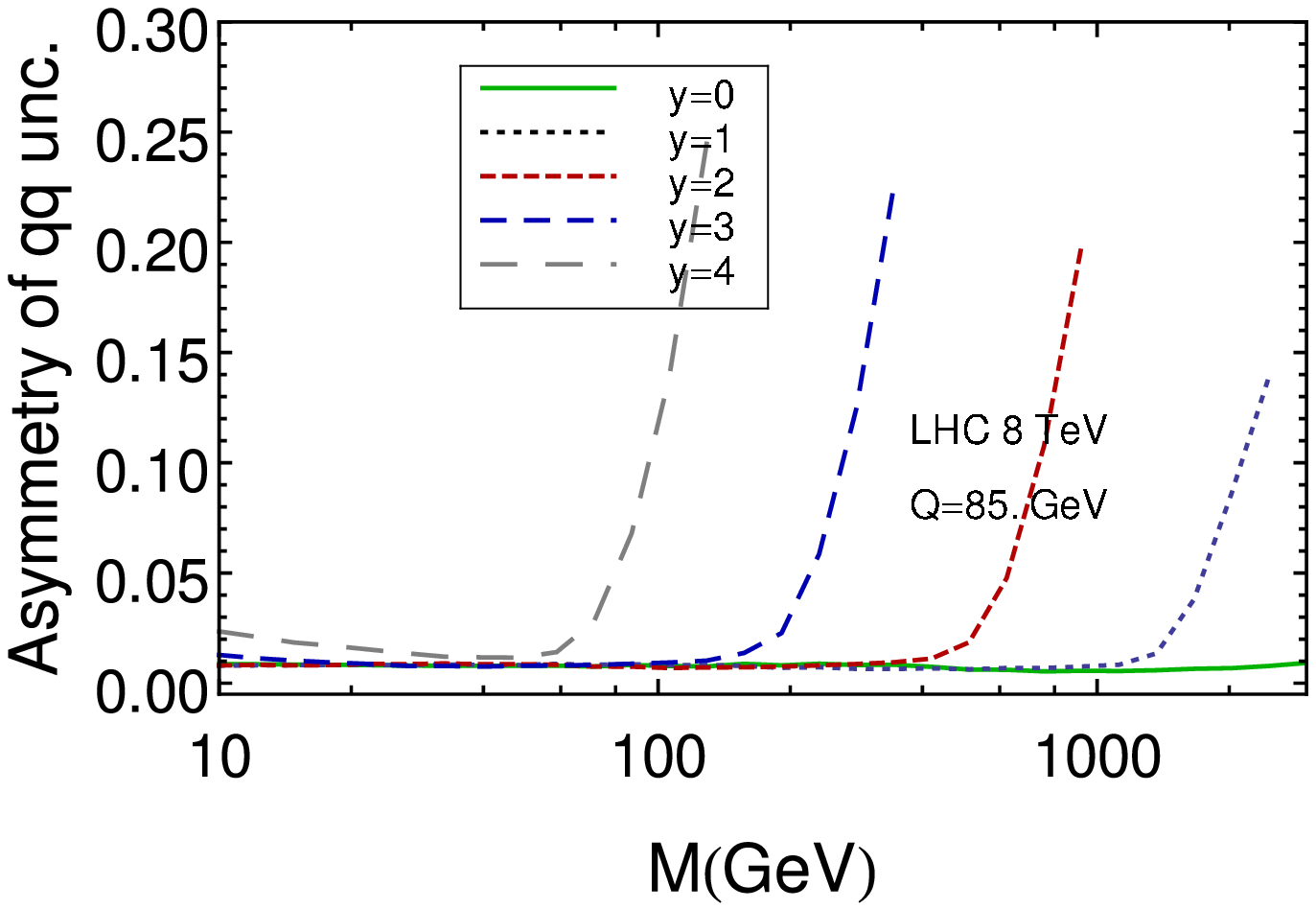}\hspace{10pt}
  \includegraphics[width=0.48\textwidth]{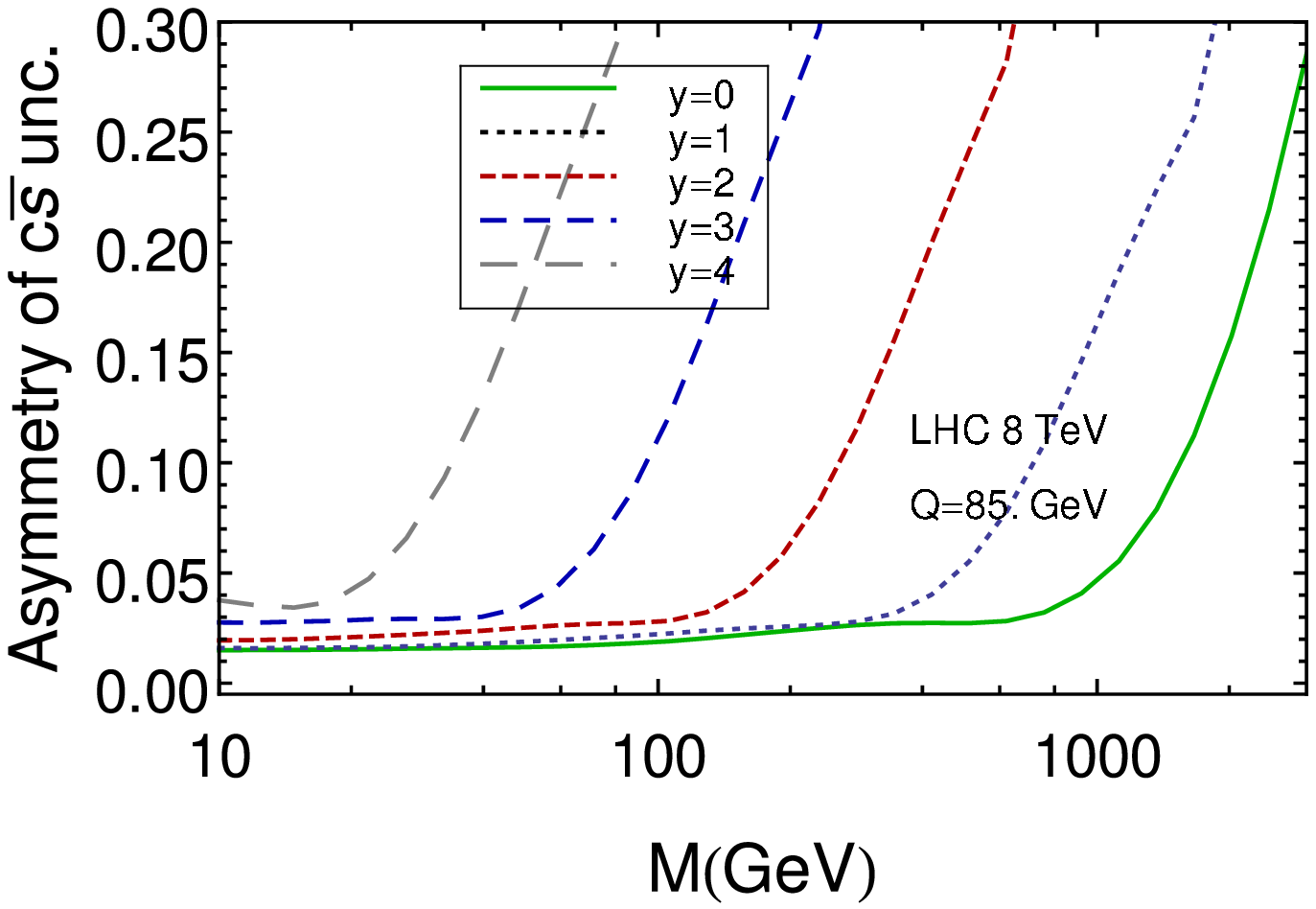}
  \end{center}
  \vspace{-1ex}
  \caption{\label{fig:lin2} Asymmetry of PDF errors of parton luminosities
  for the META PDFs.}
\end{figure}

The Hessian method of error propagation relies on the assumption of
the approximately linear dependence of QCD observables on the PDF
parameters in the vicinity of the central fit. Uncertainties of the
META PDF parameters are symmetric by definition, but the
uncertainties of the PDF functional forms are generally not. We
verified the validity of the linear approximation by first checking
the asymmetry $|\delta^H_+-\delta^H_-|/(\delta^H_++\delta^H_-)$ of
the uncertainties on the PDFs and PDF luminosities, where
$\delta^H_{\pm}$ indicate the asymmetric errors according to
Eq.~(\ref{HessianMinusError2}). For example, in Fig.~\ref{fig:lin1}
we plot the asymmetry of PDF errors of different flavors, which
gives an estimation of the non-linear behavior of the PDFs due to
both the parametrization and the PDF evolution. The asymmetries
observed in the figure are very small, at the level of a few percent
in the region with $x < 0.1$. They can exceed 15\% in the large $x$
region, where the PDFs are not well constrained. As $Q$ increases
from 8 to 10000 GeV, the asymmetry slowly spreads from large toward
smaller $x$ as a consequence of the PDF evolution. In
Fig.~\ref{fig:lin2} we further show the asymmetry of the PDF
uncertainty of the parton luminosities in different kinematic
regions at the LHC 8 TeV~\cite{Campbell:2006wx}, as functions of the
invariant mass and rapidity of the produced final state. These
asymmetries are also small except for very large invariant masses or
rapidities that correspond to the $x$ values close to 1. By
investigating the asymmetry of the PDFs as well as of sample LHC
cross sections, we conclude that the linear dependence of most LHC
observables on small variations of the PDF parameters is well
satisfied.

\begin{figure}[htb]
\begin{centering}
\includegraphics[width=0.32\textwidth]{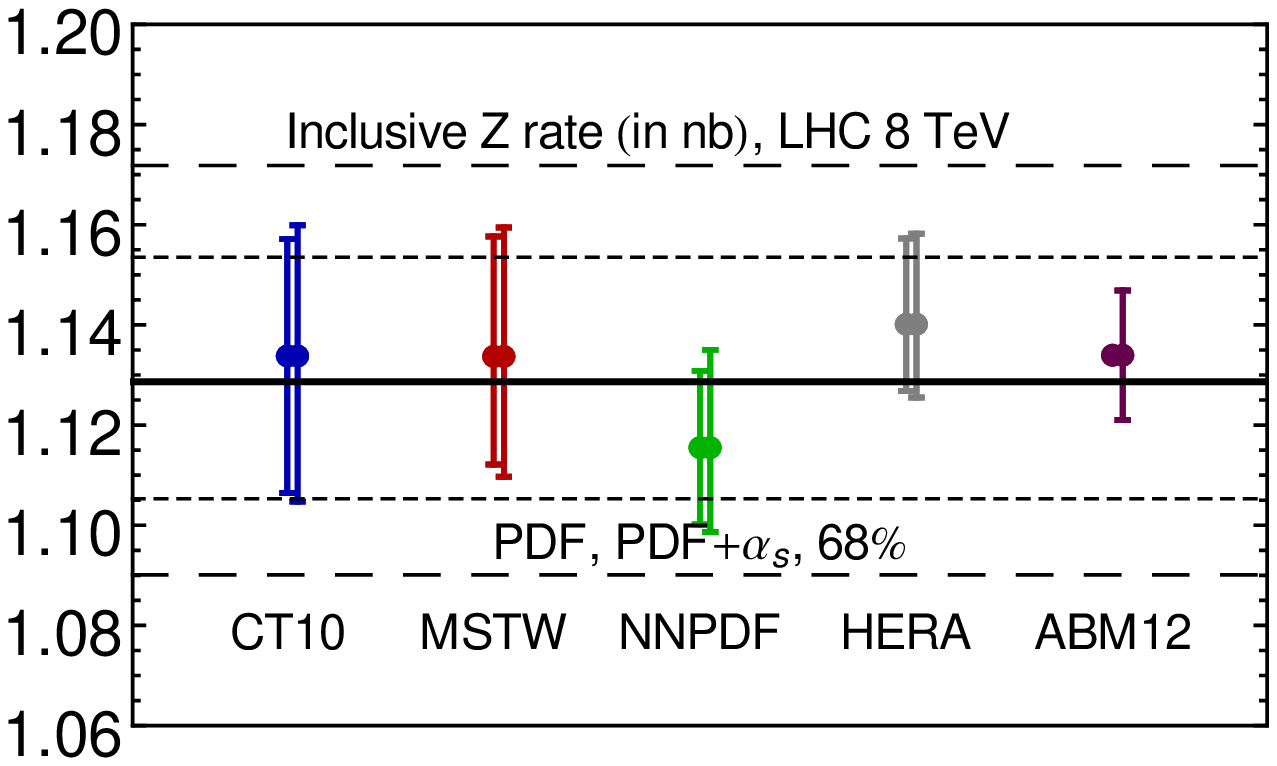} \includegraphics[width=0.32\textwidth]{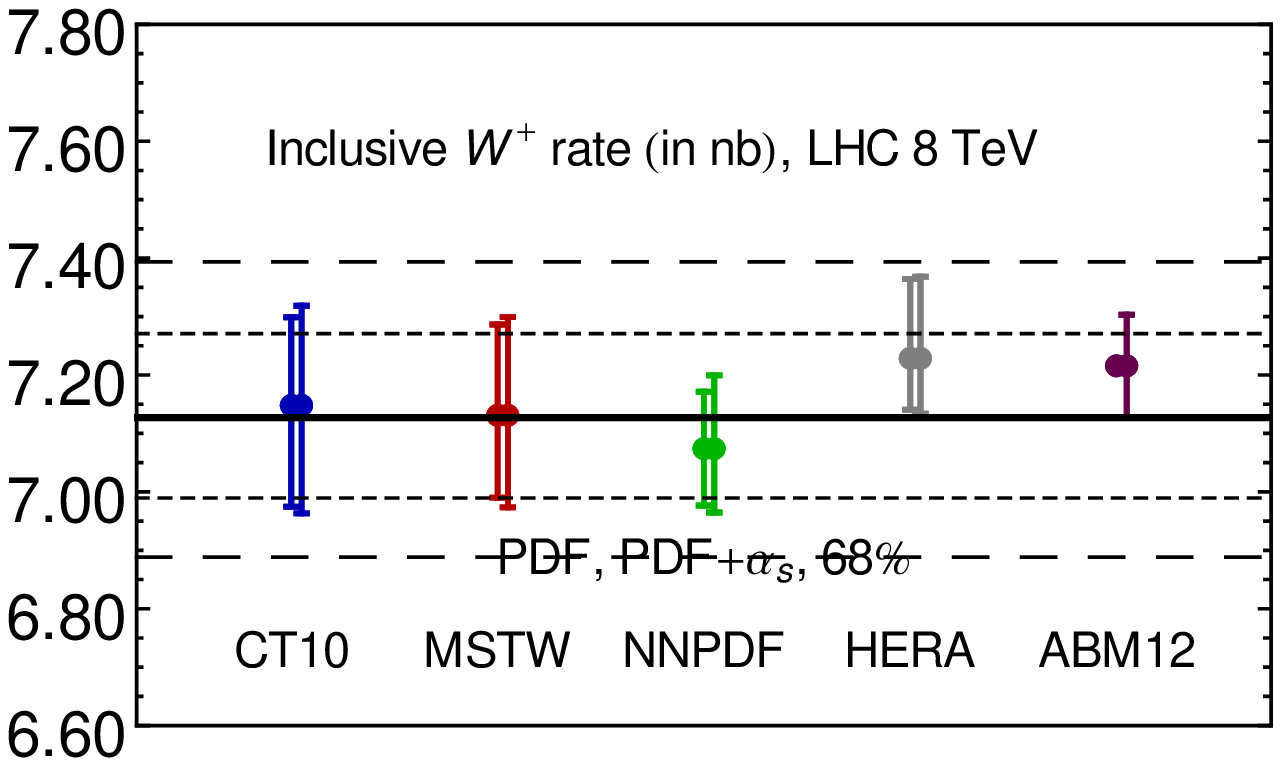}
\includegraphics[width=0.32\textwidth]{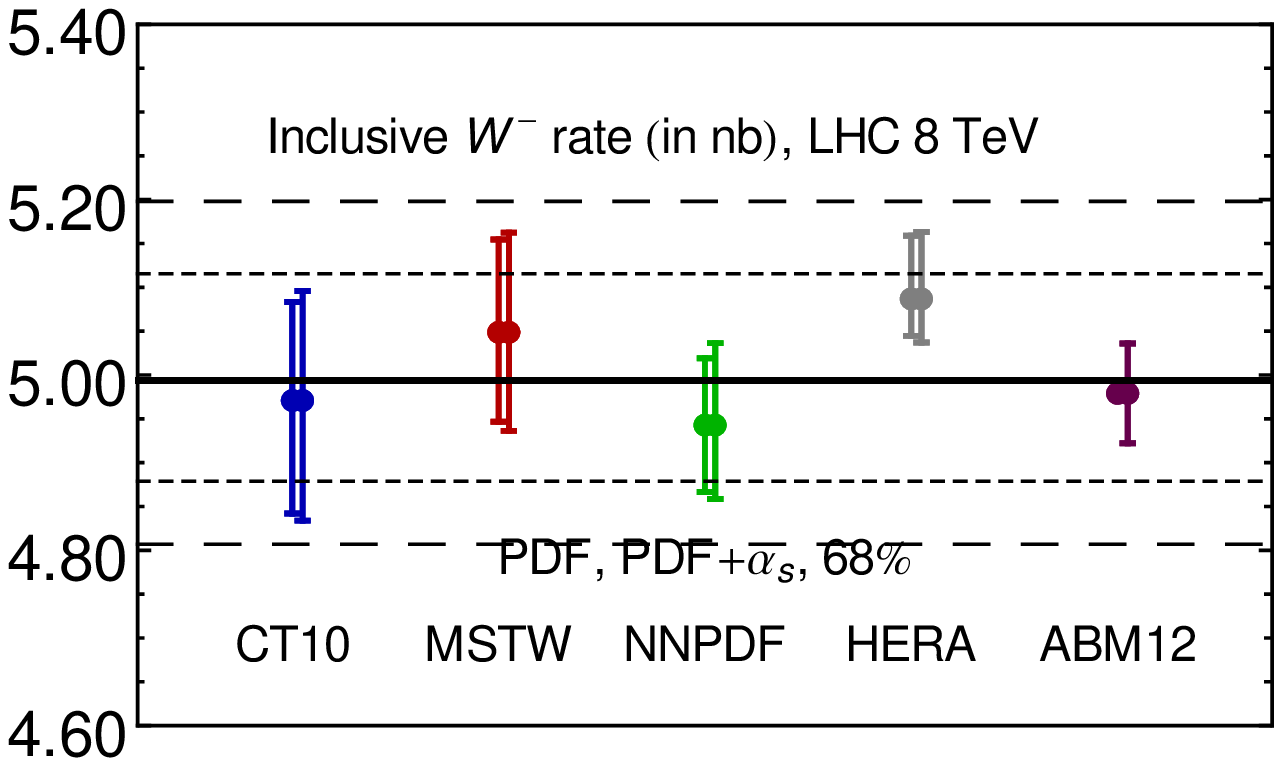}\\
 \includegraphics[width=0.32\textwidth]{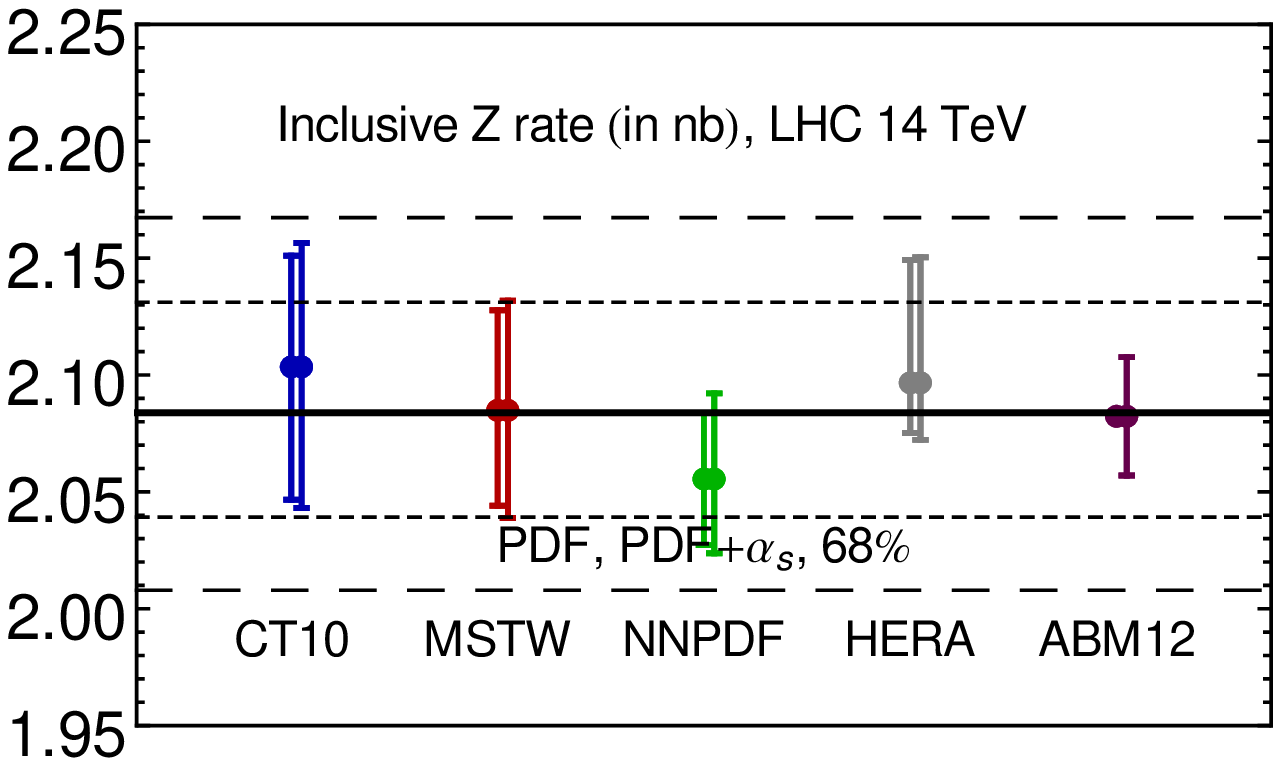} \includegraphics[width=0.32\textwidth]{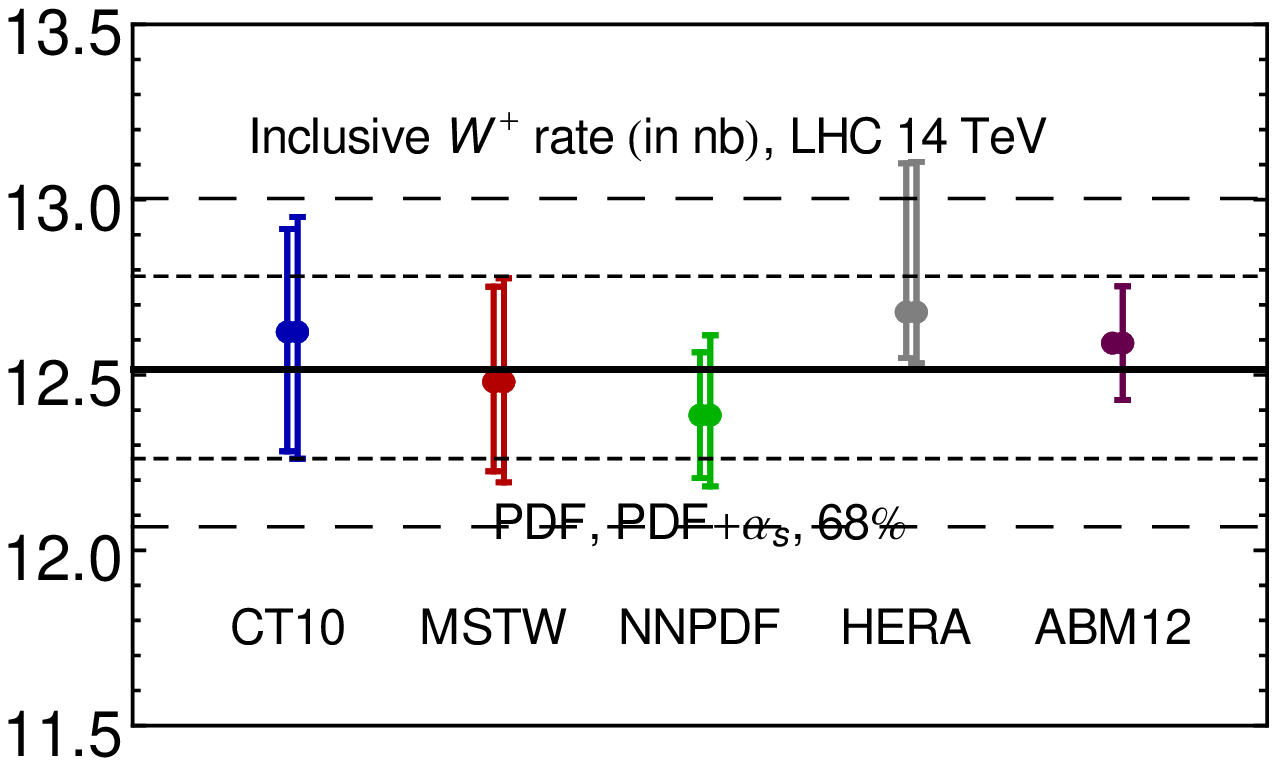}
\includegraphics[width=0.32\textwidth]{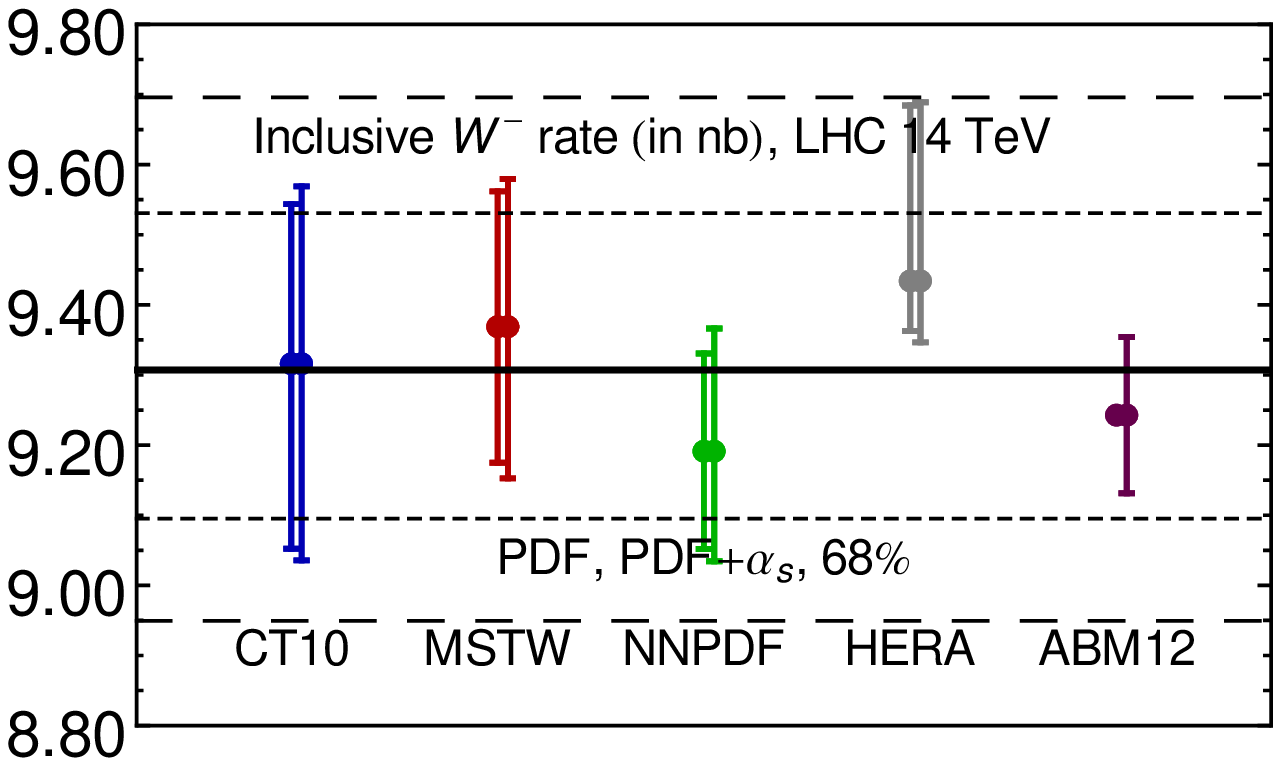}
\par\end{centering}

\vspace{-2ex}
 \caption{\label{fig:mxsec1} Comparison of inclusive $W^{\pm}$ and $Z$ cross
sections at NNLO for various PDF ensembles. The PDF and
PDF+$\alpha_s$ errors of META PDFs (at 68\% c.l.) are shown by the
short-dashed and long-dashed lines.}
\end{figure}

\begin{figure}[htb]
\begin{centering}
\includegraphics[width=0.32\textwidth]{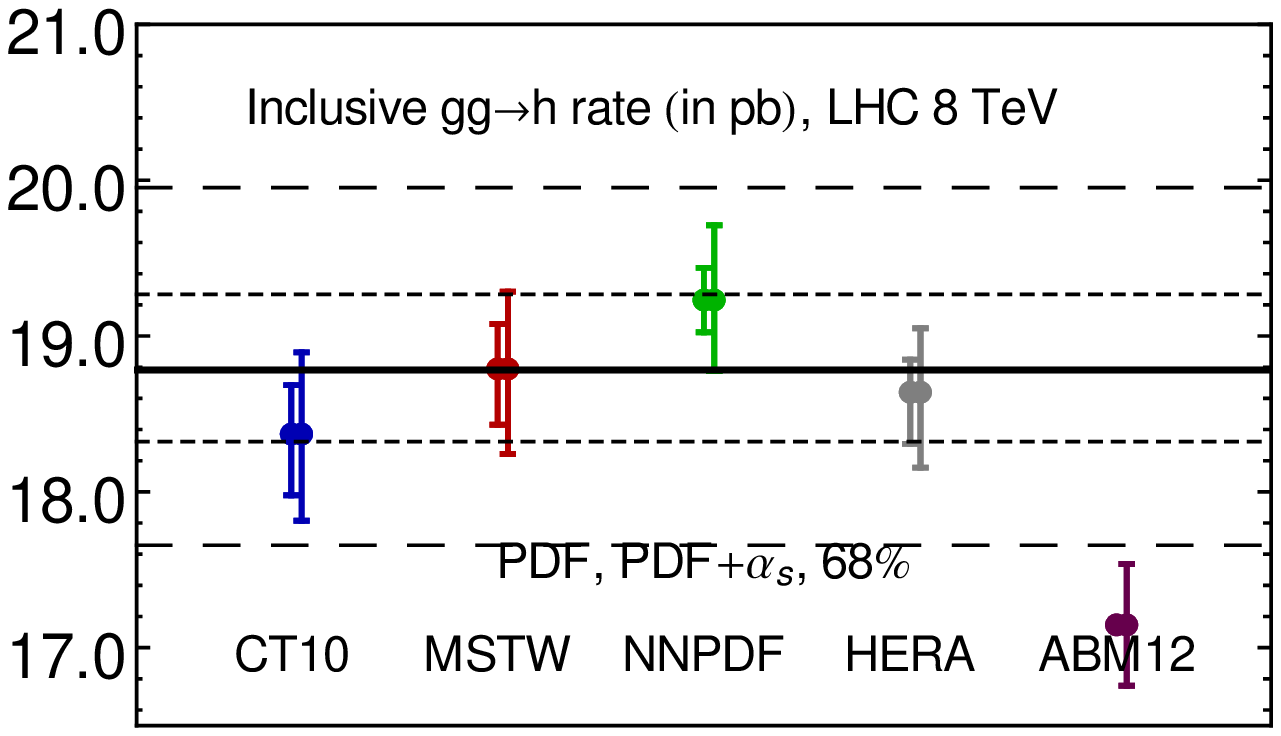} \includegraphics[width=0.32\textwidth]{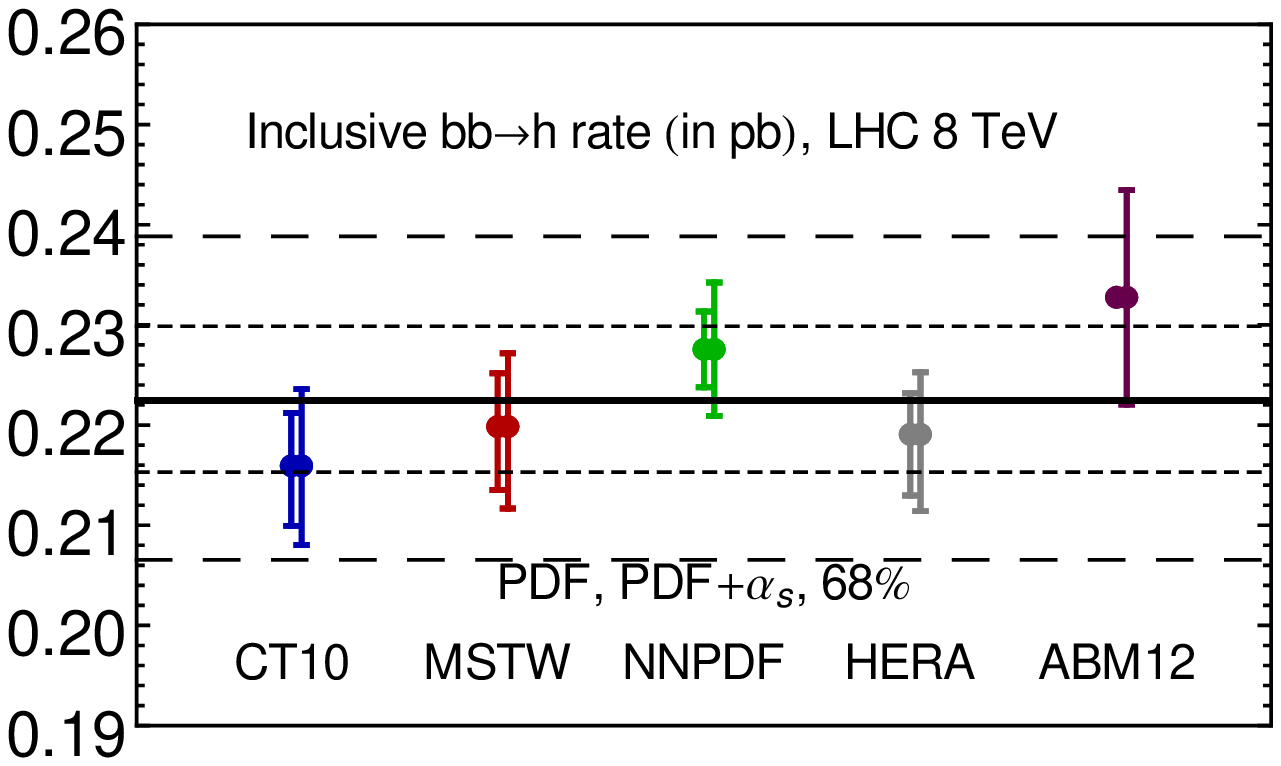}
\includegraphics[width=0.32\textwidth]{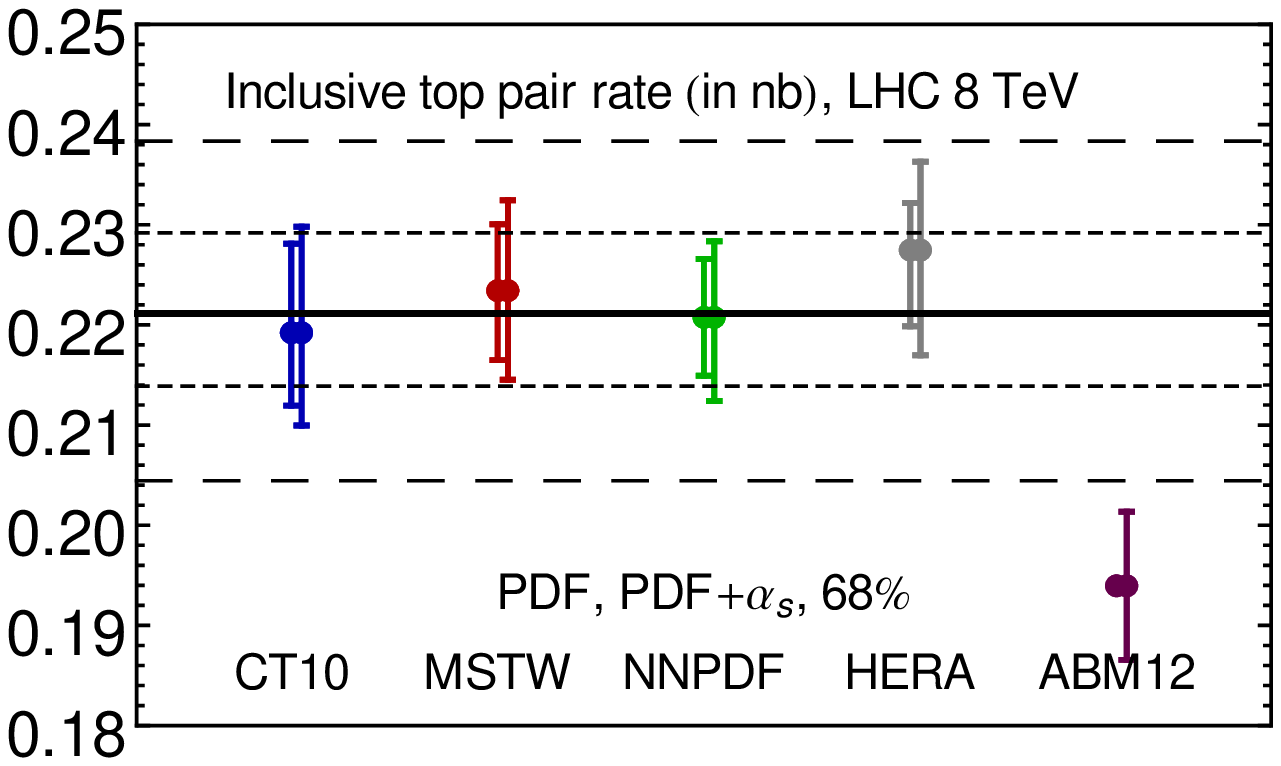}\\
 \includegraphics[width=0.32\textwidth]{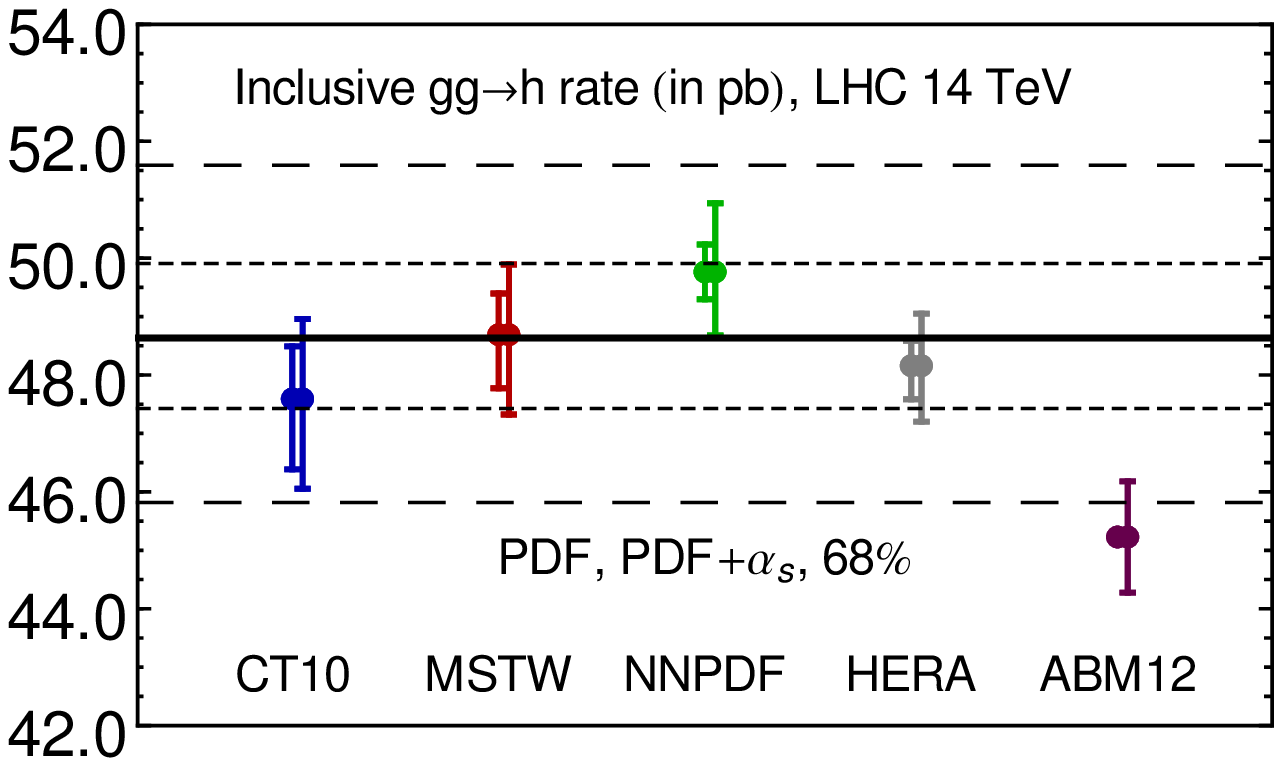} \includegraphics[width=0.32\textwidth]{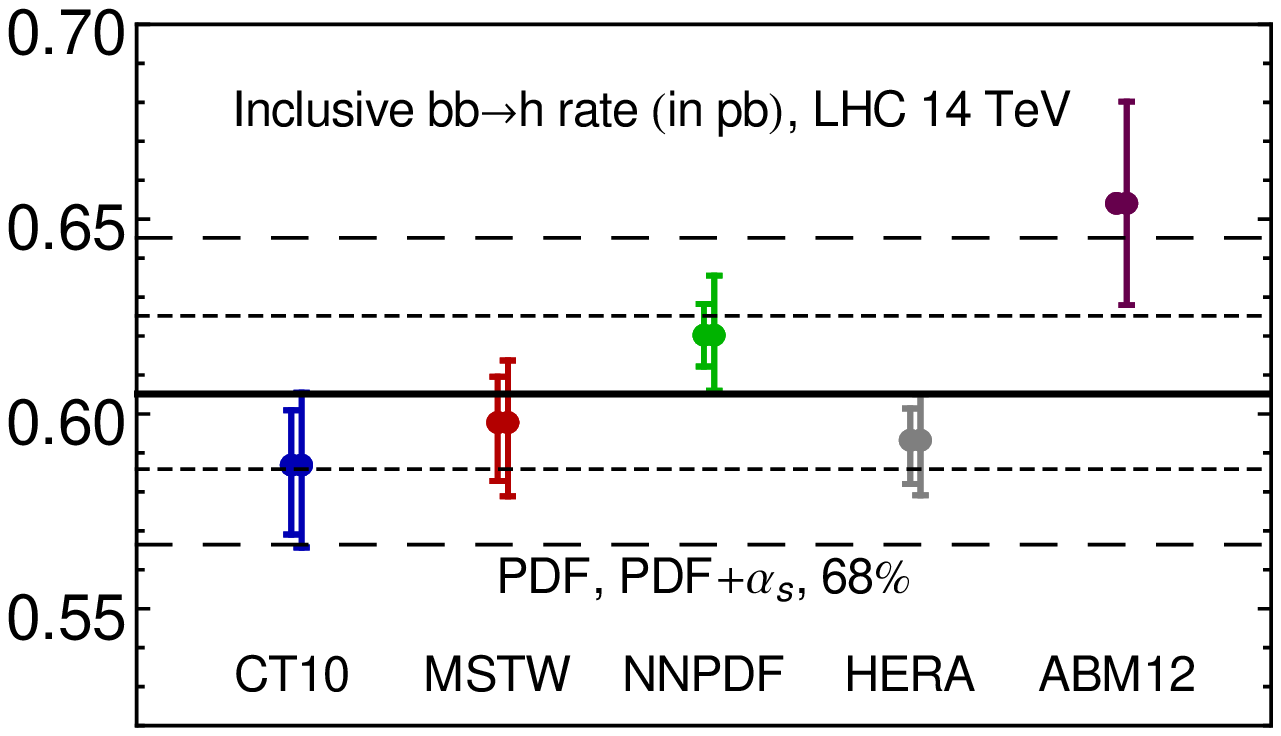}
\includegraphics[width=0.32\textwidth]{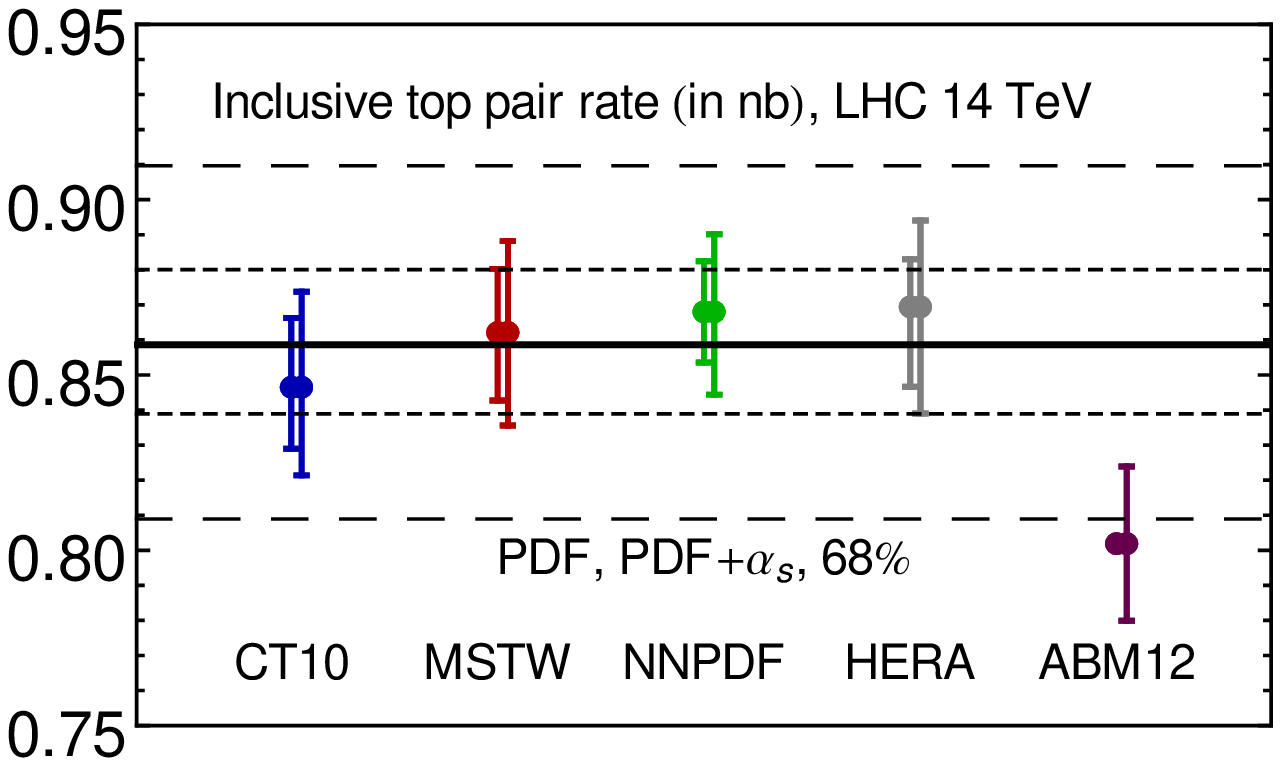}
\par\end{centering}

\vspace{-2ex}
 \caption{\label{fig:mxsec2} Comparison of inclusive Higgs boson and top quark
pair production cross sections at NNLO for various PDF ensembles. The PDF
and PDF+$\alpha_s$ errors of META PDFs (at 68\% c.l.) are shown by the
short-dashed and long-dashed lines.}
\end{figure}

\begin{figure}[htb]
\begin{centering}
\includegraphics[width=0.32\textwidth]{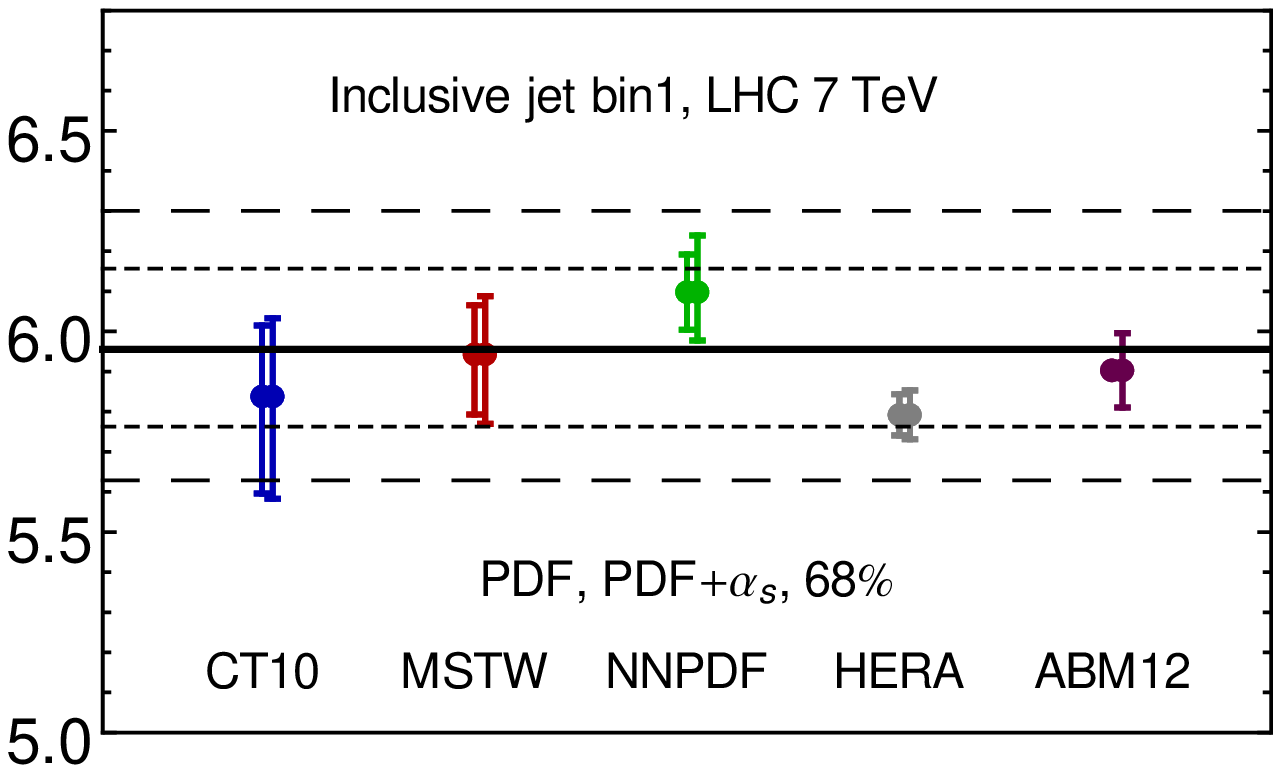} \includegraphics[width=0.32\textwidth]{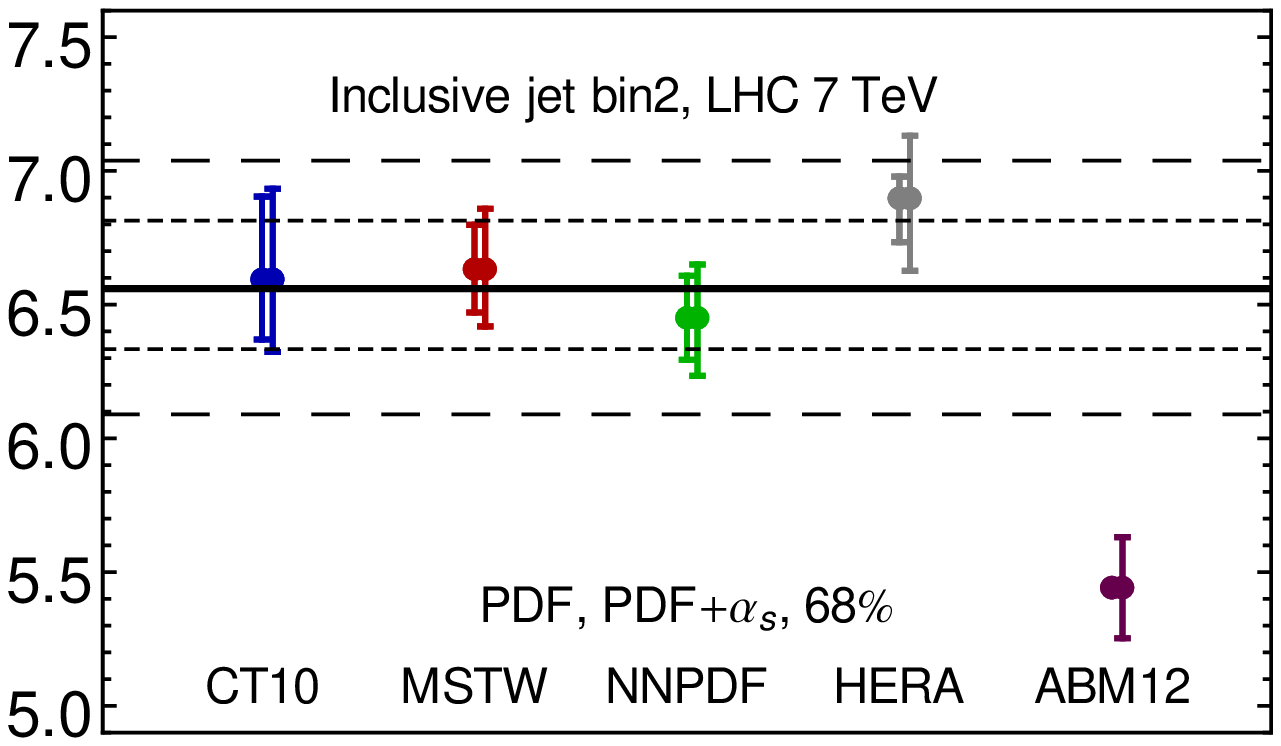}
\includegraphics[width=0.32\textwidth]{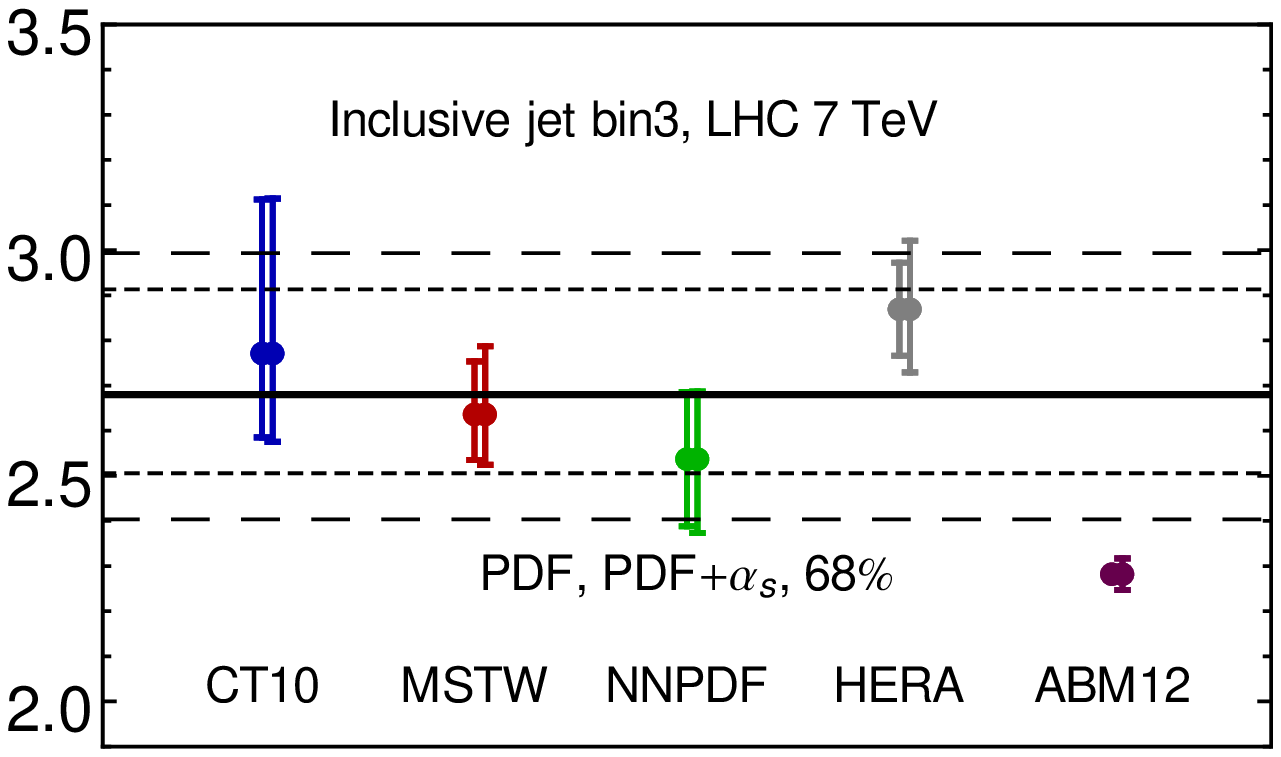}\\
 \includegraphics[width=0.32\textwidth]{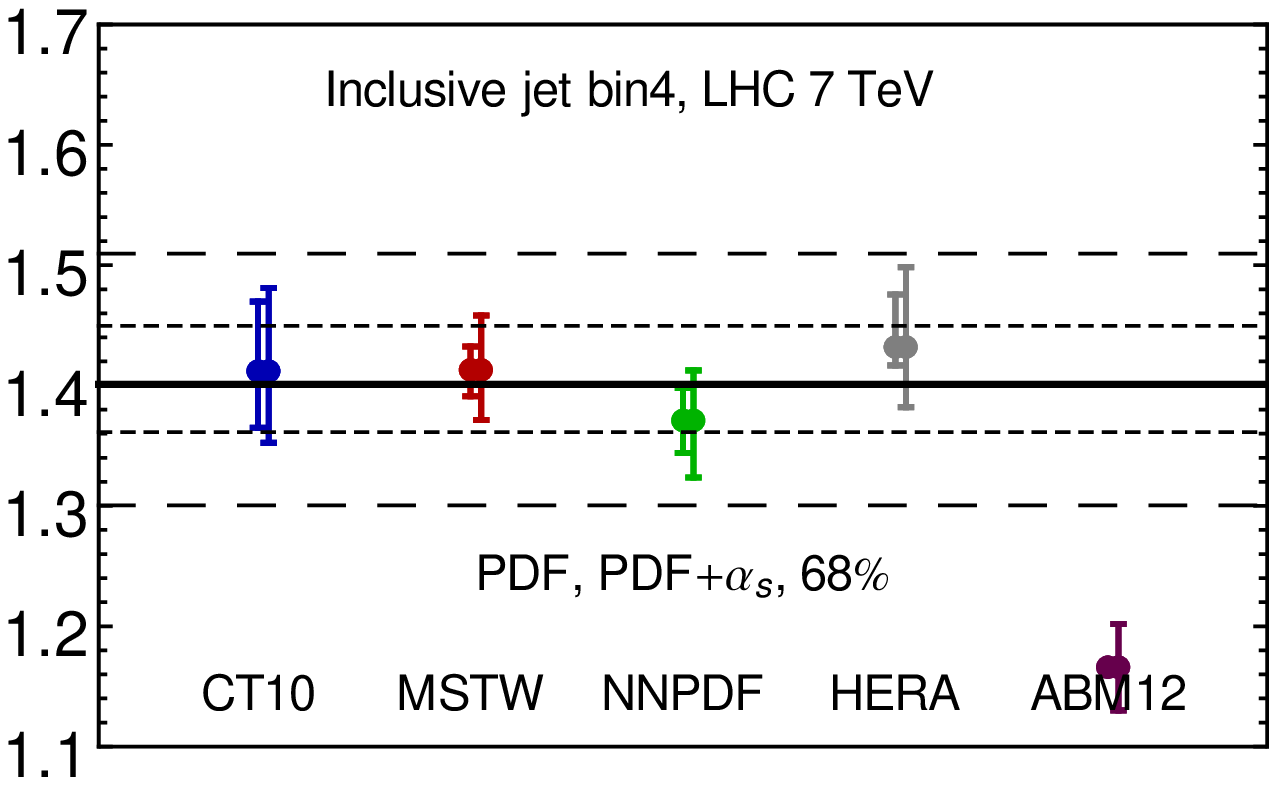} \includegraphics[width=0.32\textwidth]{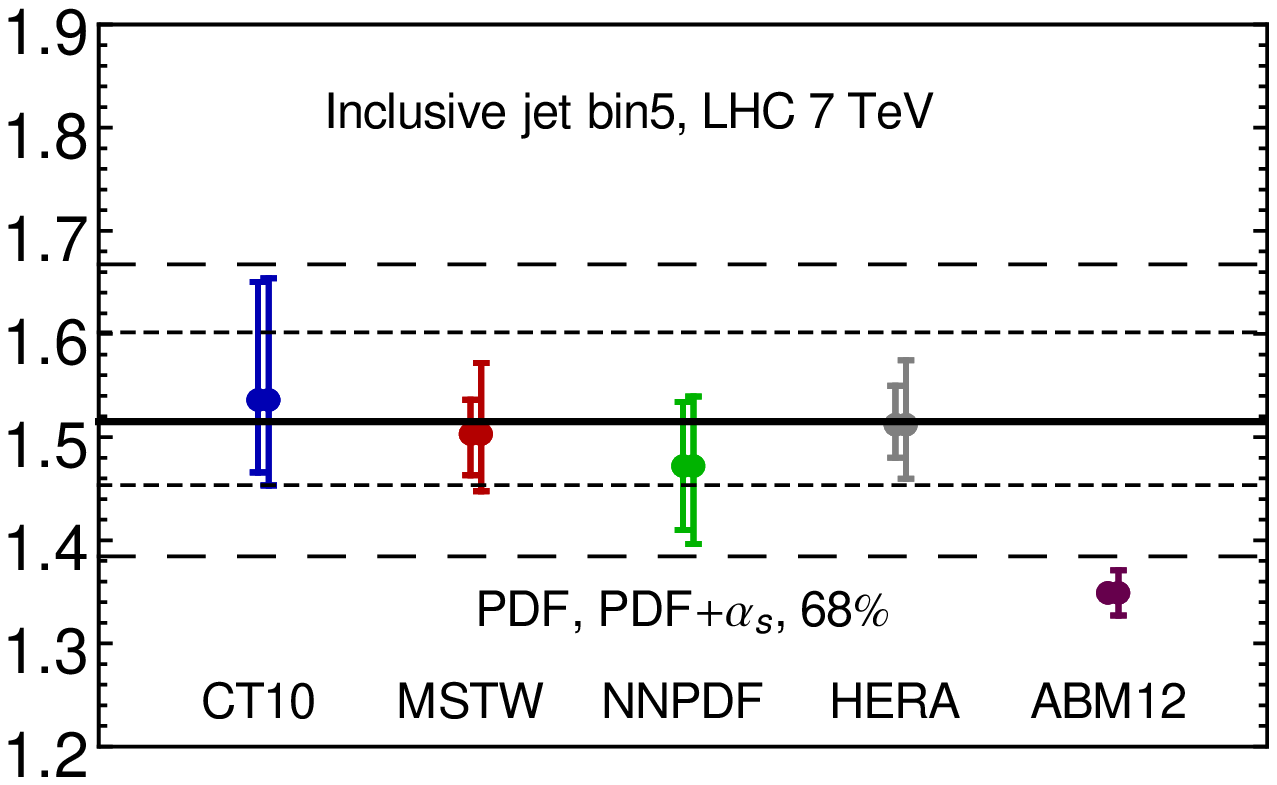}
\includegraphics[width=0.32\textwidth]{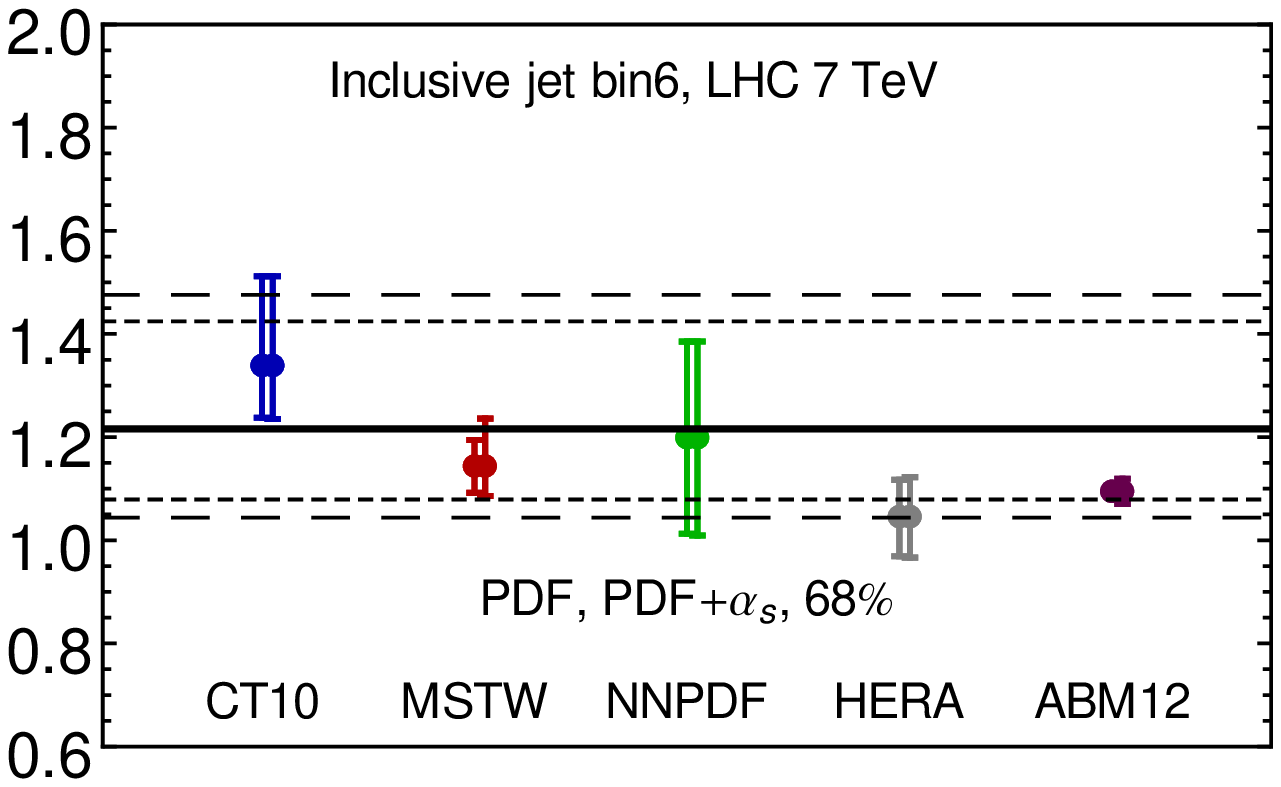}
\par\end{centering}

\vspace{-2ex}
 \caption{\label{fig:mxsec3} Comparison of inclusive jet production
   cross sections (in arbitrary unit) at NLO for various PDF ensembles.
The PDF and PDF+$\alpha_s$
errors of META PDFs (at 68\% c.l.) are shown by the short-dashed and
long-dashed lines.}
\end{figure}

Using the eigenvector sets and $\alpha_{s}$ series of the META
ensemble, we can calculate central predictions and uncertainties of
the cross sections at the LHC and compare them with the results from
individual groups. As an example,
Figs.~\ref{fig:mxsec1}-\ref{fig:mxsec3} show NNLO predictions for
$W,$ $Z$, SM Higgs, $t\bar{t}$, and NLO predictions for inclusive
jet production, using the same settings as in
Figs.~\ref{fig:xsec1}-\ref{fig:xsec3}. The solid horizontal lines
indicate the central prediction from the Hessian META PDF ensemble,
the dotted and dashed ones correspond to the 68\% c.l. PDF
uncertainty and PDF$+\alpha_{s}$ uncertainty of the META ensemble,
respectively.  For every group except ABM, we show the central
predictions with $\alpha_{s}(M_{Z})=0.118$ and the 68\% c.l. PDF
(PDF$+\alpha_{s}$) uncertainties as the left (right) error bar. 
The rightmost points in each inset correspond to the central predictions 
and 68\% c.l. PDF+$\alpha_s$ uncertainties for the ABM12 PDF ensemble
with 5 active flavors~\cite{Alekhin:2013nda}, computed according 
to their convention for $\alpha_s(M_Z)=0.1132$ and 
top quark pole mass of 171 GeV. As we expect, 
the META PDFs work as an average of CT10, MSTW2008, and
NNPDF2.3 PDFs, as can be noticed by comparing the central
predictions. The PDF or PDF+$\alpha_{s}$ uncertainties of the META
ensemble are slightly smaller than the envelope prescription in the
benchmarking study~\cite{Ball:2012wy}. For instance, the NNLO Higgs
cross section through gluon fusion is $18.75\pm1.24\,{\rm pb}$ for
LHC 8 TeV according to the envelope prescription, while it is
$18.78\pm1.15\,{\rm pb}$ for the META PDF.

\begin{figure}[htb]
\begin{centering}
\includegraphics[width=0.24\textwidth]{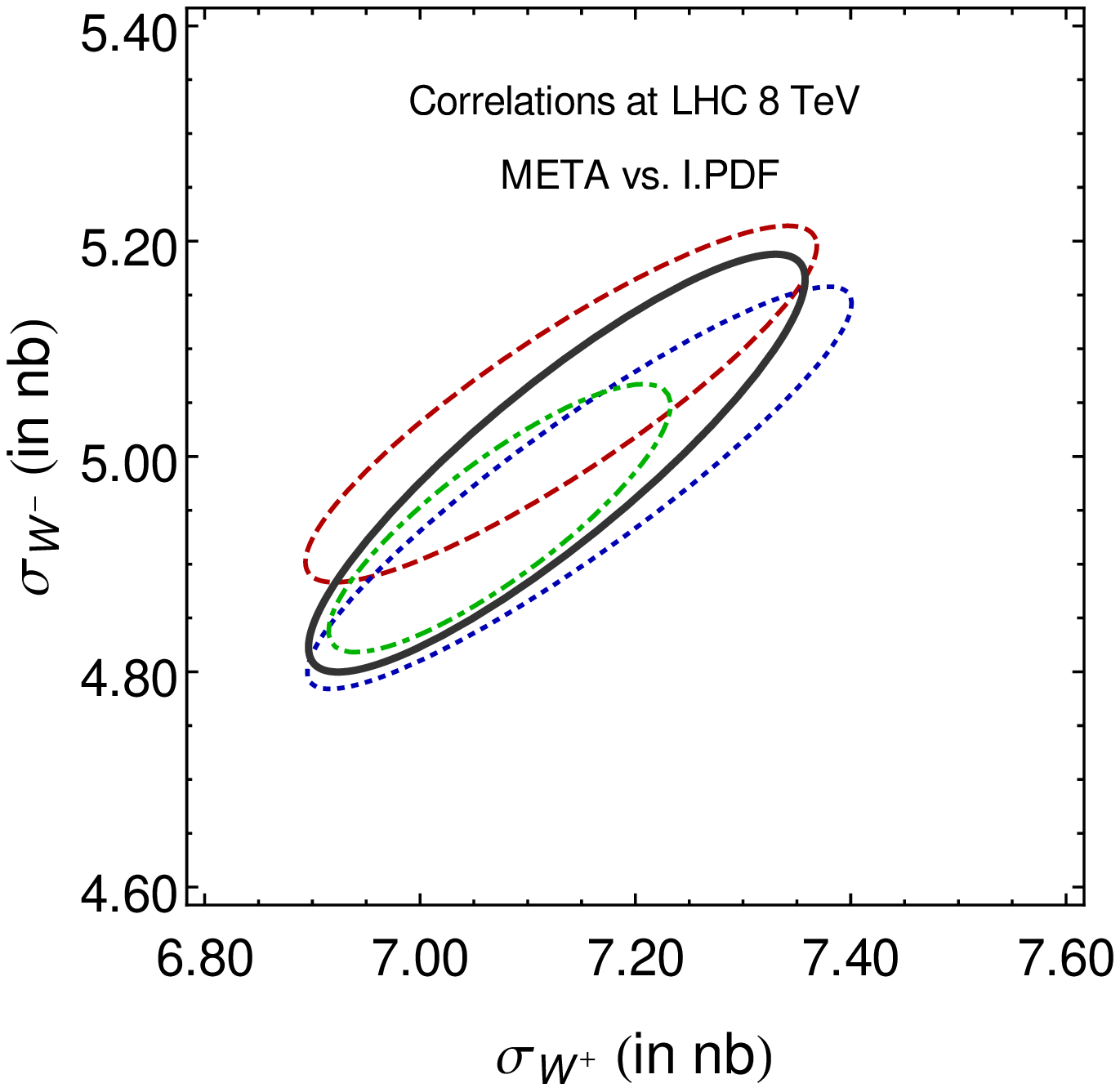} \includegraphics[width=0.24\textwidth]{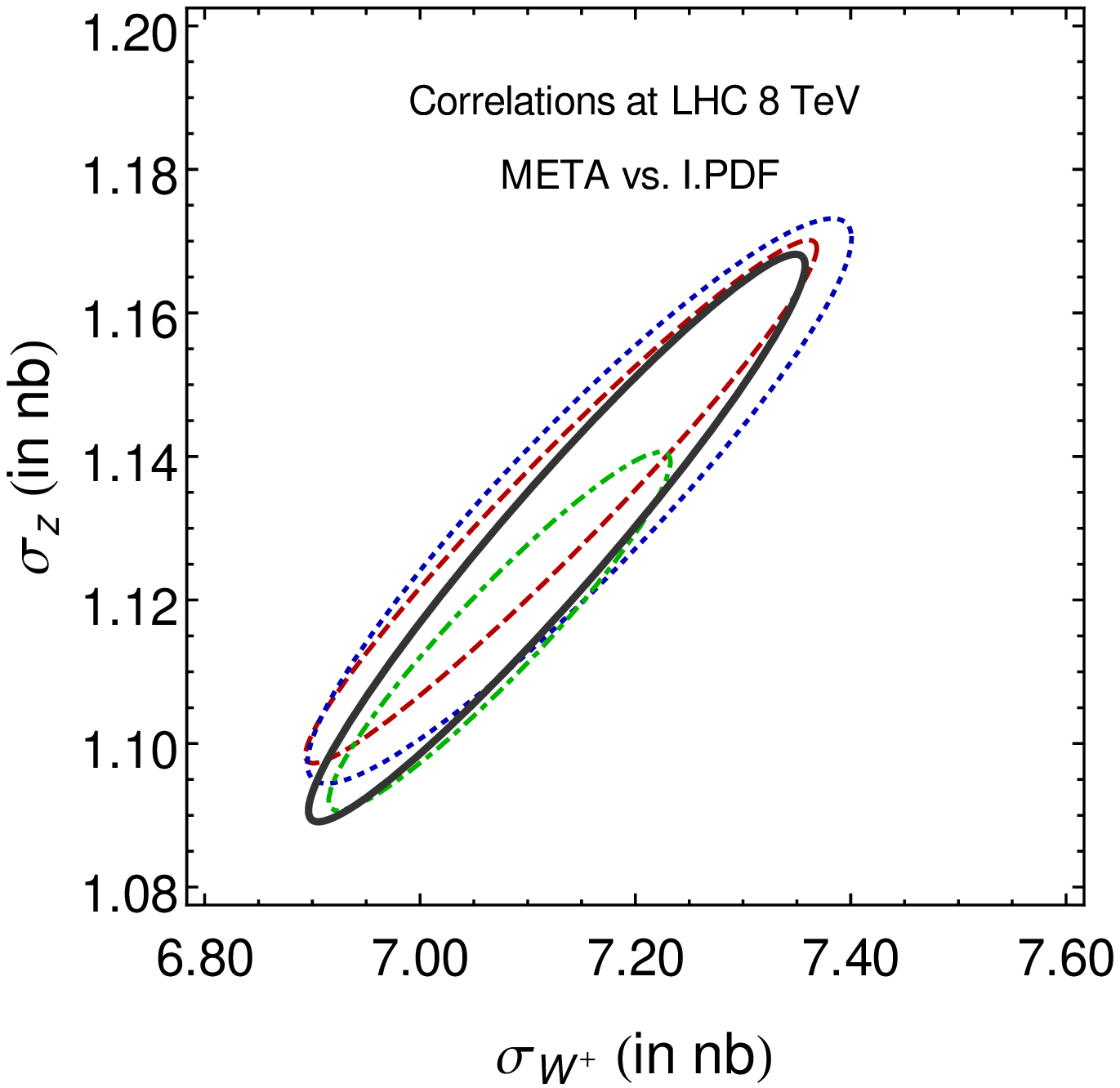}
\includegraphics[width=0.24\textwidth]{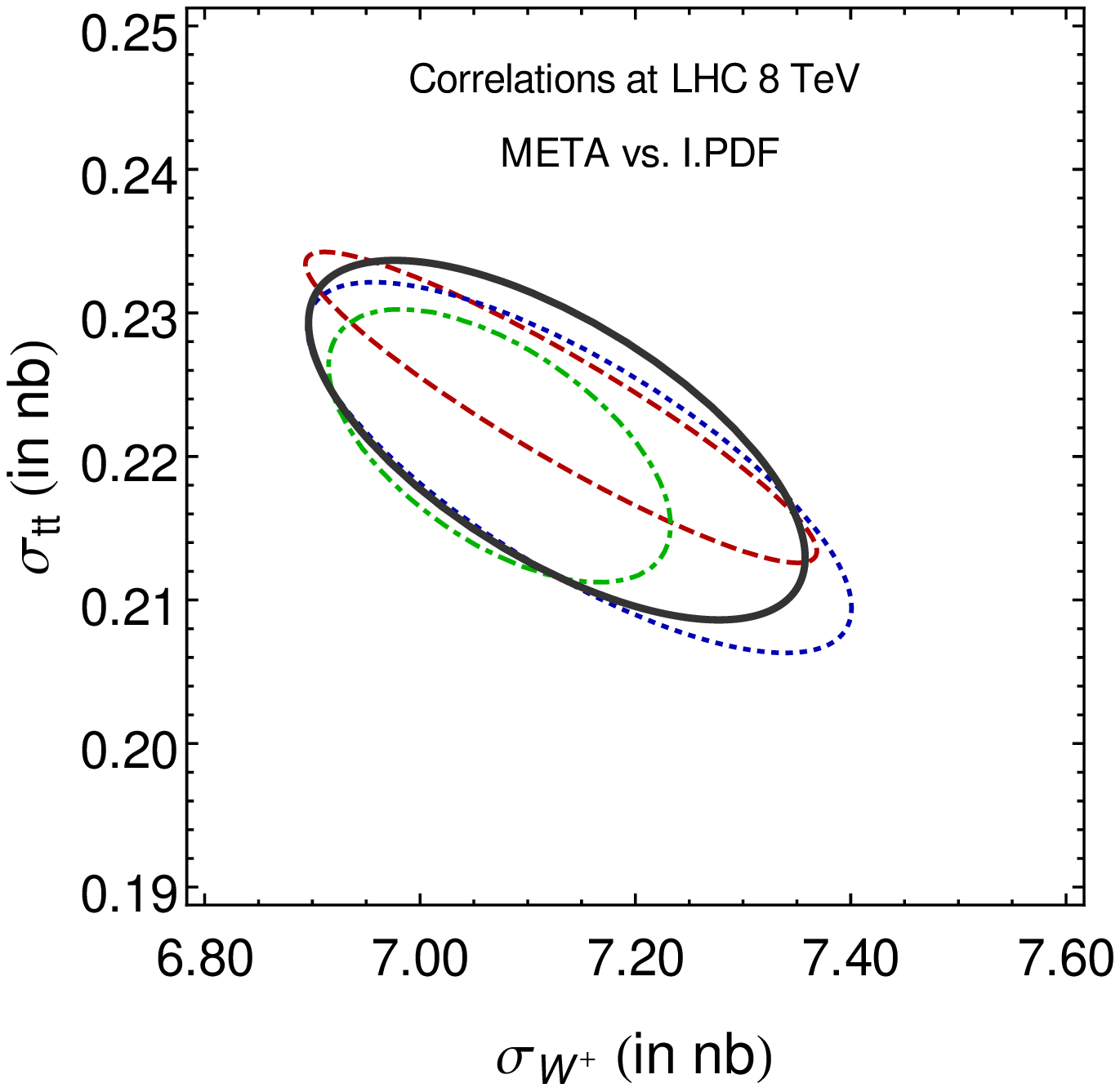} \includegraphics[width=0.24\textwidth]{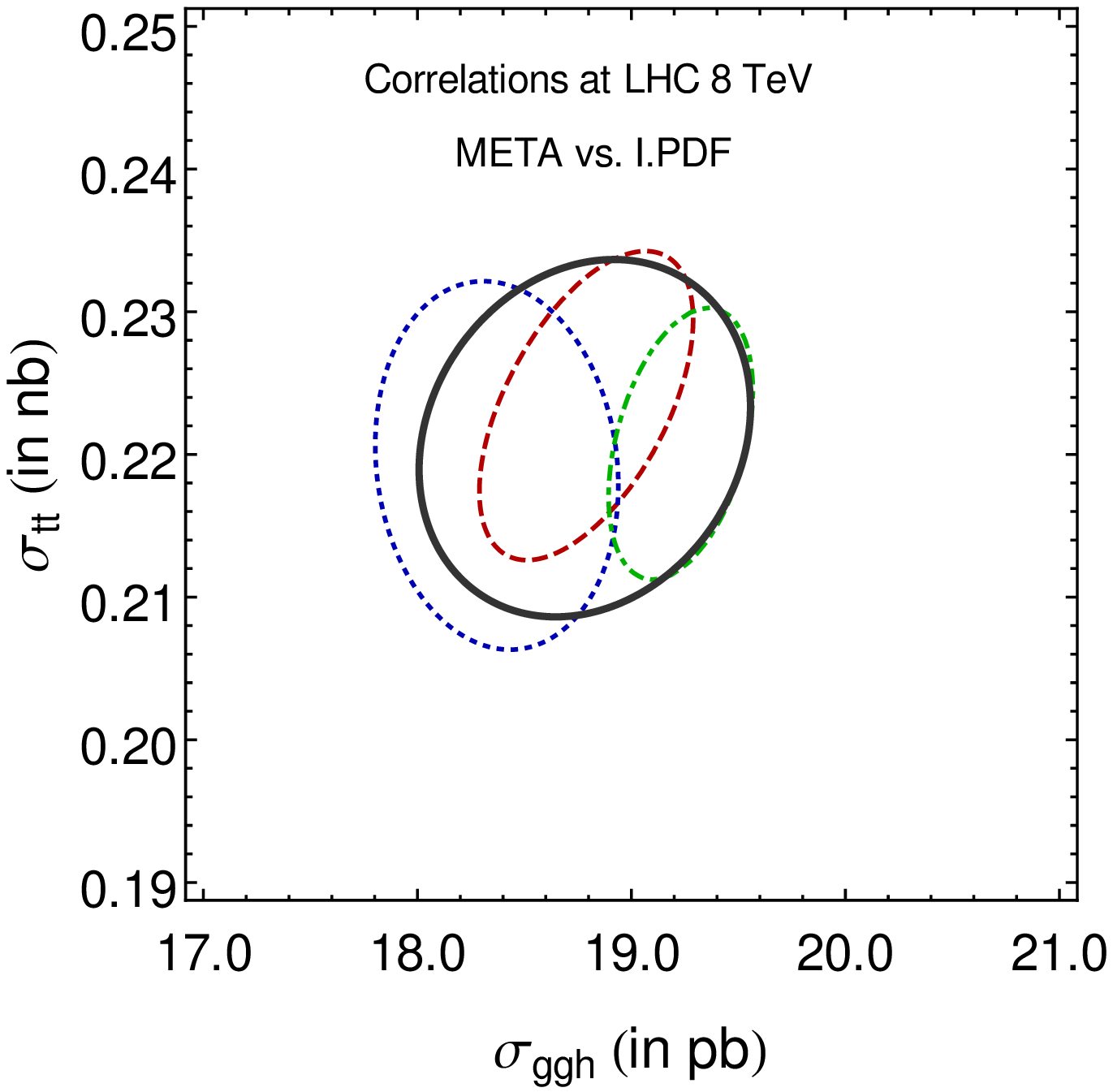}\\
 \includegraphics[width=0.24\textwidth]{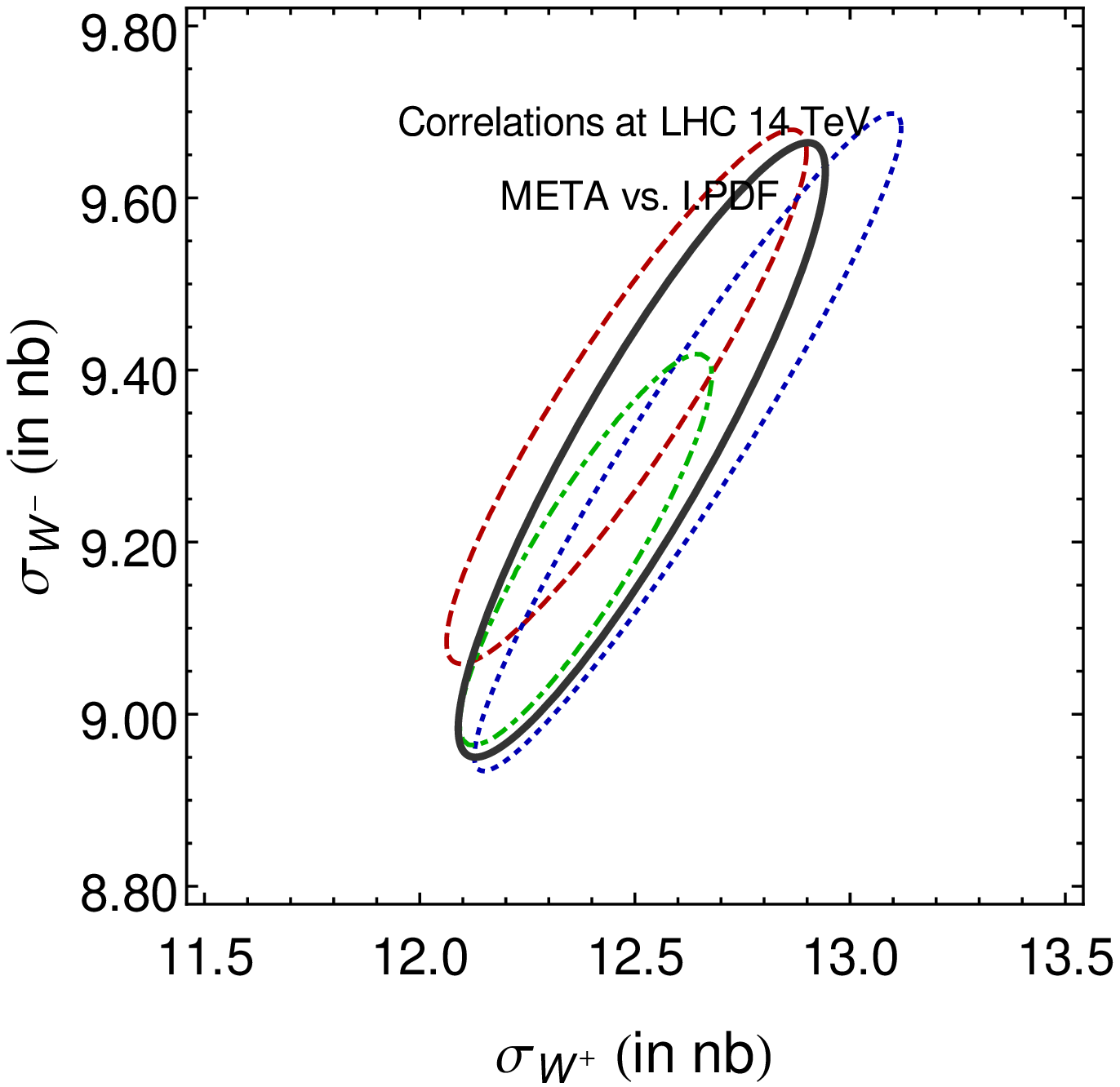} \includegraphics[width=0.24\textwidth]{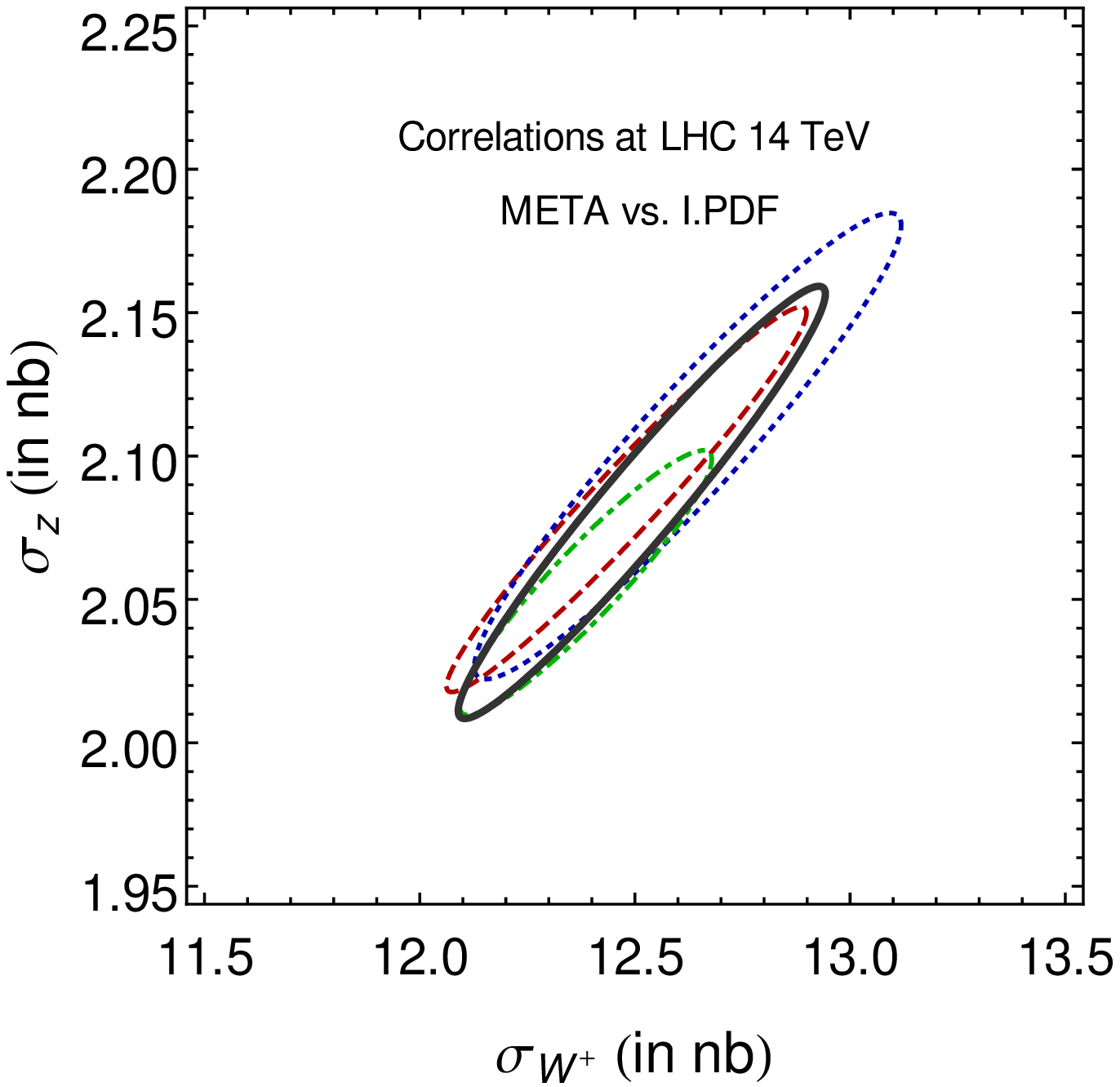}
\includegraphics[width=0.24\textwidth]{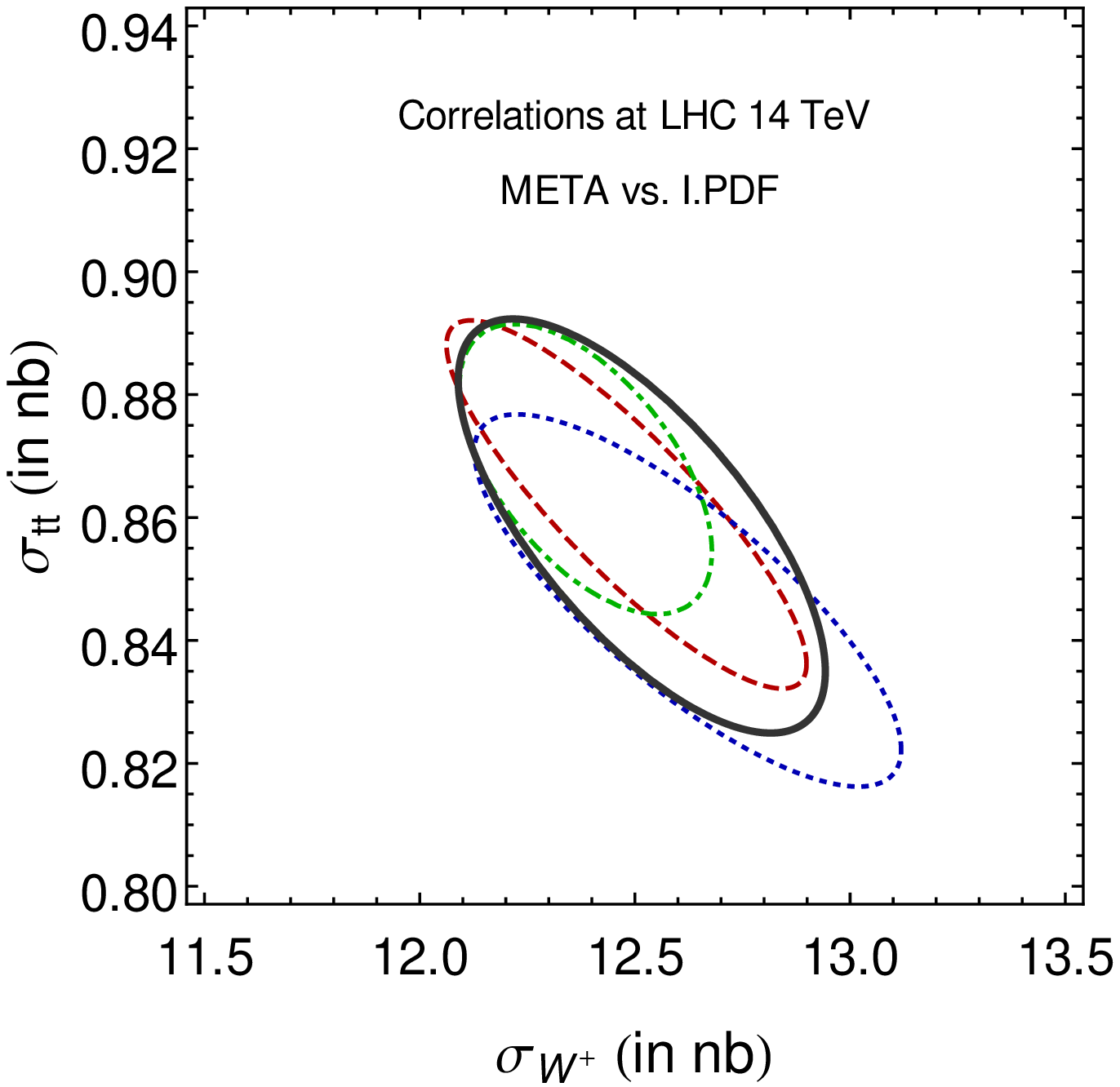} \includegraphics[width=0.24\textwidth]{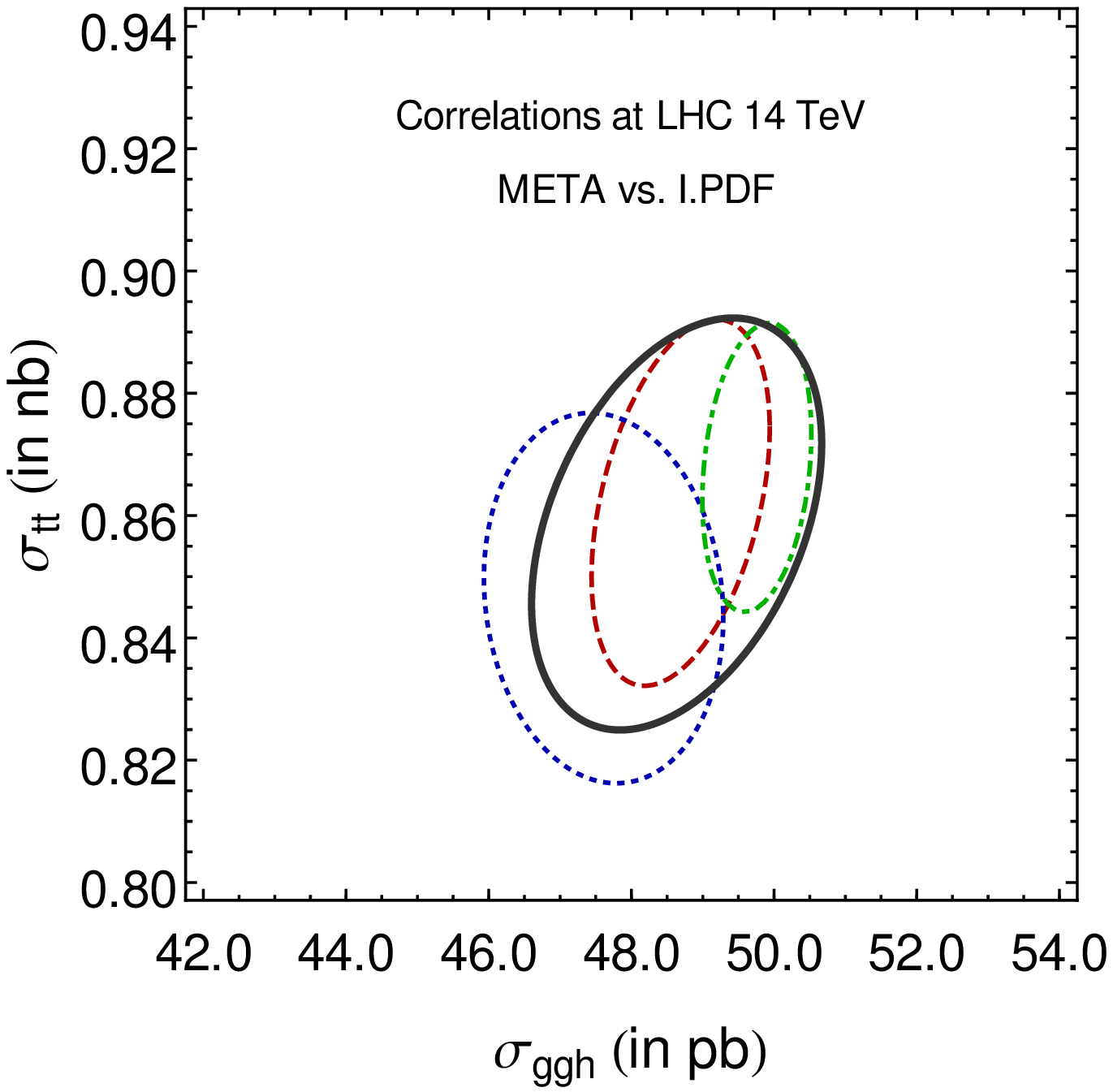}
\par\end{centering}

\vspace{-1ex}
 \caption{\label{mxsec4} Comparison of 90\% error ellipses of NNLO cross sections
using various ensembles: META PDF(black solid), CT10(blue dotted), MSTW(red dashed), NNPDF(green dot-dashed).}
\end{figure}

Fig.~\ref{mxsec4} shows the 90\% c.l. tolerance ellipses for pairs
of cross sections computed using either the META PDFs or original
PDF ensembles with $\alpha_{s}(M_{Z})=0.118$. The ellipses are
computed according to Eq.~(\ref{eq:ell0}) and correspond to the
probability density contours of 90\%.  The ellipses of the META PDFs
preserve the PDF-induced (anti-)correlations observed in the three
input PDF ensembles.

\begin{figure}[htb]
\begin{centering}
\includegraphics[width=0.32\textwidth]{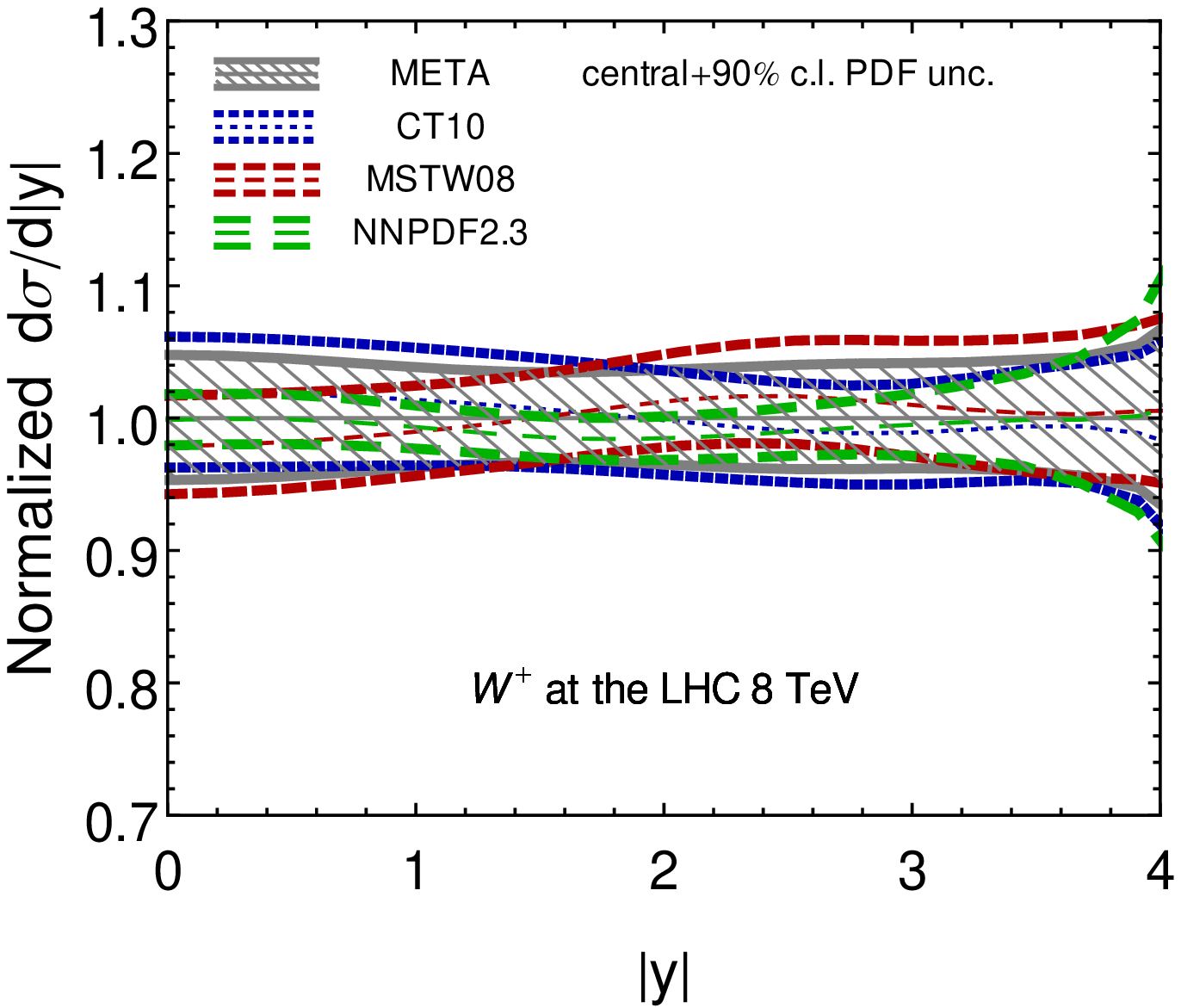} \includegraphics[width=0.32\textwidth]{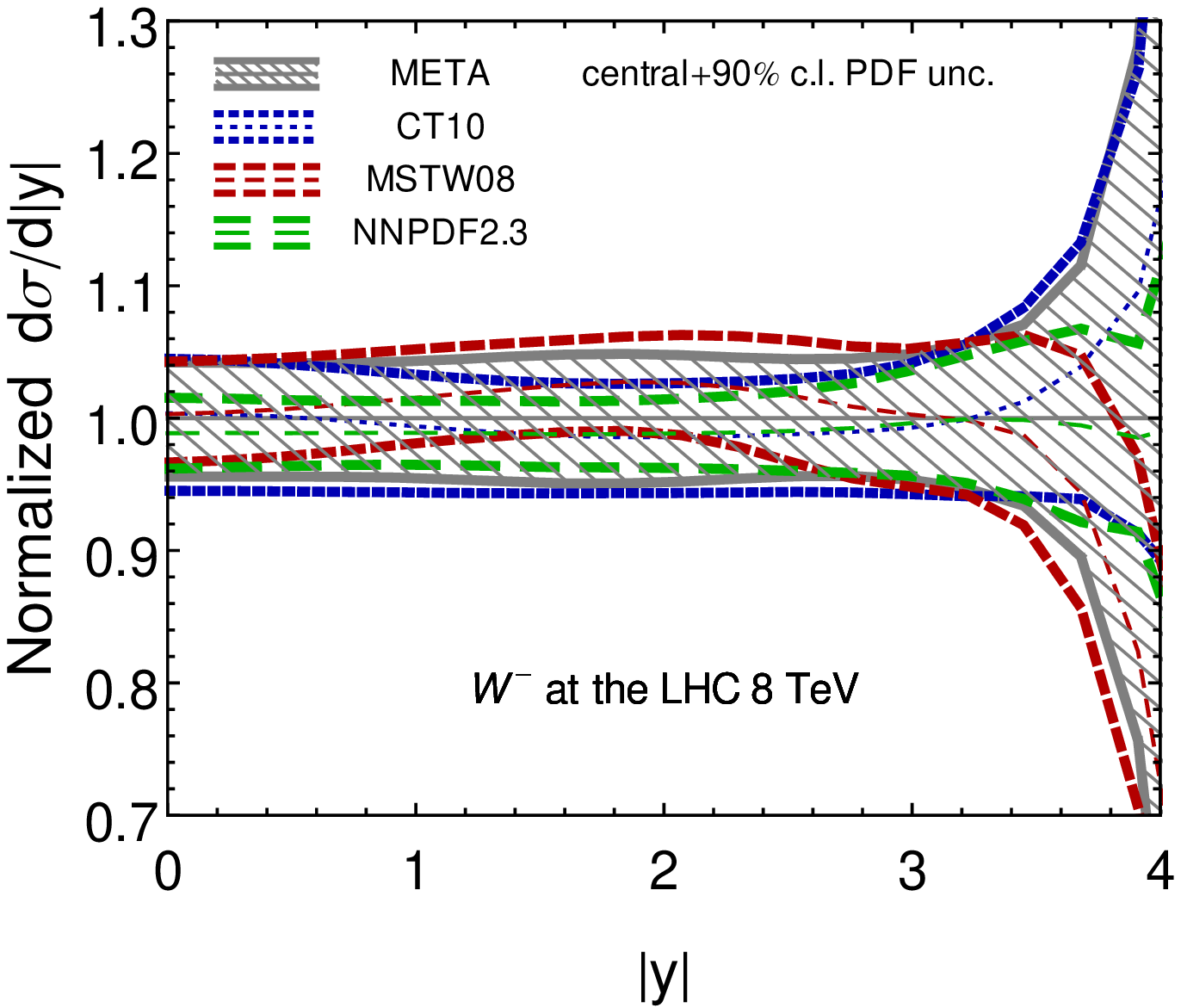}
\includegraphics[width=0.32\textwidth]{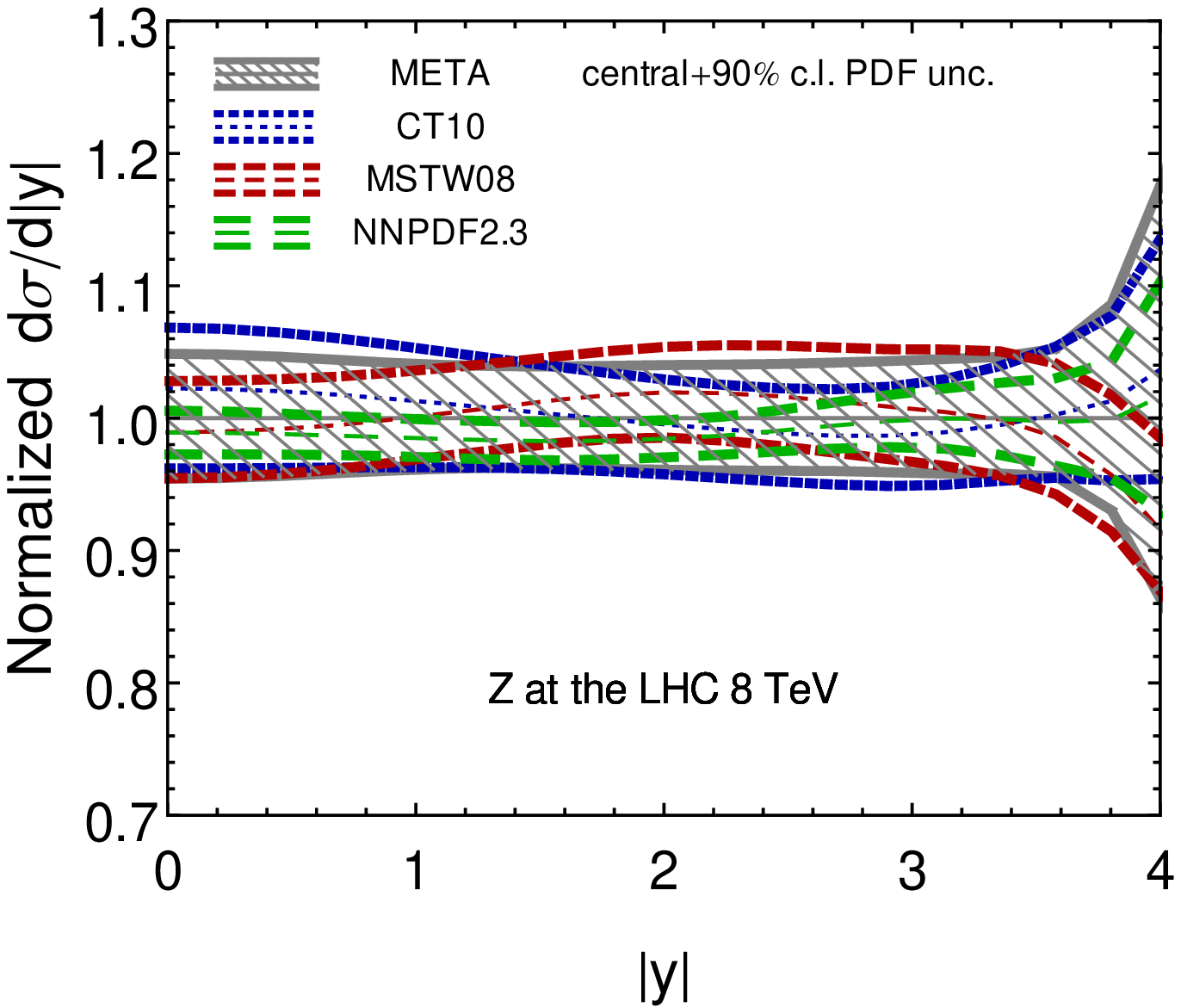}\\
 \includegraphics[width=0.32\textwidth]{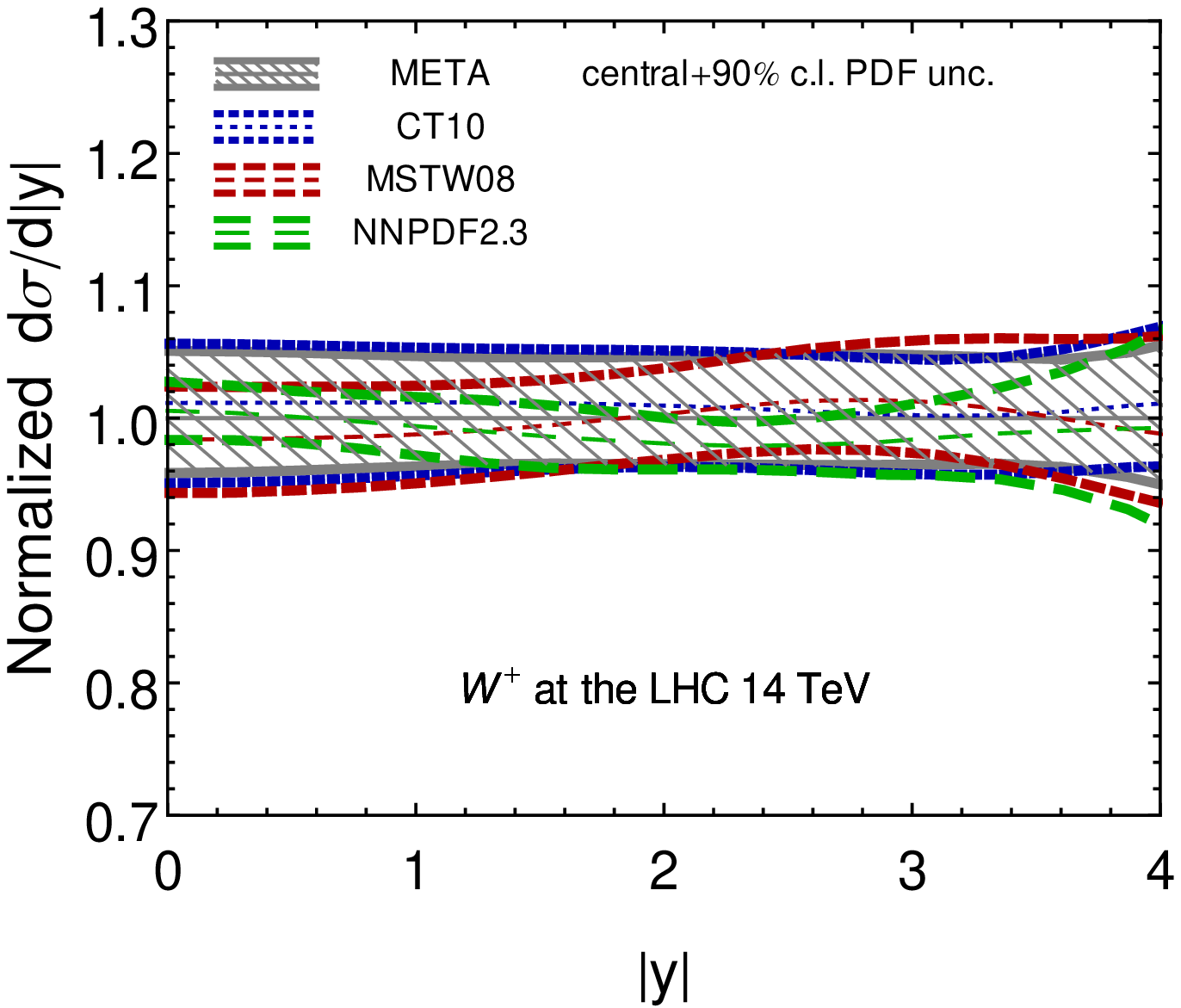} \includegraphics[width=0.32\textwidth]{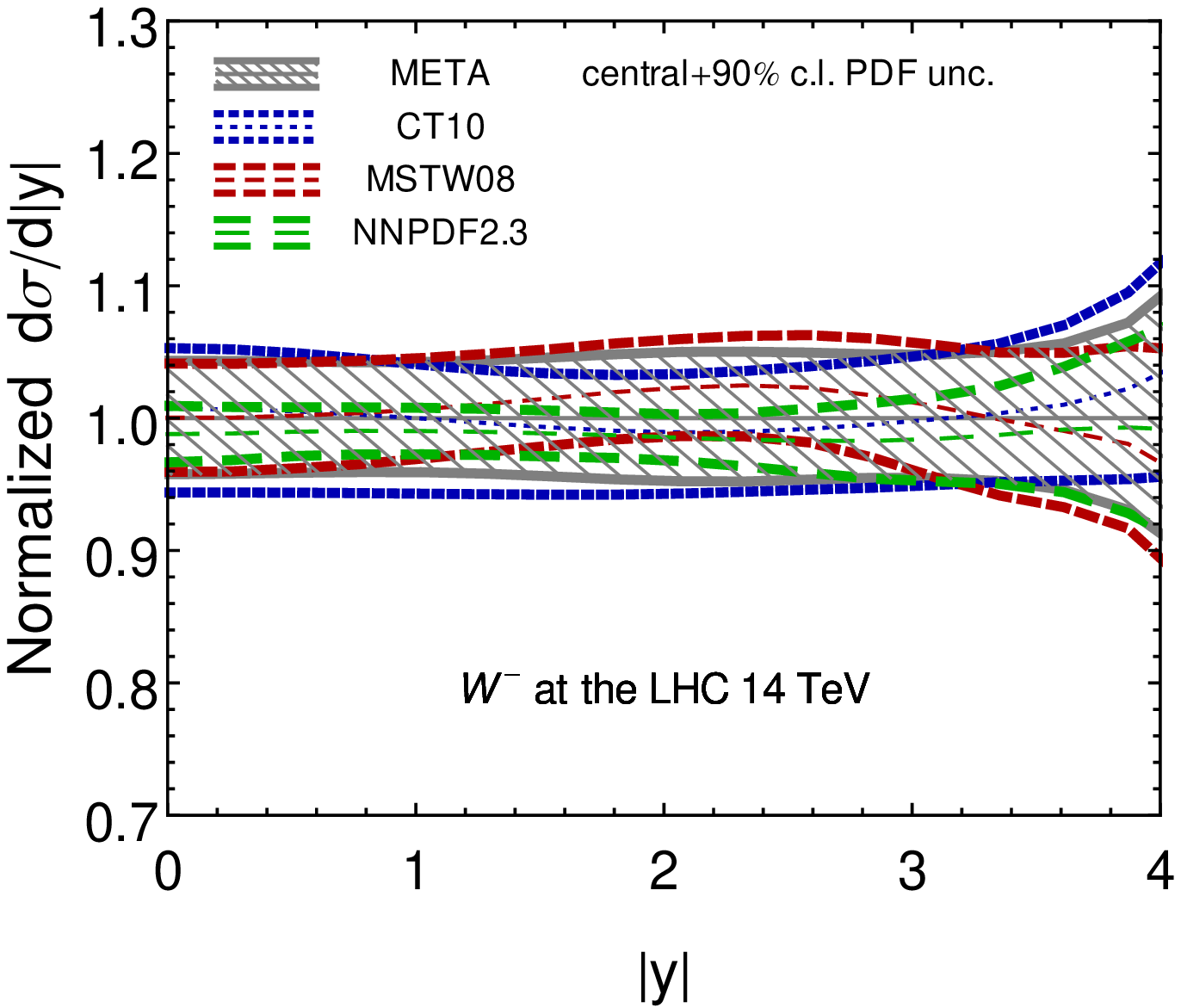}
\includegraphics[width=0.32\textwidth]{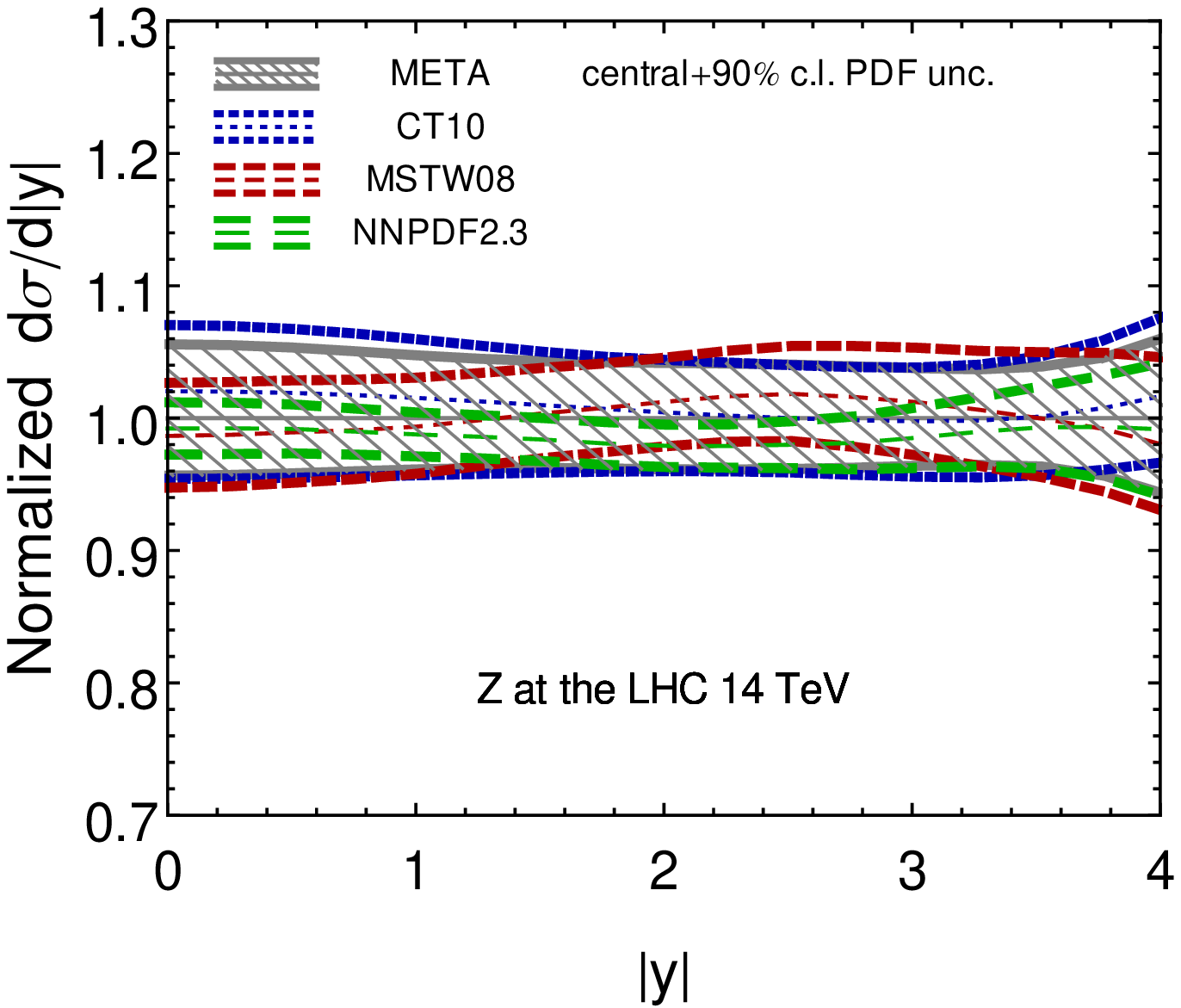}
\par\end{centering}

\vspace{-2ex}
 \caption{\label{fig:diffrap} Comparison of rapidity distributions of the W and
 Z boson produced at the LHC. All distributions are normalized to the central
 predictions of the META PDFs.}
\end{figure}

In a follow-up paper, we demonstrate that the META PDF approach 
performs as well for various differential observables that we
explored. The META PDFs provide an average of three input PDF
ensembles for various distributions, such as NNLO rapidity
distributions of  W and Z bosons produced at the LHC in Fig.~\ref{fig:diffrap},
obtained with Vrap~\cite{Anastasiou:2003ds} at the NNLO.
The rapidity
distributions are normalized to the central predictions of the META
PDFs, the bands indicate 90\% c.l. PDF uncertainties. The central
predictions of the META PDFs represent the average of the three global
sets across a wide rapidity range. The META uncertainties cover
the spread of the input PDF ensembles.

\section{Summary\label{sec:concl}}
Faithful estimation of the PDF dependence in theoretical predictions
is relevant practically and challenging conceptually. Diversity of
PDF ensembles provided by several groups reflects a variety of
factors impacting the PDFs in modern QCD calculations. Comparisons
of PDFs deal with their miscellaneous assumptions and disparate
formats, as well as with large numbers of member sets included in
the PDF ensembles. To estimate the net uncertainty in QCD cross
sections due to the PDF inputs, theoretical predictions are
currently calculated on a process-by-process basis for multiple PDF
sets and combined at the final stage based on a certain
prescription, {\it e.g.}, the PDF4LHC recommendation 
\cite{Botje:2011sn,Alekhin:2011sk,Ball:2012wy} or the replica
combination method  \cite{Forte:2013wc,Forte:2010dt}. These approaches,
while providing a reasonable uncertainty estimate, handle the PDFs
inefficiently and promote lengthy computations. Much of the
information contained in the input PDF sets is discarded in the
final combination. Development of fast interpolation interfaces for
(N)NLO cross sections, such as ApplGrid~\cite{Sutton:2010zz} or
FastNLO~\cite{Kluge:2006xs,Wobisch:2011ij}, speeds up the
computation, but does not eliminate the key hurdle of inefficient
information processing, especially when simulations involve many
scattering processes.

The meta-analysis described in this study follows an
alternative approach in which a variety of PDFs are combined into a
single PDF ensemble {\it before}
the individual theoretical predictions are computed. Such ensemble
is suitable for most LHC applications and consists of a relatively
small number of error PDFs that reproduce the total PDF+$\alpha_s$
uncertainty provided by an arbitrary number of the input PDF ensembles.
In the specific example that we constructed, a META ensemble is
comprised of 100 Hessian eigenvector sets
for evaluating the 68\% c.l. combined uncertainty of
CT10, MSTW'2008, and NNPDF2.3 at NNLO.
The three ensembles are selected because they can be
combined with minimal obstacles. They correspond to close values of
the QCD coupling strength $\alpha_s(M_Z)$ (compatible with the world
average of 0.118) and, although they follow different heavy-quark
schemes, these differences can be absorbed into their parametrizations
at the initial evolution scale of META PDFs chosen to
be above the $b$ quark mass (at 8 GeV).

Each error set of the input ensembles is approximated by an auxiliary
functional form (a meta-parametrization) in the kinematic range
typical for LHC studies, taken to be $x>10^{-5}$ and $8 <Q <10000$ GeV.
As the meta-parametrizations of all error sets share the same
functional form, distributions of their free parameters (66 in total)
can be easily compared. We determine these parameters at a common
coupling $\alpha_s(M_Z)$ of 0.118. Small differences in the input $\alpha_s$
values are compensated for by minor shifts of the
respective meta-parameters found using the $\alpha_s$ series
of the input PDF sets.

Then we combine the meta ensembles from CT10, MSTW2008 and NNPDF2.3
by generating Monte-Carlo replicas for each Hessian input ensemble
and diagonalizing the covariance matrix constructed from the
Monte-Carlo replicas. The final result consists of a central META
PDF set, equivalent to the unweighted average of three input central
sets, and 100 eigenvector META sets obtained in the Hessian method
\cite{Pumplin:2001ct,Pumplin:2002vw}. The PDF uncertainty for the
META ensemble is computed according to the master formulas in
Eq.~(\ref{HessianMinusError2}). We also provide a PDF $\alpha_{s}$
series of the META PDFs for calculating the PDF+$\alpha_{s}$
uncertainty, including the correlations, by adding the PDF and
$\alpha_s$ uncertainties in quadrature \cite{Lai:2010nw}. The META
PDFs are stored in a tabulated format together with an interface for
their numerical interpolation at~\cite{metapdfweb}.

The META ensemble predicts about the same PDF+$\alpha_s$ uncertainty
in the key LHC observables as the PDF4LHC recommendation, and in
addition provides a natural way to estimate correlations of
different observables. In comparison to massive computation efforts
characteristic of LHC simulations, the construction of the META
ensemble is relatively simple and was carried out entirely by using
standard functions for statistics and data analysis in {\it
Mathematica 8}.

Our results present an initial attempt to combine
LHC predictions from different PDF groups independently of the processes
studied. We note several avenues for further developments
of this approach. The simple combination procedure that was tried
(equivalent to using an unweighted average for finding expectation
values) would be inappropriate for including too
disparate PDF ensembles or for non-Gaussian
parameter distributions. In our case, the three global ensembles that were
combined are similar in their fitted experimental
samples, $\alpha_s(M_Z)$, central PDFs, and nominal
PDF uncertainties. We also examined the meta-parametrizations for
ABM'11 and HERAPDF1.5 ensembles and observed that their central PDF sets
for $\alpha_s(M_Z)=0.118$
sometimes lie outside of the 90\% c.l. error bands of the META
ensemble, see Figs.~\ref{fig:bench2} and \ref{fig:bench3}.
We could not fully consider the ABM'11 error sets in the
current combination at a fixed $\alpha_s(M_Z)=0.118$, as
$\alpha_s(M_Z)$ in the ABM error sets varies around
a low central value of $\alpha_s(M_Z)=0.1134$.
For HERAPDF1.5, a part of the difference with the global sets
can be attributed to a larger PDF uncertainty on some combinations
of the HERA PDF parameters, given their smaller experimental data
set. Such an ensemble could be added by averaging the META parameters
with statistical weights that account for different sizes of the
experimental data samples. 

The error PDFs of the future META ensembles can be obtained using
either the Monte Carlo (MC) sampling or Hessian methods. In the MC
approach \cite{Forte:2013wc,Forte:2010dt}, 
probability distributions of arbitrary complexity can be
in principle described, while selection of replicas is simplified via
using a shared meta-parametrization form. 
Constraints from new experiments can be
imposed using the PDF reweighting
technique~\cite{Giele:1998gw,Ball:2010gb,Sato:2013ika} in either MC
or Hessian approaches. The method of data set
diagonalization~\cite{Pumplin:2009nk} can be applied in the case of
the Hessian representation to find a small number of eigenvector
sets that dominate the PDF uncertainty of given observables. While
the 100 META eigenvector sets are designated for a wide range of LHC
theoretical predictions, a much smaller number (6-10) 
of eigenvector sets dominates the uncertainty in various important 
processes, such as massive electroweak boson production
\cite{GaoMETA2014}. The framework of the meta-analysis 
is well-suited for exploring these possibilities.

\section*{ACKNOWLEDGMENTS}

This work was supported by the U.S. DOE Early Career Research Award
DE-SC0003870 and by Lightner-Sams Foundation. We appreciate insightful
discussions with Sayipjamal Dulat, Joey Houston, Jon Pumplin, Carl
Schmidt, Dan Stump, Robert Thorne, Graeme Watt, and C.-P. Yuan.

\bibliographystyle{apsrev}
\bibliography{METApdf}

\appendix

\section*{\label{appendix:QCDEvolution}Appendix. A test of QCD evolution}

In this appendix, we verify that the dependence of $\alpha_{s}(Q)$
and PDFs $\Phi(x,Q)$ on the scale $Q$
in the input ensembles is compatible and allows their
combination. We compared the tabulated evolution of $\alpha_{s}(Q)$ and
$\Phi(x,Q)$ for five ensembles from  the LHAPDF library \cite{lhapdfweb}
with the numerical evolution by the HOPPET program  \cite{Salam:2008qg}
from the initial scale $Q_{0}=8$ GeV up to $10000$ GeV. In all cases,
we assume 3 QCD loops and 5 flavors in the QCD beta-function and
splitting kernels.

\begin{figure}[htb]
\begin{centering}
\includegraphics[width=0.35\textwidth]{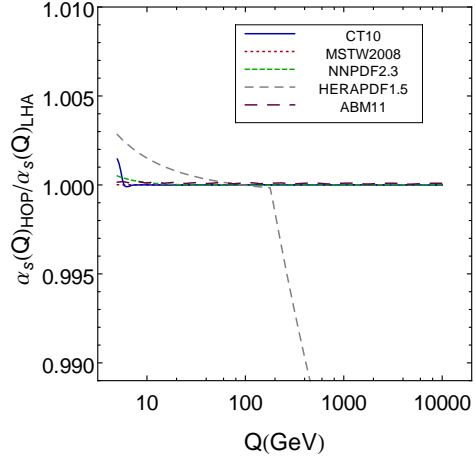}
\par\end{centering}

\vspace{-1ex}
 \caption{\label{fig:alf3} Comparison of the QCD coupling strengths
   $\alpha_{s}(Q)$ from five PDF ensembles as tabulated in
LHAPDF (LHA) or evolved with $N_{f}=5$ by HOPPET (HOP).}
\end{figure}

Fig.~\ref{fig:alf3} shows the ratio of $\alpha_{s}(Q)$ that we
get from LHAPDF to the corresponding values from HOPPET using the
original $\alpha_{s}(M_{Z})$ values
of the input ensembles listed in Table~\ref{tab:Input-PDF-ensembles}.
The agreement of the tabulated and evolved $\alpha_{s}(Q)$ is perfect
for all ensembles except for HERAPDF, where a kink at the top quark mass
occurs because HERAPDF uses the 6-flavor $\alpha_{s}(Q)$ above the
top quark threshold.

\begin{figure}[htb]
\begin{centering}
\includegraphics[width=0.32\textwidth]{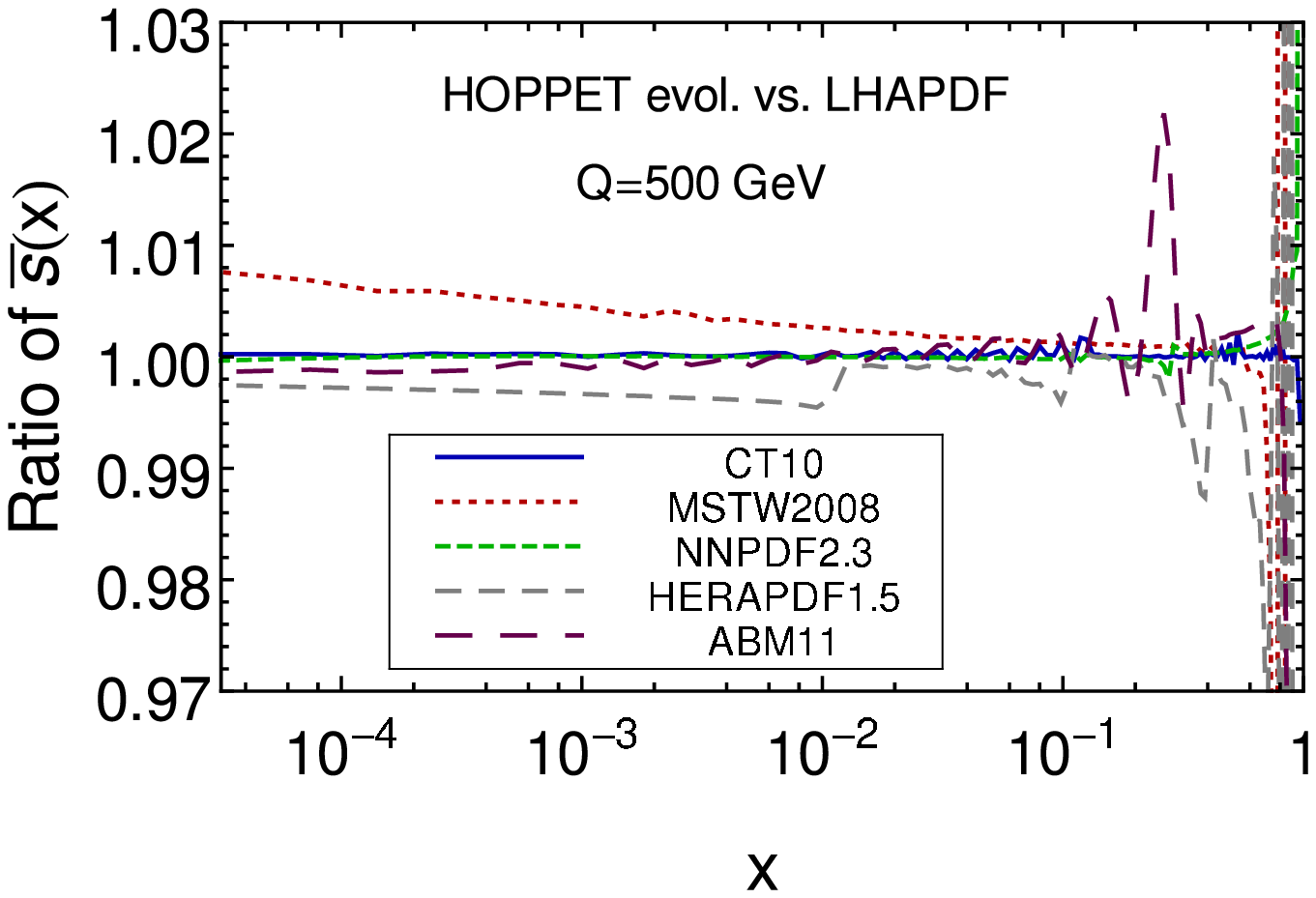} \includegraphics[width=0.32\textwidth]{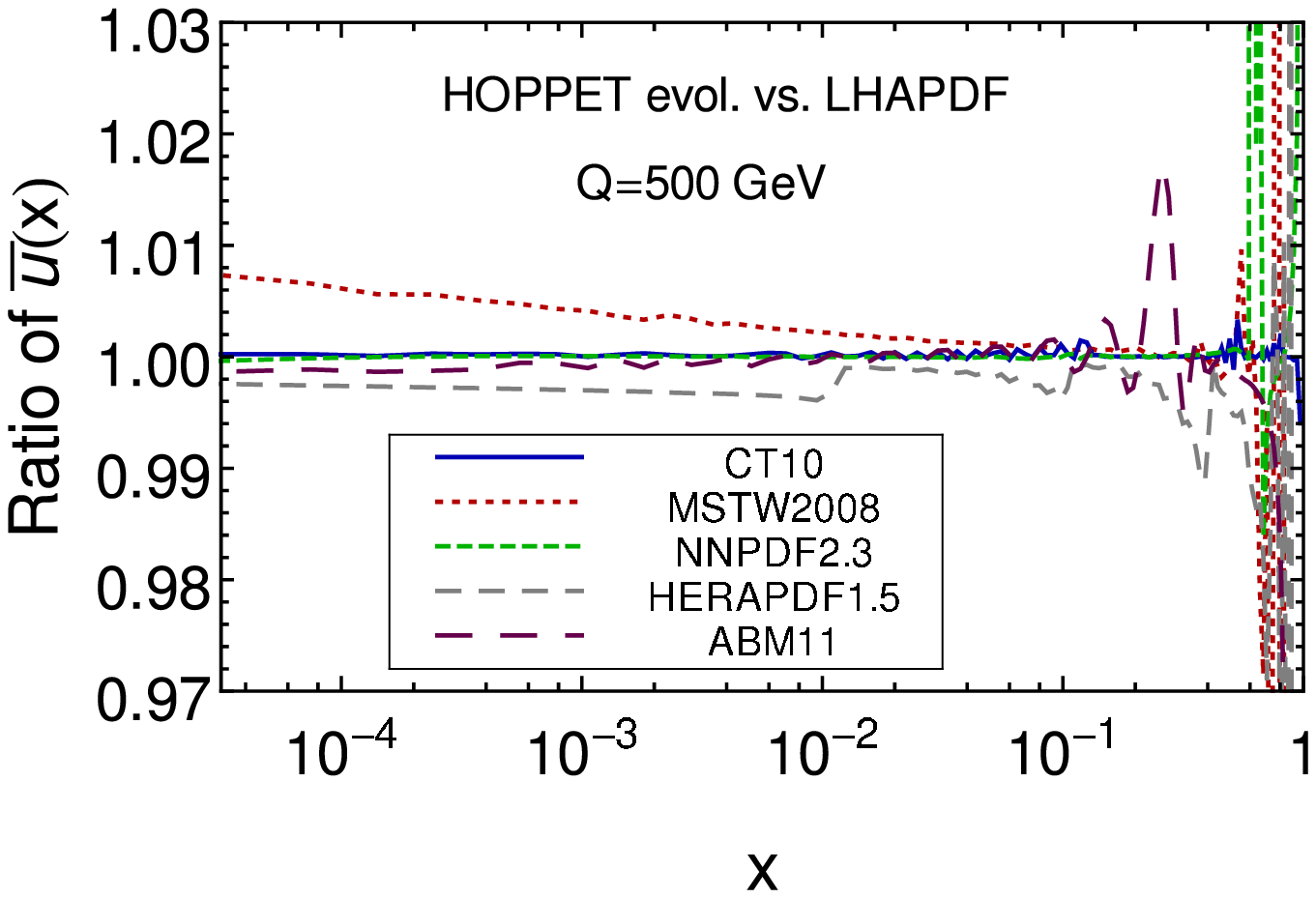}
\includegraphics[width=0.32\textwidth]{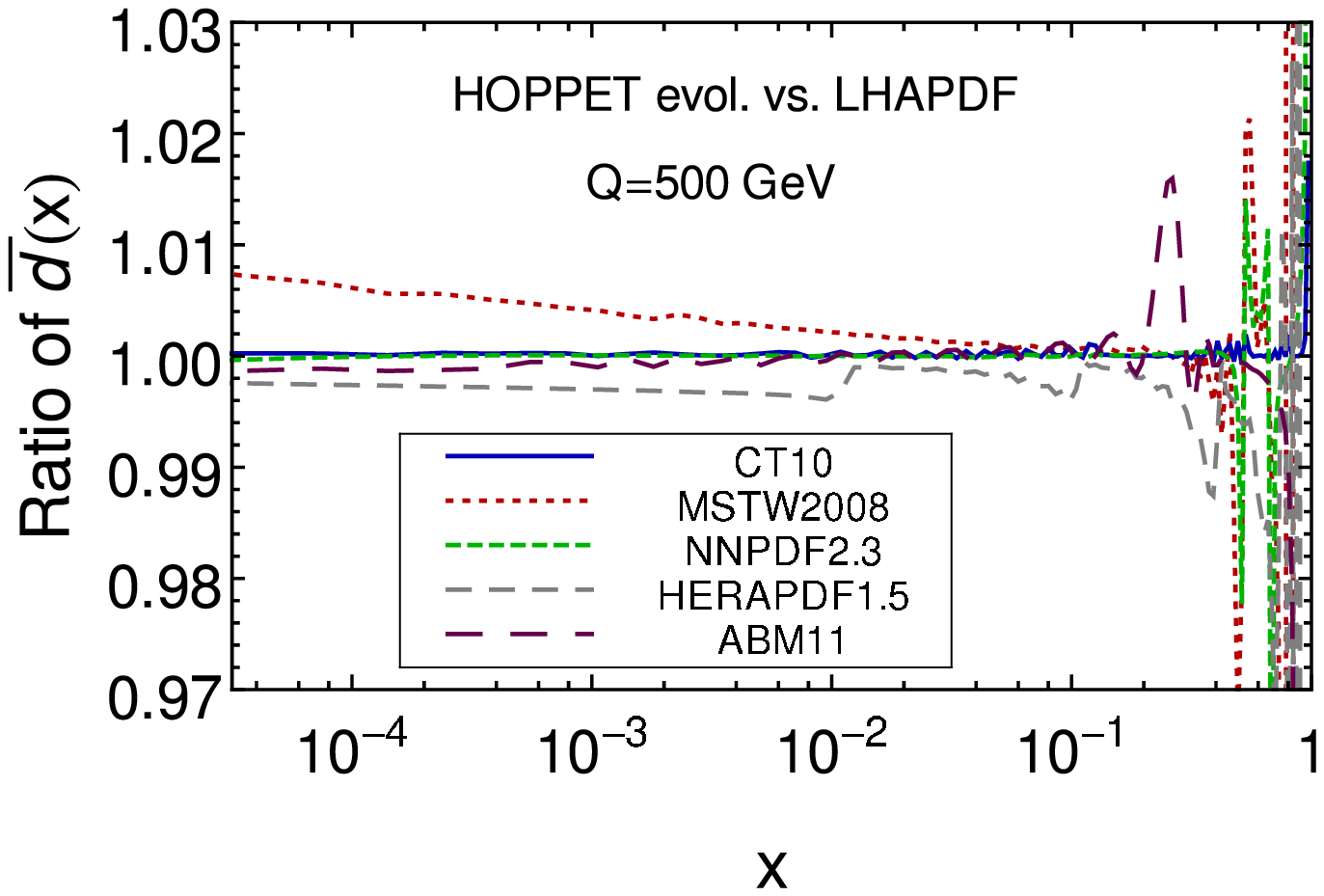} \\
 \includegraphics[width=0.32\textwidth]{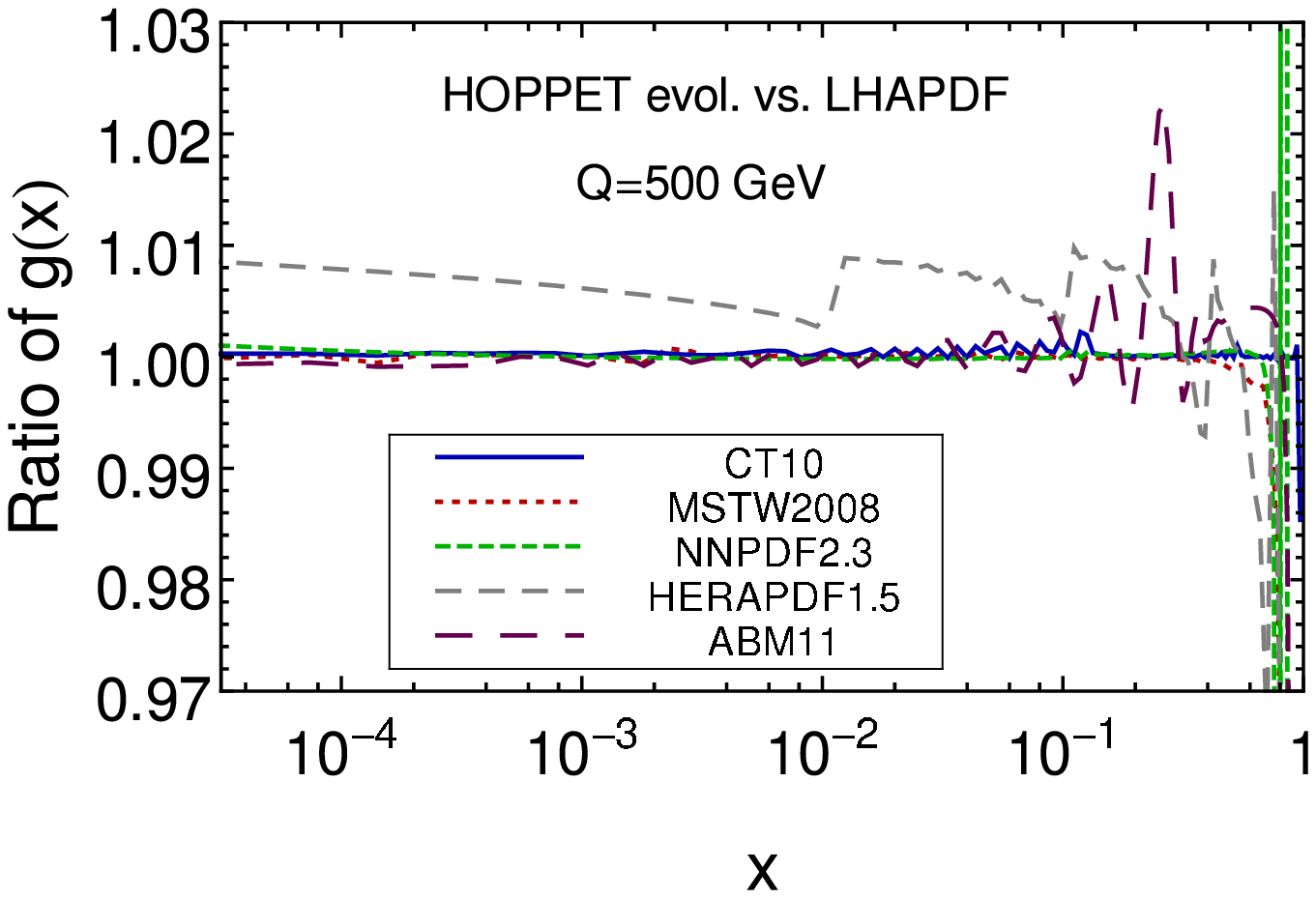} \includegraphics[width=0.32\textwidth]{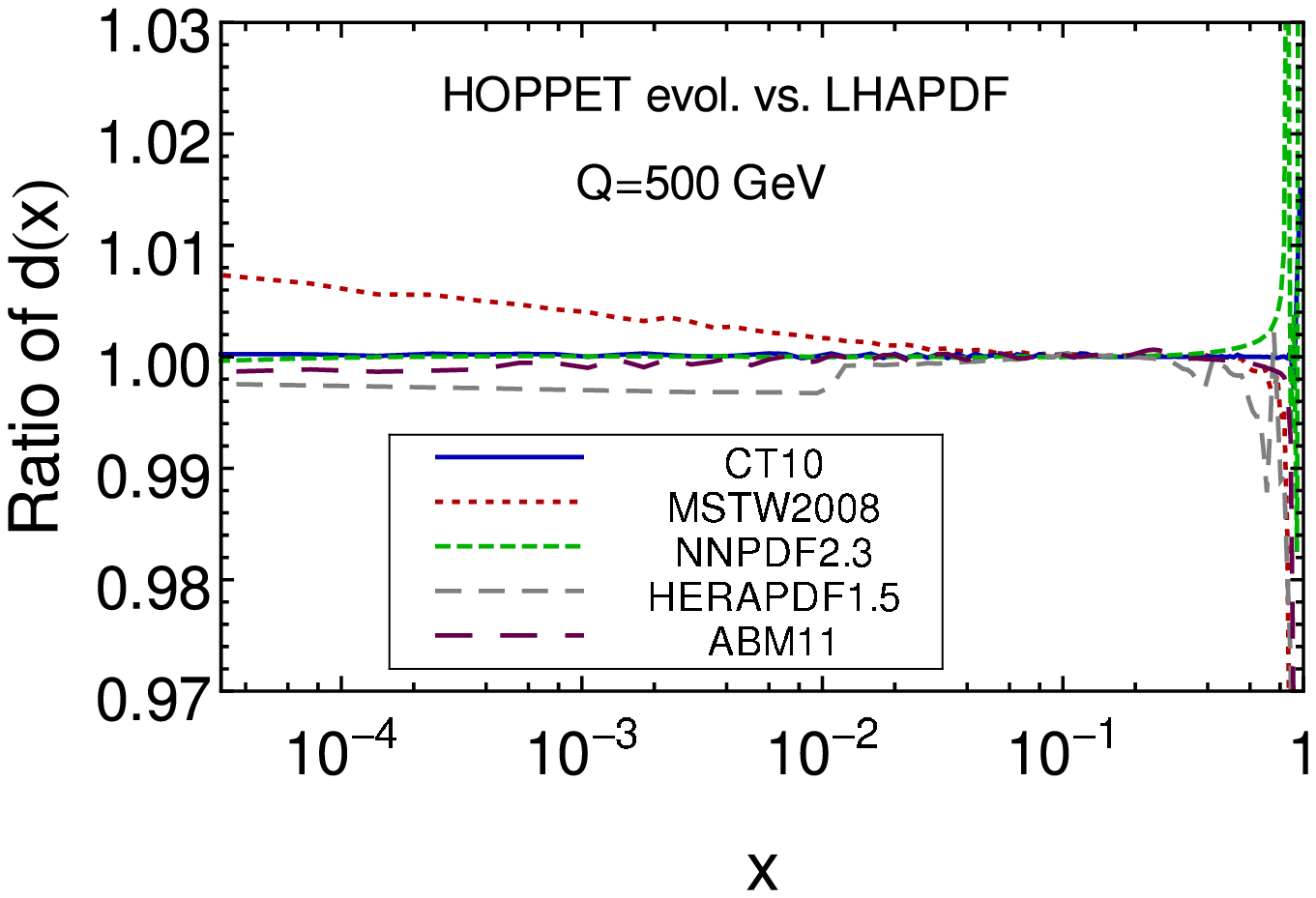}
\includegraphics[width=0.32\textwidth]{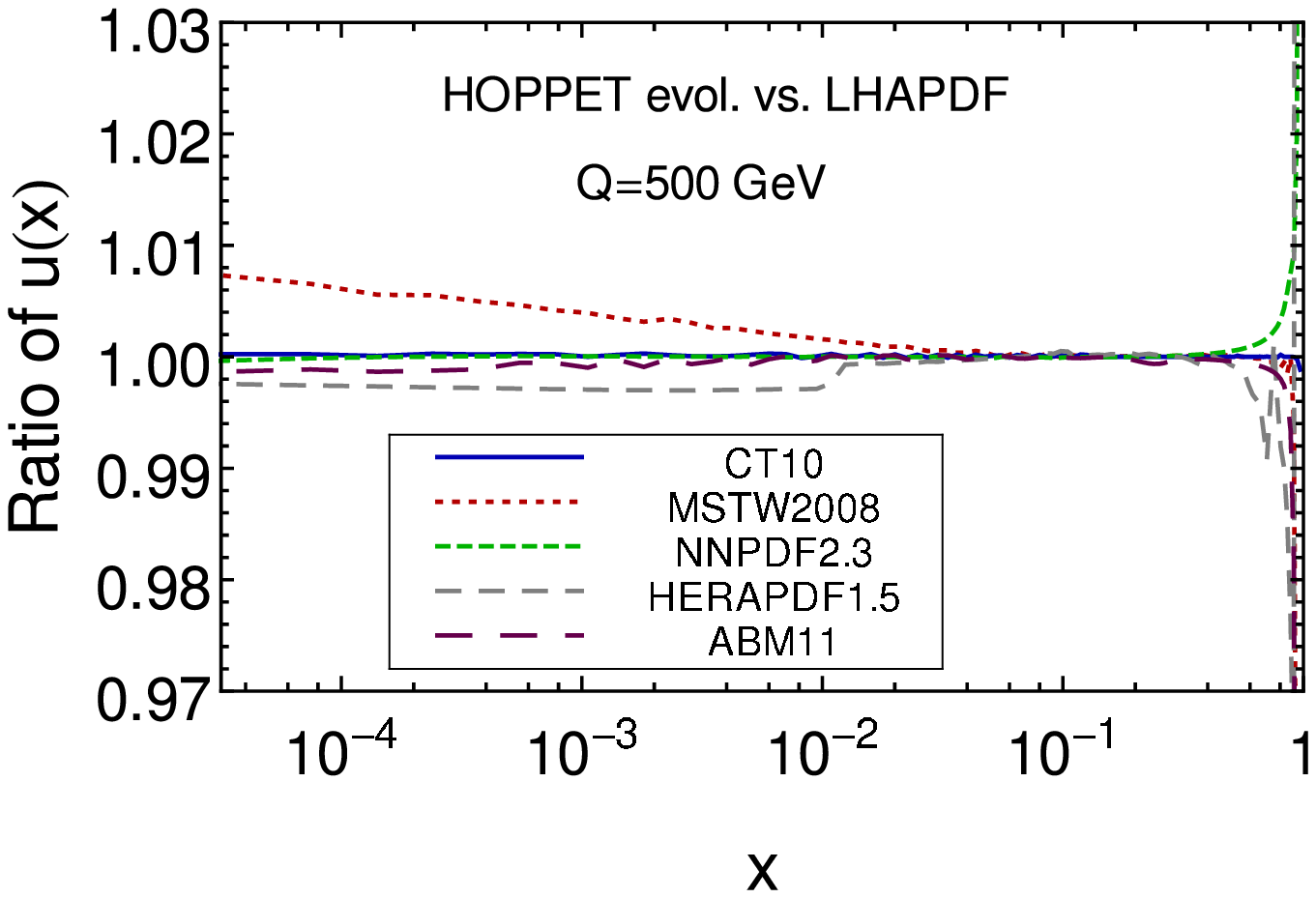} \\
 \includegraphics[width=0.32\textwidth]{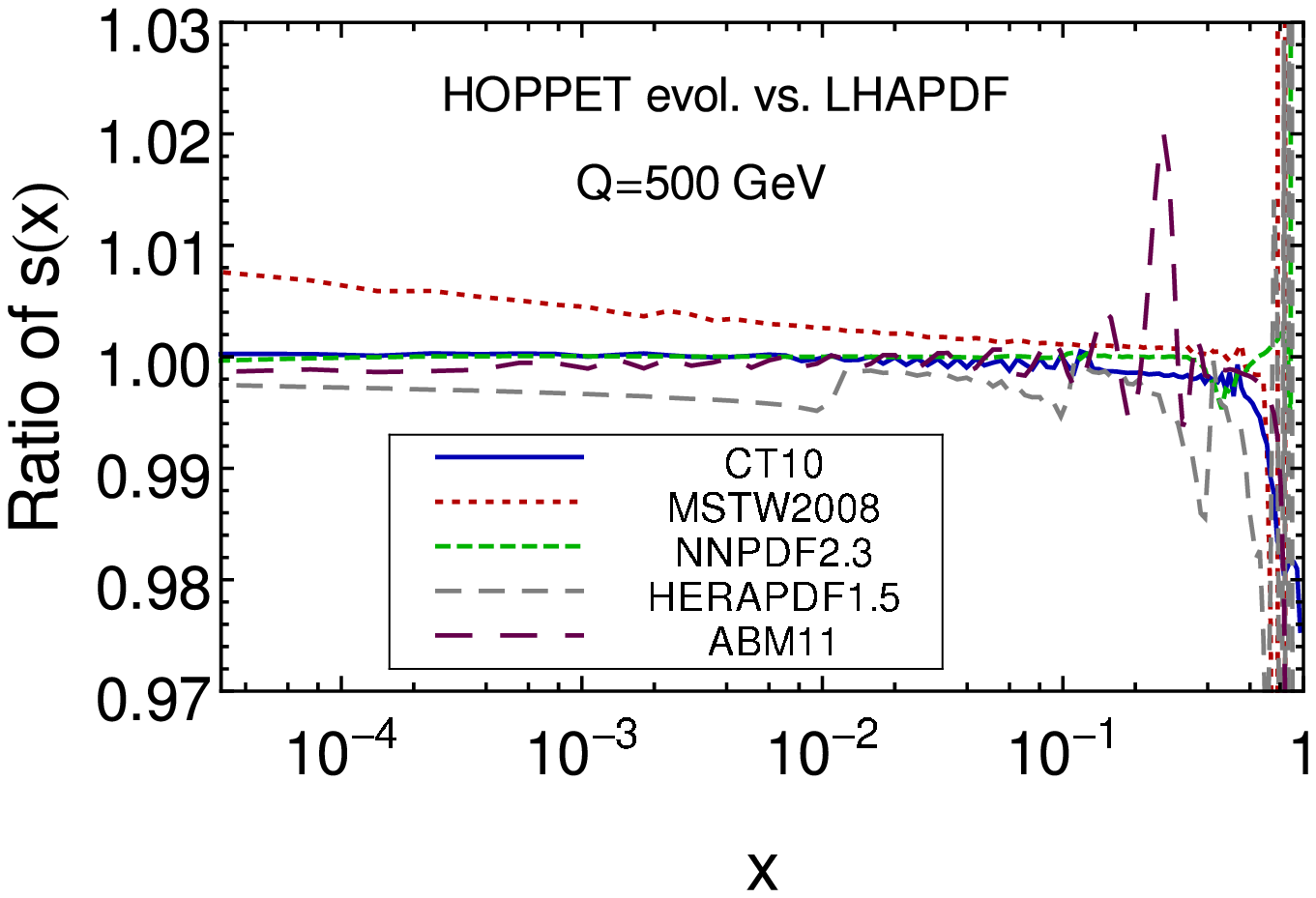} \includegraphics[width=0.32\textwidth]{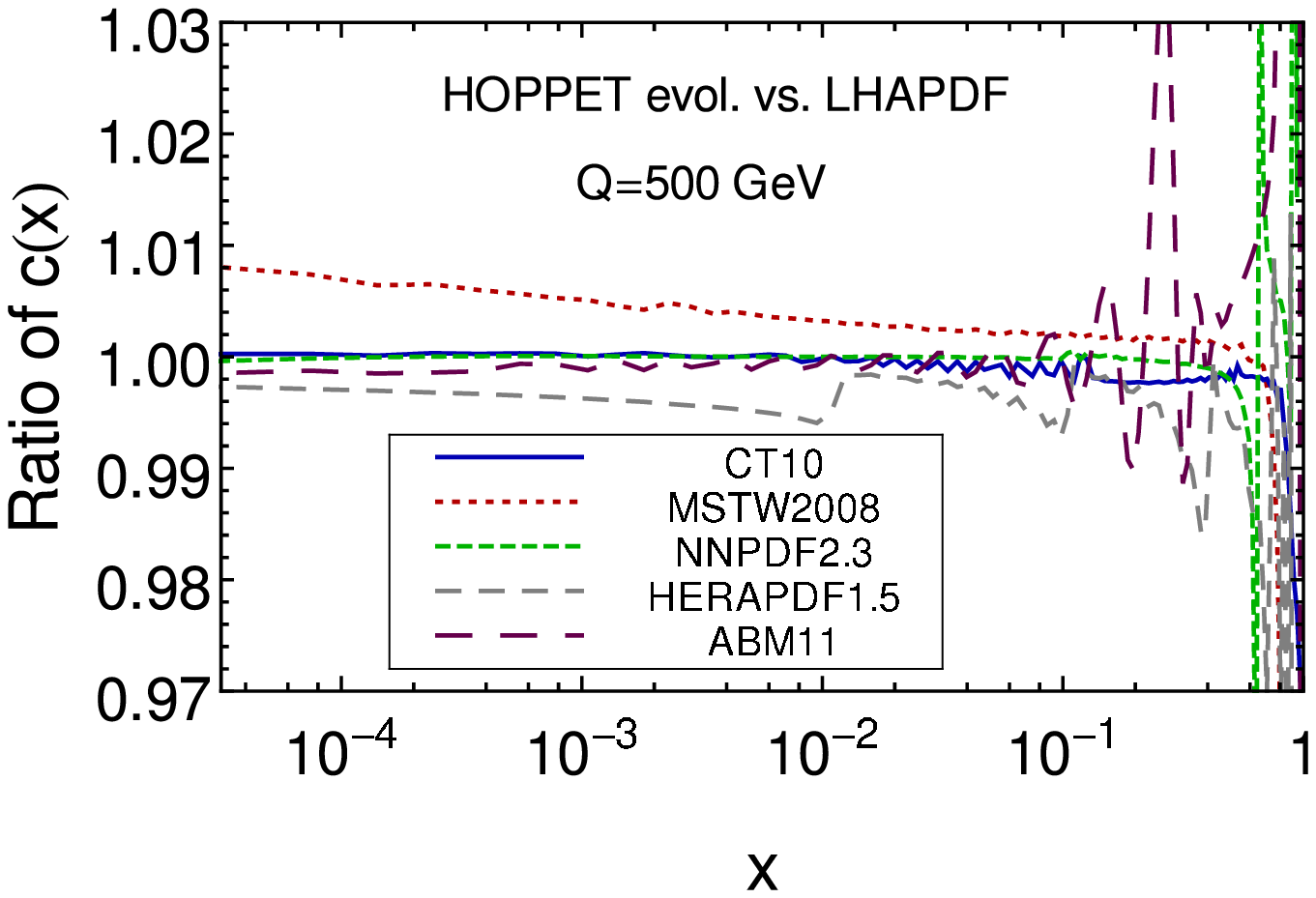}
\includegraphics[width=0.32\textwidth]{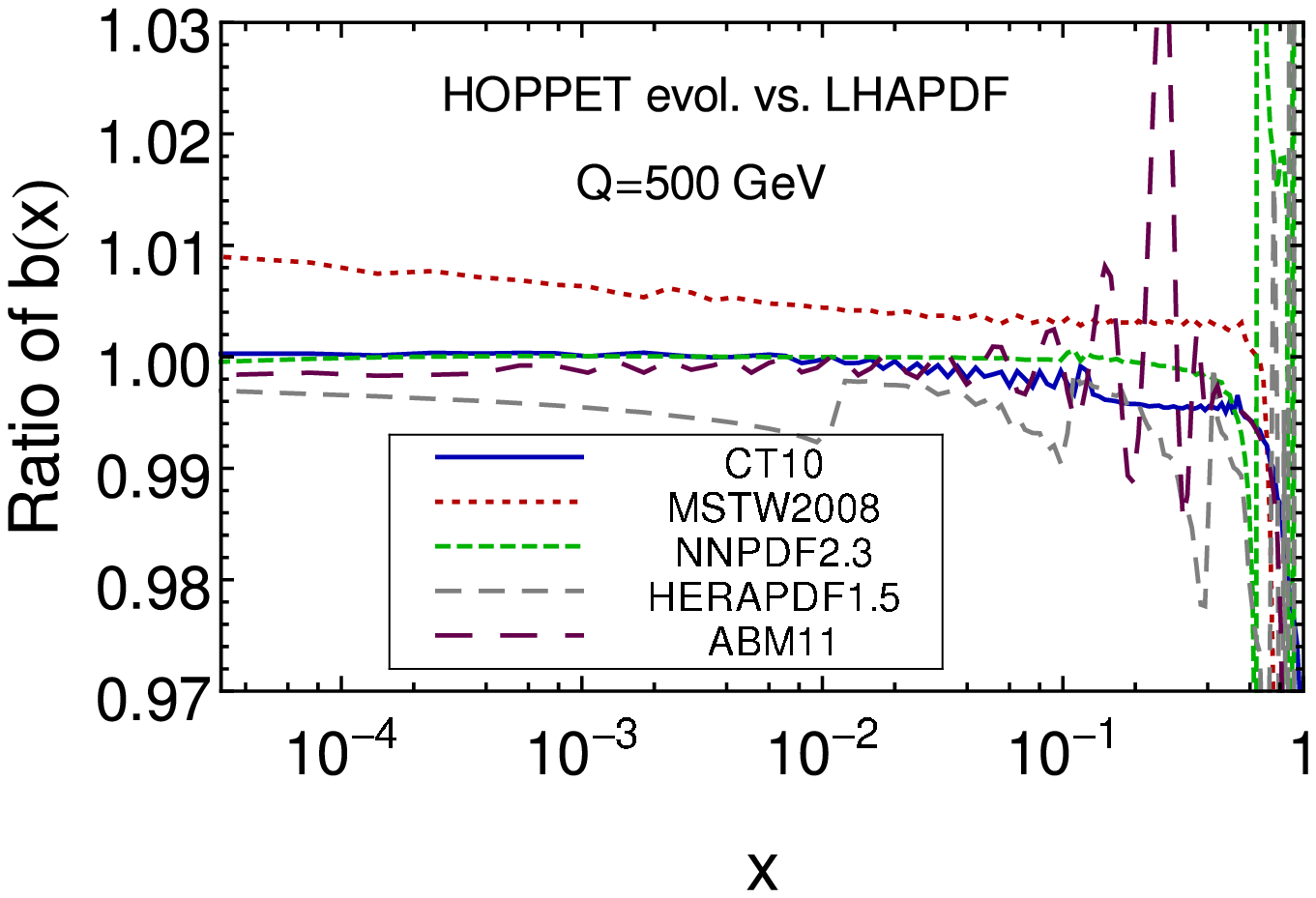}
\par\end{centering}

\vspace{-1ex}
 \caption{\label{evol2} Comparison of the PDF evolution from different best-fit
PDF sets with HOPPET for $Q=500\ {\rm GeV}$.}
\end{figure}

The PDF evolution is examined by computing ratios of $\Phi(x,Q)$
returned by LHAPDF and HOPPET at several $Q$ values above 8 GeV.
The $x$-dependent ratios at $Q=50$0 GeV shown in Fig.~\ref{evol2}
are representative of the general trend at other $Q$ values.
We observe excellent
agreement at the level of 0.1\% between LHAPDF and HOPPET for both
CT10 and NNPDF almost across the entire shown region of $10^{-5}\leq x<1$,
except for $x\gtrsim0.5$, where differences in the $Q$ dependence
can be a few percent for all sea PDFs. In the case of ABM, 
fluctuations for all sea PDFs are observed at somewhat smaller values of $x$
of about 0.3, but again they happen in the region where the sea PDFs
are small, and the accuracy of LHAPDF interpolation is sufficient
compared to the  PDF uncertainties.  The tabulated and
numerical evolution of quark PDFs from MSTW shows a systematic
discrepancy reaching 1\% at $x<0.05$,
while the tabulated and numerical evolution of the
MSTW gluon is in a perfect agreement. The tabulated and evolved HERA
PDFs differ by $\approx0.5\%$ across most $x$, which is expected at
$Q=500$ GeV given the 5-flavor evolution in
HOPPET and 6-flavor evolution of the HERAPDF's tabulated $\alpha_s(Q)$.

\end{document}